\documentclass{abelianclass}

\def\Nb{N}

\newcommand{\sungsik}[1]{ { \color{red} [(\textsf{SUNG-SIK}) \textsf{\textsl{#1}} ]}}
\newcommand{\old}[1]{ { \color{purple} (\textsf{old}) \textsf{\textsl{#1}} }}
\newcommand{\bqa}{\begin{eqnarray}} 
\newcommand{\eqa}{\end{eqnarray}}
\newcommand{\nn}{\nonumber \\}
\newcommand{\eq}[1]{Eq. \eqref{#1}}
\newcommand{\fig}[1]{Fig. \ref{#1}}
\newcommand{\ub}{\mathcal{U}}
\def\SC{\text{SC}}
\def\CDW{\text{CDW}}

\begin{document}
\title{
Dynamical kinetic energy quenching\\ 
in the antiferromagnetic quantum critical metals 
}
        \author{Anton Borissov$^{1,2}$ }
        \email{aborissov@uwaterloo.ca}
        \author{Vladimir Calvera$^{3}$}
        \email{fvcalver@stanford.edu}
        \author{Sung-Sik Lee$^{1,2}$
        \footnote{AB and VC are equal first authors of the present work.}}
        \email{slee@mcmaster.ca}
        \affiliation{$^{1}$Department of Physics \& Astronomy, McMaster University, Hamilton ON L8S 4M1, Canada}
        \affiliation{$^{2}$Perimeter Institute for Theoretical Physics, Waterloo ON N2L 2Y5, Canada}
        \affiliation{$^{3}$Department of Physics, Stanford University, Stanford, CA 94305, USA}
        \date{\today}
\begin{abstract}
We study the dynamics of  
critical spin fluctuations 
and hot electrons 
at the metallic
antiferromagnetic quantum critical points
with $Z_2$ and $O(2)$ spin symmetries,
building upon earlier works on the $O(3)$ 
 symmetric theory.
The interacting  theories in $2+1$ dimensions 
 are approached from $3+1$-dimensional theories 
 in the $\epsilon$-expansion that tunes
 the co-dimension of Fermi surface as a control parameter. 
The low-energy physics of the $Z_2$ and $O(2)$ theories qualitatively differ from each other and also from that of the $O(3)$ theory.
The difference is caused by higher-order quantum corrections beyond the one-loop order that are important even to the leading order in $\epsilon$.
The naive loop-expansion breaks down 
 due to dynamical quenching of kinetic energy:
the speed of the  collective mode  ($c$) 
and the Fermi velocity perpendicular to the magnetic ordering vector ($v$)
 become vanishingly small at low energies.
What sets the three
theories apart is the hierarchy that 
 emerges between the quenched kinetic terms.
At the infrared fixed point,
$c/v$ becomes $0$, $1$ and $\infty$ 
in the $Z_2$, $O(2)$
and $O(3)$ theories, respectively.
At intermediate energy scales,
the slow renormalization group (RG) flows of 
$c$ and $v$ toward their fixed point 
 values create approximate scale invariance controlled by approximate marginal parameters.
%
The manifold of those {\it quasi-fixed points} and the RG flow therein determines crossovers from scaling behaviours with transient critical exponents at intermediate energy scales
 to the universal scaling in the low-energy limit.
If the symmetry group is viewed as a tuning parameter,
the $O(2)$ theory corresponds to a multi-critical point 
 which has one additional quasi-marginal parameter than the other two theories.
\end{abstract}
\maketitle
\tableofcontents

\newcounter{pgnum} 

\newpage

\section{Introduction}

Given that it is challenging to theoretically understand the collective behaviours of condensed matter systems in the deep quantum regime,
theories that can be understood in controlled expansions 
are important beacons in the landscape of quantum matter theories.
Capturing quantum effects in a controlled manner is not only for improving our understanding at the quantitative level but also for delineating subtle dynamical patterns that quantum fluctuations generate.
The real value of a controlled expansion lies in the fact that 
 dynamics revealed through it can 
 point to general 
 organizing principles
 applicable even non-perturbatively\cite{
Hooft1980,SEIBERG1993469,SUNGSIKREVIEW}.

In the traditional perturbative approach, quantum corrections are organized by the number of loops in Feynman graphs. 
In weakly coupled theories, quantum corrections are usually suppressed with increasing number of loops.
However,
not all theories admit the loop expansion in the presence of small parameters.
One class of examples in which the naive loop-expansion breaks down are 
 the large-$N$ matrix models\cite{THOOFT} 
and non-Fermi liquids with $N$ vector flavours\cite{SSLEE}.
In those examples,
infinitely many graphs
remain important even in the large $N$ limit due to quantum fluctuations enhanced by many internal degrees of freedom or an extended manifold of gapless modes.
The other mechanism through which the naive perturbative expansion breaks down 
is {\it dynamical 
 kinetic energy quenching}.
If particles are slowed down due to interactions, systems become more susceptible to quantum fluctuations due to the increased density of states at low energies.
In metallic states that arise at the antiferromagnetic quantum critical points\cite{
ABANOV1,
ABANOV3,
ABANOV2,
HARTNOLL,
ABRAHAMS,
PhysRevB.87.045104,
DECARVALHO,
PATEL,
PATEL2,
VARMA2,
MAIER,
VARMA3,
MAX2,
MAX1,
LIHAI2,
SCHATTNER2,
GERLACH,
LIHAI,
WANG2},
kinetic energies of low-energy degrees of freedom are dynamically suppressed in some directions
as a result of interactions that tend to localize particles\cite{
ABANOV1,
MAX2,
SHOUVIK,LUNTS,SCHLIEF}.
In those cases, quantum corrections are organized in qualitatively new ways as the zero-th order theory describes partially localized particles\cite{SHOUVIK3,LUNTS}. 
This requires including certain higher-loop effects even to the leading order in the `perturbative' limit.

Antiferromagnetic quantum critical metals are among the most important examples of non-Fermi liquids\cite{
HERTZ,
MILLIS,
VARMALI,
POLCHINSKI2,
PLEE1,
ALTSHULER,
YBKIM,
ABANOV2,
SSLEE,
MAX0,
MAX2,
NAYAK2,
STEWART,
SCHOFIELD,
DENNIS,
SENTHIL,
MROSS,
FITZPATRICK,
SHOUVIK2,
SCHLIEF,
PhysRevX.11.021005,
PhysRevB.107.165152,
berglederer,
BORGES2023169221,
2024arXiv240509450K}
that lie beyond the scope of Landau Fermi liquids\cite{
LANDAU,
PhysRevB.42.9967,
SHANKAR,
POLCHINSKI1},
as magnetic quantum phase transitions arise commonly in strongly correlated materials such as
the high-temperature cuprates \cite{Cuprates2010Example}, iron pnictides \cite{IronPnic2012Example} and heavy fermion compounds \cite{HeavyFermion2006Example}. 
In the antiferromagnetic quantum critical metals, the interaction between gapless spin fluctuations and electrons drives both the speed of the collective mode ($c$) and the nesting angle 
 ($v$) toward zero at low energies. 
Here, $c$ denotes the speed of the antiferromagnetic spin fluctuations and $v$, the Fermi velocity perpendicular to the magnetic ordering vector,
both of which are measured in the unit of the Fermi velocity parallel to the ordering vector 
(see Fig. 		\ref{fig:VF}).
As the patches of the Fermi surface connected by the ordering wave-vector become parallel at low energies, 
$v$ flows to zero under the renormalization group (RG) flow.
At the same time, the emergent nesting makes the collective mode more dispersionless, making $c$ to flow to zero.
Interestingly, a hierarchy arises between $c$ and $v$ as the kinetic energies of both electrons and the magnetic collective mode are dynamically quenched.
For the $O(3)$ theory,
$v$ vanishes faster than $c$ such that $c/v$ diverges at low energies\cite{LUNTS}.
In this theory, this emergent hierarchy makes it possible to identify the strongly interacting fixed point in two dimensions\cite{SCHLIEF,BORGES2023169221}.
While the exact dynamical critical exponent is $z=1$ at the fixed point,
a theory with a non-zero bare nesting angle exhibits dynamical critical exponent $z = 1+O(\sqrt{v/\log (1/v)})$ that depends on the bare nesting angle $v$ at an intermediate energy scale due to the slow flow of $v$\cite{SCHLIEF,BORGES2023169221}.
A recent large-scale Monte Carlo simulation\cite{Lunts:2023un},
which uses a sign-problem-free lattice regularization
\cite{berg2012signfreemontecarlo,
PhysRevResearch.2.023008,
Xu_2019},
has observed $z$ that is smaller than $2$ and decreases with decreasing bare nesting angle.

\begin{figure}
    \centering
    \begin{subfigure}{0.5\textwidth}
        \centering
        \includegraphics[width=0.8\textwidth]{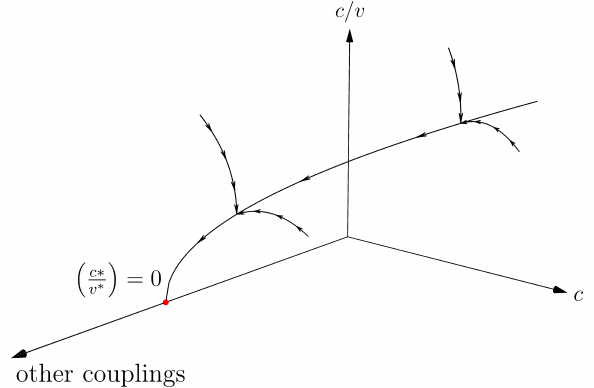}
        \caption{$Z_2$}
        \label{fig:1a}
    \end{subfigure}
    \begin{subfigure}{0.6\textwidth}
        \centering
        \includegraphics[width=0.8\textwidth]{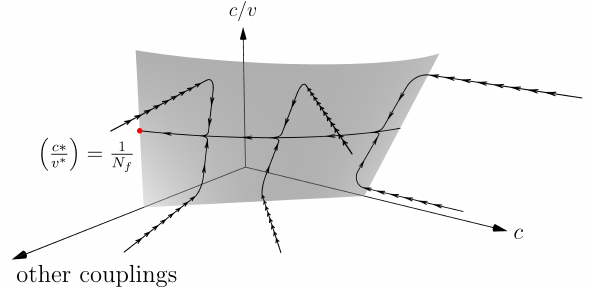}
        \caption{$O(2)$}
        \label{fig:1b}
    \end{subfigure}
    \begin{subfigure}{0.6\textwidth}
        \centering
        \includegraphics[width=0.8\textwidth]{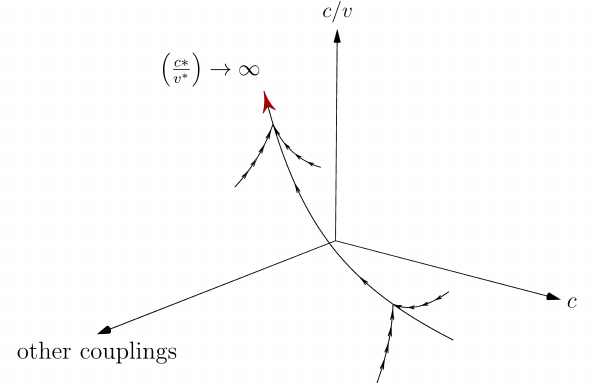}
        \caption{$O(3)$}
        \label{fig:1c}
    \end{subfigure}
    \caption{
Schematic renormalization group flow.
In all theories, the nesting angle ($v$)
and the speed of the collective mode ($c$) flow to zero in the low-energy limit.
However, their ratio $s\equiv c/v$ becomes $0$, $O(1)$ and $\infty$ for the $Z_2$, $O(2)$ and $O(3)$ theories, respectively.
In the $Z_2$ and $O(3)$ theories, the UV theory is first attracted to an one-dimensional manifold of quasi-fixed points before it flows to the true fixed point at a much lower energy scale.
In the $O(2)$ theory, the UV theory is initially attracted to a two-dimensional manifold of quasi-fixed points.
At a lower energy scale, the theory flows to an one-dimensional sub-manifold before it flows to the true fixed point in the lowest energy scale.
These multi-stage renormalization group flows are reflected in the distinct crossover behaviours of physical observables, which is summarized in
Eqs. \eqref{eq:AchiZ2}-\eqref{eq:AchiO3}.
}
\label{fig:RGflowSchematic}
\end{figure}

In this work, we study the antiferromagnetic quantum critical metals with $Z_2$ and $O(2)$ symmetries. 
These theories have been studied via quantum Monte Carlo simulations\cite{SCHATTNER2,
2023arXiv230506421T},
but a systematic field-theoretic study is still lacking.
Those theories haven't been studied much compared to their $O(3)$ symmetric counterpart, partly due to the expectation that the variation in the symmetry group may not lead to qualitatively new behaviours.
However, we find that this is not true.
The $Z_2$ and $O(2)$ theories exhibit qualitatively different hierarchies of dynamical kinetic energy quenching.
In contrast to the $O(3)$ theory,
as $v$ and $c$ flow to zero  in the low-energy limit,
$c/v$ becomes $0$ and 
$\mathcal{O}(1)$
in the $Z_2$ and $O(2)$ theories, respectively.

The theories with different symmetries also exhibit distinct crossovers from intermediate-energy scaling to universal low-energy scaling behaviours.
In the $Z_2$ theory, a UV theory is first attracted to a one-dimensional manifold of theories at an intermediate energy scale under the RG flow. 
Within the one-dimensional manifold parameterized by $s=c/v$,
the theory does not flow much over a sizable window of energy scales
due to a slow flow of $s$.
Only at much lower energy scales,
the theory logarithmically flows to the true fixed point located at $c=v=0$ with $s=0$.
The manifold of theories in which the RG flow is stagnant for a finite but large window of energy scale acts as a line of approximate fixed points.
They are referred to as {\it quasi-fixed points}
and the parameter that labels the location within the manifold is called {\it quasi-marginal coupling}.
The size of the energy window in which physical observables exhibit scaling behaviors governed by a quasi-fixed point
is determined by the quasi-marginal couplings of the effective theory.
For an effective theory that enters the manifold of quasi-fixed points closer to the true fixed point,
it takes longer RG time for the qusi-marginal coupling to change appreciably within the manifold.
Therefore, the closer the quasi-fixed points are to the true fixed point, 
the larger the window of quasi-fixed point scaling becomes. 
The existence of the quasi-marginal coupling manifests itself in the scaling behaviour of physical observables.
At intermediate energy scales,
correlation functions obey scaling forms controlled by exponents that depend on the value of the quasi-marginal coupling.
In the low-energy limit, the correlation functions
cross over to the universal forms.
The $O(3)$ theory has
a similar one-dimensional manifold of quasi-fixed points with the difference being that the true fixed 
 lies at $s=\infty$\cite{LUNTS}.
The $O(2)$ theory exhibits a more intricate crossover behaviour
due to the existence of an additional quasi-marginal parameter.
In the $O(2)$ theory,
a UV theory is first attracted to a two-dimensional manifold of quasi-fixed points.
That manifold is labelled by both $c$ {\it and} $v$ 
which are two quasi-marginal parameters.
At a parametrically lower energy scale, the theory within the two-dimensional plane of quasi-fixed points flows into a one-dimensional sub-manifold in which $c$ and $v$ are locked into a universal relation.
In the ultimate low-energy limit, the theory finally flows to the true fixed point located at $v=c=0$ with $c/v \sim \mathcal{O}(1)$.
This multi-stage RG flow creates multiple crossovers in the scaling behaviours of physical observables.
The presence of the extra quasi-marginal parameters is attributed to the fact that the $O(2)$ theory corresponds to a multi-critical point 
if the symmetry group is viewed as a tuning parameter\footnote{
The $O(2)$ theory also supports vortices as topological excitations in $2+1$ dimensions\cite{VARMA2}, which, combined with the multi-critical nature,
makes it a rather special theory.
}.
The schematic RG flow is illustrated in Fig. \ref{fig:RGflowSchematic}.
The consequences of these crossovers for physical observables are summarized in Sec. V.

Here is the outline of the rest of the paper.
In Sec. II, we set the stage by introducing the hot-spot field theory for the antiferromagnetic quantum critical metal with a general symmetry group 
that can be $Z_2$, $O(2)$ or $O(3)$.
The theory in $2+1$ dimensions is then extended to a general theory that describes the $1$-dimensional Fermi surface embedded in the $d$-dimensional momentum space,
 where  $\epsilon \equiv 3-d$ is used as a control parameter.
In Sec. III, we present the one-loop analysis of the theory.
At the one-loop order, all three theories exhibit  qualitatively similar behaviours.
However, the dynamical kinetic energy quenching makes the one-loop analysis uncontrolled for any non-zero $\epsilon$ in all three theories.
In Sec. IV, we systematically include higher-loop corrections that are important to the leading order in the $\epsilon$ expansion.
This reveals the symmetry-dependent hierarchy of the quenched kinetic energies
and 
crossover that arises from the quasi-marginal parameters.
In Sec. V, we discuss the physical consequences of the emergent hierarchies for physical observables.
In Sec. VI, we conclude with discussions and open questions.
The extensive proof of the control of the $\epsilon$-expansion is given in the appendices.

\section{
The low-energy effective theories 
with \texorpdfstring{$Z_2$}{Z2}, \texorpdfstring{$O(2)$}{O2} and \texorpdfstring{$O(3)$}{O3} symmetries
}
\label{sec:Section2}

\begin{figure}[h]
	\centering
	\begin{subfigure}[h]{0.4\textwidth}
		\centering
		\includegraphics[width=.95\textwidth]{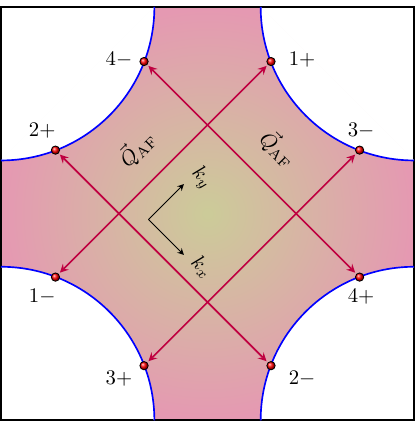}
		\caption{\label{fig:Hotspots} }
	\end{subfigure}%
	\begin{subfigure}[h]{0.35\textwidth}
		\centering
		\includegraphics[width=.8\textwidth]{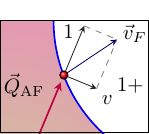}
		\caption{\label{fig:VF} }
	\end{subfigure}%
\caption{\label{fig:FS}\small{
(a) 
A Fermi surface with the four-fold rotation and reflection symmetries.
The shaded (white) region represents occupied 
 (unoccupied) states.
The (red) dots denote the hot spots on the Fermi surface connected by a commensurate antiferromagnetic ordering vector 
$ \vec{Q}_{\mathrm{AF}}$ (red arrows) 
with $ 2 \vec{Q}_{\mathrm{AF}} =0$  modulo the reciprocal vectors.
(b) 
The Fermi velocity at the hot spots can be decomposed into the components parallel and perpendicular to 
$\vec{Q}_{\mathrm{AF}}$.
The parallel component is set to be one, 
and the perpendicular component is denoted as $v$.
}
}	
\end{figure}

The low-energy degrees of freedom in the antiferromagnetic quantum critical
metal are the collective mode that describes critical antiferromagnetic spin fluctuations and the electrons near the Fermi surface.
For spin symmetry group $G = Z_2$, $O(2)$ and $O(3)$,
the number of collective modes that are gapless at the critical point is $N=1$, $2$ and $3$, respectively.
We consider a collinear antiferromagnetic ordering with a commensurate wave vector as shown in Fig. \ref{fig:FS}\footnote{
The axes have been chosen such that $\vec{Q}_\mathrm{AF}=\pm\sqrt{2}\pi\hat{k}_x,\pm\sqrt{2}\pi\hat{k}_y$ up to reciprocal lattice vectors.}.
While it is necessary to include the entire Fermi surface to understand the low-energy physics in $2+1$ dimensions\cite{BORGES2023169221},
in this study we consider the hot spot theory that only consider the collective mode and electrons near the hot spots on the Fermi surface connected by $\vec{Q}_\mathrm{AF}$. 
We consider a Fermi surface with the fourfold rotation and reflection symmetries as is shown in Fig. \ref{fig:FS}.
The hot-spot theory in $2+1$ dimensions with symmetry group $G$ is written as
$S=S_\phi+S_\psi+S_{g}+S_{u}$\cite{Sur2015Quasilocal},
where 
\begin{align}\label{eq:Action2D}
S_\phi&=\frac{1}{2}\int \frac{d^3 q}{(2\pi)^3} \left[  q_0^2+c^2|\vec{q}|^2 \right] \phi(q)\phi(-q),\\
S_\psi&=
\sum_{j=1}^{N_f}
\sum_{n=1}^{4}\sum_{m=\pm}\sum_{\sigma=\uparrow,\downarrow}\int \frac{d^3 k}{(2\pi)^3} \psi_{n,\sigma,j}^{(m)*}(k)  \left[ ik_0+e_n^m (\vec{k};v)\right] \psi_{n,\sigma,j}^{(m)}(k) ,\\
S_{g}&=g_0
\sum_{j=1}^{N_f}
\sum_{\alpha\in S_G}\sum_{n=1}^{4}\sum_{\sigma,\sigma'=\uparrow,\downarrow}\int \frac{d^3 k}{(2\pi)^3} \frac{d^3 q}{(2\pi)^3}  \left(  \phi^\alpha(q) \psi_{n,\sigma,j}^{(+)*}(k+q) \tau^{\alpha}_{\sigma,\sigma'} \psi_{n,\sigma',j}^{(-)}(k)  +h.c.\right), \\
S_{u}&=
u_0 \sum_{\alpha,\beta\in S_G}\int \frac{d^3 p_1}{(2\pi)^3} \frac{d^3 p_2}{(2\pi)^3}\frac{d^3 q}{(2\pi)^3}  \phi^\alpha(p_1) \phi^\alpha(p_2) \phi^\beta(q-p_1) \phi^\beta(-q-p_2). 
\end{align}
$k = (k_0,\vec{k})$ denotes
Matsubara frequency and two-dimensional spatial momentum.
$\psi_{n,\sigma,j}^{(m)}(k)$ is the electron field near one of the eight hot spots labelled by $n=1,2,3,4$ and $m=\pm$ (see Fig. \ref{fig:Hotspots})
with spin $\sigma=\uparrow,\downarrow$, flavour $j=1,2,..,N_f$ and momentum $\vec k$ measured relative to the hot spot\footnote{
For generality, 
we consider electrons with $N_f$ flavours, but 
 we don't need $N_f$ to be large for our analysis.}.
The collective mode that carries spatial momentum 
$ 
\vec{Q}_{\mathrm{AF}} 
+\vec{q}
$ 
is denoted as $\phi^\alpha(q)$ 
with $\alpha\in S_G$, 
where
$S_{Z_2} = \{z\}$,
$S_{O(2)} = \{x,y\}$
and
$S_{O(3)} = \{x,y,z\}$
represent the components of the critical collective mode.
In the hot-spot theory,
the energy dispersion of electron is linearized near the hot spots.
In the unit in which the component of the Fermi velocities parallel to $\vec{Q}_\mathrm{AF}$ is $1$ 
(see Fig. \ref{fig:VF}),
the dispersion near hot spot $(1,+)$  is written as
$e_{1}^{\pm}(\vec{k};v)=v k_x \pm k_y$.
The dispersions near other hot spots are fully determined through the $C_4$  and reflection symmetries :
$e_{3}^{\pm}(\vec{k};v)=-(v k_x \pm k_y)$ and
$e_{2}^{\pm}(\vec{k};v)=-e_{4}^{\pm}(\vec{k};v)= \pm k_x+vk_y$. 
Here, $v$ represents the Fermi velocity perpendicular to $\vec Q_{AF}$ measured in the unit of the component that is parallel to $\vec Q_{AF}$.
$v$ is referred to as the nesting angle because the hot spots connected by the ordering wave vector become perfectly nested at $v=0$.
Unlike the cases in which $\vec Q_{AF}$ coincides with $2 \vec k_F$\cite{
PhysRevB.99.195102,
PhysRevB.104.125123,
PhysRevB.107.165152}, $v$ is generally non-zero. 
Even if $v$ dynamically flows to zero at low energies, 
one can still ignore quadratic terms in the dispersion of electrons if $v$ flows to zero much more slowly compared to a power-law decay in energy,
which is indeed the case, as will be shown later.
The Yukawa interaction 
 $g_0$ scatters electrons across a pair of hot spots connected by $ \vec{Q}_{\mathrm{AF}}$.
$\tau^{a}$ denote Pauli matrices.
$u_0$ is the quartic boson coupling.

Under the Gaussian scaling 
 that keeps the kinetic terms of the action invariant,
 both $ g_0 $ and $ u_0 $ are relevant in $(2+1)$-dimensions. 
To access the low-energy physics of the theory in a controlled way, 
the theory is extended to general dimensions.
As a first step, 
we combine a pair of fermionic fields in the anti-podal points on the Fermi surface into a 2-component spinors,
$\Psi_{1,\sigma,j}(k)=(\psi_{1,\sigma,j}^{(+)}(k),\psi_{3,\sigma,j}^{(+)}(k))^{\top}$,$\Psi_{2,\sigma,j}(k)=(\psi_{2,\sigma,j}^{(+)}(k),\psi_{4,\sigma,j}^{(+)}(k))^{\top}$, $\Psi_{3,\sigma,j}(k)=(\psi_{1,\sigma,j}^{(-)}(k),-\psi_{3,\sigma,j}^{(-)}(k))^{\top}$ and $\Psi_{4,\sigma,j}(k)=(\psi_{2,\sigma,j}^{(-)}(k),-\psi_{4,\sigma,j}^{(-)}(k))^{\top}$\cite{SHOUVIK}.
We also introduce the gamma matrices that generate the Clifford algebra,
$\gamma_0=\sigma^y$, $\gamma_1=\sigma^x$\footnote{$ \sigma^a$'s act on spinor indices while $\tau^\alpha$'s act on spin indices.} 
with $\bar{\Psi}=\Psi^{\dag}\gamma_0$.
In the spinor form,
the electronic action can be rewritten as 
\begin{equation}
\begin{split}
S_\psi&=
\sum_{j=1}^{N_f}
\sum_{n=1}^{4}\sum_{\sigma=\uparrow,\downarrow}\int \frac{d^3 k}{(2\pi)^3} \bar{\Psi}_{n,\sigma,j}(k)  \left[ ik_0\gamma_{0}+i\varepsilon_n(\vec{k};v)\gamma_1\right] \Psi_{n,\sigma,j}(k) ,\\
S_{g}&=g_0
\sum_{j=1}^{N_f}
\sum_{n=1}^{4}\sum_{\sigma,\sigma'=\uparrow,\downarrow} \sum_{\alpha\in S_G}\int \frac{d^3 k}{(2\pi)^3} \frac{d^3 q}{(2\pi)^3}   \phi^\alpha(q)\tau^{\alpha}_{\sigma,\sigma'} \bar{\Psi}_{\bar{n},\sigma,j}(k+q) (i\gamma_1){\Psi}_{n,\sigma',j}(k).
\end{split}
\end{equation}
Here, 
$(n,\bar{n})$ are pairs of hot spots connected by $\vec{Q}_\mathrm{AF}$ 
with $\bar{1}=3$, $\bar{2}=4$, $\bar{3}=1$ and $\bar{4}=3$. 
The dispersion for the spinors is written as 
$\varepsilon_{1}(\vec{k};v)=e_{1}^{(+)}(\vec{k},v)$, $\varepsilon_{2}(\vec{k};v)=e_{2}^{(+)}(\vec{k},v)$, $\varepsilon_{3}(\vec{k};v)=e_{1}^{(-)}(\vec{k},v)$ and $\varepsilon_{4}(\vec{k};v)=e_{2}^{(-)}(\vec{k},v)$. 
The theory that describes the one-dimensional Fermi surface embedded in general space dimension $d$ is given by
$S = S_b + S_f + S_{g} + S_{u}$\cite{Lee2013DimReg,Sur2015Quasilocal} with
\begin{subequations}\label{eq:ActiondDim}
\begin{align}
S_b &= \frac12\sum_\alpha\int \frac{d^{d+1}q}{(2\pi)^{d+1}}
\left(\textbf{Q}^2 + c^2 |\vec q|^2\right) \phi^\alpha(q) \phi^\alpha(-q),\\
S_f &= 
\sum_{j=1}^{N_f}
\sum_{n=1}^4 \sum_{\sigma=\uparrow,\downarrow}
\int \frac{d^{d+1}k}{(2\pi)^{d+1}}
\overline\Psi_{n,\sigma,j}(k)
\left[ i \textbf{K} \cdot {\bm\Gamma} + i\varepsilon_n(\vec{k},v) \gamma_{d-1} \right]
\Psi_{n,\sigma,j}(k),
\\
S_{g} &= \frac{g\mu^{(3-d)/2}}{\sqrt{N_f}}
\sum_{j=1}^{N_f}
\sum_{n=1}^4 \sum_{\sigma,\sigma'=\uparrow,\downarrow} \sum_{\alpha\in S_G} \int \frac{d^{d+1}k}{(2\pi)^{d+1}} \frac{d^{d+1}q}{(2\pi)^{d+1}}
\phi^\alpha(q)
\overline\Psi_{\bar n,\sigma,j}(k+q)
\tau^\alpha_{\sigma,\sigma'}
(i\gamma_{d-1})
\Psi_{n,\sigma',j}(k),
\\
S_{u} &= 
u \mu^{3-d}
\sum_{\alpha,\beta\in S_G}
\int \prod_{j=1}^4 dp_j 
\delta\left(\sum_{i=1}^4p_i\right)
\phi^\alpha(p_1)\phi^\alpha(p_2)
\phi^\beta(p_3)\phi^\beta(p_4).
\end{align}
\end{subequations}
Here $q = (\vb{Q},\vec{q}\,)$, where $\vb{Q}=(q_0,q_\mu)$ is a vector composed
of one Matsubara frequency and $(d-2)$-dimensional momentum along the new spatial directions perpendicular to the one-dimensional Fermi surface. 
The theory respects the $SO(d-1)$ symmetry for general $d$ and
$\vb{Q} $ will be referred as `frequency'. 
$\vec{q}$ represents the original two dimensional spatial momenta. $\dd{q}=\frac{{d^{(d+1)}}q}{(2\pi)^{d+1}}$ 
and $(\vb{\Gamma},\gamma_{d-1})$ with
$\vb{\Gamma}=(\gamma_0,...,\gamma_{d-2})$ are a set of $(d\!-\!1)$-dimensional
$2\times2$ gamma matrices.
As we  are interested in $2\leq d \leq3$,
the gamma matrices are kept to be $2$ by $2$
and they satisfy
$\{\gamma_\mu,\gamma_\nu\}=2\iden\delta_{\mu,\nu}$. 
Both the Yukawa coupling and the quartic boson coupling become marginal at the upper critical dimension $d_c=3$.
We will use $\epsilon \equiv 3 -d$
as an expansion parameter in the following analysis.
$\mu$ is a floating energy scale.
$g$ and $u$ are dimensionless couplings.

The local counter terms needed to cancel singular parts of quantum corrections 
are written as
$\mathcal{S_{CT}}=\mathcal{S}_{b,\mathcal{CT}}+\mathcal{S}_{f,\mathcal{CT}}+\mathcal{S}_{g,\mathcal{CT}}+\mathcal{S}_{u,\mathcal{CT}}$ with
\begin{widetext}
\begin{subequations}\label{eq:ActionCT}
\begin{align}
\mathcal{S}_{f,\mathcal{CT}}&=\int \dd{k} \bar{\Psi}_{n,\sigma,j}\left[ i \mathcal{A}_{1}\vb{\Gamma \cdot K}+i \mathcal{A}_{3} \gamma_{d-1} \varepsilon_{n}\left(\vec{k}; \frac{\mathcal{A}_{2}}{\mathcal{A}_{3}}v\right)\right] \Psi_{n,\sigma,j}(k),\\
\mathcal{S}_{b,\mathcal{CT}}&=\frac{1}{2}\sum_{\alpha\in S_G}\int \dd{q} \left(\mathcal{A}_{4}|\vb{Q}|^2+ \mathcal{A}_{5} c^2 |\vec{q}|^2\right)\phi^\alpha(q)\phi^\alpha(-q),\\
\mathcal{S}_{g,\mathcal{CT}}&=\mathcal{A}_{6}\frac{g \mu^{\frac{3-d}{2}}}{\sqrt{N_f}} \sum_{\scriptsize{n,\sigma,\sigma'\!\!,j}} \sum_{\alpha\in S_G}\int \dd{q} \dd{k} \left[\bar{\Psi}_{\bar{n},\sigma j}(k+q) \tau_{\sigma,\sigma'}^{\alpha} (i\gamma_{d-1}) \Psi_{n,\sigma',j}(k) \phi^\alpha(q) \right ],\\
\mathcal{S}_{u,\mathcal{CT}}&=
\mathcal{A}_{7} 
u\mu^{3-d}
\sum_{\alpha,\beta\in S_G}\int \prod_{j=1}^{4}\dd{p_j}\delta\left(\sum_{i=1}^4 p_i\right)\phi^\alpha(p_1)\phi^\alpha(p_2)\phi^\beta(p_3)\phi^\beta(p_4).
\end{align}
\end{subequations}
With the multiplicative counter terms of the form 
$
\mathcal{A}_n=\mathcal{A}_n(v,c,g,u;\epsilon)=\sum_{m=1}^{\infty} \frac{1}{\epsilon^m}Z_{n,m}(v,c,g,u)$.
The renormalized action,
which is the sum of Eqs.
(\ref{eq:ActiondDim}) and Eqs. (\ref{eq:ActionCT}),
can be expressed in terms of bare fields and bare couplings.
The bare quantities and the renormalized ones are related to each other through

\begin{align}\label{eq:DefBareQuantities} 
\vec{k}&=\vec{k}_B,   &
\vb{K}&= \mathcal{Z}_\tau^{-1} \vb{K}_B, \notag\\
\Psi_{n,\sigma,j}(k)&= \mathcal{Z}_\psi^{-1/2}  \Psi_{B;n,\sigma,j}(k_B), &
\phi(k)&= \mathcal{Z}_\phi^{-1/2}   \phi_B(k_B),\notag\\
v&=\frac{\mathcal{Z}_3}{\mathcal{Z}_2} v_B, & c&=\sqrt{\frac{\mathcal{Z}_{\phi}\mathcal{Z}_{\tau}^{d-1}}{\mathcal{Z}_5}} c_B,\notag\\
g&=\dfrac{\mathcal{Z}_{\phi}^{1/2}\mathcal{Z}_{\psi}\mathcal{Z}_{\tau}^{2(d-1)}}{\mathcal{Z}_6 \mu^{\frac{3-d}{2}}} g_B, & u&=\frac{\mathcal{Z}_{\phi}^{2}\mathcal{Z}_{\tau}^{3(d-1)}}{\mathcal{Z}_4 \mu^{{3-d}}}  u_B.
\end{align}
Here, $\mathcal{Z}_n=1+\mathcal{A}_n$ and $\mathcal{Z}_{
\tau}=\mathcal{Z}_1/\mathcal{Z}_3$, $\mathcal{Z}_{\psi}=\mathcal{Z}_1
\mathcal{Z}_\tau^{-d}$ and $\mathcal{Z}_{\phi}=\mathcal{Z}_4
 \mathcal{Z}_\tau^{-(d+1)}$. 
From the condition that the bare parameters are independent of the floating energy scale $\mu$, 
one can express
the critical exponents 
and the beta functions in terms of the counter term coefficients\cite{Sur2015Quasilocal} as
\begin{subequations}
\begin{align}
z&=\left(1+(Z'_{1,1}-Z'_{3,1})\right)^{-1} ,\\
\eta_{\psi}&=\frac{z}{2}(-\epsilon(Z'_{1,1}-Z'_{3,1})+(2 Z'_{1,1}-3 Z'_{3,1})),\\
\eta_{\phi}&=\frac{z}{2}(-\epsilon(Z'_{1,1}-Z'_{3,1})+(4 Z'_{1,1}-4 Z'_{3,1}- Z'_{4,1})),\\
\beta_{v}&=zv(Z'_{2,1}- Z'_{3,1}),\\
\beta_{c}&=\frac{zc}{2}(2 Z'_{1,1}-2 Z'_{3,1}- Z'_{4,1}+ Z'_{5,1}),\\
\beta_{g}&=-zg\left(\frac{\epsilon}{2}+\frac{1}{2}(2 Z'_{3,1}+ Z'_{4,1}-2 Z'_{6,1})\right), \\
\beta_{u}&=-zu\left(\epsilon-(2 Z'_{1,1}-2 Z'_{3,1}-2 Z'_{4,1}+ Z'_{7,1})\right),
\label{eq:zetabeta}
\end{align}
\end{subequations}
where
 $Z'_{i,1}=\left(\frac{g}{2}\pdv{g}+u \pdv{u}\right)Z_{i,1}$. 
\end{widetext}

\section{
One-loop fixed point 
and breakdown of the loop expansion
} \label{sec:Section3}

\begin{figure*}[ht] 
	\centering
\begin{subfigure}[t]{0.2\textwidth}
\centering	\raisebox{0.8cm}{\includegraphics[width=\textwidth]{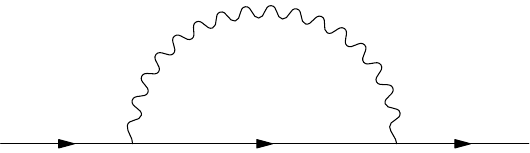}}
\caption{\label{fig:FermionSE1Loop} }
\end{subfigure}%
\begin{subfigure}[t]{0.2\textwidth}
\centering	\raisebox{0.3cm}{\includegraphics[width=\textwidth]{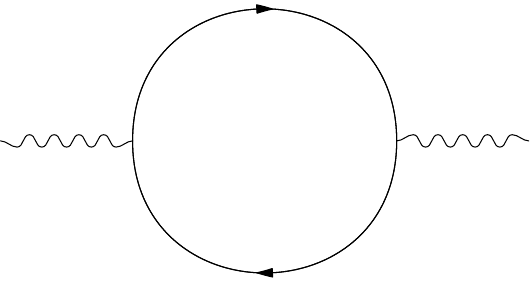}}
\caption{\label{fig:BosonSE1Loop} }
\end{subfigure}%
\begin{subfigure}[t]{0.15\textwidth}
\centering	\raisebox{0.3cm}{\includegraphics[width=\textwidth]{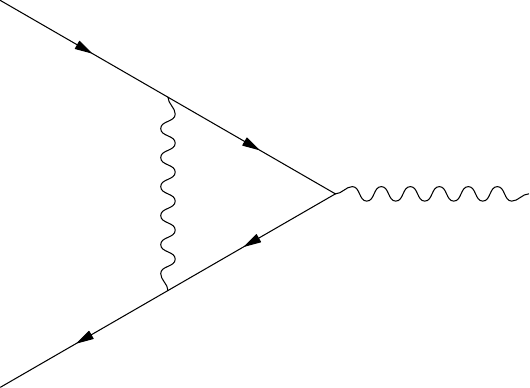}}
\caption{\label{fig:Yukawa1Loop} }
\end{subfigure}
\begin{subfigure}[t]{0.15\textwidth}
\centering	\raisebox{0.0cm}{\includegraphics[width=\textwidth]{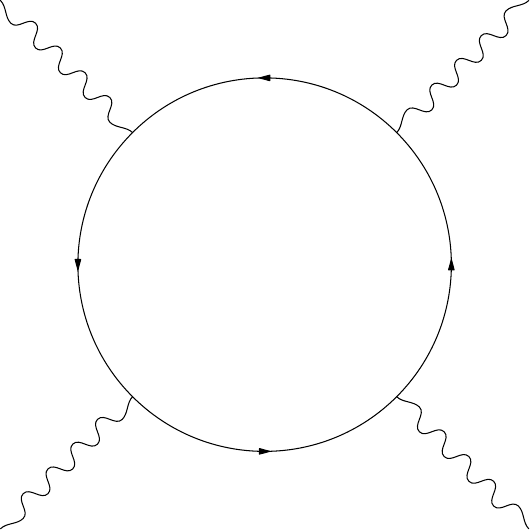}}
\caption{\label{fig:Quartic1LoopFermion} }
\end{subfigure}
\begin{subfigure}[t]{0.15\textwidth}
\centering	\raisebox{0.6cm}{\includegraphics[width=\textwidth]{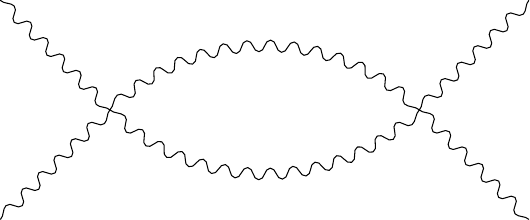}}
\caption{\label{fig:Quartic1LoopBoson} }
\end{subfigure}
	\caption{
The one-loop diagrams,
where the solid and wavy lines represent the propagators of electron 
 and collective mode, respectively.
Even in the small $\epsilon$ limit,
these one-loop quantum corrections are not sufficient due to the dynamical quenching of the kinetic energies.
} 
	\label{fig:1LoopDiagrams}
\end{figure*}

In this section, we present the `blind' one-loop analysis.
Although the one-loop analysis is not sufficient even to the leading order in $\epsilon$ as will be shown later, this is the starting point of our analysis.
The counter-terms that cancel the singular quantum corrections 
from the one-loop diagrams shown in Fig. \ref{fig:1LoopDiagrams}
are given by
(see Appendix \ref{app:quantumCorrections} for the detail)
\begin{align}
\begin{split}
Z_{1,1}&=-\frac{\Nb}{8 \pi^2 N_f} \frac{g^2}{c}h_{1}(v,c) ,\\
Z_{2,1}&=\frac{\Nb}{8 \pi^2 N_f}  \frac{g^2}{c} h_{2}(v,c) ,\\
Z_{3,1}&=-Z_{2,1},\\
Z_{4,1}&=-\frac{1}{4\pi}\frac{g^2}{v},\\
\end{split}
\begin{split}
Z_{5,1}&=0, \\
Z_{6,1}&= \frac{2-\Nb}{16 \pi^3 N_f}\frac{g^2}{c} h_3(v,c) ,\\
Z_{7,1}&=\frac{\Nb+8}{2\pi^2}\frac{u}{c^2},
\\
\end{split}
\label{eq:Zs}
\end{align}
where $\Nb = 1,2,3$ 
for the $Z_2$, $O(2)$ and $O(3)$ theories, respectively, and
\begin{equation}\label{eq:hFunctions}
	\begin{split}
		h_1(v,c)=\frac{1}{v_F} \frac{d_0(\hat{c})-\hat{c}}{1-\hat{c}^2},
  \quad \quad\quad
 h_2(v,c)=\frac{2 \hat{c}}{v_F} \frac{1-\hat{c}d_0(\hat{c})}{1-\hat{c}^2}, 
  \quad \quad\quad
h_3(v,c)=\frac{4\pi}{v_F}\frac{d_0(\hat{c})-(v/c) d_0(v \hat{c}/c)}{1-(v/c)^2}
	\end{split}
\end{equation}
with 
$ d_0(z)=\frac{\arcsin\sqrt{1-z^2}}{\sqrt{1-z^2}}$,
$\hat{c}=c/v_F$
and
$v_F=\sqrt{1+v^2}$. 
The dynamical critical exponent $z$ 
and the beta functions are
 readily obtained to be
 \begin{subequations}\label{eq:flowEquations1L}
\begin{align}
z^{-1}&=1-\frac{\Nb g^2}{8 \pi^2 c N_{f}}(h_1(v,c)-h_2(v,c)),\label{eq:z1Loop}\\
\frac{dv}{d\ell} 
&= - \frac{z\Nb g^2}{4\pi^2 N_f} \frac{v}{c}h_2(v,c),\label{eq:v1Loop}\\
\frac{dc}{d\ell} 
&= - \frac{zg^2}{8\pi^2}\left(\frac{c\pi}{v} - \frac{\Nb}{N_f} (h_1(v,c) - h_2(v,c)) \right),\label{eq:c1Loop}\\
\frac{dg}{d\ell} 
&= - \frac{zg}{2}\left( \frac{g^2}{4\pi v} - \epsilon + \frac{g^2}{8\pi^3 c N_f}((2-\Nb ) h_3(v,c) + 2\pi \Nb h_2(v,c))\right),\label{eq:g1Loop}\\
\frac{du}{d\ell} 
&= uz\left(\epsilon - \frac{g^2}{2\pi v} - \frac{u(\Nb +8)}{2c^2\pi^2}+\frac{\Nb g^2}{4\pi^2 c N_f}(h_1(v,c) - h_2(v,c)))\right).\label{eq:u1Loop}
\end{align}
\end{subequations}
Here 
$\ell=\ln\left(\frac{\Lambda}{\mu}\right)$ is the logarithmic scale associated with the floating energy scale $\mu$ measured in the unit of a UV cutoff scale $\Lambda $.
The anomalous dimensions of the fields are given by
\begin{equation}
\begin{split}
\eta_{\psi}=&\frac{\Nb zg^{2}}{16\pi^2 c N_{f}}\left((3-\epsilon) h_2(v,c) - (2-\epsilon) h_1(v,c)\right), \\
\eta_{\phi}=&\frac{z g^{2}}{16\pi^2 c N_{f}}\left(\Nb(4-\epsilon)(h_2-h_1)+\frac{2\pi N_f c}{v}
\right).
\end{split}
\end{equation}

Within the parameter space with $g=0$,
there are two fixed points: 
1) the free fixed point with $u=0$, and 
2) the Wilson-Fisher fixed point with $u \neq 0$
\cite{WilsonFisher1972WFBoson}.
At both fixed points, $v$ and $c$ are exactly marginal parameters.
Once the boson-fermion coupling $g$ is turned on, 
it grows under the RG flow as it is a relevant perturbation in $d<3$\footnote{ At the free fixed point, the scaling dimension of $g$ is $[g]=\frac{1}{2} \epsilon$. At the Wilson-Fisher fixed point, the anomalous dimension of the boson field lowers it to $[g]=\frac{1}{2} \epsilon - \frac{N+2}{2(N+8)^2} \epsilon^2 +O(\epsilon^3)$, which is clearly positive for small $\epsilon$. The constraints from the conformal bootstrap implies that $g$ remains relevant even at the Wilson-Fisher fixed point for all $N$ in $2+1$ dimensions \cite{RevModPhys.91.015002}.
}.
With a non-zero fermion-boson coupling,
electrons connected by the magnetic ordering vector are mixed.
The self-energy correction from the mixing makes $v$ flow to zero as the Fermi surface tends to become nested near the hot spots\cite{
ABANOV1,
MAX2,
SHOUVIK,
LUNTS,
SCHLIEF}. 
As the fermions start to become more dispersionless in the direction perpendicular to $\vec Q_{AF}$,
the boson,
which mixes with 
particle-hole 
 excitations,
 also becomes dispersionless.
The emergent nesting also makes the density of low-energy particle-hole states with momenta close to $\vec Q_{AF}$ to increase. 
This enhances the screening and renormalizes $g$ to smaller values at low energies. 
Namely, the interaction quenches the kinetic energies through dynamical nesting 
while the quenched kinetic energies suppress the interaction through screening.
The net result of this competition between the kinetic and interaction energies is that all parameters of the theory ($v$, $c$, $g$, $u$) flow to zero in the low-energy limit under the one-loop beta function\cite{SHOUVIK}.
This is neither a non-interacting theory nor a theory of completely localized particles because the ratio between the interaction and velocities stay constant as they flow to zero,
and
the anomalous dimensions are controlled by the ratios.
Therefore, it is convenient to use the following ratios to characterize the IR fixed point,
\begin{align} \label{eq:RatiosDef}
s\equiv \frac{c}{v}, \quad   y\equiv  \frac{g^2}{c}, \quad \kappa\equiv  \frac{u}{c^2}.
\end{align}
Here,
$s$ is the speed of the boson measured in the unit of the Fermi velocity perpendicular to $\vec Q_{AF}$\footnote{
Alternatively, one can use $w\equiv v/c$, $\lambda \equiv g^2/v$ as in Ref. \cite{LUNTS}.
}.
$y$ and $\kappa$ represent the Yukawa coupling and the quartic boson coupling 
measured in the unit of $c$.
%
%
%
%
The 
beta functions for the new parameters become
\begin{subequations}\label{eq:BetaFunctions1Loop}
\begin{align}
\frac{ds}{d\ell}
&=-\frac{z \,s\,y}{8 \pi}\left( s -  8\pi\alpha N(h_1(c/s,c)+h_2(c/s,c))  \right),\label{e13b} \\
\frac{dy}{d\ell}
&=z  \, y \left( \epsilon-\frac{s y}{8\pi } - \alpha y N\left( h_1(c/s,c)+h_2(c/s,c)+ \frac{2-N}{N}\frac{h_3{(c/s,c)}}{\pi} \right)\right), \label{e13c}\\
\frac{d\kappa}{d\ell}
&=z\kappa\left(\epsilon-\frac{s y }{4 \pi }-(\Nb+8)\gamma \kappa\right),\label{e13d}
\end{align}
\end{subequations}
where $\alpha=\frac{1}{8 \pi^2 N_f}$ and $\gamma=\frac{1}{2\pi^2}$.
To understand the RG flow,
we first note that the beta functions for $s, y, c$ are independent of $\kappa$ at the one-loop order.
This allows us to understand the RG flow in the three-dimensional space of $s, y, c$ irrespective of the flow of $\kappa$.
Let us first  assume that (a) fixed points arise in the region with $c \ll 1$ and (b) the flow of $c$ is much slower compared to the flow of $(s,y)$.
The validity of these assumptions will be confirmed from the solution obtained from them.
In this case, we can first focus on the RG flow in the subspace of  
$(s,y)$
for a fixed $c \ll 1$.
For a fixed $c$, $(s,y)$ 
 flow to a $c$-dependent quasi-fixed point,\footnote{
This is not a real fixed points, but acts as a line of fixed points parameterized by $c$ to the extent that the flow of $c$ can be ignored\cite{LUNTS}}
\begin{align}
\label{eq:1Lqfp}
s(c)^* = \frac{\Nb}{2N_f} + \mathcal{O}(c)^2,
~~
    y(c)^* = 
    \frac{8 \pi N_f \left(
    \Nb+2 N_f\right)}{
    N(4+2N_f-N)
    } 
    \epsilon   
    +
    \frac{32  (2-\Nb)  
    N_f \left(2 N_f+\Nb\right)^2}{N^2 \left(4+2N_f-N\right)^2} 
    \epsilon  
    c
    + \mathcal{O}(c)^2.
\end{align}
%
Here we use the fact that
the $h_i$'s can be approximated as $h_1(c/s,c)\approx \pi/2-2c$, $h_2(c/s,c)\approx 2c$ and $h_3(c/s,c)\approx \frac{2\pi^2s}{1+s}-4 \pi c$
in the small $c$ limit with a fixed $s$ 
(see Appendix \ref{app:quantumCorrections}).
This represents an attractive quasi-fixed point in the space of $(s,y)$ because
the linearized beta function near \eq{eq:1Lqfp} is
$\frac{d \vec X}{d \ell} = M \cdot \vec X$,
where
$\vec X = 
\Bigl(
s - s_i(c)^*,
y - y_i(c)^*
\Bigr)$
and 
$M=
\left(
\begin{array}{cc}
 -\frac{2N_f+N}{4+2N_f-N} \frac\epsilon2 & 0  \\
 0 & -\epsilon  
\end{array}
\right)$
to the leading order in $\epsilon$.
This shows that $s$ and $y$ flow to \eq{eq:1Lqfp} exponentially in $\ell$ as long as the flow of $c$ is much slower.
Within the one-dimensional manifold of $(s,y,c)= \Bigl(
s^*(c),y^*(c),c 
\Bigr)$ parameterized by $c$,
the flow of $c$ is governed by
\begin{equation}
\dv{c}{\ell}=-\frac{z(c)^{*} {y(c)^{*}}\Nb}{2\pi^2 N_{f}} c^2 +\mathcal{O}(c^3),
\end{equation}
where $z(c)^*=1+\mathcal{O}(\epsilon)$ and $y(c)^*=\mathcal{O}(\epsilon)$.
At low energies, $c$ flows to zero as
$1/\ell$.
This confirms that (a) $c$ indeed flows to $0$
and (b) the flow of $(s,y)$ is much faster than $c$ near the fixed point with $c=0$, justifying the previous assumptions that led to this solution.
Since $s$ is $O(1)$ at the fixed point, $v$ flows to zero at the same rate that $c$ flows under the one-loop RG flow.

\begin{figure}
	\centering
	\begin{subfigure}{0.5\textwidth}
		\includegraphics[width=\textwidth]{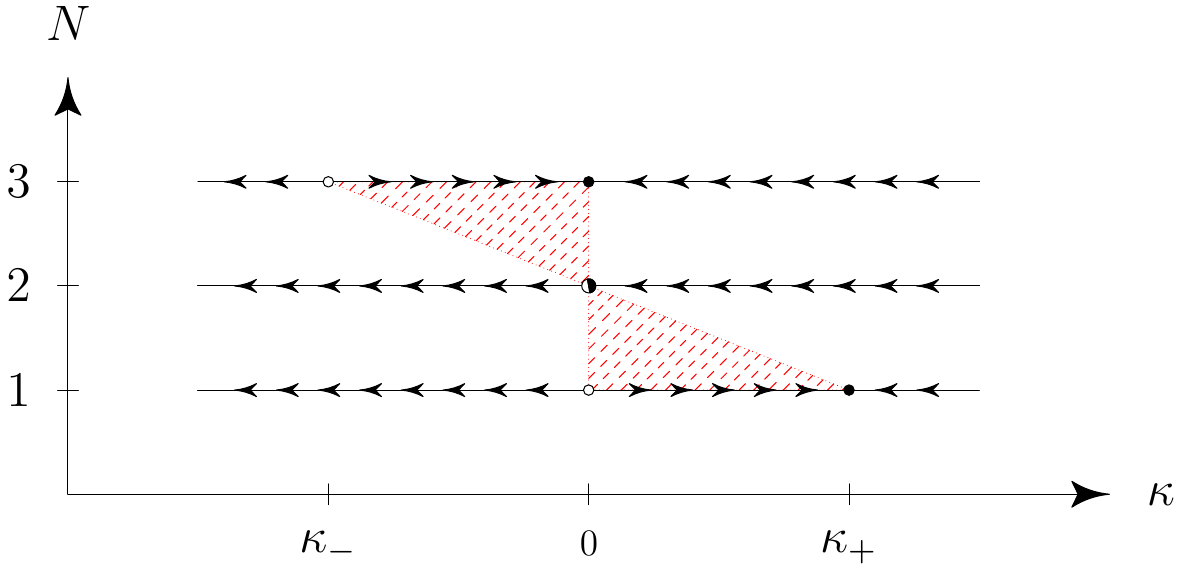}
	\end{subfigure}
		\caption{ 
The one-loop RG flow of the quartic coupling with decreasing energy scale.
As $N$ increases from $1$ to $3$, 
the stable and unstable fixed points collide to create a fixed point 
which is marginally stable from one side 
and unstable from the other side at $N=2$.
  }.
  \label{fig:1Lkappaflow}
\end{figure}

Now let us turn to the beta function of $\kappa$.
Near $(s,y,c)=
\Bigl(
s^*(0), y^*(0),0 
\Bigr)$,
the beta function for $\kappa$ becomes
\begin{equation}
\dv{\kappa}{\ell}=z\kappa\left(
\frac{2 (2-\Nb)}{
2N_f + 4 - \Nb 
}
\epsilon
-(\Nb+8)\gamma \kappa \right).
\label{e13d2}
\end{equation}
The RG flow of $\kappa$ is depicted in \fig{fig:1Lkappaflow}.
For $N \neq 2$, there are two fixed points : 
$\kappa_1^* = \frac{4 \pi ^2 (2-\Nb)  }{\left(\Nb+8\right) \left(
4+2N_f-\Nb \right)}
\epsilon
$
and $\kappa_2^*=0$.
For $N=2$, 
the quartic boson coupling stay marginal even in $d<3$,
which leads to
$\kappa_1^*=\kappa_2^*=0$.
To have a global picture,
it is instructive to pretend that $\Nb$ is a continuous tuning parameter.
For $\Nb=1$ 
the fixed point $\kappa_1^*$ is stable 
while  $\kappa_2^*$ is  unstable.
The two fixed points approach
as $\Nb$ increases 
and collide at $\Nb=2$.
The merged fixed point is  
 marginally attractive (repulsive) on the positive (negative) side of the fixed point.
As $\Nb$ increases above $2$, the two fixed points re-emerge with stability switched :  
$\kappa_2^*$ becomes stable and $\kappa_1^*$ is unstable for $\Nb=3$.

At the one-loop fixed point, both $v$ and $c$ 
 dynamically flow to zero,
 and one has to worry about possible infrared singularity caused by the vanishingly small kinetic energies.
In $d=3$, 
the one-loop analysis is 
 still sufficient because 
the couplings flow to zero faster than the velocities 
($c$ flows to zero only as $(\log{l})^{-1}$ and $\kappa$ and $y$ flow as $1/l$). 
The low-energy physics is then described by a stable quasi-local fixed point with logarithmic corrections to the free theory\cite{SHOUVIK}
similar to the marginal Fermi Liquid\cite{Varma1996MarginalFermiLiquid}. 
In $d<3$, however, the quenched kinetic energies make  certain higher-loop diagrams important at low energies no matter how small $\epsilon$ is as long as $\epsilon > 0$ because $y$ does not flow to zero.
In the $O(3)$ theory, it has been shown that the higher-loop corrections qualitatively change the nature of the fixed point\cite{Sur2016Anisotropic,Lee2018Review,Lunts2017SU2ControlParameter, Sur2016Anisotropic}.
In the rest of the paper, we analyze the effect of the higher-loop corrections that become important due to the quenched kinetic energy for the $Z_2$ and  $O(2)$ theories.

\section{Beyond the one-loop}

Due to the quantum fluctuations enhanced by the dynamically quenched kinetic energy,
one should take into account certain higher-loop graphs even to the leading order in the  $\epsilon$-expansion.
The challenge is the presence of infinitely many higher-loop graphs that are divergent at the one-loop fixed point for any $\epsilon \neq 0$. 
Here, we adopt the 
 following strategy\cite{SHOUVIK3,LUNTS} : 
(1) we first include the leading divergent quantum corrections, that is, the quantum corrections that are divergent in $1/c$ but least suppressed in $\epsilon$,
(2) identify the new fixed point that arises from the leading divergent quantum correction,
and (3) check if other quantum corrections that used to be divergent at the one-loop fixed point become finite in the small $c$ limit and remain suppressed  with $\epsilon$.
If the leading divergent quantum correction is not enough to suppress other higher-loop corrections, we include the next leading divergent quantum correction until the control is achieved.
Not knowing a priori which contributions should be included is awkward,
but this is a working strategy as will be demonstrated in the followings.
We emphasize that attaining the control through the inclusion of higher-loop corrections is not for 
 improving the quantitative accuracy of predictions.
Rather, the higher-loop corrections are crucial for understanding the organizing principles which determines 
 the qualitative behaviours of the theory at low energies\cite{SHOUVIK3,LUNTS,SCHLIEF}.
As we will see, 
the outcomes of this exercise are quite different for the theories with different 
symmetry groups.
In this section,
we first review the results that have been already obtained for the $O(3)$ theory\cite{
LUNTS,
SCHLIEF,
BORGES2023169221},
and discuss new results for the $Z_2$ and $O(2)$ theories one by one.

\subsection{\texorpdfstring{$O(3)$}{O(3)} theory (review) }\label{sec:Section3.5}

For the $O(3)$ theory, 
\fig{fig:BosonSE2LoopFermion} 
is the only diagram that needs to be included beyond the one-loop graphs 
 to the leading order in $\epsilon$\cite{LUNTS}.
Its counter term is written as
 \begin{align}
\label{eq:O32LBSE}
Z_{5,1}^{O(3)}&=-\frac{2}{N_f}\frac{s^2 y^2}{c^2}
h_5(v,c).
\end{align} 
The small $c$ limit of $h_5(v,c)$ is a function of $s$ only and is given by
\begin{align}
\tilde h_5(s) &= \lim_{c\to 0} h_5(c/s,c) = \frac{s^2}{128\pi^2}\rho(s),
\label{eq:h5}
\end{align}
where
\begin{align}
\rho(s) 
= 
\begin{cases}
\frac{3-3s^2-\sqrt{1-s^2}(2+s^2)\log\left( \frac{1+\sqrt{1-s^2}}{s} \right)}{2(s^2-1)^3}& 0 < s < 1,\\
\frac{3-3s^2+\sqrt{s^2-1}(2+s^2)\arcsec(s)}{2(s^2-1)^3}& s>1.
\end{cases}
\label{eq:rho}
\end{align}
See Ref. \cite{LUNTS} for the definition of $h_5(v,c)$ for general $v$ and $c$.
The fixed point obtained with the inclusion of the vertex correction to the boson self-energy is referred to as the modified one-loop (M1L) fixed point.
While $v$ and $c$ still flow to zero at low energies,
the vertex correction  speeds up the boson 
to the extent that 
$s$ diverges in the low energy limit.
Remarkably, $w \equiv 1/s$ can be used as a new dynamically generated control parameter even when $\epsilon$ is not small.
The theory is exactly solvable in the small $w$ limit in any $2 \leq d \leq 3$\cite{SCHLIEF2}.
This leads to the identification of the strongly interacting fixed point in two space dimensions and the exact critical exponent at the critical point\cite{SCHLIEF}.
In $2+1$ dimensions, 
the interacting fixed point with $v=0$ has relevant perturbations and 
theories with non-zero bare nesting angles flow to superconducting states before flowing to the fixed point\cite{BORGES2023169221}.
However, theories with small $v$ and weak bare four-fermion interactions
must go through the bottleneck region in the space of couplings with a slow RG flow due to the proximity to the fixed point.
This gives rise to a family of quasi-fixed points 
labelled by the bare nesting angle.
The quasi-fixed points control the scaling behaviour in the normal state and the pathway to the superconducting state\cite{BORGES2023169221}.

\subsection{\texorpdfstring{$Z_2$}{Z2} theory}\label{sec:Section4}

\begin{figure*}[ht] 
	\centering
	\begin{subfigure}[t]{0.3\textwidth}
		\centering
		\includegraphics[width=\textwidth]{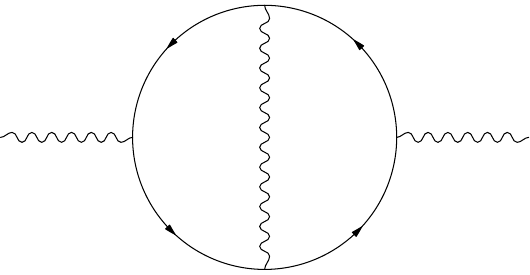}
		\caption{\label{fig:BosonSE2LoopFermion}}
	\end{subfigure}%
	\begin{subfigure}[t]{0.3\textwidth}
		\centering
		\includegraphics[width=\textwidth]{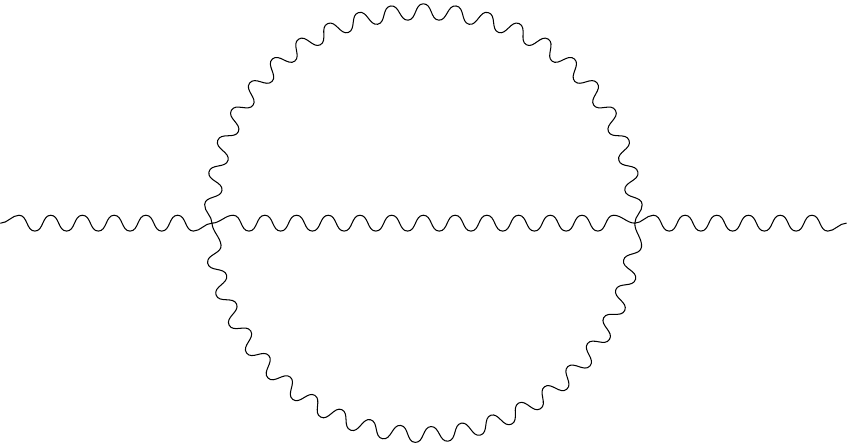}
		\caption{\label{fig:BosonSE2LoopBoson} }
	\end{subfigure}%
	\caption{\label{fig:2LoopDiagrams}
Two-loop diagrams that remain important even to the leading order in $\epsilon$ for the $Z_2$ theory. 
 }
\end{figure*}


For the $Z_2$ theory ($N=1$), the two-loop boson self-energy in Fig.~\ref{fig:BosonSE2LoopFermion} gives rise to the counter term 
that is the opposite of \eq{eq:O32LBSE},
\begin{align}
\label{eq:CountertermsM1Lwrong}
Z_{5,1}^{Z_2} = - Z_{5,1}^{O(3)}.
\end{align} 
%
%
This difference in sign is caused by the fact that the vertex correction inside the self-energy is anti-screening for the $O(3)$ theory due to the non-Abelian nature of the group while it is screening for the $Z_2$ group.
Namely, the generator of the $O(3)$ group anti-commute on average,
 $\sum_{\alpha=1}^3 \tau^\alpha \tau^\beta \tau^\alpha = - \tau^\beta$ inside the vertex correction,
 while the $Z_2$ theory has only one generator that commutes.
%
%
Consequently, the two-loop boson self-energy has the opposite effect on the speed of the collective mode.
For the $O(3)$ theory, the collective mode speeds up relative to fermions so that $s$ diverges 
as $c$ and $v$ individually flow to zero
in the low energy limit. 
In the $Z_2$ theory, 
 the collective mode slows down further,
 and $s$ vanishes in the low energy limit. 
It is useful to use a combination of $s$ and $c$ that becomes finite in characterizing the fixed point.
For this purpose, we introduce
\begin{align}
\chi\equiv \frac{y s^4\rho_0(s)}{c^2}
\end{align} 
with 
\begin{align}\rho_0(s)=\ln\left( \frac{2 e^{-3/2}}{s} \right)\end{align}
being the small $s$ limit of 
\eq{eq:rho}.
With the addition of the contribution from Fig.~\ref{fig:BosonSE2LoopFermion}, the new beta functions become
\begin{subequations}\label{eq:flowEquationsM1Lstep1}
\begin{align}
z^{-1}&=1-\alpha y \left(h_{1}\left(\frac{c}{s},s\right)-h_{2}\left(\frac{c}{s},s\right)\right),\\
    \frac1{zs}\frac{ds}{d\ell} &=
 \alpha y\left(h_{1}\left(\frac{c}{s},s\right) -\chi \frac{16 \pi^2 h_{5}\left(\frac{c}{s},c\right)}{s^2 \rho_{0}(s)}+h_{2}\left(\frac{c}{s},s\right)-\frac{s}{8\pi\alpha}\right),  \\
    \frac1{zy}\frac{dy}{d\ell} &=
-\frac1{zs}\frac{ds}{d\ell}
    +\epsilon-\frac{y}{4\pi}\left(s+4\alpha h_3\left(\frac{c}{s},c\right)\right),   \\
    \frac1{z\kappa}\frac{d\kappa}{d\ell} &=
\left[\alpha y\chi \left(\frac{32\pi^2  h_5\left(\frac{c}{s},c\right)}{s^2 \rho_{0}(s)}\right)-9\gamma\kappa\right] -\frac{sy}{4\pi}+\epsilon , \\
    \frac1{z\chi}\frac{d\chi}{d\ell} &=
-\frac1{zy}\frac{dy}{d\ell}
    +\frac{s\rho_{0}'(s)}{\rho_{0}(s)}
    \frac1{zs}\frac{ds}{d\ell}
    +2\epsilon- \frac{y}{2\pi}\left(s-8\pi\alpha h_2\left(\frac{c}{s},c\right)+4\alpha h_3\left(\frac{c}{s},c\right)\right).
\end{align}
\end{subequations}
The only stable fixed point of these beta functions is
\begin{align}
\label{eq:wrong fixed point}
    s        \to 0,           \qquad
    \chi^*   = 4\pi,          \qquad
    y^*      = \frac{\sqrt{8\gamma}}{(4\alpha+\gamma)}\frac\epsilon{s} \rightarrow \infty \qquad
    \kappa^* = \frac{2\alpha}{9\gamma(4\alpha+\gamma)}\frac\epsilon{s}\rightarrow \infty .
\end{align}
%
%
%
However, this is not a controlled fixed point due to the divergent $\kappa$ and $y$
in the small $s$ limit.
In particular, \fig{fig:BosonSE2LoopBoson} 
 contributes to $Z_{4,1}$ at order $\kappa^2$,
 which is larger than the vanishingly small contributions to $Z_{4,1}$ from the one-loop diagram and \fig{fig:BosonSE2LoopFermion}.

This forces us to include both diagrams in 
Fig.~\ref{fig:2LoopDiagrams}.  
Now we show that this gives rise to a stable and controlled fixed point in the small $\epsilon$ limit.
We refer to this as the modified one-loop (M1L) fixed point for the Ising theory.
The full counter-terms are given by
\begin{align}\label{eq:CountertermsM1L}
\begin{split}
Z_{1,1}&= -\alpha h_{1}(v,c) y ,\\
Z_{3,1}&=-Z_{2,1}=-\alpha h_{2}(v,c) y,\\
Z_{4,1}&=-\frac{1}{4 \pi}s y-\frac{3}{4}\gamma^2\kappa^2,
\end{split}
\begin{split}
Z_{5,1}&=-\frac{3}{4}\gamma^2\kappa^2+\frac{2}{N_f}\frac{s^2 y^2}{c^2}h_5(v,c),\\
Z_{6,	1}&=\alpha y \frac{h_3(v,c)}{2 \pi},\\
Z_{7,1}&=9 \gamma\kappa,
\end{split}
\end{align}
where $\alpha=\frac{1}{8 \pi^2 N_f}$, $ \gamma=\frac{1}{2\pi^2}$.
%
%
%
The beta functions are modified into
\begin{subequations}
\label{eq:flowEquationsM1L}
\begin{align}
\frac1{zs}\frac{ds}{d\ell}
    =&
 \alpha y\left(
h_{1}\left(\frac{c}{s},s\right)
 -\chi \frac{16 \pi^2 h_{5}\left(\frac{c}{s},c\right)}{s^2 \rho_{0}(s)}
 +h_{2}\left(\frac{c}{s},s\right)
 -\frac{s}{8\pi\alpha}
\right),  \\
\frac1{zy}\frac{dy}{d\ell}
    =&
-\frac1{zs}\frac{ds}{d\ell}
+\left(\epsilon-\frac{3}{2}\gamma^2\kappa^2\right)
-\frac{y}{4\pi}\left(
s+4\alpha h_3
\left(
\frac{c}{s},c
\right)
\right),   \\
\frac1{z\kappa}\frac{d\kappa}{d\ell}
    =&
\left[
\alpha y\chi 
\left(
\frac{32\pi^2  h_5\left(
\frac{c}{s},c
\right)}
{s^2 \rho_{0}(s)}
\right)
-9\gamma\kappa
\right] 
-\frac{sy}{4\pi}
+\epsilon -3\gamma^2\kappa^2 , \\
\frac1{z\chi}\frac{d\chi}{d\ell}
    =&
-\frac1{zy}\frac{dy}{d\ell}
    +\frac{s\rho_{0}'(s)}{\rho_{0}(s)}\frac1{zs}\frac{ds}{d\ell}
    +2\left(\epsilon-\frac{3}{2}\gamma^2\kappa^2\right)
- \frac{y}{2\pi}
\left(
s-
8\pi\alpha h_2\left(\frac{c}{s},c\right)+
4\alpha h_3\left(\frac{c}{s},c\right)
\right).
\end{align}
\end{subequations}
For $c\ll s \ll 1$,
the expressions for the dynamical critical exponent and the beta functions simplify as
\allowdisplaybreaks
\begin{subequations}\label{eq:flowEquationsM1LSmallcs}
\begin{align}
z^{-1}&=1-\alpha \pi y/2 ,
\\
    \frac1{zs}\frac{ds}{d\ell}
    =&
\alpha y\left(\frac{\pi}{2}
-\frac{\chi}{8} 
\right),  \\
\frac1{zy}\frac{dy}{d\ell}
    =&
-
\frac1{zs}\frac{ds}{d\ell}
+\left(\epsilon-\frac{3}{2}\gamma^2\kappa^2\right)
,   \\
\frac1{z\kappa}\frac{d\kappa}{d\ell}
    =&
\left[
\frac
{\alpha y\chi }
{4}
-9\gamma\kappa
\right] 
+\epsilon - 3\gamma^2\kappa^2, \\
    \frac1{z\chi}\frac{d\chi}{d\ell}
    =&
-
    \frac1{zy}\frac{dy}{d\ell}
+2\left(\epsilon-\frac{3}{2}\gamma^2\kappa^2\right).
\end{align}
\end{subequations}
The stable fixed point of the beta functions is
\begin{equation}
\label{eq:FixedPointValuesM1L}
s^{*}=0, \quad
\alpha y^{*}=\frac{1}{\pi} 
\left(
3\sqrt{6 \epsilon}+\epsilon
\right)
,\quad \gamma\kappa^{*}=\sqrt{\frac{2\epsilon}{3}}, \quad \chi^{*}=4\pi.
\end{equation}
Since 
$\chi^*  \sim O(1)$,
$c \sim  s^2 \log^{1/2}(1/s) \ll s$ in the small $s$ limit.
The stable fixed point indeed arises in the region with
$c \ll s \ll 1$ in which
\eq{eq:flowEquationsM1LSmallcs} is valid.
It is noted that the coupling constants $y^*$ and $\kappa^*$, which control the anomalous dimensions, is proportional to $\sqrt{\epsilon}$ to the leading order in $\epsilon$.
This is another manifestation of the breakdown of the conventional loop-expansion, 
which is caused by the dynamical quenching of the kinetic energy.

At the modified one-loop fixed point in \eq{eq:FixedPointValuesM1L},
there is no quantum correction that is divergent.
This is shown in Appendix \ref{app:ControlGraphs}. 
All other higher-order corrections are suppressed by powers of $s$, $\sqrt{\epsilon}$ or both relative to the modified one-loop corrections except for one exception.
The exception arises for a class of diagrams for the fermion self-energy whose contributions to $Z_{2,1}$ and $Z_{3,1}$ vanish at the fixed point,
but vanish more slowly  than the modified one-loop contributions in the small $s$ limit.
This is due to the fact that for the one-loop fermion self-energy the external momentum can be directed to flow through the boson propagator only, which results in a suppression by $c$ for the momentum-dependent self-energy corrections.
Those higher-order quantum corrections that are larger than the modified one-loop correction are written as
(see  App. \ref{app:Z2} for details)
\begin{align}
    Z_{2,1}=Z_{3,1}=Ay^3s,
    \label{eq:AZ23}
\end{align}
where $A$ is a constant of order unity.
Because these corrections vanish at $s=0$, the fixed point in \eq{eq:FixedPointValuesM1L} is not affected by them.
The main reason for considering these contributions is to make sure that the stability of the fixed point is not affected by them.

To understand the stability of the fixed point, we have to examine the RG flow of the coupling constants for a small but non-zero $s$.
%
With the inclusion of \eq{eq:AZ23},
\eq{eq:CountertermsM1L} is modified as
\begin{align}\label{eq:CountertermsM1L_extra}
\begin{split}
Z_{1,1}&= -\alpha h_{1}(v,c) y ,\\
Z_{2,1}&=\alpha h_{2}(v,c) y + A y^3s,\\
Z_{3,1}&=-\alpha h_{2}(v,c) y + A y^3s,\\
Z_{4,1}&=-\frac{1}{4 \pi}s y-\frac{3}{4}\gamma^2\kappa^2,
\end{split}
\begin{split}
Z_{5,1}&=-\frac{3}{4}\gamma^2\kappa^2+\frac{2}{N_f}\frac{s^2 y^2}{c^2}h_5(v,c),\\
Z_{6,	1}&=\alpha y \frac{h_3(v,c)}{2 \pi},\\
Z_{7,1}&=9 \gamma\kappa,
\end{split}
\end{align}
and 
the beta functions become
\begin{subequations}\label{eq:flowEquationsComplete}
\begin{align}
\frac1{zs}\frac{ds}{d\ell}
    =&
 \alpha y\left(
h_{1}\left(\frac{c}{s},s\right)
 -\chi \frac{16 \pi^2 h_{5}\left(\frac{c}{s},c\right)}{s^2 \rho_{0}(s)}
 +h_{2}\left(\frac{c}{s},s\right)
 -\frac{s}{8\pi\alpha}
\right)+3A s y^3,  \\
\frac1{zy}\frac{dy}{d\ell}
    =&
-\frac1{zs}\frac{ds}{d\ell}
+\left(\epsilon-\frac{3}{2}\gamma^2\kappa^2\right)
-\frac{y}{4\pi}\left(
s+4\alpha h_3
\left(
\frac{c}{s},c
\right)
\right)
 + 6A s y^3,   \\
\frac1{z\chi}\frac{d\chi}{d\ell}
    =&
-\frac1{zy}\frac{dy}{d\ell}
    +\frac{s\rho_{0}'(s)}{\rho_{0}(s)}\frac1{zs}\frac{ds}{d\ell}
    +2\left(\epsilon-\frac{3}{2}\gamma^2\kappa^2\right)
- \frac{y}{2\pi}
\left(
s-
8\pi\alpha h_2\left(\frac{c}{s},c\right)+
4\alpha h_3\left(\frac{c}{s},c\right)
\right)
 + 12A s y^3,
\end{align}
\end{subequations}
where the beta function for $\kappa$ remains the same as \eq{eq:flowEquationsM1L}.
Near the fixed point, 
the beta function for $s$ is proportional to $(s-s^*)(\chi-\chi^*)$ while  other beta functions are linear in the deviation away from the fixed point.
This implies that the flow of $s$ is much slower than the flow of 
$y$, $\kappa$ and $\chi$ near the fixed point.
Sufficiently close to the fixed point, we can then consider the $s$-dependent quasi-fixed points 
$\left(  y^*(s), \kappa^*(s), \chi^*(s) \right)$
at which the beta functions for  $y$, $\kappa$ and $\chi$ vanish,
\begin{subequations}
    \label{QuasiFixedPointManifold_Full}
\begin{align}
    y^*(s)      &= \frac{\sqrt{54\epsilon}+\epsilon}{\pi\alpha} - \frac{s \left(8 \pi ^2 \alpha  \left(4 \epsilon +15 \sqrt{6} \sqrt{\epsilon }+54\right)+3
   \sqrt{6} \sqrt{\epsilon }+54\right)}{8 \pi ^3 \alpha ^2} + \frac{8748}{(\pi\alpha)^4}As\epsilon,
    \\
    \kappa^*(s) &= \sqrt{\frac{2\epsilon}{3\gamma^2}} -\frac{\left(8 \pi ^2 \alpha +1\right) s \left(\sqrt{6} \sqrt{\epsilon }+18\right)}{24 \pi ^2 \alpha  \gamma }
    +
    \frac{972}{(\pi\alpha)^3\gamma}As\epsilon
    ,
   \\
    \chi^*(s)   &= 4\pi - \frac{s}{\alpha\pi}
    +
    \frac{1296}{\pi^2\alpha^3}As\epsilon.
\end{align}
\end{subequations}
We use 
$\{ s^*, y^*, \kappa^*,\chi^* \}$ 
to denote the true fixed point
in \eq{eq:FixedPointValuesM1L},
and 
$\{  y^*(s), \kappa^*(s), \chi^*(s) \}$ 
to denote the $s$-dependent quasi-fixed points.
It is noted that the contributions from $Z_{2,1}$ and $Z_{3,1}$, which is of order $y^3s$, are sub-leading compared with other contributions.
Therefore, the contributions beyond the modified one-loop order can be indeed ignored.

The linearized beta function of 
$\mathbb{V}=
\left(y-y^{*}(s),
\kappa-\kappa^{*}(s),
\chi-\chi^{*}(s)\right) $
for a fixed $s$ reads
$\frac{1}{z^*(s)}\dv{\mathbb{V}}{\ell}=
M\mathbb{V}$,
where
$z^*(s) = 1+ \mathcal{O}(\sqrt{\epsilon})$ is the dynamical critical exponent evaluated at the $s$-dependent quasi-fixed point
and
$M=\left[
\begin{array}{ccc}
0
& -4\pi^{-1}\epsilon N_f(18+\sqrt{6\epsilon}) 
& N_f(\sqrt{54\epsilon}+\epsilon)^2 \\
\frac\pi{N_f}\frac{\sqrt\epsilon}{\sqrt{24}}
& -\sqrt{54\epsilon}-4\epsilon
& \pi\epsilon(3+\sqrt{\epsilon/6})\\
0
& - \pi^{-1}\sqrt{24\epsilon}
& -\frac12(\sqrt{54\epsilon}+\epsilon)
\end{array}
\right] + O(s)$.
The eigenvalues of $M$ are 
$\left\{-3 \sqrt{6} \sqrt{\epsilon },-\frac{\epsilon }{2}-3 \sqrt{\frac{3}{2}} \sqrt{\epsilon }, -4\epsilon\right\} 
+ \mathcal{O}(\epsilon^{3/2})
+ \mathcal{O}(s)$.
For a fixed $\epsilon > 0$, all eigenvalues are negative definite in the small $s$ limit,
which implies that $\{y,\kappa,\chi\}$
flow to the $s$-dependent quasi-fixed-point around length scale $\ell^{Z_2}_1 \sim \epsilon^{-1}$ provided that the flow of $s$ can be ignored. 
To understand the flow of $s$ within the manifold of quasi-fixed points, we can directly solve the beta function for $s$ within the one-dimensional manifold.
However, it is more convenient to find $c(\ell)$ and 
use the relation 
$\chi = ys^4\rho_0(s)/c^2 = 1/4\pi$ 
satisfied within the manifold 
to extract $s(\ell)$.
Within the one-dimensional manifold of the quasi-fixed points, the beta function for $c$ reads
$\frac{\beta_c}{zc} =  \left( 2 - \frac12\frac{s\rho_0'(s)}{\rho_0(s)} \right) \Bigl( Z_{2,1}'  - Z_{3,1}' \Bigr)$,
which in the small $c,s$ limit becomes
\begin{align}
    \frac{dc}{d\ell} = -8\alpha y^* c^2.
\end{align}
Its solution is 
$c(\ell) = \frac{1}{c_1^{-1} + (8\alpha y^* \Delta \ell)}$.
Here, $\Delta \ell = \ell - \ell_1^{Z_2}$ 
with $\ell_1^{Z_2}$ being the logarithmic length scale 
at which 
$y$, $\kappa$ and $\chi$ reach their $s$-dependent quasi fixed points.
$c_1 = \left[ y^*s_1^4\log(
 \frac{2 e^{-3/2}}{s_1} 
 )/\chi^* \right]^{1/2}$ is the value of $c$ 
at scale $\ell_1^{Z_2}$
expressed in terms of $s_1 \equiv s(\ell_1^{Z_2})$.
From this, we readily obtain the flow of $s(\ell)$ to be
\begin{align}
    s(\ell) &= 
    \frac{(\chi^*/y^*)^{1/4}}{ \left[ c_1^{-1} + (8\alpha y^* \Delta \ell) \right]^{1/2} \log^{1/4}\left[ 2e^{-3/2}(y^*/\chi^*)^{1/4} (c_1^{-1} + (8\alpha y^* \Delta \ell))^{1/2} \right]}.
\end{align}
This shows that $s$ indeed flows to zero in the low-energy limit
and the speed at which $s$ flows to its true fixed point value is much smaller than that of the rest of remaining couplings flowing to the $s$-dependent quasi-fixed points.
This justifies our earlier assumption of ignoring the flow of $s$ when we identify the $s$-dependent quasi-fixed point in the small $s$ limit.
The theory flows to the true fixed point 
only in length scales much larger than 
$\ell_0^{Z_2} = \ell_1^{Z_2} +  \frac{1}{ 8\alpha y^* c_1}$.
For $s_1 \ll 1$, 
$c_1 \ll 1$ and
$\ell_0^{Z_2} \gg \ell_1^{Z_2}$.
Due to this hierarchy,
a UV theory that is initially close to the fixed point is first attracted to an $s$-dependent quasi-fixed point around scale $\ell_1^{Z_2}$
before
it rolls down to the true fixed point along the one-dimensional manifold of quasi-fixed points
at much larger length scales
$\ell_0^{Z_2}$.
This is illustrated in \fig{fig:1a}.

\subsection{\texorpdfstring{$O(2)$}{O2} theory}\label{sec:Section5}

\begin{figure}[htpb!]
\centering
\begin{subfigure}{0.2\textwidth}
    \includegraphics[width=\textwidth]{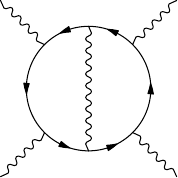}
    \caption{\label{fig:2Lphi4}}
\end{subfigure}
\begin{subfigure}{0.3\textwidth}
    \centering
    \raisebox{0.45cm}{\includegraphics[width=\textwidth]{Fig/13.pdf}}
    \caption{\label{eq:2Lphi2}}
\end{subfigure}
\caption{Two-loop diagrams that remain   important even to the leading order in $\epsilon$ for the $O(2)$ theory.}
\end{figure}
    
In this section, we turn to the $O(2)$ theory.
We recall that at the one-loop order
the stable fixed point arises at
$s^* = 
\frac{1}{N_f}$,
$y^* =  4\pi \epsilon N_f$, 
$\kappa^* = 0$
and 
$c^*=0$.
One peculiar feature of the $O(2)$ theory is that \fig{fig:BosonSE2LoopFermion} vanishes.
This is due to a cancellation of the vertex correction caused by a balance between commuting and anti-commuting vertices:  $\sum_{\alpha=1,2} \tau^\alpha \tau^\beta \tau^\alpha = 0$ for $\beta=1,2$.
Furthermore, all non-vanishing boson self-energy corrections to $\mathcal A_5$ are regular in the small $c$ limit.
In those non-vanishing diagrams, all virtual fermions in the loops can stay right on the Fermi surface irrespective of the external momentum.
Consequently, the momentum-dependent self-energy correction is solely generated from the dispersion of the boson, which is suppressed by a factor of $c^2$ 
(see Appendix \ref{app:O2} for the proof).
This is in stark contrast to the behaviour of the $Z_2$ theory discussed in the previous section.
Nonetheless, there are still higher-order quantum corrections for  $\mathcal{A}_{7}$ that diverge at the one-loop fixed point.
The leading divergent quantum correction that arises from the two-loop four-boson vertex correction in \fig{fig:2Lphi4} is given in the $c\ll1$ limit by
\begin{align}
    Z_{7,1} &= -
    \frac{N}{4N_f^2}
    \frac{g^6}{v^2u}\tilde h_5(s).
    \label{eq:2LZ7}
\end{align}
While suppressed by a higher power of $\epsilon$, 
it diverges at the one-loop fixed point.
With \eq{eq:2LZ7} included, the beta functions are modified to
\begin{subequations}
\begin{align}
    \frac{dc}{d\ell} &= zc\pi \alpha y\left(1 - N_f s- \frac{8c}{\pi}\right),\\
    \frac{ds}{d\ell} &= zs \pi \alpha y\left(1-N_f s\right),\\
    \frac{dy}{d\ell} &= zy\left(\epsilon -\pi \alpha y(1+N_f s)\right),\\
    \frac{d\kappa}{d\ell} &= z\kappa \left(
    \epsilon 
    - \frac{sy}{4\pi} 
    +\frac{s^2y^3 \tilde h_5(s)}{c\kappa N_f^2}
	-10\gamma\kappa
	\right).
\end{align}
    \label{eq:41}
\end{subequations}
The stable fixed point of 
 Eqs. \eqref{eq:41} arises at   
    \begin{align}
        c\to0, \qquad
        s^* = \frac{1}{N_f},   \qquad
        y^* = 4\pi N_f\epsilon,\qquad
        \kappa^* = \frac{(\pi\epsilon)^{3/2}\sqrt{\rho(N_f^{-1})}}{\sqrt{10c}N_f^{3/2}} 
        \rightarrow \infty
    \end{align}
        to the leading order in $\epsilon$,
        where $\rho(s) = 128\pi^2 s^{-2} \tilde h_5(s)$.

This fixed point is not yet controlled because $\kappa$
diverges at the fixed point.
This makes all quantum corrections that involve $\kappa$ singular at the fixed point.
Those divergent higher-loop corrections are expected to be tamed once the leading divergent quantum correction is included. 
Specifically, the two-loop boson self-energy in \fig{eq:2Lphi2},
which is order of $\kappa^2$,
introduces a positive anomalous dimension to the boson field,
weakening the quartic boson coupling and the Yukawa coupling.
The counter-term for the two-loop boson self-energy reads
    \begin{align}
    Z_{4,1;2L;b} = Z_{5,1;2L;b} = - \frac{u^2 \left(N+2\right)}{(2\pi)^4 c^4 }.
    \label{eq:2LZ4}
    \end{align}
Now we show that including both \eq{eq:2LZ7} and \eq{eq:2LZ4} to the one-loop quantum corrections gives rise to a stable and controlled fixed point in the small $\epsilon$ limit. 
We refer to this as the modified one-loop (M1L) fixed point for the $O(2)$ theory. 
Altogether, the counterterms for the $O(2)$ theory are given by (cf. \eqref{eq:Zs}, \eqref{eq:2LZ7}, \eqref{eq:2LZ4})
\begin{align}\label{eq:CountertermsM1L_O2}
\begin{split}
Z_{1,1}&= -2\alpha h_{1}(v,c) y ,\\
Z_{3,1}&=-Z_{2,1}=-2\alpha h_{2}(v,c) y,\\
Z_{4,1}&=-\frac{1}{4 \pi}s y-\gamma^2\kappa^2,
\end{split}
\begin{split}
Z_{5,1}&=-\gamma^2\kappa^2,\\
Z_{6,1}&=0,\\
Z_{7,1}&=10\gamma\kappa-\frac{2}{4N_f^2}\frac{s^2y^3}{\kappa c}\tilde h_5(s),
\end{split}
\end{align}
in the $c\ll1$ limit. The IR beta functions read
\begin{subequations}
\begin{align}
    \frac{dc}{d\ell} &= zc\pi \alpha y\left(1 - N_f s- \frac{8c}{\pi}\right),\\
    \frac{ds}{d\ell} &= zs \pi \alpha y\left(1-N_f s\right),\\
    \frac{dy}{d\ell} &= zy\left(\epsilon -\pi \alpha y(1+N_f s) 
    -2\gamma^2\kappa^2
    \right),\\
    \frac{d\kappa}{d\ell} &= z\kappa \left(\epsilon - \frac{sy}{4\pi} 
    +\frac{s^2y^3 \tilde h_5(s)}{c\kappa N_f^2}
    -10\gamma\kappa
    -4\gamma^2\kappa^2
    \right).
    \label{eq:42betaO2}
\end{align}
\end{subequations}
The beta functions exhibit a stable fixed point at
\begin{align}
c^*=0, \qquad
s^*=\frac{1}{N_f}, \qquad
\frac{y^*}{\sqrt[3]{c^*}} = \left(\frac{5\gamma^3\epsilon}{256\alpha^4\tilde h_5(4\alpha/\gamma)}\right)^{1/3},
\qquad
\kappa^* = 
\sqrt{\frac{\epsilon}{2\gamma^2}}.
\label{eq:O2truefp}
\end{align}
It is noted that both $y^*$ and $c^*$ vanish at the new fixed point with 
$ \frac{y^*}{ \sqrt[3]{c^*}}$ fixed to be $O(\epsilon^{1/3})$.

Now, we analyze the stability of the fixed point.
As is shown in Appendix.~\ref{app:O2},
there is a group of higher-order fermion self-energy corrections 
and cubic vertex corrections
that are zero at the fixed point but vanish more slowly than the modified one-loop corrections in the small $c$ limit.
Their contributions can be written as
\begin{align}
    Z_{2,1}=Z_{3,1}=Ay^3s, ~~
    Z_{6,1}=By^2,
\end{align}
where $A$ and $B$ are constants of order unity.
While they do not affect the fixed point, 
they can in principle affect the stability of the fixed point.
%
In the small $c$ limit, 
the complete set of counter-terms that include all leading order contributions is given by
\begin{align}\label{eq:CountertermsM1L_O2_extra}
\begin{split}
Z_{1,1}&= -2\alpha h_{1}(v,c) y ,\\
Z_{2,1}&=2\alpha h_{2}(v,c) y + Ay^3,\\
Z_{3,1}&=-2\alpha h_{2}(v,c) y + Ay^3,\\
Z_{4,1}&=-\frac{1}{4 \pi}s y-\gamma^2\kappa^2,
\end{split}
\begin{split}
Z_{5,1}&=-\gamma^2\kappa^2,\\
Z_{6,1}&=By^2,\\
Z_{7,1}&=10\gamma\kappa-\frac{2}{4N_f^2}\frac{s^2y^3}{\kappa c}\tilde h_5(s).
\end{split}
\end{align}
The resulting beta functions become
\begin{subequations}
\begin{align}
    \frac{dc}{d\ell} &= zc\pi \alpha y\left(1 - N_f s- \frac{8c}{\pi} + \frac{3Ay^2}{\pi\alpha}\right),\\
    \frac{ds}{d\ell} &= zs \pi \alpha y\left(1-N_f s+ \frac{3Ay^2}{\pi\alpha}\right),\\
    \frac{dy}{d\ell} &= zy\left(\epsilon -\pi \alpha y(1+N_f s) 
    -2\gamma^2\kappa^2
    + 3Ay^3
    - 4By^2
    \right),
\end{align}
\end{subequations}
where the beta function for $\kappa$, which is not affected by $A$ and $B$, is still given by \eq{eq:42betaO2}.

Let us show that the fixed point in \eq{eq:O2truefp} is attractive.
If the beta functions are 
 expanded around the fixed point, one immediately notices that
the beta functions for both $s$ and $c$ vanish to the linear order in 
$\delta s = s-s^*$ and $\delta c = c - c^*$ while the beta functions for $\kappa$ and $y$ do not.
This implies that the flow of $s$ and $c$ can be ignored in understanding the flow of $\kappa$ and $y$ near the fixed point.
Namely, $y$ and  $\kappa$ first flow to a quasi-fixed point that depends on the bare values of $s$ and $c$
at an intermediate energy scale before $s$ and $c$ begin to flow at lower energy scales.
The $s$,$c$-dependent quasi-fixed points of $y$ and $\kappa$ are given by
    \begin{align}
&y^*(\delta s,c) = \left(\frac{c\gamma^3\epsilon(5\sqrt{2}+\sqrt\epsilon)}{256\sqrt2\alpha^4\tilde h_5(4\alpha/\gamma)}\right)^{1/3}
\left(1 - \delta s \left[ \frac{\gamma}{6\alpha} + 
\frac{\tilde h_5'(4\alpha/\gamma)}{3\tilde h_5(4\alpha/\gamma)} \right] \right)
+
\mathcal{O}(c^{2/3}),
\\
&\kappa^*(\delta s,c) = 
\sqrt{\frac{\epsilon}{2\gamma^2}}-
\left(\frac{c(5\sqrt{2}+\sqrt\epsilon)}{2048\alpha\gamma^{3/2}\tilde h_5(4\alpha/\gamma)\sqrt{2\epsilon}}\right)^{1/3}
\left(1 - \delta s \left[ \frac{\gamma}{24\alpha} + 
\frac{\tilde h_5'(4\alpha/\gamma)}{3\tilde h_5(4\alpha/\gamma)} \right] \right)
+ \mathcal{O}(c^{2/3}).
\label{eq:O2twoparamqfp_modified}
    \end{align}
Here, 
$\{ c^*, s^*, y^*, \kappa^* \}$ 
denotes the true fixed point
in \eq{eq:O2truefp},
and 
$\{  y^*(\delta s,c), \kappa^*(\delta s,c) \}$ represents
the quasi-fixed points that depend on $c$ and $\delta s \equiv s-s^*$.
We note that $A$ and $B$ only affect the quasi-fixed points at the order of $c^{2/3}$ and higher.
Since their contributions are suppressed compared with the leading order corrections, we can ignore them.
$\tilde h_5'(s)$ is the derivative of $\tilde h_5(s)$ with respect to $s$.

The fact that the flows of $y$ and $\kappa$ are indeed 
 much faster than those of $s$ and $c$ 
can be checked from the linearized beta functions for 
    $\delta y = y- y^*(s,c)$
    and
    $\delta \kappa = \kappa - \kappa^*(s,c)$. 
The linearized beta function for $\delta y$ and $\delta \kappa$ becomes
            \begin{align}
            \begin{pmatrix}
                \dot{\delta y}\\
                \dot{\delta \kappa}
            \end{pmatrix}
            &=
            \begin{pmatrix}
                -\frac{y^*}{4\pi N_f} 
                & -4\gamma^2 y^*\kappa^*\\ 
                \frac{3(y^*)^2\tilde h_5(N_f^{-1})}{cN_f^4}
                -\frac{\kappa^*}{4\pi N_f}
                & \epsilon
                -20\gamma\kappa^*
                -12\gamma^2(\kappa^*)^2
                -\frac{y^*}{4\pi N_f}
            \end{pmatrix}
            \begin{pmatrix}
                \delta y\\
                \delta \kappa
            \end{pmatrix}.
            \end{align}
This has eigenvalues
$            \left\{
            -\sqrt{200\epsilon},
            -3 \epsilon \right\}
$
to the leading order in $\epsilon$ and $c$.
As long as $\epsilon > 0$, 
the eigenvalues remain definitely negative in the small $c$ limit.
As a result, $y,\kappa$  flow to  \eq{eq:O2twoparamqfp_modified}
below a logarithmic length scale $\ell_2^{O(2)} \sim \epsilon^{-1}$. 
Now, we examine the flow of $s$ and $c$.
Within the two-dimensional manifold of quasi-fixed points,
the beta functions for $\delta s = s - s^*$ and $c$ read
$\dot \delta s \sim - c^{1/3} \delta s$
and $\dot c \sim - c^{7/3}$.
Interestingly, $| {\dot \delta s}/\delta s| \gg | \dot c/c |$ in the small $c$ limit.
This additional layer of hierarchy between the RG speeds of $s$ and $c$ allows one to understand the flow of $s$ first without considering the flow of $c$ within a window of length scales.
Within the two-dimensional manifold of quasi-fixed points,
the flow of $s$ is governed by
\begin{align}
\frac{ds}{d\ell} &= -\delta s\left(
	\frac{\gamma^{3/2}(5\sqrt2+\sqrt\epsilon)c\epsilon}{2^{10}\alpha \tilde h_5(4\alpha/\gamma)}
\right)^{1/3},
\end{align}
where $c$ is still regarded as a constant. 
Its solution is given by
\begin{align}
\delta s(\ell) &= \delta s_2 \exp\left( -(\ell-\ell_2^{O(2)} ) \left[ 
	\frac{\gamma^{3/2}(5\sqrt2+\sqrt\epsilon)c_2\epsilon}{2^{10}\alpha \tilde h_5(4\alpha/\gamma)}
\right]^{1/3} \right),
\end{align}
where $\ell_2^{O(2)} $ is the logarithmic length scale at which $y$ and $\kappa$ flow into the two-dimensional quasi-fixed points,
$\delta s_2 = s( \ell_2^{O(2)}) - s^*$
and $c_2 = c( \ell_2^{O(2)})$.
The RG `time' $\ell_1^{O(2)} $ needed for $\delta s$ to become small is
$\ell_1^{O(2)} \sim \ell_2^{O(2)}  + [\epsilon c_2]^{-1/3}$. 
It is noted that $\ell_1^{O(2)} $ is indeed much larger than
$\ell_2^{O(2)} $ for $c_2 \ll 1$, which confirms the fact that the flow of $s$ can be
ignored for the flow of $\kappa, y$.
At length scales $\ell \sim \ell_1^{O(2)} $, the theory
flows into the one-dimensional family of quasi-fixed point,
\begin{align}
    s &=\frac{1}{N_f}, &
    y^*(c) &= \left(\frac{\gamma^3\epsilon(5\sqrt{2}+\sqrt\epsilon)}{256\sqrt2\alpha^4\tilde h_5(4\alpha/\gamma)}\right)^{1/3},
            &
            \kappa^*(c) &= \sqrt{\frac{\epsilon}{2\gamma^2}}-
            \left(\frac{c(5\sqrt2+\sqrt\epsilon)}{2048\alpha\gamma^{3/2}\tilde h_5(4\alpha/\gamma)\sqrt{2\epsilon}}\right)^{1/3},
\end{align}
which is a sub-manifold of \eq{eq:O2twoparamqfp_modified}.
Finally, the flow of $c$ 
 within 
 this one-dimensional manifold of quasi-fixed points is governed by
\begin{align}
    \frac{dc}{d\ell} &= - c^{7/3} \left[ 
    \frac{\gamma^3\epsilon(10+\sqrt{2\epsilon})}{\alpha \tilde h_5(4\alpha/\gamma)}
    \right]^{1/3}
\end{align}
and its solution is 
\begin{align}
c(\ell) = \left[ c_1^{-4/3} + 
\frac43     \left(\frac{\gamma^3\epsilon(10+\sqrt{2\epsilon})}{\alpha \tilde h_5(4\alpha/\gamma)}\right)^{1/3} (\ell-\ell_1^{O(2)} )
\right]^{-3/4},
\end{align}
where $c_1 \equiv c(\ell_1^{O(2)} ) \approx c_2$.
The flow of $c$ is indeed negligible for $\ell \ll \ell_0^{O(2)} $,
where $\ell_0^{O(2)} \sim \ell_1^{O(2)}  + c_1^{-4/3} \epsilon^{-1/3}$.

For a theory with $c_2 \ll 1$, there exists the
hierarchy of length scales: $\ell_0^{O(2)}  \gg \ell_1^{O(2)}  \gg
\ell_2^{O(2)} $. This creates three distinct stages of RG
flow. 
In $\ell_1^{O(2)}  \gg  \ell \gg \ell_2^{O(2)} $, the theory
flows into a point within the two-dimensional
manifold of quasi-fixed points labeled by $s$ and
$c$. 
In $\ell_0^{O(2)}  \gg  \ell \gg \ell_1^{O(2)} $, the flow of
$c$ is still negligible but the flow of $s$ takes
the theory to the one-dimensional manifold of
quasi-fixed points labeled only by $c$. 
In the
largest length scale  with $\ell \gg \ell_0^{O(2)} $, the
theory flows to the true infrared fixed point as
all couplings take the universal value at
\eq{eq:O2truefp}. This hierarchy of the length
scales manifests itself in crossover behaviours of
physical observables as will be discussed in the
next section.

This shows that
\eq{eq:O2truefp} is indeed a stable fixed point.
At the fixed point,
all higher-order diagrams which are not included in our analysis
are systematically suppressed either by $\epsilon$ or $c$.
We refer the readers to Appendix \ref{app:ControlGraphs} for details.

\section{Physical observables}

In this section, we derive the scaling forms of the electron spectral function and the dynamical spin structure factor.
We also examine fluctuations in the particle-hole and particle-particle channels that are enhanced in the non-Fermi liquid state.

\subsection{The spectral function and dynamical spin structure factor}

\begin{table}[t]
\centering
\begin{tabular}{|c|c|c|c|c|} 
    \hline
               &  quasi-marginal & $z$ & $\eta_\psi$ & $\eta_\phi$  \\
               & parameters &&& \\ 
   \hline
    $Z_2$ & $s$ & $1+3 \sqrt{\frac{3}{2}} \sqrt{\epsilon }-27 s \left(N_f+1\right)$ & $ -3 \sqrt{\frac{3}{2}} \sqrt{\epsilon } + 27 s \left(N_f+1\right) $  & $ -3 \sqrt{6} \sqrt{\epsilon } + 54 s \left(N_f+1\right) $ \\
  %
    \hline $O(2)$   & $\delta s = s-s^*$, &  
    $1 + 
    \left( \frac{5c\epsilon N_f}{256\pi\tilde h_5(N_f^{-1})} 
    \right)^{1/3} 
    \times $ 
    & $
    -\left(  \frac{5c\epsilon N_f}{256\pi\tilde h_5(N_f^{-1})} 
    \right)^{1/3} 
    \times $ 
    & $  \frac\epsilon2 -
    \left( \frac{5c\epsilon N_f}{32\pi \tilde h_5(N_f^{-1})} 
    \right)^{1/3}
    \times $   \\
     & $c$ 
     & $  \left( 1 - \delta s\left[ \frac{2N_f}{3}+\frac{\tilde h_5'(N_f^{-1})}{3\tilde h_5(N_f^{-1})} \right] \right)
     $ 
     & $  \left( 1 - \delta s\left[ \frac{2N_f}{3}+\frac{\tilde h_5'(N_f^{-1})}{3\tilde h_5(N_f^{-1})} \right] \right)
     $ 
     & $  \left( 1 - \delta s\left[ \frac{11N_f}{12}+\frac{\tilde h_5'(N_f^{-1})}{3\tilde h_5(N_f^{-1})} \right] \right)
     $    \\
   \hline
    $O(3)$ & $w=1/s$ &
    $1+ \frac{3 }{4 N_f} \epsilon w$
   &
   $ -\frac{3}{4 N_f} \epsilon w $ 
   &
   $\frac{\epsilon}{2}
   -\frac{1}{N_f} \epsilon w$ \\   
   \hline 
\end{tabular}
\caption{
Critical exponents expressed to the leading order in $\epsilon$ and the magnitude of the quasi-marginal parameters deformed away from the true fixed points.
In the $Z_2$, $O(2)$ and $O(3)$ theories, 
 the quasi-marginal parameters consist of 
 $\{ s = c/v \}$,
 $\{ s, c \}$
 and
 $\{ w =v/c\}$, respectively.
The true fixed points of the $Z_2$, $O(2)$ and $O(3)$ theories are located at $\{ s^* = 0  \}$, $\{ s^*=1/N_f, c^*=0 \}$ and $\{ w^* =0\}$, respectively.
For the $O(2)$ theory, $\delta s \equiv s-s^*$.
}
\label{table2}
\end{table}%

The electron spectral function and the dynamical spin structure factor can be extracted from the two-point functions.
The electron Green's function at the hot spots 
and the spin-spin correlation function at the antiferromagnetic ordering vector can be written as\cite{LUNTS}
\begin{align}
G(\omega)  \sim &
~ 
e^{
-\int_0^{\log \frac{1}{\omega}} 
d l
\left(
\frac{2\eta_\psi(l)+z(l)(2-\epsilon)-(3-\epsilon)}{z(l)}
\right)
},
\label{eq:Aomega}
\\
D(\omega)  \sim &
~ 
e^{
-\int_0^{\log \frac{1}{\omega}} 
dl
\left( \frac{2\eta_\phi(l)+z(l)(2-\epsilon)-(4-\epsilon)}{z(l)} \right)
}.
\label{eq:chiomega}
\end{align}
Here, $\omega = | {\bf K} |$ is the frequency. 
$l$ denotes the logarithmic energy scale, which is related to the logarithmic length scale $\ell$ through $d\ell = \frac{dl}{z(l)}$.
$z(l)$, $\eta_\psi(l)$ and $\eta_\phi(l)$ 
are the dynamical critical exponent, and the anomalous dimensions of the fermion and the boson fields, respectively.
They depend on scale through the scale dependent couplings
 as is shown in Table \ref{table2}.
We focus on frequencies that are low enough that irrelevant couplings are fixed by the set of quasi-marginal couplings. 
Nonetheless, we consider  a window of frequency that is large enough that the quasi-marginal couplings can take general values before they take the universal values in the low-frequency limit.
In this range of frequency that covers an intermediate energy scale to the low-energy limit, we need to consider critical exponents as functions of the quasi-marginal couplings.
%
Within a window of intermediate energy scales where the flow of quasi-marginal parameter(s) can be ignored,
the exponents are essentially independent of $l$ ($\ell$).
In this case, the correlation functions
exhibit simple power-law behaviours in frequency with exponents that are functions of the quasi-marginal parameter(s) within the manifold of quasi-fixed points.
As the frequency is lowered below a crossover scale, the theory flows toward the true infrared fixed point.
In the low-energy limit, the physical observables exhibit power-law behaviours controlled by the universal scaling exponents of the fixed point with logarithmic corrections associated with the slowest flowing quasi-marginal couplings.
Below we discuss the scaling form of the spectral function and the dynamical spin structure factor, which can be obtained  from Eqs. \eqref{eq:Aomega} and \eqref{eq:chiomega} through the analytic continuation of the thermal frequency.
For simplicity, we assume that the theory defined at an intermediate energy scale $\Lambda$ is close to the fixed point.

In the $Z_2$ theory, 
the theory flows to $s=s_1$ within the one-dimensional manifold of quasi-fixed points at energy scale  
$\omega_1^{Z_2} = \Lambda 
\exp[-\int_0^{\ell_1^{Z_2}} z(\ell) d\ell]$.
For $s_1 \ll 1$, 
the flow of $s$ can be largely ignored at energies above the crossover energy scale given by
$\omega^{Z_2}_0 
 =  \omega^{Z_2}_1 
\exp[ -\int_{\ell_1^{Z_2}}^{\ell_0^{Z_2}} 
z(\ell) d\ell] 
\sim
      \omega^{Z_2}_1  \exp\left( - \frac{\pi}{144(6N_f^2\epsilon^3)^{1/4}} \frac1{s_1^2\sqrt{\log(1/s_1)}} \right)$.
Therefore, $s_1$ plays the role of a quasi-marginal coupling even though the true infrared fixed point is at $s^*=0$.
Across the crossover energy scale $\omega_0^{Z_2}$,
the scaling behaviour of physical observables change due to the flow of $s$.
This leads to the following crossover behaviours of physical observables,
\begin{align}
    &\mathcal A_0(\omega)
    \sim
    \begin{cases}
        \omega^{ -1+ \sqrt{27\epsilon/2} -27 s_1 \left(N_f+1\right) }
        & \omega^{Z_2}_0 \ll \omega < \omega^{Z_2}_1 
\\
        \omega^{-1+\sqrt{27\epsilon/2}}
        \exp\left( a_\psi \sqrt{\frac{\log(1/\omega)}{\sqrt{\log\log(1/\omega)}}}    \right)
        & \omega \ll \omega^{Z_2}_0
    \end{cases},
    \nn   
    &\chi''_0(\omega)
    \sim
    \begin{cases}
        \omega^{ -2+\epsilon-s_1 \sqrt{216\epsilon}}
        & \omega^{Z_2}_0 \ll \omega < \omega^{Z_2}_1 
\\
        \omega^{-2+\epsilon} 
        \exp\left( a_\phi \sqrt{\frac{\log(1/\omega)}{\sqrt{\log\log(1/\omega)}}}    \right)
        & \omega \ll \omega^{Z_2}_0
    \end{cases},
    \label{eq:AchiZ2}
\end{align}
where 
$\mathcal A_0(\omega)$ is
the frequency dependent spectral function of electrons at the hot spots 
and 
$\chi''_0(\omega)$ is the dynamical spin structure factor measured at the ordering wave vector $\vec Q_{AF}$.
$a_\psi = 27(N_f+1)\left(\frac{\pi ^2}{648 \sqrt{6} N_f \epsilon ^{3/2}}\right)^{1/4}$
and
$a_\phi = \left(\frac{12 \sqrt{6} \pi ^2 \sqrt{\epsilon }}{N_f}\right)^{1/4}$.
In $\omega^{Z_2}_0 \ll \omega < \omega^{Z_2}_1$, the exponents that control the scaling behaviour of observables depend on $s_1$.
If a UV theory is not already near the manifold of quasi-fixed points, the precise value of $s_1$ that the theory reaches within the manifold can be different from the UV value of $s$.
In this case, $s_1$ can be determined through other observables.
For example, the dynamical spin structure factor at momenta $\vec Q_{AF}+\vec q$ takes a more general scaling form $\chi_{\vec q}''(\omega) 
\sim \chi''_0(\omega) f_\theta\left(
\frac{\omega}{q^{z(s)}}
\right)$ in the intermediate energy scale,
where $\theta = \tan^{-1} q_y/q_x$ and 
$z(s)$ is the $s$-dependent dynamical critical exponent given by
Table \ref{table2}\cite{Lunts:2023un}.
$s_1$ can be fixed through the 
 relative scaling between $\omega$ and $q$
in $\omega^{Z_2}_0 \ll \omega < \omega^{Z_2}_1$.
Once $s_1$ is known, the scaling behaviours of $\mathcal
A_0(\omega)$ and $\chi''_0(\omega)$ are fully
determined through \eq{eq:AchiZ2}.
In the low-energy limit ($\omega
\ll \omega^{Z_2}_0$), the observables obey the
universal scaling forms dictated by the fixed
point. 
The super-logarithmic deviation from the simple power-law
behaviour is due to the slow convergence of the
quasi-marginal coupling to the true fixed-point
value.
\begin{figure}[htpb!]
\centering
\begin{subfigure}{0.4\textwidth}
    \includegraphics[width=\textwidth]{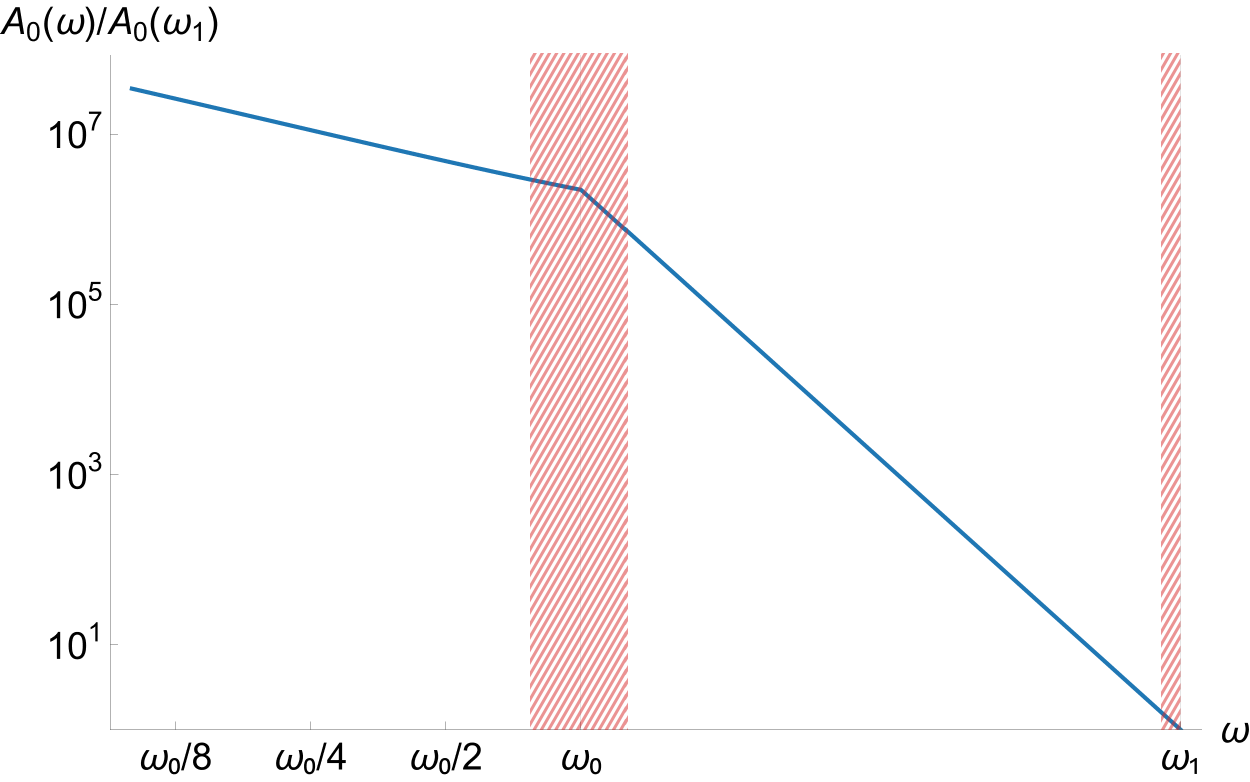}
    \caption{\label{fig:new_37}}
\end{subfigure}
\begin{subfigure}{0.4\textwidth}
    \centering
    \includegraphics[width=\textwidth]{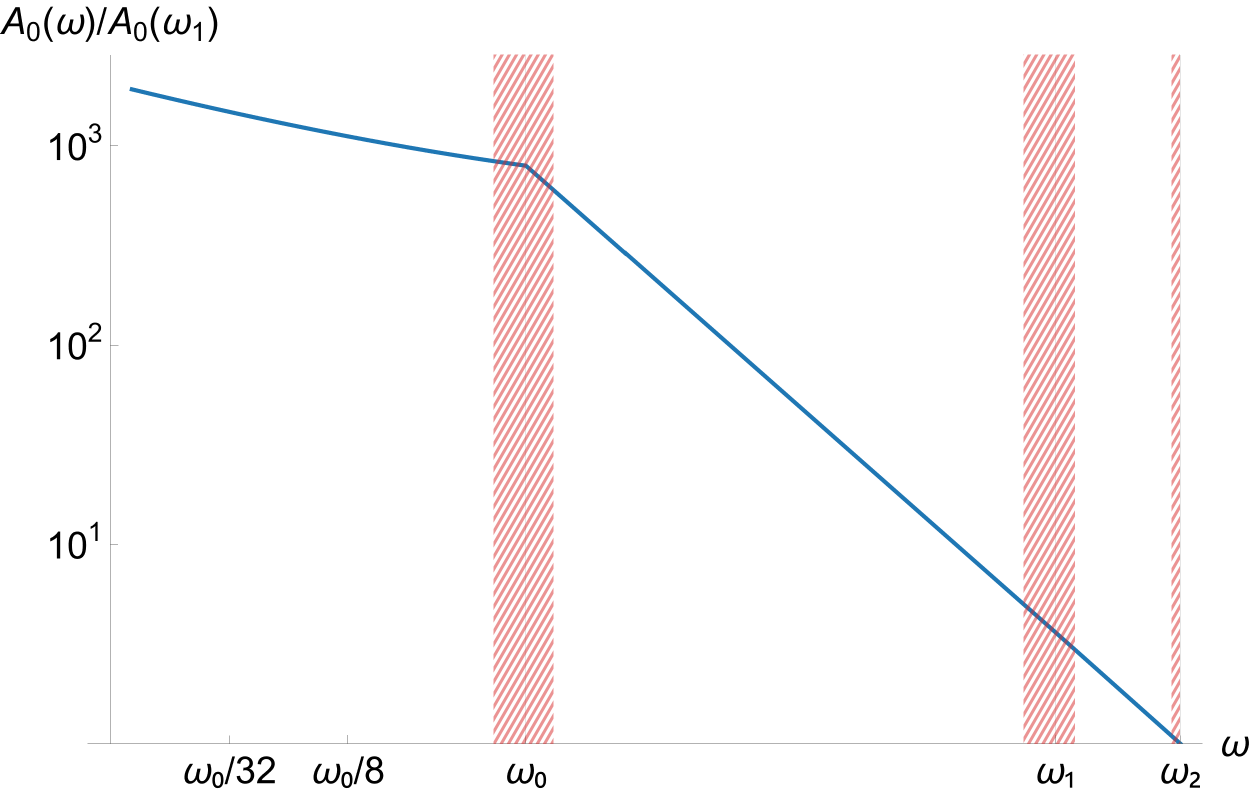}
    \caption{\label{fig:new_38}}
\end{subfigure}
\begin{subfigure}{0.4\textwidth}
    \includegraphics[width=\textwidth]{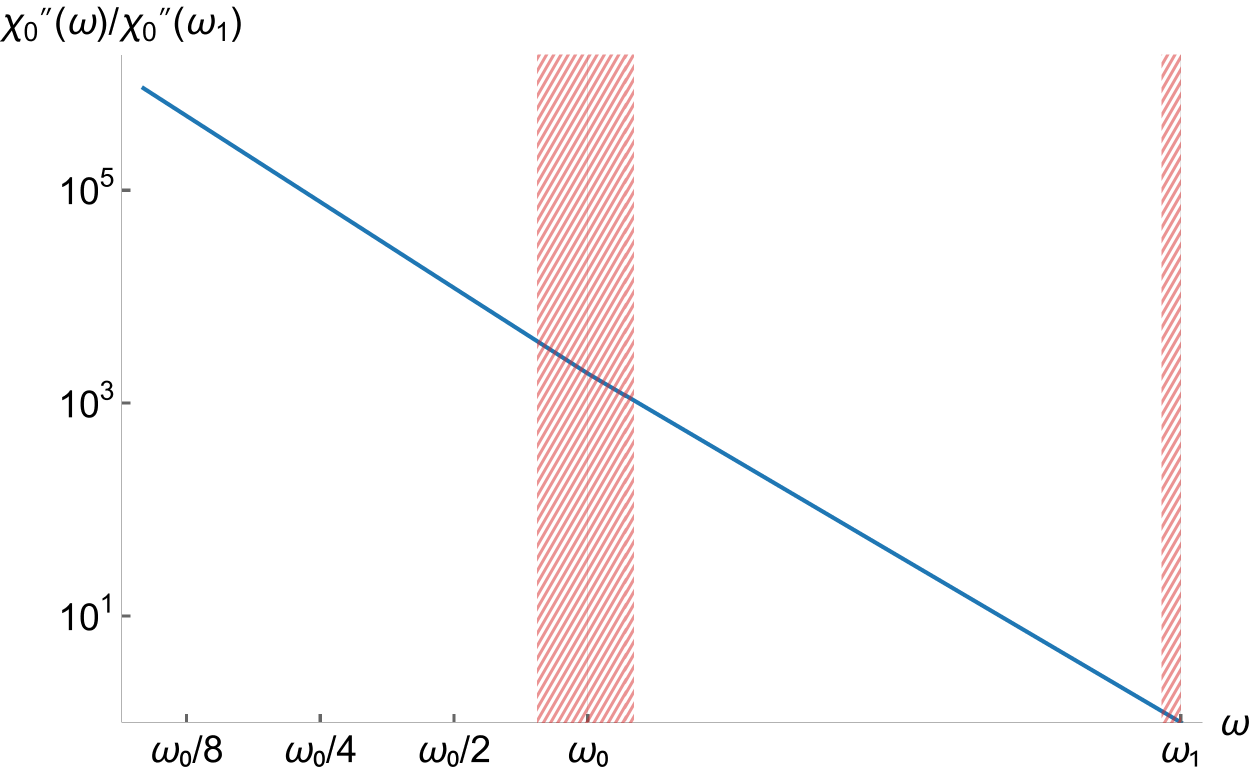}
    \caption{\label{fig:new_39}}
\end{subfigure}
\begin{subfigure}{0.4\textwidth}
    \centering
    \includegraphics[width=\textwidth]{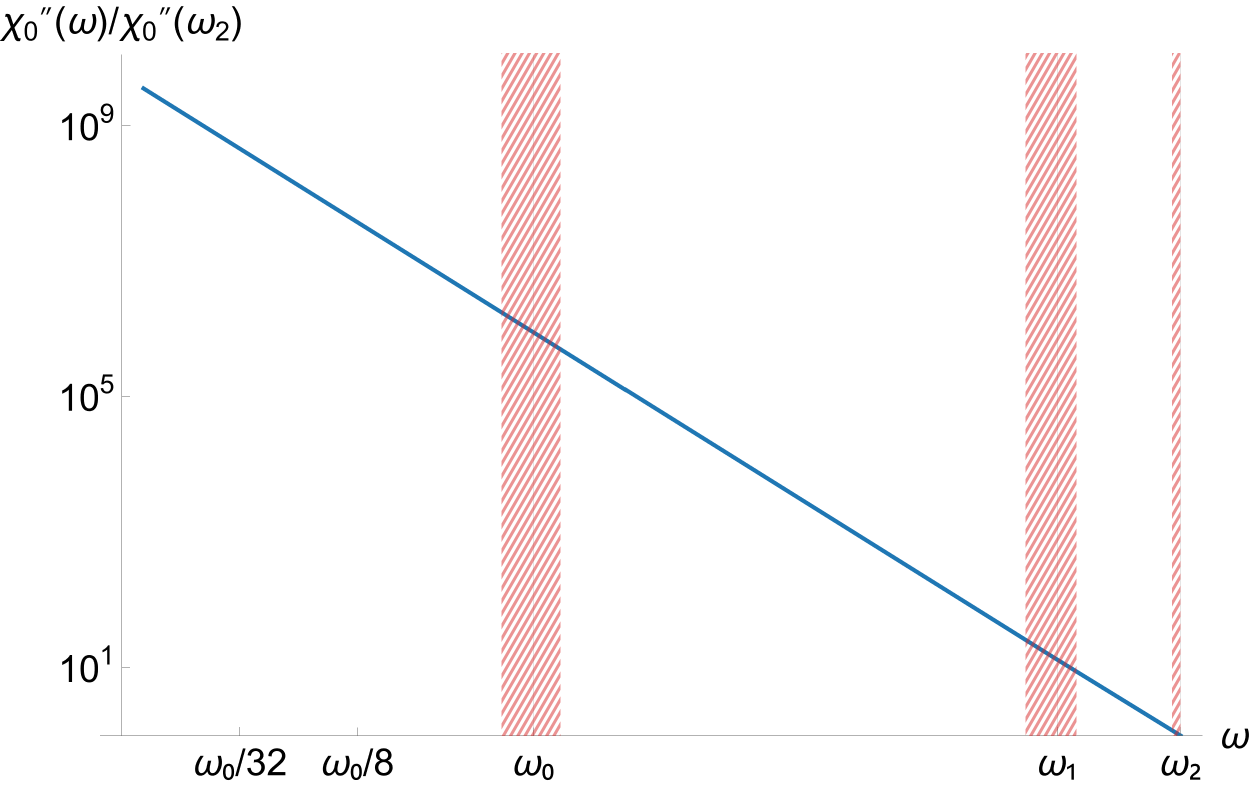}
    \caption{\label{fig:new_40}}
\end{subfigure}
\caption{
\label{fig:spec_scaling}
Schematic profiles of the electron spectral function at the hot spot momentum and 
the dynamical spin susceptibility at the magnetic ordering wavevector 
plotted in the log-log scale 
as functions of frequency for $\epsilon=0.2$ and $N_f=1$. 
(a,c) are for the $Z_2$ theory
and (b,d) are for the $O(2)$ theory.
For the $Z_2$ theory,
we choose  $s_1=0.1$,
and
for the $O(2)$ theory,
we choose $c_2=0.055$, $s_2=1.1$.
The shaded regions indicate the crossover regimes.
Outside the shaded region,
we use Eqs. 
    \eqref{eq:AchiZ2}
    and
    \eqref{eq:AchiO2},
    which are valid far away from the crossover scales.
}
\end{figure}

In the $O(2)$ theory, 
the manifold of quasi-fixed points is two-dimensional.
Suppose the theory is at $(c_2, s_2)$
within the manifold with two-dimensional quasi-fixed points
at energy scale 
$\omega^{O(2)}_2 
 =  \Lambda
\exp[ -\int_{0}^{\ell_2^{O(2)}} 
z(\ell) d\ell]$.
For $c_2 \ll 1$,
the renormalization group speeds of $s$ and $c$ are significantly different.
This hierarchy is captured by two distinct crossover scales.
The first is 
$\omega^{O(2)}_1 
=\omega^{O(2)}_2 
\exp[ -\int_{
\ell_2^{O(2)} 
}^{\ell_1^{O(2)}} 
z(\ell) d\ell]
\sim
\omega^{O(2)}_2 e^{ -\left(  \frac{256\sqrt2 \pi \tilde h_5(N_f^{-1})}{N_f c_2\epsilon(5\sqrt2+\sqrt\epsilon)} \right)^{1/3} }$
and the second is
$\omega^{O(2)}_0  = 
\omega^{O(2)}_1 
\exp[ -\int_{
\ell_1^{O(2)} 
}^{\ell_0^{O(2)}} 
z(\ell) d\ell]
\sim
\omega^{O(2)}_1 e^{
    -\left( \frac{27\pi^4 \tilde h_5(N_f^{-1})}{640 N_f c_2^4\epsilon} \right)^{1/3}
}$.
It is noted that 
$\omega^{O(2)}_0 \ll \omega^{O(2)}_1 \ll
\omega^{O(2)}_2$
for $c_2 \ll 1$.
In $\omega^{O(2)}_1 \ll \omega < \omega^{O(2)}_2$,
the flows of both $s$ and $c$ are negligible,
and the physical observables exhibit simple power-law  behaviours with exponents that depend both on $s_2$ and $c_2$.
In $ \omega^{O(2)}_0 \ll \omega \ll \omega^{O(2)}_1$,
$\delta s = s - s^*$ becomes vanishingly small but the flow of $c$ can be still ignored.
In this range of frequencies, the 
 physical observables exhibit power-law behaviours with exponents that depend only on $c_2$.
In the low-energy limit with $\omega \ll \omega^{O(2)}_0$,
the physical observables take the universal power-law forms with corrections associated with the logarithmic flow of $c$.
The crossover behaviour 
of $\mathcal A_0(\omega)$ and $\chi''_0(\omega)$ in the $O(2)$ theory is
summarized as
\begin{align}
    &\mathcal A_0(\omega)
    \sim
    \begin{cases}
	\omega^{-1+b c_2^{1/3}\left[1 - b'\delta s_2\right]}
        & \omega^{O(2)}_1 \ll \omega < \omega^{O(2)}_2 \\
    \omega^{-1+b c_2^{1/3}}
        & \omega^{O(2)}_0 \ll \omega \ll \omega^{O(2)}_1 \\
    \frac{\omega^{-1}}
    {\exp\left(\alpha \log^{3/4}(1/\omega)\right)
    \log^{1/4}(1/\omega)}
        & \omega \ll \omega^{O(2)}_0
    \end{cases}
    \nn
    &\chi''_0(\omega)
    \sim
    \begin{cases}
\omega^{-2+\epsilon+c_2^{1/3}\delta s_2 bN_f} 
		& \omega^{O(2)}_1 \ll \omega < \omega^{O(2)}_2 \\
\omega^{-2+\epsilon} 
        & 
        \omega \ll \omega^{O(2)}_1
    \end{cases},
    \label{eq:AchiO2}
\end{align}
where 
$\delta s_2 = s_2 - s^*$,
$b = \left( \frac{(5\sqrt2+\sqrt\epsilon) \epsilon  N_f}{256\sqrt2 \pi  \tilde h_5(N_f^{-1})}  \right)^{1/3}$, 
$b' = \frac{\tilde h_5'(N_f^{-1})}{3 \tilde h_5(N_f^{-1})}+\frac{2 N_f}{3}$
and
$\alpha = \left( \frac{N_f\epsilon(5\sqrt2+\sqrt\epsilon)}{864\sqrt2 \tilde h_5(N_f^{-1})} \right)^{1/4}$.
For $\chi''_0(\omega)$, the scaling behaviour remains the same in all $\omega \ll \omega_1^{O(2)}$.

For comparison, we also show the predictions for the $O(3)$ theory.
While the non-perturbative solution is available  everywhere in $2 \leq d \leq 3$\cite{SCHLIEF,SCHLIEF2,BORGES2023169221} for this theory, 
we use the expressions obtained within the $\epsilon$ expansion\cite{LUNTS} 
so that the results can be compared with those of the $Z_2$ and $O(2)$ theories on the equal footing.
Similar to the $Z_2$ theory,
there is only one crossover associated with one quasi-marginal coupling,  $w=v/c$.
Suppose the theory is within the one-dimensional manifold of quasi-fixed points labeled with $w_1 \ll 1$ at energy scale $\omega_1^{O(3)}$.
The spectral function and the dynamical susceptibility become\footnote{
It is noted that non-perturbative expressions valid to all orders in $\epsilon$ are available for the $O(3)$ theory\cite{SCHLIEF,BORGES2023169221}, but here we show the expressions valid only to leading order in 
 $\epsilon$ to compare with the results of the $O(2)$ and $Z_2$ theories for which non-perturbative solutions are unavailable.}
\begin{equation}
\begin{aligned}
\mathcal{A}_0(\omega)
\sim & 
\left\{
\begin{array}{lll}
\omega^{
-1+3w_1 \epsilon/(4N_f)
} 
& 
 \mbox{for} & ~~ 
 \omega^{O(3)}_0
 \ll \omega < \omega^{O(3)}_1, \\
\frac{1}{
\omega ~
\exp\left(\frac{ 
 3^\frac{4}{3}}{2^\frac{14}{3}(h^{*}_5)^\frac{1}{3}}\left(\log\frac{1}{\omega}\right)^\frac{1}{3}\right) 
 ~\left[ \log\frac{1}{\omega} \right]^{\frac{2}{3}}
 }
  & 
 \mbox{for} &~~ \omega \ll 
 \omega^{O(3)}_0
 \end{array}
\right., \\
\chi^{''}_0(\omega) \sim & 
\left\{ 
\begin{array}{lll}
\omega^{
-2+\epsilon+w_1\epsilon/N_f
}
 & 
 \mbox{for} & ~~
 \omega^{O(3)}_0 \ll \omega < \omega^{O(3)}_1, \\
\frac{1}{
 \omega^{2-\epsilon}~
\exp\left(\frac{ 
 3^\frac{1}{3}}{2^\frac{8}{3}(h^{*}_5)^\frac{1}{3}}\left(\log\frac{1}{\omega}\right)^\frac{1}{3}\right) ~
}
& \mbox{for} & ~~ \omega \ll 
 \omega^{O(3)}_0
\end{array}
\right..
\end{aligned}
    \label{eq:AchiO3}
\end{equation}
Here, 
$\omega^{O(3)}_0 = \omega_1^{O(3)} e^{- \frac{ (2 N_f)^{3/2}}{96 \sqrt{2 h_5^*} \epsilon^{3/2} } w_1^{-\frac{3}{2}}}$ is the crossover energy scale below which the flow of $w$ becomes significant
and
$h_5^* = \lim_{s\to\infty} s\tilde h_5(s) = (512\pi)^{-1}$. 

To illustrate 
the crossovers of physical observables,
we plot the electron spectral function at the hot spots and 
the dynamical spin susceptibility
at the magnetic ordering vector as functions of frequency
in \fig{fig:spec_scaling}.
At high frequencies,
the scaling behaviours are controlled by exponents that depend on the quasi-marginal couplings.
In the low-energy limit, the scaling behaviour is governed by the universal exponents associated with the true fixed point.
The crossovers occur at scales around which the renormalization group flow of $s$ or $c$ changes.

\subsection{Enhanced fluctuations}

In this section, we examine fluctuations enhanced at the quantum critical point, closely following the discussion of Ref.~\cite{Sur2015Quasilocal}. 
We emphasize that this analysis, which only includes electrons near the hot spots,
does not necessarily determine the actual instabilities 
 of the full system
because electrons away from the hot spots also play important roles for instabilities of the entire Fermi surface.
The goal of the current analysis is to understand relative enhancement  between superconducting (SC) and charge density wave (CDW) fluctuations
of hot electrons.
To quantify the enhancement of fluctuations in these channels,
we insert fermion bi-linear operators of the following form,
\begin{equation}
\mathcal{S}_{\rho}[V_{\rho},\Theta]=V_{\rho} \mu \int\dd{k}
{\bf O}_k^{(\rho)}
+ h.c. .
\end{equation}
Here, $V_\rho$ 
with $\rho=\SC$ and $\CDW$ 
denotes the infinitesimal symmetry breaking field coupled to superconducting and charge density wave order parameters
${\bf O}_k^{(\rho)}
=\tilde{\Psi}^{(\rho)}_{n,\sigma,j}(k) \Theta_{n,\sigma,j;m,\tilde{\sigma},l}\Psi_{m,\tilde{\sigma},l}(k)$,
where
$ \tilde{\Psi}^{(\SC)}_{n,\sigma,j}(k) ={\Psi}^{\top}_{n,\sigma',j}(-k) 
(i\tau_{\sigma'\sigma}^{y})$
and 
$ \tilde{\Psi}^{(\CDW)}_{n,\sigma,j}(k) =\bar{\Psi}_{n,\sigma,j}(k) $.
$\Theta_{n,\sigma,j; n',\sigma',j'}$ represents the wavefunction of two particles or a particle-hole pair in the space of hot spots ($n,m$),
spin ($\sigma$, $\sigma'$)
and flavour ($j$, $l$).
To be concrete, we consider the following wavefunction,
\begin{equation}
    \Theta_{n,\sigma,j; n',\sigma',j'} = b_n \d_{n,n'} [\vec{a}\cdot \vec{\tau}]_{\s,\s'} \delta_{j,j'}\hat{\Omega},
\end{equation}
where $\vec{b}= (b_1,b_2,b_3,b_4)$ 
and
$\vec{a}=(a_0,a_x,a_y,a_z)$ 
denote the wavefunctions in the space of hot spots and spin, respectively
with
$\vec{\tau} = (1,\tau^x,\tau^y,\tau^z)$
being the $2 \times 2$ identity matrix and the Pauli matrices acting on the spin indices.
$\hat \Omega$, which is a $2 \times 2$ matrix acting on the spinor indices, determines the orbital wavefunction together with $\vec b$.
Its RG flow equation takes the form of 
$ \dv{V_\rho}{l}=V_\rho(1+\gamma_{\rho})$
to the linear order in $V_\rho$,
where
$\gamma_\rho$ represents the anomalous dimension.
The anomalous dimension can be decomposed as 
$\gamma_{\rho} = \gamma_{0} + z \lambda_\rho$, 
where $\gamma_0$ is a channel-independent one 
and $\lambda_\rho$ is the channel-dependent one.
The bigger the anomalous dimension of the source is, 
the stronger the 
 enhancement of the fluctuations is in that channel. 
The most strongly enhanced channel depends on spin-symmetry group $G$.
Technical details for the computation of $\lambda_\rho$ can be found in App.~\ref{app:Instabilities}. 
The results for enhanced SC and CDW fluctuations are summarized in Table~\ref{table34}.
In the table, we characterized a channel by the spin ($\vec a \cdot \vec \tau)$,
the momentum and the orbital wavefunction of fermion bi-linears.

In the superconducting channel,
the most enhanced vertex corresponds to a spin-triplet in the vector representation of $D_4$ (i.e. p-wave) for $G=Z_2$ and $O(2)$.
For $G=O(3)$, the spin-singlet pairing with $s$ or $d$-wave symmetry is dominant.
While the pairing vertex with zero center of mass momentum and momentum $2k_F$ are degenerate for electrons at the hot spots in the $O(3)$ theory, 
the pairing channel with zero momentum prevails once the full Fermi surface is taken into account\cite{BORGES2023169221}.
In the charge density wave channel, those with momentum $2k_F$ are most strongly enhanced due to the strongest nesting.
The most enhanced vertex transforms as a vector under $D_4$ but can be either spin triplet or singlet in the $Z_2$ and $O(2)$ theories.

We emphasize that these tables only indicate the enhancement of fluctuations associated with electrons close to the hot spots.
To understand the actual instability of the system,
one has to include all gapless electronic degrees of freedom on the Fermi surface.
For the $O(3)$ theory, the functional renormalization group analysis of the full Fermi surface reveals that the leading instability is the d-wave superconductivity\cite{BORGES2023169221}.
The instability is driven mainly by `lukewarm' 
 electrons that are 
 close enough to the hot spots that they are subject to the strong attractive interaction medicated by spin fluctuations but far enough to be coherent at low energies.
Hot electrons right at the hot spots do not drive the superconducting instability because of the strong pair breaking effect caused by incoherence.
It would be of great interest to use 
 the functional renormalization group analysis to understand the instabilities of the Fermi surface in the 
$Z_2$ and $O(2)$ theories.

It is noted that the $Z_2$ and $O(2)$ theories in two space dimensions have been studied through the quantum  Monte Carlo simulation \cite{SCHATTNER2,2023arXiv230506421T}.
In those numerical studies, the leading order superconducting instability was found to be in the $d$-wave channel. 
In our work, superconducting fluctuations are enhanced most in the $p$-wave channel, although the $d$-wave channel is also enhanced by spin fluctuations 
(c.f. Appendix~\ref{app:Instabilities}) .  
In part, this discrepancy can be attributed to the fact that `lukewarm' electrons away from hot spots, which are not included in the present study, play an important role in driving superconducting instabilities\cite{BORGES2023169221}.

\begin{table}[t]
\begin{subtable}{0.45\textwidth}
\begin{tabular}{|c|c|c|c|c|}
\hline
$G$   & $\lambda_\rho$     & $\vec a \cdot \vec \tau$ &   Momentum & Orbital\\ \hline
\multirow{2}{*}{$Z_2$} & \multirow{2}{*}{$\frac{2g^2}{16\pi cN_f}$}   & $\tau^{x/y}$ & $0$ & $p_x/p_y$           \\ 
\cline{3-5} 
     &   & $\tau^{z}$      & $0$ & $p_{x'}/p_{y'}$\\ \hline
$O(2)$   & $\frac{4g^2}{16\pi cN_f}$       & $\tau^z$     & $0$ & $p_x/p_y$           \\ \hline
\multirow{2}{*}{$O(3)$} & \multirow{2}{*}{$\frac{3g^2}{16\pi cN_f}$} &  \multirow{2}{*}{1}                           & 0     & $s/d_{xy}$                 \\ \cline{4-5} 
 &    & & $2k_F$ & $s/d_{xy}$                \\ \hline
\end{tabular}
\label{table3}
\caption{Pairing channel}
\end{subtable}
\begin{subtable}{0.45\textwidth}
\begin{tabular}{|c|c|c|c|c|}
\hline
$G$                     & $\lambda_\rho$                           & $\vec a \cdot \vec \tau$  &   Momentum & Orbital\\ \hline
\multirow{2}{*}{$Z_2$} & \multirow{2}{*}{$\frac{2g^2}{16\pi cN_f}$}   & $\tau^{x},\tau^y$ & $2k_F$ & $p_{x'},p_{y'}$           \\ \cline{3-5} 
  &   & $1,\tau^{z}$      & $2k_F$ & $p_{x},p_{y}$ \\ \hline
\multirow{2}{*}{$O(2)$} & \multirow{2}{*}{$\frac{4g^2}{16\pi cN_f}$} &1  & $2k_F$     & $p_x, p_{y}$  \\ \cline{3-5} 
&     &     $\tau^z$ & $2k_F$ & $p_{x'},p_{y'}$                \\ \hline
$O(3)$   & $\frac{6g^2}{16\pi cN_f}$  & $1$     & $2k_F$ & $p_x,p_y$           \\ \hline
\end{tabular}
\label{table4}
\caption{Charge-density wave}
\end{subtable}
\caption{ 
The anomalous dimension ($\lambda_\rho$) of the most enhanced fermion bi-linear composed of hot electrons in the antiferromagnetic quantum critical metal with spin symmetry $G=Z_2, O(2)$ and $O(3)$.
$\vec a \cdot \vec \tau$ is the representation of the fermion bi-linear under $G$.
The center of mass momentum is either $0$ or $2k_F$ in the particle-particle channel, and $2k_F$ in the particle-hole channel.
$p_{x'}$ and $p_{y'}$ correspond to the 
 $p$-wave orbital aligned along $k_y$ and $k_x$ directions of 
\fig{fig:Hotspots}, respectively.
$p_x$ and $p_y$ represent 
the orbitals aligned along
$(k_x+ k_y )/\sqrt{2}$
and
$(-k_x+ k_y )/\sqrt{2}$, respectively.
}
\label{table34}
\end{table}%

\section{Conclusion}\label{sec:sectiontion6}

We have studied the 
 hot spot theory for the antiferromagnetic quantum critical metal with $Z_2$ and $O(2)$ spin symmetries 
 in the $\epsilon$-expansion that uses the co-dimension of Fermi surface as a control parameter.
The naive loop expansion breaks down due to dynamical kinetic energy quenching where some components of velocities are renormalized to zero at low energies.
As a result, higher-loop quantum corrections become qualitatively important 
even to the leading order in the $\epsilon$-expansion.
Interestingly, the difference in the spin symmetry creates drastically different hierarchies among the quenched velocities :
for the $Z_2$, $O(2)$ and $O(3)$ symmetric theories,
the ratio between the speed of the collective mode and the Fermi velocity perpendicular to the magnetic ordering vector flows to 
$0$, $1$ and $\infty$, respectively, under the renormalization group flow.
Furthermore, theories with different spin symmetries exhibit distinct crossover behaviours with different numbers of quasi-marginal parameters that act as marginal parameters within finite windows of energy scales.

The reason why the symmetry group alters the nature of the infrared fixed points so drastically is two-folded.
Firstly, the dynamical kinematic energy quenching makes the theory highly susceptible to the vertex correction,  which would have been unimportant at small $\epsilon$ otherwise.
Secondly, the nature of the vertex correction depends on the symmetry group.
In the $Z_2$ ($O(3)$) theory, the vertex correction is screening (anti-screening), 
which makes $c/v$ flow to zero (infinity).
In the $SU(N)$ theories with $N \geq 2$, the vertex correction is anti-screening 
and $c/v$ flows to infinity as is the case for the $O(3)$ theory\cite{SCHLIEF,SCHLIEF2}.
The $O(2)$ theory is the unique case in that the vertex correction is absent
and $c/v$ is stabilized at an $O(1)$ value.

We conclude with some open questions.
First, it would be of great interest to see if the ratio between quenched velocities can be used as a control parameter 
to gain a non-perturbative access to the strongly coupled theory in $2+1$ dimensions for the $Z_2$ and $O(2)$ theories as it was done for the $O(3)$ theory\cite{SCHLIEF}.
Second, the present hot spot theory only captures the electrons near the hot spots and the critical spin fluctuations.
In order to understand the effects of large-angle scatterings\cite{2024arXiv240509450K}
or disorder\cite{
PhysRevB.103.235157,
JANG2023169164},
one has to go beyond the hot spot theory and consider the full Fermi surface within the low-energy theory\cite{
BORGES2023169221}.
Finally, it is desirable to find a systematic way of studying quantum field theories with dynamically quenched kinetic energy.
Considering higher-order corrections ad hoc until the control is achieved is not ideal because one does not know a priori the extent to which the kinetic energy is quenched 
and the magnitudes of higher-order diagrams that are determined by the ratio between the interaction and the quenched velocities.
For such theories, it may be interesting to use the quantum renormalization group\cite{lee:2014uf} that promotes couplings to dynamical variables, allowing them to 
find their renormalization flow based on the action principle.

\onecolumngrid
\appendix

\section{Quantum Corrections}\label{app:quantumCorrections}

In this Appendix, 
we provide the details  for the computation of  the counter terms.

\subsection{One-loop boson self-energy}
Fig. \ref{fig:BosonSE1Loop}
gives rise to the one-loop quantum correction to the boson kinetic energy,
\begin{equation}\label{eq:BosonSE1}
\delta\mathcal{S}_{b} = \int \dd q \Upsilon_{(0,2)}(q) \sum_{\alpha\in S_G} \phi^\alpha(q)\phi^\alpha(-q),
\end{equation}
where 
\begin{align}\label{eq:BosonSE2}
\Upsilon_{(0,2)}(q) &=     g^2\mu^{3-d}\sum_{n}\int \dd k \Tr[i\gamma_{d-1}G_{\bar{n}}(q+k)i\gamma_{d-1}G_{n}(k)].
\end{align}
By tracing over the $\gamma$-matrices we obtain
\begin{subequations}
\begin{align}\label{eq:BosonSE3}
\Upsilon_{(0,2)}(q)
&= 2g^2\mu^\epsilon\sum_{n}\int \dd k \frac{\epsilon_{\bar{n}}(\vec{q}+\vec{k})\epsilon_{n}(\vec{k})-\vb{(Q+K)}\cdot\vb{K}} {(\epsilon_{\bar{n}}(\vec{q}+\vec{k})^2+\vb{|Q+K|}^2)(\epsilon_{{n}}(\vec{k})^2+\vb{|K|}^2)}.
\end{align}
After integrating over the parameters $\epsilon_n(\vec k)$ and $\epsilon_{\bar n}(\vec k)$, we may introduce Feynman parameters to express the integral as
\begin{align}\Upsilon_{(0,2)}(q)
&=-\frac{g^2\mu^\epsilon}{4v}\frac{\Gamma{(1)}}{\Gamma{\left(\frac{1}{2}\right)}^2}  \int_0^1 \frac{\dd x}{\sqrt{x(1-x)}} \int \frac{\dd \vb{K}^{d-1}}{(2\pi)^{d-1}} \frac{\vb{(Q+K)}\cdot\vb{K}}{x\vb{|Q+K|}^2+(1-x)\vb{|K|}^2}.
\end{align}
A shift $\vb{K} \to \vb{K}+x\vb{Q}$ leads to
\begin{align}\Upsilon_{(0,2)}(q)
&=-\frac{g^2\mu^\epsilon}{4v}\frac{\Gamma{(1)}}{\Gamma{\left(\frac{1}{2}\right)}^2}  \int_0^1 \frac{\dd x}{\sqrt{x(1-x)}} \int \frac{\dd \vb{K}^{d-1}}{(2\pi)^{d-1}} \frac{\vb{|K|}^2-x(1-x)|\vb{Q}|^2}{\vb{|K|}^2+x(1-x)\vb{|Q|}^2}.
\end{align}
The remaining integrations gives rise to the $\epsilon$ pole,
\begin{align}\Upsilon_{(0,2)}(q)
&=\frac{g^2\vb{Q}^{2}}{8\pi v\epsilon}+\mathcal{O}(\epsilon^0).
\end{align}
\end{subequations}
The counter-term action is obtained to be
\begin{equation}
\mathcal{S}_{b;\mathcal{CT}}=-\frac{g^2}{4\pi v\epsilon }\left[ \frac{1}{2}\int \dd{q} |\vb{Q}|^2\phi(q)\phi(-q) \right].
\end{equation}
From this, we obtain 
\begin{align}
Z_{4,1} &= - \frac{g^2}{4\pi v}.
\label{eqA5}
\end{align}


\subsection{One-loop fermion self-energy }
\label{app:FermionSE}
The quantum correction from the diagram in Fig. \ref{fig:FermionSE1Loop} reads
\begin{equation}\label{eq:FermionSE1}
\delta\mathcal{S}_{f}=-\frac{g^2 \mu^{3-d}N}{N_f}\sum_{n,\sigma,j}\int \dd{k} \bar{\Psi}_{n,\sigma,j}(k) \Upsilon^{1L}_{f,(n)}(k)\Psi_{n,\sigma,j}(k),
\end{equation}
where
\begin{align}
	\Upsilon^{1L}_{f,(n)}(k)&=\int \dd{q} i\gamma_{d-1}G_{\bar{n}}(q)i\gamma_{d-1}D(q-k).\notag\\
	&=\int \dd{q}\frac{i\gamma_{d-1} \varepsilon_{\bar{n}}(\vec{q})-i{\vb{Q\cdot\Gamma}}}
	{(|\vb{Q}|^2+\varepsilon_{\bar{n}}(\vec{q})^2)(\vb{|Q-K|}^2+c^2(\vec{q}-\vec{k})^2)} 
	\end{align}
We perform the change of variables
$(q_x,q_y)\rightarrow(q_{\parallel},q_{\perp})$, 
where
$q_{\parallel}=\varepsilon_{\bar{n}}(\vec{q})$ and
$q_{\perp}=\varepsilon_{\tilde{n}}(\vec{q})$, where $\tilde 1 = 2, \tilde 2 = 1, \tilde 3 = 4, \tilde 4 = 3$.
Notice that with this definition, the coordinates are perpendicular so that
$\vec{q}\,^2={v_F^{-2}}\left(q_{\parallel}^2+q_{\perp}^2\right)$, where
$v_F^2=1+v^2$.
Changing variables causes the integration measure to transform as
$\dd{q}\rightarrow v_F^{-2}\dd{q'}$, where
$\dd{q'}=\frac{\dd{q_\parallel}\dd{q_\perp}\dd{^{d-1}\vb{Q}}}{(2\pi)^{d+1}} $.
It is also useful to define $\hat{c}=\frac{c}{v_F}$. The resulting integral is
\begin{align}
	\Upsilon^{1L}_{f,(n)}&=\int \frac{\dd{q'}}{v_{F}^2}\frac{i\gamma_{d-1} q_{\parallel}-i{\vb{Q\cdot\Gamma  }}}{(|\vb{Q}|^2+q_{\parallel}^2)(\vb{|Q-K|}^2+\hat{c}^2(q_{\parallel}-k_{\parallel})^2+\hat{c}^2(q_{\perp}-k_{\perp})^2)} \notag\\
	&=\frac{1}{2\hat{c}v_{F}^2} \int \frac{\dd{^{d-1}\vb{Q}}\dd{q_\parallel} }{(2\pi)^{d}}
	\frac{i\gamma_{d-1} q_{\parallel}-i{\vb{Q\cdot\Gamma  }}}{(|\vb{Q}|^2+q_{\parallel}^2)({\vb{|Q-K|}^2+\hat{c}^2(q_{\parallel}-k_{\parallel})^2)}^{1/2}} \notag\\	
	&=\frac{1}{4\hat{c}v_{F}^2}
	\int_0^1 \frac{dx}{\sqrt{x}}
	\int \!\! \frac{\dd{^{d-1}\vb{Q}}\dd{q_\parallel} }{(2\pi)^{d}}
\frac{i\gamma_{d-1} q_{\parallel}-i{\vb{Q\cdot\Gamma  }}}{((1-x)\vb{Q}^2+x(\vb{Q-K})^2+ (1-x) q_{\parallel}^2+x\hat{c}^2(q_{\parallel}-k_{\parallel})^2))^{3/2}}.
\end{align}	
Shifting the integration variables as 
$\vb{Q}\rightarrow
\vb{Q}+x\vb{K}$ and $q_{\parallel}\rightarrow q_{\parallel}+\frac{x
\hat{c}^2}{1-x+x\hat{c}^2}k_{\parallel}$, 
we obtain
\begin{align}
\Upsilon^{1L}_{f,(n)}	&=\frac{1}{4\hat{c}v_{F}^2}
\int_0^1\frac{\dd{x}}{\sqrt{x}}
\int    \frac{\dd{^{d-1}\vb{Q}}\dd{q_\parallel} }{(2\pi)^{d}} 
\frac{i\gamma_{d-1}\frac{x\hat{c}^2}{1-x+x\hat{c}^2}k_{\parallel}-i{x\vb\Gamma\cdot\vb{K}}}
	{\left(
		\vb{Q}^2+(1-x) x_2 \vb{K}^2+(1-x+x\hat{c}^2 ) q_\parallel^2 + \frac{(1-x) x \hat{c}^2}{1-x+ x \hat{c}^2}k_\parallel^2
		\right)^{3/2}} \label{eqA90} \\
	&=
\frac{1}{4\hat{c}v_{F}^2}
\frac{1}{(4\pi)^{d/2}}\frac{\Gamma((3-d)/2)}{\Gamma(3/2)}
\int_0^1\dd{x}\sqrt{\frac{x}{1-x+x\hat{c}^2 }}
	\frac{
		\left(
		i\gamma_{d-1}\frac{\hat{c}^2}{1-x+x\hat{c}^2}k_{\parallel}-{i\vb{\Gamma}\cdot\vb{K}}
		\right)
	}
	{\left((1-x)x \left(\vb{K}^2+ \frac{ \hat{c}^2}{1-x+x\hat{c}^2}k_\parallel^2\right)\right)^{(3-d)/2}}
 \label{eqA91}\\
	&=
	\frac
	{\Gamma{\left(\frac{3-d}{2}\right)}}
	{\hat{c}v_{F}^2(4\pi)^{\frac{d+1}{2}}}
	\int_0^1\dd{x}\sqrt{\frac{x}{1-x+x\hat{c}^2 }}
	\frac{
		\left(
			i\gamma_{d-1}\frac{\hat{c}^2}{1-x+x\hat{c}^2}k_{\parallel}-\vb\Gamma\cdot\vb{K}
		\right)
	}
	{\left(
		(1-x) x \left(\vb{|K|}^2+ \frac{ \hat{c}^2}{1-x+x\hat{c}^2}k_\parallel^2
				\right)
	\right)^{(3-d)/2}} \label{eqA92} \\
	&= 
	\frac{1}{8\pi^2 c \epsilon}
	\left(
	h_2(v,c)i\gamma_{d-1}\varepsilon_{\bar{n}}(\vec{k})
	-h_1(v,c) i\Gamma\cdot\vb{K}
	+\mathcal{O}(\epsilon)
	\right).
 \label{eq:A10}
\end{align}	
Here, $h_1(v,c)$ and $h_2(v,c)$ are defined as
\begin{equation}\begin{split}\label{eq:h12}
h_1(v,c)&=\frac{1}{v_F}        \int_0^1{\dd{x}}\sqrt{\frac{x}{1-x+x\hat{c}^2 }}=\frac{1}{v_F} \frac{d_0(\hat{c})-\hat{c}}{1-\hat{c}^2},  \\
h_2(v,c)&=\frac{\hat{c}^2}{v_F}\int_0^1{\dd{x}}\sqrt{\frac{x}{(1-x+x\hat{c}^2)^{3} }}=\frac{2 \hat{c}}{v_F} \frac{1-\hat{c}d_0(\hat{c})}{1-\hat{c}^2}
\end{split}
\end{equation}
with 
$ d_0(z)=\frac{\arcsin(\sqrt{1-z^2})}{\sqrt{1-z^2}}$. 
 It is useful to express the small $c$ limit of the functions $h_1(v,c)$ and $h_2(v,c)$ for fixed $s = c/v$ as

\begin{align}
    \lim_{c\to0}h_1(c/s,c) &= \frac\pi2 - 2 c + \mathcal{O}(c^2),\\
    \lim_{c\to0}h_2(c/s,c) &= 2 c + \mathcal{O}(c^2).
\end{align}

\noindent The counter-term is then given by
\begin{equation*}\label{eq:FermionSE3}
\mathcal{S}_{f,\mathcal{CT}}=\frac{-g^2N}{8\pi^{2} c N_f \epsilon}\sum_{n,\sigma,j}\int \dd{k} \bar{\Psi}_{n,\sigma,j}(k) 
\left[
ih_1(v,c)   \vb\Gamma\cdot\vb K
-ih_2(v,c)  \gamma_{d-1}\epsilon_{\bar{n}}(\vec{k}) 
\right]
\Psi_{n,\sigma,j}(k).
\end{equation*}
The counter-term coefficients are given by
\begin{align}
    Z_{1,1} &= \frac{-g^2N h_1(v,c)}{8\pi^{2} c N_f}, &
    Z_{2,1} &= \frac{+g^2N h_2(v,c)}{8\pi^{2} c N_f}, &
    Z_{3,1} &= \frac{-g^2N h_2(v,c)}{8\pi^{2} c N_f}.
\end{align}


\subsection{One-loop Yukawa vertex correction}

The contribution from the diagram in Fig. \ref{fig:Yukawa1Loop} reads
\begin{align}
\delta \mathcal{S}_{g}=\frac{g^3\mu^{3\epsilon/2}}{N_f^{3/2}}
\sum_{\substack{n,j,\alpha\\\sigma,\sigma'}}
\int \dd{q}\,\dd{k}  
\left[  
	\phi(k_1-k_2) \bar{\Psi}_{\bar{n},\sigma,j}(k_1)
	\left(\sum_{\beta\in S_G}\tau^\beta\tau^\alpha\tau^\beta\right)_{\sigma,\sigma'} \Upsilon_{g,(n)}^{1L}(k_1,k_2)
	(i\gamma_{d-1})\Psi_{n,\sigma',j}(k_2) 
\right],
\label{eqA12}
\end{align}
where
\begin{align}
\Upsilon_{g,(n)}^{1L}(k_1,k_2)=\int \dd{q} i\gamma_{d-1}G_{\bar{n}}(k_1+q) i\gamma_{d-1}G_{n}(k_2+q)D(q).
\end{align}
As we are interested in the divergent part in the $\epsilon\rightarrow0$
limit, let us choose $\vb{K}_1=\vb{K}_2=\vb{K}$ and simplify the calculation by setting $\vec{k}_1 = \vec{k}_2 = 0$,
\begin{align}
\Upsilon_{g,(n)}^{1L}\left(\vb{K}\right)=
\int \dd{q} 
\frac{\varepsilon_{{n}}(\vec{q})\varepsilon_{\bar{n}}(\vec{q})-\vb{Q}^2}
{
	(\vb{Q}^2+\varepsilon_{\bar{n}}(\vec{q}+\vec{k}))
	(\vb{Q}^2+\varepsilon_{{n}}(\vec{q}+\vec{k}))
	(\vb{|Q-K|}^2+c^2\vec{q}\,^2).
}
\end{align}
Introducing two Feynman parameters, $x,y$, we may combine the denominators into one that is bilinear in $\vb{Q}$ and $\vb{K}$.
After shifting $\vb{Q} \to \vb{Q}-\vb{K}(1-x-y)$, the integral can be written as 
\begin{align}
\Upsilon_{g,(n)}^{1L}\!\left(\vb{K}\right)=
2
\int_0^1 \dd x \int_0^{1-x} \dd y
\int \dd{q} 
\frac
{
	\vec{v}_{n}\cdot\vec{q}\,\vec{v}_{\bar{n}}\cdot\vec{q}-(\vb{Q}-\vb{K}(1-x-y))^2
}
{
	(\vb{Q}^2+\vec{q}\cdot \underline{X} \cdot \vec{q}+M^2)^3
}
,
\end{align}
where $M^2=(x+y)\left(1-x-y\right)\vb{|K|}^2$ and $\underline{X}=x \vec{v}_{n}
\vec{v}_{n}^{\,T}+y \vec{v}_{\bar{n}} \vec{v}_{\bar{n}}^{\,T}+(1-x-y)c^2 \iden$
is a symmetric matrix. After $\vec{q}$ is rescaled as $\vec{q}\rightarrow
\underline{X}^{-1/2}\cdot\vec{q}$, the frequency and momentum integrals are
performed. The result in the small $\epsilon$ limit is
\begin{align}
\Upsilon_{g,(n)}^{1L}(q,k)
&=
\frac
{-1}
{16\pi^3 c\epsilon} h_{3}(v,c) +\mathcal{O}(\epsilon^0),
\label{eqA16}
\end{align}
where
\begin{align}\label{eq:h3Def}
h_{3}(v,c)=\pi c
\int_0^1 \dd x \int_0^{1-x} \dd y
\frac1{\sqrt{\det(\underline{X})}} \left( 2+ \frac{(1-x-y)c^2(1-v^2)}{\det(\underline{X})} \right).
\end{align}
The determinant is
$\det\left(\underline{X}\right)=(c^2\zeta_{3}+\zeta_{1})(c^2
\zeta_{3}+v^2\zeta_{1})-v^2\zeta_{2}^2$, where 
$\zeta_{1}=x+y$,
$\zeta_{2}=x-y$ and 
$\zeta_{3}=1-x-y$. 
%
%

\noindent
The integration over $\zeta_2$ gives
\begin{align}
h_3(v,c)
= \frac{4\pi c}{v} \int_0^1\dd{\zeta_3}
	\left[
\frac{2}{v}\arcsin\left(  \frac{  1-\zeta_3   }{(1-a\zeta_3)(1-b\zeta_3)}  \right)
+
\frac{2 (1-\zeta_3) \zeta_3 \left(\frac{a}{1-a \zeta_3}-\frac{b}{1-b \zeta_3}\right)}{c v^2 \sqrt{v^2+1} (a-b)
\sqrt{\zeta_3 \left(1-\left(1-\hat{c}^2\right) \zeta_3\right)}}
   \right],
\end{align}
where $a=1-c^2$ and $b=1-s^2$. The first term can be integrated by parts to simplify the integrand:
\begin{align}
h_3(v,c) 
&=
\int_0^1 \dd\zeta_3
\frac{2 \pi  (1-b) \zeta_3}{\sqrt{v^2+1} (1-b \zeta_3) \sqrt{\zeta_3 \left(1-\left(1-\hat{c}^2\right) \zeta_3\right)}}
\\
&=
\frac{4\pi}{\sqrt{1+v^2}}
\frac
{d_0(\hat c) - s^{-1}d_0(s^{-1}\hat c)}
{1-s^{-2}}
\end{align}
In the small $c$ limit with $c \ll s$, this is simplified to
\begin{align}
h_3(c/s, c) = \frac{2\pi^2 s}{1+s} - 4\pi c + \mathcal O(c^2).
\end{align}

Returning back to the calculation of the quantum correction in Eq.~\eqref{eqA12}, we note 
that the discrete sum over the Pauli index $\beta$ vanishes in the $G=O(2)$ theory because
\begin{align}  
	\left(\sum_{\beta\in S_G}\tau^\beta\tau^{\alpha}\tau^\beta\right)_{\sigma,\sigma'} 
	= (2-N) \tau^\alpha_{\sigma,\sigma'}.
	\label{eqA23}
\end{align}
Substituting the $\epsilon$-pole found in Eq.~\eqref{eqA16} together with Eq.~\eqref{eqA23} into Eq.~\eqref{eqA12} we obtain
\begin{align}
\delta \mathcal{S}_{g}=
\left(\frac{-g^2(2-N)h_3(v,c)}{16\pi^3cN_f\epsilon}\right)
\frac{g}{N_f^{1/2}}
\sum_{\substack{n,j\\\alpha\sigma\sigma'}}
\int \dd{q}\,\dd{k}  
\left[  
	\phi^\alpha(k_1-k_2) \bar{\Psi}_{\bar{n},\sigma,j}(k_1) \tau^{\alpha}_{\sigma,\sigma'}
	(i\gamma_{d-1})\Psi_{n,\sigma',j}(k_2) 
\right].
\end{align}
We obtain 
$Z_{6,1}$ to be
\begin{align}
Z_{6,1} = \frac{g^2(2-N)h_3(v,c)}{16\pi^3cN_f}.
\end{align}


\subsection{One-loop correction to the quartic coupling}
\label{app:FourBoson}

The one-loop corrections to the quartic interaction arise from Figs. \ref{fig:Quartic1LoopFermion} and \ref{fig:Quartic1LoopBoson}. 
The contribution Fig. \ref{fig:Quartic1LoopFermion} reads
\begin{align}
\delta\mathcal{S}_{u,a}
=
-\frac{g^4\mu^{2\epsilon}}{2N_f}\sum_{\alpha_j\in S_G}
\int 
	\prod_{j=1}^{4}\dd{p_j}
	\delta\left(\sum_{i=1}^4 p_i\right)
\Upsilon_{u,a}^{1L}(p_1,p_2,p_3) \prod_{j=1}^4\phi^{\alpha_j}(p_j),
\end{align}
where
\begin{align}
\Upsilon_{u,a}^{1L}(p_1,p_2,p_3)
=-
\sum_{n}
\int \dd{k}\Tr[
i\gamma_{d-1}{G}_{n}      (k_1)\cdot 
i\gamma_{d-1}{G}_{\bar{n}}(k_2)\cdot 
i\gamma_{d-1}{G}_{n}      (k_3)\cdot 
i\gamma_{d-1}{G}_{\bar{n}}(k_4)].
\end{align}
In the above expression,
we use $k_1=k$, $k_2=k_1+p_1$, $k_3=k_2+p_2$ and $k_4=k_3+p_3$. 
For the diverging part of quantum correction in the small $\epsilon$ limit, 
we can set all the external frequencies to zero.
By tracing over the $\gamma$-matrices,
we obtain
\begin{align}
\Upsilon_{u\!,a}^{1L}
=
\sum_{n}\int\dd{k}
\frac
{-1}
{	(\varepsilon_{n}(\vec{k}_1)-i \vb{|K|})
	(\varepsilon_{\bar{n}}(\vec{k}_2)-i \vb{|K|})
	(\varepsilon_{n}(\vec{k}_3)-i \vb{|K|})
	(\varepsilon_{\bar{n}}(\vec{k}_4)-i \vb{|K|})
}
+
c.c..
\label{eq:A30}
\end{align}
Integrating over the variables
($\varepsilon_{n}(\vec{k})$,$\varepsilon_{\bar{n}}(\vec{k})$), we see that the
integral vanishes because both poles of each integration lie in the upper half
plane (similarly for the complex conjugate). Therefore, there is no contribution
to the quantum effective action coming from this diagram.



The contribution from Fig. \ref{fig:Quartic1LoopBoson} is
\begin{align}
\delta\mathcal{S}_{u,\mathcal{CT}}^{1L}
&=
    -\frac{u^2\mu^{2(3-d)}}{2} 
\sum_{\alpha,\beta\in S_G}
\int \prod_{i=1}^4 \dd{k_i} \delta\left(  \sum_{j=1}^4 k_j \right)
\phi^\alpha(k_1)\phi^\alpha(k_2)\phi^\beta(k_3)\phi^\beta(k_4)
\nonumber\\
&\hspace{5cm}\times
\left[
(8N+32) \Upsilon^{(1L)}_{u,b}(k_3+k_4) + 32 \Upsilon^{(1L)}_{u,b}(k_1+k_4)
\right],
\label{eqA29}
\end{align}
where
\begin{align}
\Upsilon^{(1L)}_{u,b}(k)
&= \int \dd{q} D(q)D(q-k).
\end{align}

\noindent By using the variables $\tilde q = (\vb{Q},c\vec q)$ and the parameter $\tilde k =(\vb{K},c\vec{k})$,
we can rewrite the integration as
\begin{align}
\Upsilon^{(1L)}_{u,b}(k)
&= \frac{1}{c^2} \int \frac{\dd^{d+1}\tilde q}{(2\pi)^{d+1}} \frac{1}{\tilde q^2}\frac{1}{(\tilde q-\tilde k)^2}.
\end{align}

\noindent 
From the identity,
\begin{align}
\int \frac{\dd^{D}q}{(2\pi)^{D}} \frac{1}{(q^2)^{n_1}}\frac{1}{((q+p)^2)^{n_2}}
&= 
\frac{\pi ^{D/2}}{\left(p^2\right)^{-\frac{D}{2}+n_1+n_2}}
\frac{\Gamma \left(-\frac{D}{2}+n_1+n_2\right)}{\Gamma (n_1) \Gamma (n_2)}
\frac{\Gamma \left(\frac{D}{2}-n_1\right) \Gamma \left(\frac{D}{2}-n_2\right)}{\Gamma (D-n_1-n_2)},
\label{eqGrozin}
\end{align}
we obtain
\begin{align}
\Upsilon^{(1L)}_{u,b}(k)
&= 
\frac{1}{8\pi^2c^2 \epsilon} + \mathcal{O}(\epsilon^0).
\end{align}
\noindent Substituting this back into Eq.~\eqref{eqA29}, we obtain
\begin{align}
\delta\mathcal{S}_{u}^{1L}
&=
    -\frac{(8N+64)u^2}{16\pi^2c^2 \epsilon}
\sum_{\alpha,\beta\in S_G}
\int \prod_{i=1}^4 \dd{k_i} \delta\left(  \sum_{j=1}^4 k_j \right)
\phi^\alpha(k_1)\phi^\alpha(k_2)\phi^\beta(k_3)\phi^\beta(k_4)
\end{align}
%
%
and the counter term coefficient,
\begin{align}
Z_{7,1;(1L)} = \frac{N+8}{2\pi^2}\frac{u}{c^2}.
\end{align}



\subsection{Two-loop boson self-energy}
\label{app:BSE_2L_FL}

The two-loop boson self-energy arise from Figs. 
\ref{fig:BosonSE2LoopFermion} 
and
\ref{fig:BosonSE2LoopBoson}.
The contribution from Fig. \ref{fig:BosonSE2LoopFermion} reads
\begin{align}\label{eq:Yukawa1Loop}
    \delta\mathcal{S}_{\phi^2,a}^{2L}=\frac{g^4\mu^{2(3-d)}(2-N)}{N_f}\int \dd k \Upsilon_{(0,2),a}^{2L,\alpha}(k)\phi^\alpha(k)\phi^\alpha(-k),
\end{align}
where 
\begin{align}\label{eq:BosonSE2LoopFermionDiagram}
\Upsilon_{(0,2),a}^{2L,\alpha}(k)
=\sum_{n=1}^4
\int\!\dd{p}\!\dd{q} 
\Tr\left[
      i\gamma_{d-1}{G}_{n}(p+k)
\cdot i\gamma_{d-1}{G}_{\bar{n}}(p)
\cdot i\gamma_{d-1}{G}_{n}(q)
\cdot i\gamma_{d-1}{G}_{\bar{n}}(q+k)
\right]
D(p-q).
\end{align}
As we are only interested in the spatial part of the self-energy, $\vb{K}$ is
set to zero. Let's proceed to calculate the contribution from hot spot 1, as the
other contributions are related by the $C_4$ symmetry. 
Its contribution is
\begin{align}
\centering
\Upsilon_{(0,2),a}^{2L,\alpha,n=1}(\vec{k})
&= \int \dd{p}\!\dd{q} \Tr \left[
\left(
  \frac{\mathbb{P_{+}}}{(\varepsilon_1(p+k)-i|\vb{P}|)(\varepsilon_3(p)-i|\vb{P}|)}+
  \frac{\mathbb{P_{-}}}{(\varepsilon_1(p+k)+i|\vb{P}|)(\varepsilon_3(p)+i|\vb{P}|)}
\right)\right.
\nonumber\\
&\hspace{1cm}\times
\left.
\left(
  \frac{\mathbb{Q_{+}}}{(\varepsilon_1(q)-i |\vb{Q}|)(\varepsilon_3(q+k)-i|\vb{Q}|)}+
  \frac{\mathbb{Q_{-}}}{(\varepsilon_1(q)-i |\vb{Q}|)(\varepsilon_3(q+k)-i|\vb{Q}|)}
\right)
\right]
D(p-q).
\label{eq:A40}
\end{align}
We perform the change of variables $\vec{q}\to\vec{q}+\vec{p}$ and $(p_x,p_y)\rightarrow(\varepsilon_1(p),\varepsilon_3(p))$. 
The determinant of the Jacobian corresponding to the transformation is $(2v)^{-1}$. 
After integrating over $(\varepsilon_1(p),\varepsilon_3(p))$, only two terms survive due to the location of the poles. \footnote{Also notice that $\Tr[\mathbb{P}_+ \mathbb{Q}_-]=\Tr[\mathbb{P}_- \mathbb{Q}_+]$.}
With this information, we can write
\begin{align}
\Upsilon_{(0,2),a}^{2L,\alpha,n=1}(\vec{k})
&=\frac{1}{2v}
\int
\frac{\dd^{d-1} \vb{P}}{(2\pi)^{d-1}} 
\frac{\dd^{d+1} q}{(2\pi)^{d+1}}
\frac{\mathrm{Tr}\left[\mathbb{P}_+ \mathbb{Q}_-\right]}{c^2 \vec{q}^2+(\vb{Q}-\vb{P})^2}
\frac
{2 \left(\varepsilon_1(k-q)\varepsilon_3(k+q)+(|\vb{P}| +|\vb{Q}| )^2\right)}
{
\left(\varepsilon_1(k-q)^2+(|\vb{P}|+|\vb{Q}|)^2\right)
\left(\varepsilon_3(k+q)^2+(|\vb{P}|+|\vb{Q}|)^2\right)
}.
\end{align}

\noindent
The $cq_y$ dependence of the boson propagator will be neglected as the dependence on $q_y$ is dominated by the remnants of the fermionic propagators. 
After the $q_y$ integral is performed, we obtain
\begin{align}
\Upsilon_{(0,2),a}^{2L,\alpha,n=1}(\vec{k})
&=\frac{1}{2v}
\int
\frac{\dd^{d-1} \vb{P}}{(2\pi)^{d-1}} 
\frac{\dd^{d-1} \vb{Q}}{(2\pi)^{d-1}} 
\frac{\dd q_x}{2\pi}
\frac{\mathrm{Tr}\left[\mathbb{P}_+ \mathbb{Q}_-\right]}{c^2 q_x^2+(\vb Q-\vb P)^2}
\frac{|\vb{P}|+|\vb{Q}|}{(vq_x-k_y)^2+(|\vb{P}|+|\vb{Q}|)^2}.
\end{align}


\noindent Integrating over $q_x$ gives
\begin{align}
\Upsilon_{(0,2),a}^{2L,\alpha,n=1}(\vec{k})
&=\frac{1}{8v^2}
\int
\frac{\dd^{d-1} \vb{P}}{(2\pi)^{d-1}} 
\frac{\dd^{d-1} \vb{Q}}{(2\pi)^{d-1}} 
\frac{1-\hat{\vb P}\cdot \hat{\vb Q}}{|\vb Q-\vb P|}
\frac{s(|\vb{P}|+|\vb{Q}|)+|\vb Q-\vb P|}{(sk_y)^2+(s(|\vb{P}|+|\vb{Q}|)+|\vb Q-\vb P|)^2},
\label{eqA43}
\end{align}
where 
$\hat{\vb P} = \vb P / |\vb P|$,
$\hat{\vb Q} = \vb Q / |\vb Q|$, and
$s = c/v$. This integration can be done exactly using a shift to an elliptico-cyllindrical coordinate system where
$\vb P$ and $\vb Q$ parameterize two foci in a $2(d-1)$ dimensional space. More explicitly, we may take the origin of the new coordinates
at $(\vb P + \vb Q)/2$, with the foci located at $\pm \vb T/2 = \pm(\vb Q - \vb P)/2$. To make this manifest,
let us introduce the parameters
$\vb T = \vb Q - \vb P$,
$\xi   = (|\vb P|+|\vb Q|)/(|\vb Q-\vb P|) \in (1,\infty)$,
$\eta  = (|\vb P|-|\vb Q|)/(|\vb Q-\vb P|) \in (-1,1)$. The cylindrical integration factors out to give
\begin{align}
\Upsilon_{(0,2),a}^{2L,\alpha,n=1}(\vec{k})
&=\frac{1}{8v^2}
\int
\frac{\dd^{d-1} \vb{T}}{(2\pi)^{d-1}} 
\int_1^\infty  \dd\xi
\int_{-1}^1    \dd\eta
\left[
	\int_{S_{d-3}} \dd\Omega_{d-3}
\right]
\frac{\sqrt{1-\eta ^2}}{8\pi^2\sqrt{\xi^2-1}}
\frac{|\vb T|^{d-1}
(s\xi+1)
}{(sk_y)^2+\vb T^2 (\xi  s+1)^2}.
\end{align}

\noindent Next, we peform the integration over $\eta$ and the cylindrical coordinate $\Omega_{d-3}$.
Subtracting the mass renormalization,
we obtain the momentum-dependent self-energy,
\begin{align}
\Upsilon_{(0,2),a}^{2L,\alpha,n=1}(\vec{k})
-\Upsilon_{(0,2),a}^{2L,\alpha,n=1}(0)
&=
\frac{-(sk_y)^2}{64\pi v^2}
\int_1^\infty  \dd\xi
\int \frac{\dd^{d-1} \vb{T}}{(2\pi)^{d-1}} 
\frac{|\vb T|^{d-3}
(s\xi+1)
}{\left(\frac{sk_y}{1+s\xi}\right)^2+\vb T^2}
\frac{1}{\sqrt{\xi^2-1}(1+s\xi)^3}
\\
&= 
    \frac{-(sk_y)^2}{256\pi^2 v^2\epsilon} \rho(s),
\label{eqA44}
\end{align}
where
\begin{align}
        \rho(s)
    &=\frac{-3 s^2-\sqrt{1-s^2} \left(s^2+2\right) \log \left(\frac{\sqrt{1-s^2}+1}{s}\right)+3}{2\left(s^2-1\right)^3}.
\end{align}



The small $s$ limit of $\rho(s)$ is defined to be
\begin{align}
    \rho_0(s) = \log\frac{2}{e^{3/2}s}.
\label{eq:rho0}
\end{align}

\noindent Using Eq.~\eqref{eqA44} we can compute the quantum correction for all hot spot indices $n=1,\ldots,4$, which gives the full correction to the boson self-energy,
\begin{align}
\delta\mathcal{S}_{\phi^2,a}^{2L}
= -\frac{g^4 (2-N)}{64 \pi ^2 N_f v^4 \epsilon } \rho(s)
\left[\frac12
\sum_{\alpha\in S_G}\int \dd k \,c^2\,(k_x^2 + k_y^2)\phi^\alpha(k)\phi^\alpha(-k)\right]
\end{align}
and the associated
multiplicative 
 counterterm,
\begin{align}
    Z_{5,1} &= \frac{2(2-N)}{\pi ^2 N_f}\frac{g^4}{v^2 c^2} \tilde h_5(s),
\end{align}
    where
    $\tilde h_5(s)$ is defined in
\eq{eq:h5}.

The contributions 
from Fig. \ref{fig:BosonSE2LoopBoson} is
\begin{align}
\delta\mathcal{S}_{b}
=
-\frac{u^2\mu^{2(3-d)}\left( 32N+64 \right)}{2} 
\sum_{\alpha\in S_G}
\int \dd{k} \dd{p} \dd{q}
\phi^{\alpha}(-q) \phi ^{\alpha}(q)\Upsilon^{2L}_b(q),
\end{align}
where
\begin{align}
\Upsilon^{2L}_b(q) = \int \dd{k} \dd{p} D(k) D(p) D(k+p+q) =   \frac{1}{(2\pi)^{2(d+1)}c^4}\int \frac{\dd{\tilde k} \dd{\tilde p}}{\tilde k^2 \tilde p^2 (\tilde k + \tilde p + \tilde q)^2},
\end{align}
and $\tilde q = (\vb Q, c\vec q)$. 
This gives rise to 
\begin{align}
\Upsilon^{2L}_b(q) = -\frac{\vb Q^2 + c^2 \vec q^2}{512 \pi^4 c^4 \epsilon}
\end{align}
and the counter-term coefficient,
\begin{align}
Z_{4,1;2L;b} = Z_{5,1;2L;b} = - \frac{u^2 \left(N+2\right)}{(2\pi)^4 c^4}.
\label{eqA56}
\end{align}


\subsection{Two-loop vertex correction to the quartic coupling}
\label{app:QuarticTwoLoop1}

The contribution of the diagram in Fig. 
\ref{fig:2Lphi4}
to the quantum effective action is
\begin{align}\label{eq:BosonSE2LoopFermion}
\delta\mathcal{S}_{u}^{2L}
&=
\frac{g^6\mu^{3(3-d)}}{2N_f^2}
\sum_{\alpha_1,\ldots,\alpha_4\in S_G}
\int \dd{k_1} \dd{k_2} \dd{k_3}
\phi^{\alpha_1}(k_1) \phi ^{\alpha_2}(k_2) \phi ^{\alpha_3}(k_3) \phi ^{\alpha_4}(-k_1-k_2-k_3)	
\nonumber\\
&\hspace{4cm}\times
\sum_{\alpha\in S_G}\sum_{n=1}^4
	\Tr(\tau^{\alpha_1}\tau^{\alpha_2}\tau^{\alpha}\tau^{\alpha_3}\tau^{\alpha_4}\tau^{\alpha})
	\Upsilon^{2L}_n(k),
\end{align}
where 
\begin{align}
\Upsilon^{2L}_n(k)
&=
\int \dd{p} \dd{q} 
      D(q-p-k_1-k_4)
\Tr\left[
      i\gamma_{d-1}{G}_{n}      (p)
\cdot i\gamma_{d-1}{G}_{\bar{n}}(p-k_2)
\cdot i\gamma_{d-1}{G}_{n}      (q+k_3)
	\right.
	\nonumber
	\\
	&\hspace{4cm}\left.
\cdot i\gamma_{d-1}{G}_{\bar{n}}(q)
\cdot i\gamma_{d-1}{G}_{n}      (q-k_1-k_2-k_3)
\cdot i\gamma_{d-1}{G}_{\bar{n}}(p+k_1)
\right].
\end{align}

\begin{figure}[htpb!]
    \centering
    \includegraphics[width=0.2\textwidth]{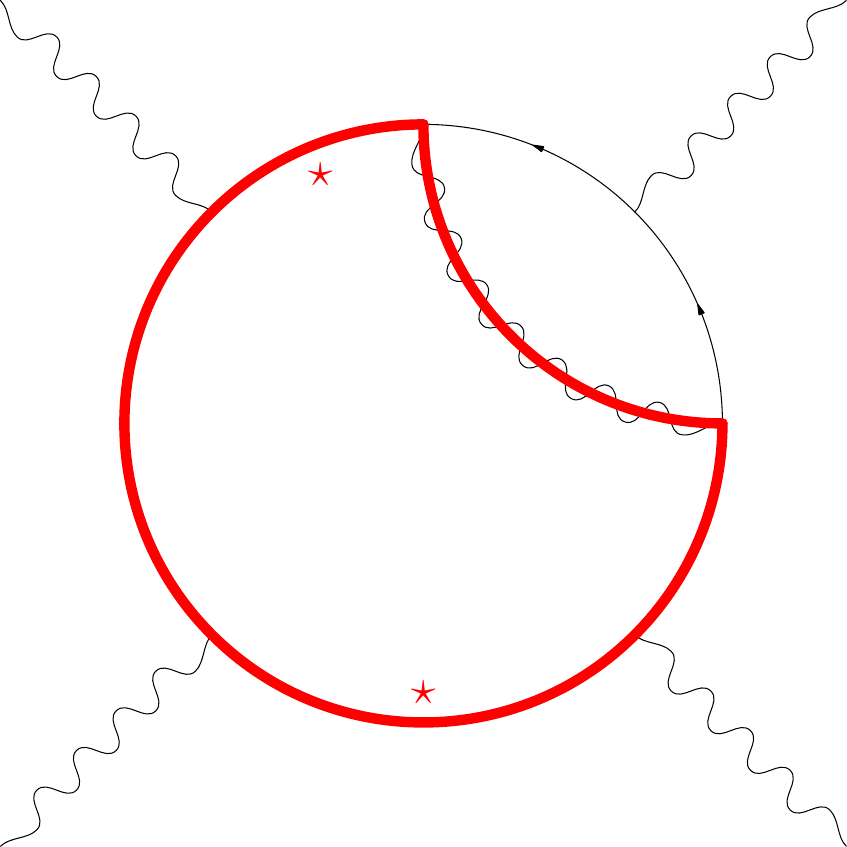}
    \caption{
    A quantum correction to the quartic vertex that has an additional suppression of $s$ due to a double pole. 
    The fermion propagators with stars carry the same energy when the external momenta are zero.
    If we ignore the dispersion of the boson in the small $c$ limit, 
    the integration of the spatial momentum in the loop denoted as the thick (red) lines vanishes due to a double pole.
    }
\label{fig:4phi2Lextra}
\end{figure}

\noindent
To calculate the $\epsilon$ pole, we will consider all frequency set to zero, and set $\vec k_3=\vec k_4=0$.
We first consider the case with $n=1$ as other cases are related through the  $C_4$ symmetry. 
With $p_x \to p_x + \frac{1}{2v} e_4(\vec k)$ and $p_y \to p_y + \frac{1}{2}e_4(k)$,
we redirect all external momentum into the internal boson propagator. 
If we define a vector $\mathfrak{K} = (0,e_1(k)/2v,e_1(k)/2)$, then we can express the integration succinctly as
\begin{align}
\Upsilon^{2L}_{n=1}(k)
&=
\int 
\frac{\dd{p} \dd{q}}{(\vb Q - \vb P)^2 + (cq_x-cp_x-ce_1(k)/2v)^2 + (cq_y - cp_y - ce_1(k)/2)^2}
\nonumber\\
&\hspace{1cm}\times
  \Tr\left[
  	\left( \frac{\mathbb{P_{+}}}{(\varepsilon_3(\vec p)-i|\vb{P}|)^2(\varepsilon_1(\vec p)-i|\vb{P}|)}+ c.c. \right)
  	\left( \frac{\mathbb{Q_{+}}}{(\varepsilon_1(\vec q)-i|\vb{Q}|)^2(\varepsilon_3(\vec q)-i|\vb{Q}|)}+ c.c. \right)
  \right].
  \label{eq:59}
\end{align}
Next, we may neglect the $cq_y$ and $cp_y$ dependence in the boson propagator as the $q_y$ and $p_y$ dependence of the integrand is 
dominated by the fermionic propagators. After integrating over all spatial momentum, $p_x,p_y,q_x,q_y$, we may neglect 
the IR cutoff $ce_1(k)$ compared to $ce_1(k)/2v$ as the latter is larger in the small $v$ limit to obtain
\begin{align}
\Upsilon^{2L}_{n=1}(k)
&=
\frac{s^2}{16v^2}
\int 
	\frac{\dd^{d-1}\vb P}{(2\pi)^{d-1}}
	\frac{\dd^{d-1}\vb Q}{(2\pi)^{d-1}}
	\frac{1 - \hat{\vb P}\cdot\hat{\vb Q}}{\sqrt{(\vb Q - \vb P)^2+\delta_1^2}}
	\frac{1}{(\sqrt{(\vb Q-\vb P)^2+\delta_1^2}+s(|\vb P|+|\vb Q|))^3},
\end{align}
where $\delta_1$ is an IR cutoff determined from the external momenta.
Here, we use 
$\epsilon_1(\vec p)$,
$\epsilon_3(\vec p)$,
$\epsilon_1(\vec q)$
and
$\epsilon_3(\vec q)$
as integration variables.
The factor of $1/v^2$ arises from the Jacobian of the transformation of the integration variables.
An extra factor of $s^2$ arises because 
two integrations have nothing but double poles
in the small $s$ limit.
For example,  the term that arises from the projector 
$ \mathbb{P_{+}} \mathbb{Q_{+}}$
in \eq{eq:59} have double poles for 
$\epsilon_3(\vec p)$
and
$\epsilon_3(\vec q)$
at 
$i |{\bf P}|$
and
$i |{\bf Q}|$, respectively.
For a non-zero $s$, the boson propagator gives rise to a non-zero contribution but with the $s^2$ suppression.
This kind of extra suppression by $s$ is a generic feature of diagrams that satisfy the following two conditions:
(a) it has a loop made of boson propagators and more than one fermion propagators of one type of hot spot,
(b) loop momenta can be chosen such there is no other loop momenta that  traverse through those `degenerate' fermion propagators in the loop identified in (a).
In these cases, the integration over the spatial momentum exhibits a double or higher-order pole from the fermion propagators and is suppressed by powers of $s$.
Another examples of such diagrams is shown in \fig{fig:4phi2Lextra}.

The remaining integration can be done through the elliptico-cyllindrical coordinates that was introduced for the computation of  
the two-loop boson self-energy. 
It reduces to 
\begin{align}
\Upsilon^{2L}_{n=1}(k)
&=
\frac{s^2}{256\pi^2v^2}
\int_1^\infty \frac{\dd\xi}{\sqrt{\xi^2-1}}
\int_0^\infty 
	\frac{ T^{-1-2\epsilon} \dd{T} }
	{\sqrt{1+(\delta_1/T)^2}\left( s\xi + \sqrt{1+(\delta_1/T)^2} \right)^3}.
\end{align}
Integrating over $T$ and $\xi$,
we obtain an $\epsilon$-pole for the counter-term coefficient,
\begin{align}
    Z_{7,1} = -\frac{Ng^6}{4N_f^2 v^2u} \tilde h_5(s).
\end{align}

\section{Control of Higher Order Graphs}
\label{app:ControlGraphs}

Due to the quenching of kinetic energy, the $\epsilon$-expansion is not organized by the loop-expansion.
As a result, one has to include some higher-loop diagrams even at the leading order in the $\epsilon$-expansion.
We call the leading order in $\epsilon$
the modified one-loop order.
Then, it is necessary to check that higher-loop graphs beyond the modified one-loop order are negligible at the infrared fixed points.
For the $O(3)$ theory,
this has been shown to be the case in Ref. \cite{LUNTS}. 
In this appendix, we show that the $\epsilon$-expansion is also controlled for the $Z_2$ and $O(2)$ theories.
We first derive an upper bound for general diagrams 
that applies to both $Z_2$ and $O(2)$ theories.
Afterwards, we refine the bound in each theory separately.

Here is our strategy of proof.
We first assume that the fully dressed propagators are given by the bare propagators to the leading order in $\epsilon$.
We use this to show that higher-order diagrams that do not include self-energy corrections as proper sub-diagrams,
which are henceforth referred to as {\it half-skeleton} diagrams,
are small in the small $\epsilon$ limit.
Based on this, we can put a bound on non-half-skeleton diagrams in a self-consistent manner:
since the renormalized propagators are identical to the bare ones to the leading order, 
the magnitudes of non-half skeleton diagrams are bounded by 
 the associated skeleton diagrams.
This analysis confirms that 
all self-energy corrections are indeed suppressed in the small $\epsilon$ limit,  which justifies our initial assumption. 
This further allows us to show that all higher-order quantum corrections beyond the modified one-loop order do not affect the fixed point and its stability to the leading order in $\epsilon$.
From now on, we focus on half-skeleton diagrams.

\subsection{General upper bound applicable to both \texorpdfstring{$Z_2$}{Z2} and \texorpdfstring{$O(2)$}{O2} theories}

The contribution of a one-particle irreducible diagram to the quantum effective action is written as
\begin{align}
    \mathcal{I} (v,c;\{k_i\})
 = 
\ub~
     \mathcal{F} (k_i),
 \label{eqGenericScaling}
\end{align}
where $\ub$ sets the overall magnitude of the quantum correction
and
$ \mathcal{F} (k_i)$
determines how quantum corrections depend on the external momenta $k_i$.
Due to an enhanced low-energy phase space and a loss of dispersion,
$\ub$ and  $ \mathcal{F} (k_i)$
may diverge or vanish
if $v$ and $c$ go to zero 
for fixed $g$, $u$ and $k_i$.
In order to estimate the magnitude of quantum correction at the fixed point with vanishing $v$ and $c$, it is necessary to understand how 
$\ub$ and $ \mathcal{F} (k_i)$
depend on the velocities.
In the following two subsections,
we discuss 
how $\ub$
and $ \mathcal{F} (k_i)$
depend on $v$ and $c$.

\subsubsection{
Upper bound of \texorpdfstring{$\ub$}{U}
}

$\ub$ generally diverges in the limit that $v$ and $c$ vanish because the phase space for low-energy excitations increases with decreasing velocities.
The enhancement is captured by the Jacobian of the transformation that maps the loop momenta to the energies that virtual particles carry within the loop.
To compute the Jacobian,
we follow Ref. \cite{SHOUVIK2}
to choose loop momenta, $\{q_i\}_{i=1}^L$, for $L$-loop diagrams, such that 
there exists
at least one 
propagator that carries only 
$q_i$.
%
To illustrate the algorithm for choosing such loop variables,
we begin with a diagram containing $L$ loops and perform a three-stage procedure. 
In the first stage, we start by cutting boson propagators within boson loops, that is, loops that contain only boson propagators, 
such that the number of boson loops is decreased by one after each cut. 
At the end of this stage, the remaining loops in the diagram, if any, have at least one fermion propagator. 
We call these non-boson loops mixed loops.
The next stage consists of sequentially cutting fermion propagators so that at each step the number of loops decreases by one. 
Throughout the entire algorithm, we ensure that the diagram remains connected. 
At the end of the second stage, our diagram takes the form of a tree diagram. 
The third stage consists of reattaching the previously cut propagators in the reverse order. 
At each reattachment step, we assign an internal momentum to traverse 
only the reintroduced propagator and the tree 
 which has been generated at the end of the second stage.
This guarantees that each propagator glued back in the third stage carries only one loop momentum.
 This loop momentum prescription is called an exclusive loop covering (ELC),
 and the propagators that were cut and reattached are called exclusive propagators.

Next, we identify a set of propagators in the diagram that can tame the loop integrations' power-law UV divergence. 
This is important because some directions in the space of internal momenta become flat at $v=c=0$, causing a UV divergence.
For small but non-zero $v$ and $c$, such UV divergences are turned into an enhancement by $1/v$ or $1/c$.
The choice of propagators that tame UV divergences of each loop momentum is not unique.
For our bound, we choose those propagators such that the enhancement by $1/c$ and $1/v$ is minimized.
We note that the actual magnitude of a diagram does not depend on how loop-momenta are labeled and can be smaller than our bound.
Given that one can not compute arbitrarily high-loop diagrams, however,  one choice is more convenient in establishing an upper bound for general diagrams. 
Our goal is to find a choice of propagators that gives the smallest upper bound within different choices.

The choice of loop momenta that minimize the enhancement  depends on $N$
because $Z_2$, $O(2)$ and $O(3)$ theories have different ratios between $v$ and $c$ near the fixed point. 
In the $N=1$ theory, $c \ll v$ near the fixed point. 
This means that the fermion is faster in the direction perpendicular to $\vec Q_{AF}$ compared with the boson. 
If either a boson propagator or a fermion propagator can 
 be chosen to tame the UV divergence, it is the fermion propagator that will cutoff the UV divergence with a smaller enhancement.
This is in contrast to the $O(3)$ theory in which $v \ll c$ and the UV divergence is cut off by the boson propagator\cite{LUNTS}.
In the direction parallel to $\vec Q_{AF}$, boson dispersions are suppressed by a factor of $c$ compared to the fermion dispersion in all theories. 
Therefore, we preferentially pick the fermion propagators to tame UV divergences in the small $v,c$ limit for $N=1$.
We will use the same choice for the $N=2$ theory  because in this case 
$v \sim c$
and choosing either a boson or a fermion propagator for taming UV divergence gives the same bound.
Let us first introduce some notations needed for the derivation
of the bound.
Let $L$, $L_b$ and $L_{fm} = L - L_b$ denote 
the total number of loops,
the number of boson loops 
and
the number of mixed loops,
respectively.
Similarly, let
$I_f$, $I_b$ and 
$I = I_f + I_b$ 
be the number of internal fermion propagators,
the number of internal boson propagators
and 
the total number of propagators, respectively.
We denote the kinetic energy of the $i$-th exclusive fermion propagator as $\mathcal E_{E,i}$, 
while $\mathcal E_{NE,j}$ corresponds to the energies of the other (non-exclusive) fermion propagators.
For exclusive boson propagators,
we use 
\begin{align}
\mathfrak b_{i,x}' = cq_{i,x}, ~~
\mathfrak b_{i,y}' = cq_{i,y}, 
    \label{B10}
\end{align}
to be $c$ times the $x$- and $y$- component of spatial momentum that runs in the $i$-th exclusive boson propagator. 
It is noted that the boson kinetic energy is quadratic in $\vec q$, 
and an exclusive boson propagator can tame a power-law divergence of both $q_x$ and $q_y$ in that exclusive loop.

For the bosonic loops, we can use the exclusive boson propagators to tame UV divergences.
Since $q_x$ and $q_y$ always appear with a factor of $c$, 
the integration over $q_x$ and $q_y$ will give rise to a factor of $c^{-2}$.  
This results in a factor of
 $c^{-2L_b}$ for the boson loops.
The next job is to identify the contribution of the remaining mixed loops.
For the mixed loops, we choose the exclusive fermion propagators and
some non-exclusive fermion/boson propagators to make sure that 
(1) all power-law UV divergences of the loop momenta are tamed by those propagators 
and (2) the Jacobian that results from the change of variables from the loop momenta to the energies of those propagators is minimally enhanced in the limit that $v$ and $c$ are small.
%
%
We represent
the set of momentum in the non-exclusive boson propagators 
in terms of two vectors,
$\mathcal B_{NE,x} \equiv (\mathfrak b_{1,x},\ldots,\mathfrak b_{I_b-L_b,x})^T$ 
 and
$\mathcal B_{NE,y} \equiv
(\mathfrak b_{1,y},\ldots,\mathfrak b_{I_b-L_b,y})^T$.
To tame the mixed-loop integrations,
we write down a linear relation between the variables, 
$\Bigl\{
    \mathcal{E}_{E},
    \mathcal{E}_{NE},
    \mathcal{B}_{NE,x},
    \mathcal{B}_{NE,y}\Bigr\}$,
and the loop-momentum variables chosen in the exclusive loop covering as
\begin{align}
    \begin{pmatrix}
        \mathcal{E}_{E}, &
        \mathcal{E}_{NE},&
        \mathcal{B}_{NE,x},   &
        \mathcal{B}_{NE,y}
    \end{pmatrix}^T
    =
    \mathbb{M}_0\mathbb{P}+
    \mathbb{N}_0\mathbb{Q}+
    \mathbb{S}_0\mathbb{R}.
    \label{B1}
\end{align}
Here, 
$\mathbb{P}=(p_{1,x},p_{2,x},\ldots,p_{L_{fm},x},p_{1,y},p_{2,y},\ldots,p_{L_{fm},y})^T$ 
and 
$\mathbb{Q}=(q_{1,x},q_{2,x},\ldots,q_{L_{b},x},q_{1,y},q_{2,y},\ldots,q_{L_{b},y})^T$ denote 
the momentum components of the exclusive fermion propagators and 
the exclusive boson propagators, respectively. 
$\mathbb{R}$ denotes the $x,y$-components of the external spatial momenta.
$\mathbb M_0, \mathbb N_0, \mathbb S_0$ are 
$(I_f+2I_b-2L_b) \times 2 L_{fm}$,
$(I_f+2I_b-2L_b) \times 2 L_{b}$
and
$(I_f+2I_b-2L_b) \times 2 E$ matrices, respectively,
where
$E$ is the number of external momenta.
In particular, $\mathbb M_0$ takes the form of
\begin{align}
    \mathbb{M}_0 &= 
    \begin{pmatrix}
        v \iden & \mathbb{A}\\
        v \mathbb{B}_1 & \mathbb{B}_2\\
        c \mathbb C_1  & 0\\
        0  & c \mathbb C_2
    \end{pmatrix},
\end{align}
where $\iden$ is the $L_{fm}\times L_{fm}$ identity matrix, $\mathbb{A}$ is a diagonal matrix with $\pm1$ as the diagonal elements,
$\mathbb{B}_1$, $\mathbb{B}_2$, $\mathbb{C}_1$, $\mathbb{C}_2$ are matrices with entries $\pm1$.

Let us consider the first 
$I_f+I_b-L_b$ rows of \eq{B1} for  $\mathbb Q = 0$ and  $\mathbb R = 0$,
\begin{align}
    &\begin{pmatrix}
        \mathcal{E}_{E}  \\
        \mathcal{E}_{NE} \\
        \mathcal{B}_{NE,x}
    \end{pmatrix}
    =
    \mathbb{M}\mathbb{P}
    =
    \begin{pmatrix}
        v \iden & \mathbb{A}\\
        v {\mathbb{B}}_1 & {\mathbb{B}}_2\\
        c {\mathbb{C}}_1  & 0
    \end{pmatrix}
    \begin{pmatrix}
        p_x\\
        \vdots\\
        p_y\\
        \vdots
    \end{pmatrix}.
    \label{eqB3}
\end{align}
%
%
Note that $I_f + I_b-L_b \ge 2L_{fm}$ follows from $I-L=V-1 \ge L_{fm}$.\footnote{The proof of this statement
follows from counting vertices related to the exclusive fermion propagators. Each exclusive fermion has two vertices, call them `start' and `end' according to the direction of propagation. (Note that a vertex could be both a start and an end of two different propagators.) By construction of the exclusive propagators, the number of `start' vertices in a graph is then equal to $L_{fm}$. This shows that $V\ge L_{fm}$. To show that $V-1\ge L_{fm}$ notice that if there is at least one four-boson vertex, then it is not the `start' or `end' of any fermion propagator, and thus we arrive at the inequality. If there are no four-boson vertices, then we can choose the exclusive propagator in such a way so that it's endpoint is an external vertex. This also means (cf. definition of $\bar\delta$) that the external vertex will not be a `start'-ing vertex. From here it follows that $V-1\ge L_{fm}$.} 
Let us now investigate two cases based on whether or not $\mathbb{M}$ has full rank.

{\it Case I:}
If this matrix has $2L_{fm}$ linearly independent rows, 
all of the $2L_{fm}$ columns are linearly independent.\footnote{This is a simple fact from linear algebra. It follows from the fact that multiplying a matrix by an invertible one does not change the rank of the matrix. If these invertible matrices are elementary row or column operations, then this means that the rank of a matrix is equal to the number of linearly independent rows and also equal to the number of linearly independent columns.} 
In this case,
$\mathbb{M}$ is of the full rank and has a trivial kernel:
$dK \equiv \dim \ker \mathbb M = 0$. 
Then, we find $2L_{fm}$ linearly independent rows in $\mathbb M$ that include the first $L_{fm}$ linearly independent rows.
Let us delete the other rows from $\mathbb{M}$ to form a square matrix, $\mathbb{M}_1$:
\begin{align}
    \begin{pmatrix}
        \mathcal{E}_{E}  \\
        \bar{\mathcal{E}}_{NE} \\
        \bar{\mathcal{B}}_{NE,x}
    \end{pmatrix}
    \equiv
    \mathbb{M}_1\mathbb{P} &= 
    \begin{pmatrix}
        v \iden & \mathbb{A}\\
        v \bar{\mathbb{B}}_1 & \bar{\mathbb{B}}_2\\
        c \bar{\mathbb{C}}_1  & 0
    \end{pmatrix}
    \begin{pmatrix}
        p_x\\
        \vdots\\
        p_y\\
        \vdots
    \end{pmatrix}.
    \label{eqB4}
\end{align}

\noindent 
Here, we use the notation 
$\bar{\mathcal{E}}_{NE}$, $\bar{\mathcal{B}}_{NE,x}$
to denote those non-exclusive fermion and boson dispersions that are left upon removal of some of the rows.
Also, we use the notation $\bar{\mathbb{B}}_1,\bar{\mathbb{B}}_2$ to denote matrices of size $\bar{\delta} \times L_{fm}$ and $\bar{\mathbb{C}}_{1}$ to have size $(L_{fm}-\bar\delta)\times L_{fm}$.
Among all possible choices of
$\bar{\mathbb{B}}_1$,
$\bar{\mathbb{B}}_2$, 
and $\bar {\mathbb{C}}_1$ 
that give $\det(\mathbb{M}_1) \ne 0$,
we maximize $\bar\delta$.\footnote{Physically, maximizing the number of rows coming from the fermionic sector is equivalent to saying that we have found fermions that will tame the loop momenta in the $x$-direction.}
Viewing $\mathbb{M}_1$ as a transformation matrix that maps the exclusive fermion momenta, $\mathbb P$, to the kinetic energy of internal particles, $\Bigl\{ \mathcal E_E, \bar{\mathcal E}_{NE}, \bar{\mathcal B}_{NE,x} \Bigr\}$, 
we see that $\det(\mathbb M_1)^{-1}$ gives the desired factor of enhancement in the limit that $v$ and $c$ are small.
To compute the Jacobian of this transformation, we multiply $\mathbb M_1$ with matrices where the determinant is easy to compute,
\begin{align}
    \mathcal{M}_\alpha = \begin{pmatrix}
        \iden & 0 & 0 \\
        - \bar{\mathbb{B}}_2 \mathbb{A}& \iden & 0\\
        0 & 0 & \iden
    \end{pmatrix}
    ,\hspace{2cm}
    \mathcal{M}_\beta  = \begin{pmatrix}
        \iden & 0 & 0 \\
        0&v^{-1}\iden & 0\\
        0 & 0 & c^{-1}\iden
    \end{pmatrix},
\end{align}
so that
\begin{align}
    \mathcal{M}_\beta\mathcal{M}_\alpha \mathbb{M}_1
&=
    \begin{pmatrix}
        \iden & 0 & 0 \\
        0&v^{-1}\iden & 0\\
        0 & 0 & c^{-1}\iden
    \end{pmatrix}
    \begin{pmatrix}
        \iden & 0 & 0 \\
        - \bar{\mathbb{B}}_2 \mathbb{A}& \iden & 0\\
        0 & 0 & \iden
    \end{pmatrix} 
    \begin{pmatrix}
        v \iden & \mathbb{A}\\
        v \bar{\mathbb{B}}_1 & \bar{\mathbb{B}}_2\\
        c \bar{\mathbb{C}}_1  & 0
    \end{pmatrix}
= 
    \begin{pmatrix}
        \iden & 0 & 0 \\
        0&v^{-1}\iden & 0\\
        0 & 0 & c^{-1}\iden
    \end{pmatrix}
    \begin{pmatrix}
        v\iden & \mathbb A\\
        v\bar{\mathbb{B}}_1-v\bar{\mathbb{B}}_2 \mathbb{A} & 0\\
        c \bar{\mathbb{C}}_1&0
    \end{pmatrix}
\\
&=
    \begin{pmatrix}
        v\iden & \mathbb A\\
        \bar{\mathbb{B}}_1-\bar{\mathbb{B}}_2 \mathbb{A} & 0\\
        \bar{\mathbb{C}}_1&0
    \end{pmatrix}.
\end{align}

\noindent Taking the determinant of both sides we arrive at
$    \left|
    \det(\mathbb{A}) \det
    \begin{pmatrix}
        \bar{\mathbb{B}}_1-\bar{\mathbb{B}}_2 \mathbb{A}\\
        \bar{\mathbb{C}}_1
    \end{pmatrix}
    \right|
    =
    v^{-\bar\delta} c^{-L_{fm}+\bar\delta} \left|\det(\mathbb{M}_1)\right|$,
from which we obtain 
$\det(\mathbb{M}_1) \sim v^{\bar\delta}c^{L_{fm}-\bar\delta}$.
Therefore, the factor of enhancement that arises from the Jacobian for the mixed loops becomes
\begin{align}
\det(\mathbb{M}_1)^{-1} \sim
\frac{1}{v^{\bar\delta}c^{L_{fm}-\bar\delta}}.
    \label{eqnInitUpperBound}
\end{align}
Note that for all exponents in \eqref{eqnInitUpperBound} except $\bar\delta$ can be easily read from each diagram.
Let us discuss how  $\bar\delta$ can be identified for general graphs.
Algebraically, we have introduced $\bar\delta$ in Eq.~\eqref{eqB4} as the number of additional fermionic row vectors that are linearly independent of the row vectors associated with the exclusive fermions. 
Operationally we may apply the following algorithm to find $\bar\delta$. 
For each exclusive fermion with hot spot index $n$, consider the associated loop that is formed according to the ELC. 
If there exists a non-exclusive fermion at hot spot $\bar{n}$ in this loop, then `reserve' this fermion if it has not yet been reserved, and move on to another exclusive fermion. 
The total number of such reserved fermions is then equal to $\bar\delta$. 
The row vectors of those reserved fermions take the form of
$(..,*,\mp v,*,..,*,1,*,...)$,
where $*$ represents some general entries that appear in the non-exclusive fermions.
Therefore, these vectors are linearly independent of the row vectors for the exclusive fermions which are proportional to 
$(..,0,\pm v,0,..,0,1,0,...)$.
Therefore, the row vectors for the additional reserved fermions can be added to the row vectors of the exclusive fermions to form an invertible matrix in \eq{eqB4}.
If there are Yukawa vertices connected to external boson lines, $\bar \delta \geq 1$ because two fermion propagators divided by one of those vertices are in distinct hot spots and belong to a common loop. 
This also applies to diagrams that include Yukawa vertex corrections with internal boson lines because the internal boson propagator can be viewed as an external leg from the point of view of the sub-diagram for the vertex correction.

{\it Case II:}
Now, let us consider the case in which
$\mathbb M$ has a non-trivial kernel: $dK \equiv \dim\ker\mathbb M > 0$.
In this case, the space of internal $x$-momenta cannot be fully spanned by 
$\{ \mathcal{E}_E, \mathcal{E}_{NE}, \mathcal{B}_{NE,x} \}$.
Therefore, some exclusive fermions should be used to span the $x$-momentum,
and some of the components of $\mathcal{B}_{NE,y}$ should be brought in to tame the $y$-momentum integrations.
%
We need $dK$ components of
$\mathcal{B}_{NE,y}$.
After choosing the propagators, let us pick out the set of linearly independent rows (including the first $L_{fm}$ rows that are necessarily linearly independent) of the matrix $\mathbb M$, call this matrix $\mathbb M_2$.
The kinetic energies of the corresponding propagators are related to the mixed-loop momenta through
\begin{align}
    \begin{pmatrix}
        \tilde{\mathcal{E}}_{E}    \\
        \tilde{\mathcal{E}}_{NE}   \\
        \tilde{\mathcal{B}}_{NE,x} \\
        \tilde{\mathcal{B}}_{NE,y}
    \end{pmatrix}
    =
    \mathbb{M}_2\mathbb{P} = 
    \begin{pmatrix}
        v \iden & \mathbb{A}\\
        v \tilde{\mathbb{B}}_1 & \tilde{\mathbb{B}}_2\\
        c \tilde{\mathbb{C}}_1  & 0\\
        0  & c \tilde{\mathbb{C}}_2\\
    \end{pmatrix}
    \begin{pmatrix}
        p_x\\
        \vdots\\
        p_y\\
        \vdots
    \end{pmatrix}.
    \label{B11}
\end{align}

\noindent
In this case, $\iden$ and $\mathbb{A}$ are $L_{fm}\times L_{fm}$ matrix; $\tilde{\mathbb{B}}_{1}$, $\tilde{\mathbb{B}}_{2}$ are $\bar\delta \times L_{fm}$ matrices; $\tilde{\mathbb{C}}_1$ is a $(L_{fm}-\bar\delta-dK)\times L_{fm}$; $\tilde{\mathbb{C}}_2$ is a $dK\times L_{fm}$. 
To extract the Jacobian of this matrix, we use
\begin{align}
    \mathcal{M}_\beta\mathcal{M}_\alpha \mathbb{M}_2
    &=
    \begin{pmatrix}
        \iden & 0 & 0 & 0\\
        0&v^{-1}\iden & 0 & 0\\
        0 & 0 & c^{-1}\iden & 0\\
        0 & 0 & 0 & (vc)^{-1}\iden\\
    \end{pmatrix}
    \begin{pmatrix}
        \iden & 0 & 0 & 0 \\
        - \tilde{\mathbb{B}}_2 \mathbb{A}& \iden & 0 & 0\\
        0 & 0 & \iden & 0\\
        - c \tilde{\mathbb{C}}_2 \mathbb{A} & 0 & 0 & \iden
    \end{pmatrix} 
    \begin{pmatrix}
        v \iden & \mathbb{A}\\
        v \tilde{\mathbb{B}}_1 & \tilde{\mathbb{B}}_2\\
        c \tilde{\mathbb{C}}_1  & 0\\
        0  & c \tilde{\mathbb{C}}_2
    \end{pmatrix}
    \\
    &= 
    \begin{pmatrix}
        \iden & 0 & 0 &0\\
        0&v^{-1}\iden & 0&0\\
        0 & 0 & c^{-1}\iden&0\\
        0 & 0 & 0 & (vc)^{-1}\iden
    \end{pmatrix}
    \begin{pmatrix}
        v\iden & \mathbb A\\
        v\tilde{\mathbb{B}}_1-v\tilde{\mathbb{B}}_2 \mathbb{A} & 0\\
        c \tilde{\mathbb{C}}_1&0\\
        -vc \tilde{\mathbb{C}}_2\mathbb A & 0
    \end{pmatrix}
    =
    \begin{pmatrix}
        v\iden & \mathbb A\\
        \tilde{\mathbb{B}}_1-\tilde{\mathbb{B}}_2 \mathbb{A} & 0\\
        \tilde{\mathbb{C}}_1&0\\
        \tilde{\mathbb{C}}_2&0
    \end{pmatrix}.
\end{align}

\noindent
The determinant of both sides leads to
    $v^{-\bar\delta}c^{-(L_{fm}-\bar\delta-dK)}(vc)^{-dK} \det(\mathbb{M}_2) \sim 1$.
The factor of enhancement that arises from the Jacobian for the mixed loops becomes    
\begin{align}
    \det(\mathbb{M}_2)^{-1} \sim 
    \frac{1}{v^{\bar\delta+dK} c^{L_{fm}-\bar\delta}}.
\end{align}

\begin{figure}
\centering
\begin{subfigure}{0.6\textwidth}
\includegraphics[width=\textwidth]{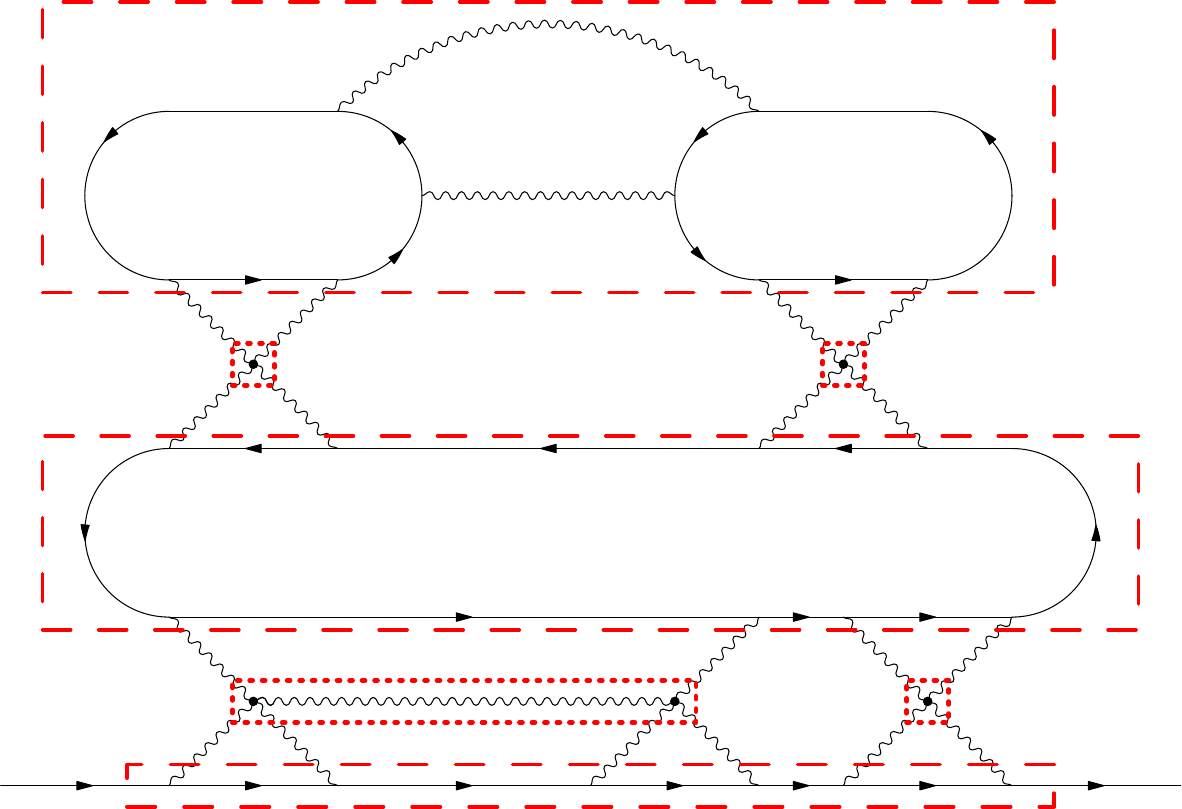}
\end{subfigure}
\caption{
The sub-diagrams enclosed by dashed and dotted boxes represent generalized fermion vertices (GFV) and generalized boson vertices (GBV), respectively.
}
\label{fig9}
\end{figure}

\begin{figure}
\centering
\begin{subfigure}{0.3\textwidth}
\includegraphics[width=\textwidth]{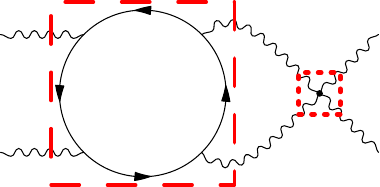}
\caption{$\alpha = 1$, $\xi = 0$}
\end{subfigure}
\begin{subfigure}{0.3\textwidth}
\includegraphics[width=\textwidth]{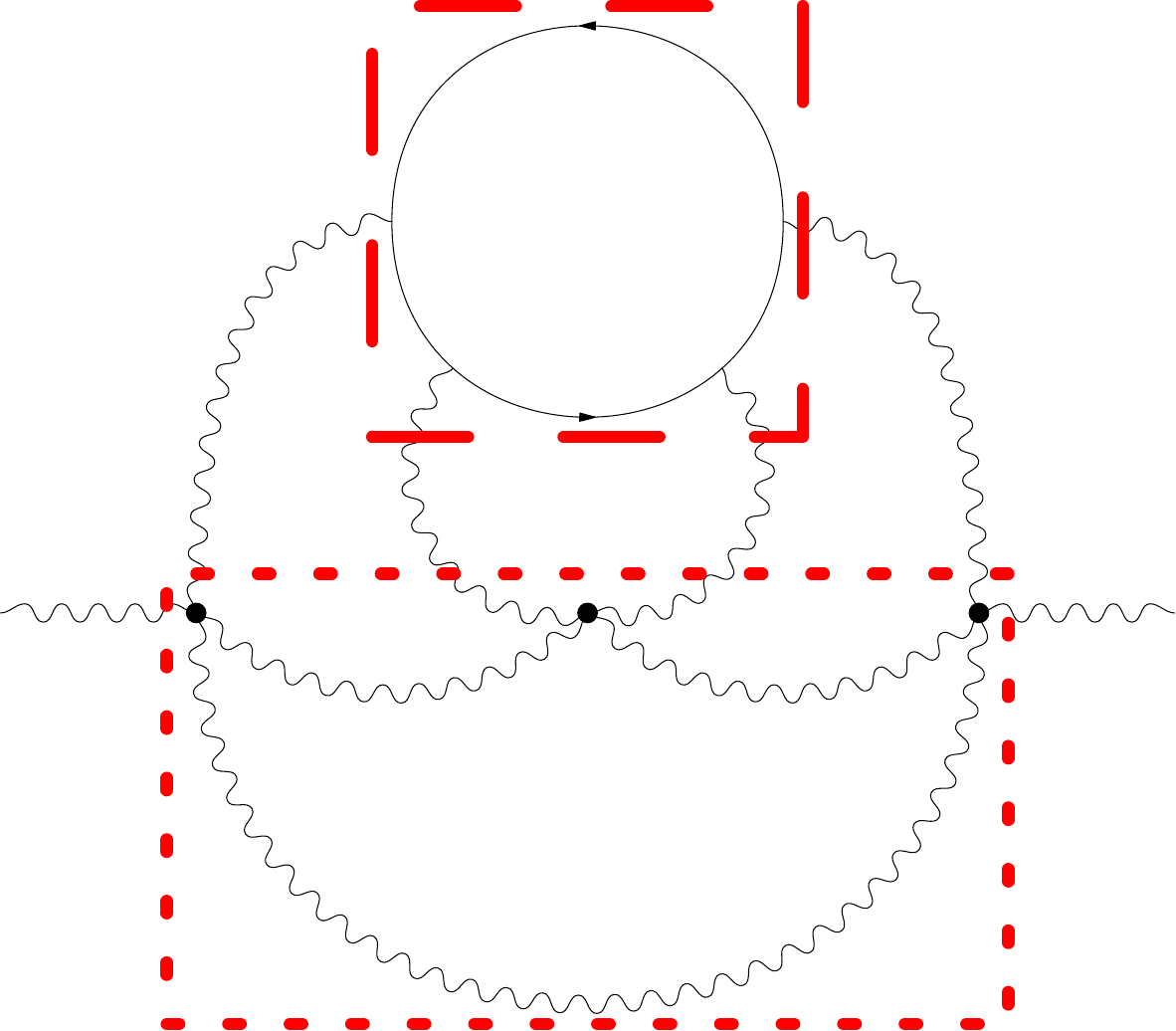}
\caption{$\alpha = 0$, $\xi = 2$}
\end{subfigure}
\begin{subfigure}{0.3\textwidth}
\includegraphics[width=\textwidth]{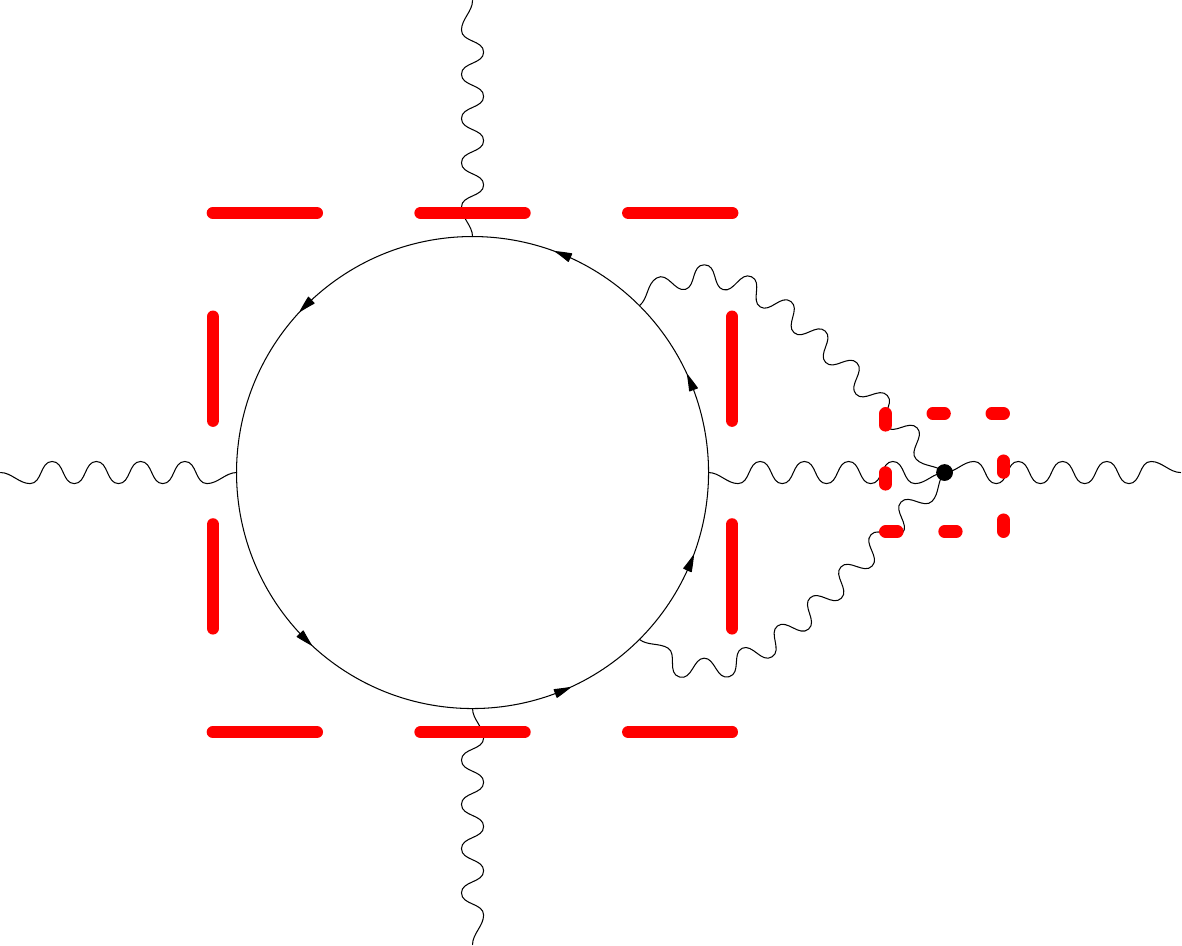}
\caption{$\alpha = 1$, $\xi = -2$}
\end{subfigure}
\caption{
Examples of $\alpha,\xi$.
$\alpha$ is the number of GFVs that are directly connected to some external legs. 
$\xi$ is the total number of legs attached to GBVs minus the total number of legs attached to GFVs. 
}
\label{fig10}
\end{figure}

To understand 
what $dK$ represents,
it is useful to introduce what are called 
generalized fermion vertices (GFV) and generalized boson vertices (GBV). 
 A GFV (GBV) represents a maximally connected subgraph of a diagram that \textit{only} includes the Yukawa vertices (four-boson vertices). 
Some examples of GFVs and GBVs are shown in \fig{fig9} and \fig{fig10}.
This is useful because 
each GFV can contribute only one independent vector to
$\ker\mathbb{M}$.
If a loop momentum is in $\ker\mathbb{M}$,
all internal fermions are placed on the Fermi surface and 
the $x$-components of 
all internal boson 
vanish.\footnote{Without loss of generality, let us assume, as we do throughout the paper, that diagrams only contain fermions from $1\pm$ vertices.}

\begin{figure}
    \centering
    \includegraphics[width=0.3\textwidth]{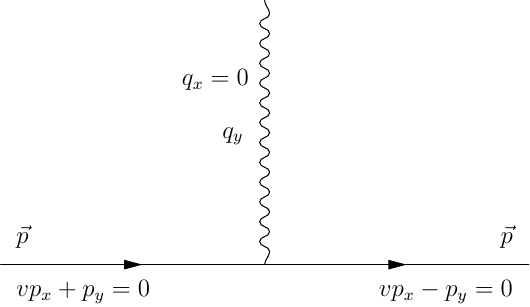}
    \caption{
    $dK$ represents the dimension of the manifold in the space of all internal momenta in which all bosons have zero $x$-momentum and all fermions are on the Fermi surface.
    Since boson has $q_x=0$, $q_y$ fixes the momenta of both incoming and outgoing fermions upto a discrete choices.
While each GFV can contribute at most one dimension to $dK$,
GFV's that are connected to external legs can not.
    }
    \label{fig:scattering}
\end{figure}

A boson with zero $x$-momentum that scatters a fermion from patch $n$ to $\bar{n}$ must carry a $y$ momentum that is fixed by the $x$-momentum of the fermions in order to keep the fermions on the Fermi surface (\fig{fig:scattering}).
That $y$-momentum is equal to the momentum difference between the two points on the Fermi surface connected by the boson. 
Therefore, all fermions in a fermion loop carry the same $x$-momenta and all bosons attached to that fermion loop carry the same $y$-momentum.
Furthermore, the $x$-momentum of the fermions and $y$-momentum of the bosons are tied to each other
from the condition that the fermions are on the Fermi surface.
If two fermion loops are attached by a boson propagator through Yukawa couplings, 
the $x$-momentum of fermions in the two loops should be equal.
Within $\ker\mathbb{M}$,
each GFV is parameterized by single parameter:
the $y$-momenta of all bosons participating in the GFV.
Namely, each GFV can supply at most one vector to $\ker\mathbb{M}$.
Therefore, $dK$ is less than or equal to the number of GFVs, $N_{GFV}$. 
Generally, $dK \neq N_{GFV}$. 
If an external line is connected to a GFV, 
the external momentum determines the $x$ momentum that runs within that fermion loop and the $y$ momentum that internal bosons carry.
Therefore, all GFVs connected to external legs do not contribute to $dK$.
This leads to an inequality for $dK$,
\begin{align}
    dK \le N_{GFV}-\alpha,
    \label{AlphaDk}
\end{align}
where $\alpha$ is the number of GFVs that are directly connected to the external momentum. 
If $V_u=0$, then 
$V_u=L_b=dK=0$ 
because the entire diagram is a GFV connected to external lines.

Using this information on $dK$, we can put a bound on $V_u-L_b-dK$ for general diagrams.
For this, we first separate all generalized boson vertices (GBVs) from graphs, by cutting all boson propagators which connect Yukawa vertices and the four-boson vertices in GBVs.
Upon doing so, the remaining graph will be a disjoint subset of GFVs and GBVs. 
The newly created external edges connect vertices of different types: GFVs with GBVs.
Let $V_u^{(i)}, I^{(i)}, E^{(i)}, L_b^{(i)}$ be the numbers of vertices, internal propagators, external legs, and internal loops, respectively, in the $i$-th GBV.
For each $i$, we have  
\begin{align}
    \left\{\begin{aligned}
        &V_u^{(i)} - I^{(i)} + L_b^{(i)} = 1\\
        &E^{(i)} + 2I^{(i)} = 4V_u^{(i)}
    \end{aligned}\right\}
    \implies
    2(V_u^{(i)}-L_b^{(i)}) = E^{(i)}-2.
    \label{eqDeltaSingleGBV}
\end{align}
Adding this over all 
generalized boson vertices,
we obtain
\begin{align}
    2(V_u-L_b) = E_{GBV} - 2N_{GBV}.
    \label{eq:B17}
\end{align}
Here, $E_{GBV}$ describes the total number of external legs to the union of all GBVs in the diagram.
We use the fact that 
$\sum_i (V_u^{(i)}-L_b^{(i)}) = V_u - L_b$
because all bosonic vertices and bosonic loops are parts of GBVs.
Using the fact that $E_{GBV} \ge 4N_{GBV}$ and $E_{GFV} \ge 4N_{GFV}$ for diagrams that do not include self-energies as sub-diagram,
and denoting
$\xi \equiv E_{GBV} - E_{GFV}$, 
we obtain the following inequality:
\begin{align}
V_u-L_b  &=\frac{1}{2} 
\left[ E_{GBV}   - 2 N_{GBV} + \frac{1}{2} \left(
    \xi - E_{GBV} + E_{GFV} \right) \right]
    \\&
=\frac{1}{4} \left( E_{GBV} - 4N_{GBV} + \xi + E_{GFV}  \right)
    \\&
    \ge N_{GFV} + \frac\xi4.
    \label{eqBeforeDeltaRelation}
\end{align}
By combining 
this inequality with Eq.~\eqref{AlphaDk}, 
we obtain
a lower bound of $V_u-L_b-dK$:
\begin{align}
    V_u-L_b - dK \ge \left\lceil{\alpha + \frac\xi4} \right\rceil.
    \label{eqDeltaRelation}
\end{align}
%
If $\alpha=0$, the diagram admits only four-boson external vertices and as a result $\xi\ge0$, so that $\alpha+\xi/4\ge0$.
When $\alpha\ge1$, then there may be external Yukawa vertices, in which case $\xi \ge -E \ge -4$,\footnote{This is true for fermion self-energies, boson self-energies, and four-boson vertex corrections of relevance to this work.} so that $\alpha+\xi/4\ge0$.
We see that in both cases, we find
\begin{align}
    V_u-L_b-dK\ge0.
    \label{nonNegativeDeltaRelation}
\end{align}
In Figs.~\ref{fig9},~\ref{fig10},
we show the values of
$N_{GFV}, N_{GBV},\alpha,\xi$ 
for a few examples.

In summary, the overall magnitude of a general diagram 
in \eq{eqGenericScaling}
is bounded by
\begin{equation}
\ub \leq    
 \frac{g^{V_g}u^{V_u}}{v^{\bar\delta+dK}c^{L_{fm}-\bar\delta+2L_b}}  
 = c^{\frac{E-2}{2}+V_u-L_b-dK}s^{\bar\delta+dK}y^{V_g/2}\kappa^{V_u}.
\label{eq:GU}
\end{equation}
Here $E$ is the number of external legs.
$V_u$ ($V_g$) is the number of quartic 
 (Yukawa) vertices.
$L_b$ is the number of boson loops.
$\bar \delta$ and $dK$ are non-negative integers that are kinematically determined 
in each diagram.

\subsubsection{
Bound on \texorpdfstring{$\mathcal{F} (k_i)$}{F(ki)}
from a decoration scheme}
\label{sec:decorscheme}

$ \mathcal{F} (k_i)$ in 
\eq{eqGenericScaling}
determines how quantum corrections depend on the external momenta $k_i$.
For $n=1,4,6,7$, 
counter terms are independent of spatial momentum, which implies
$ \mathcal{F} (k_i) \sim 1$
and
$Z_{n,1} \leq \ub$.
For $n=2,3,5$,
the counter terms are linearly or quadratically dependent on the external momentum.
In the limit that the velocities are small,
the dependence on the external momentum can be suppressed by additional powers of $v$ and $c$ as
\begin{align}
 \mathcal{F} (k_i)
& \sim  
\begin{cases}
v^{a_2} c^{b_2} k_x 
~&\mbox{for}~ n=2\\
v^{a_3} c^{b_3} k_y
~&\mbox{for}~ n=3\\
v^{a_5} c^{b_5} (k_x^2 + k_y^2)   
~&\mbox{for}~ n=5\\
\end{cases},
\label{eq:Fvc}
\end{align}
where $a_n$ and $b_n$ are exponents that are to be determined in each diagram.
Because 
$Z_{2,1}$,
$Z_{3,1}$
and
$Z_{5,1}$
are multiplicative coefficients of
$v k_x, k_y$
and
$c^2 (k_x^2 + k_y^2)$ 
in the kinetic terms,
we have
\bqa
Z_{2,1} \sim \ub v^{a_2-1} c^{b_2}, ~
Z_{3,1} \sim \ub v^{a_3} c^{b_3}, ~
Z_{4,1} \sim \ub v^{a_4} c^{b_4-2}.
\eqa
The general upper bound derived in this section does not capture potential enhancements by $\log 1/s$ that arise in some of diagrams.
In Sec. \ref{app:LogZ2}, we discuss the criterion for the presence of such logarithmic enhancements.

In this subsection,
we derive a general lower bound on $a_n$ and $b_n$ in \eq{eq:Fvc}.
This is achieved through a simple decoration scheme for Feynman graphs. 
%
For the sake of simplicity, we assume that the diagram only includes fermions from hot spots $1\pm$. 
Diagrams that include fermions from hot spots different than those that are directly connected by $\vec Q_{AF}$ are generally suppressed by higher powers of $v$.
The decoration scheme proceeds as follows: 
fermion propagators at hot spot $1+$ are drawn with solid lines, while those at hot spot $1-$ are represented by dashed lines; 
each boson propagator is represented by a double line, 
one of which is a solid line and the other is a dashed line. 
Yukawa vertices are drawn by connecting the solid (dashed) line of the fermion with the solid (dashed) line of the boson. 
The four-boson vertex is represented as a single point, where all four solid and dashed lines converge.
One crucial aspect is that each vertex, whether Yukawa or four-boson, always has an even number of solid and dashed lines emerging from it. 
This is a consequence of momentum conservation.
Graphically, 
it implies a ``conservation of line type.'' 
For example, if a solid (dashed) line enters a vertex, it must leave the vertex as a solid (dashed) line. 
Furthermore, this implies that it is always possible to enter and leave a graph via external legs only through one type of internal lines.
If a diagram has two external fermion legs, the decorated lines of these external legs are connected by a single continuous line of the same type. 
For the boson self-energy, the solid (dashed) external lines can be joined with each other only through lines of the same type.

\begin{figure}[htpb!]
	\centering
	\begin{subfigure}{0.3\textwidth}
        \includegraphics[width=\textwidth]{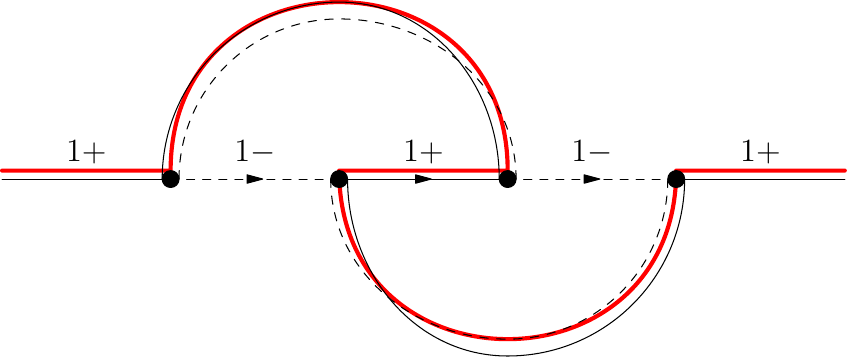}
		\caption{}
        \label{decorFSE}
	\end{subfigure}
        \begin{subfigure}{0.3\textwidth}
            \includegraphics[width=\textwidth]{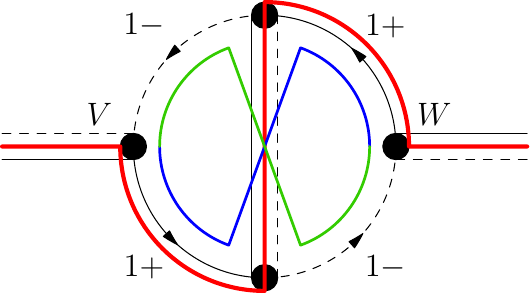}
            \caption{}
            \label{decorBSE}
	\end{subfigure}
    \caption{
A decoration scheme that `guides' the external momentum that makes it manifest how each diagram depends on the external momentum in the limit that $v$ and $c$ are small.
The thick red line denotes the path along which the external momentum flows.
In figure (b), 
the paths $\pi_1$ (blue) and $\pi_2$ (green) connect the vertices $V$ and $W$, forming a loop.
}
 \end{figure}

This decoration scheme, paired with line continuity, allows us to guide the external momentum along a particular line type. 
For instance, consider a fermion self-energy diagram where an external fermion propagator is at hot spot $1+$
(see \fig{decorFSE}). 
By tracing this solid line to its end, we reach the other external fermion leg. 
If we guide the external momentum, $k$, to follow this path, it traverses through some boson propagators and fermion propagators of hot spot $1+$ only. 
This implies 
\begin{equation}
 \mathcal{F} (k_i) \sim
ck_x, ck_y, 
\varepsilon_1(\vec k) 
\label{eq:Ffermionself}
\end{equation}
for the fermion self-energy at hot spot $1$.
$\varepsilon_1(\vec k)$ gives rise to the leading contribution to
$Z_{2,1}$ and $Z_{3,1}$ 
individually,
which are order of $\ub$.
However,  $\varepsilon_1(\vec k)$
do not contribute to the difference between
$Z_{2,1}$ and $Z_{3,1}$,
which is further suppressed by $s=c/v$ as
\bqa
Z_{2,1} - Z_{3,1} \sim s~ \ub,
\label{eq:Z2m3}
\eqa
where the leading contribution arises through 
$c k_x = s \frac{\varepsilon_1(\vec k) + \varepsilon_3(\vec k)}{2}$.

For the boson self-energy, there are two separate cases.
In the first case, the external momentum can be directed to flow only through boson propagators.
In this case, the external momentum appears only through $c \vec k$ and 
\begin{equation}
 \mathcal{F} (k_i) \sim
c^2 (k_x^2 + k_y^2).
\label{eq:Fbosonself0}
\end{equation}
In the second case, the external momentum traverses through some fermion propagators, and the quantum correction exhibits stronger dependence on the external momentum.
In this case, let
us represent two external boson legs by a pair of solid and dashed external lines
(see \fig{decorBSE}). 
Due to line continuity, the solid lines connect continuously, and so do the dashed lines. 
Now, let's label the two external vertices as $V$ and $W$. 
We denote a path from $V$ to $W$ along the solid lines as $\pi_1$ 
and a path from $W$ to $V$ along the dashed lines as $\pi_2$. 
This guarantees that $\pi_1$ ($\pi_2$) only includes fermion propagators from hot spot $1+$ ($1-$).
If $k$ denotes the external momentum, we direct $k$ to flow through $\pi_1$ only. 
Furthermore, let's consider a loop made of $\pi_1 \cup \pi_2$, from $V$ back to $V$ via $W$, and denote the loop momentum by $p$. 
With this, we end up choosing a loop in which the 
the external momentum $k$ and the loop momentum $p$ appear together 
in the fermion propagators of hot spot $1+$
while the fermion propagators of hot spot $1-$ 
 only depend on $p$ 
 and possibly other loop momenta but not on $k$.
This enables us to shift the loop momentum $p$ as 
\begin{align}
p_x \to p_x - \frac{1}{2v}\varepsilon_1(k), \hspace{1cm}
p_y \to p_y - \frac{1}{2}\varepsilon_1(k)
\end{align}
to make all fermion propagators independent of $k$.
Instead, $k$ appears only in the boson propagators through
$\Delta \equiv c \varepsilon_1(\vec k)/v$ 
and 
\begin{equation}
 \mathcal{F} (k_i) \sim
s^2 (k_x^2 + k_y^2).
\label{eq:Fbosonself}
\end{equation}

In the next sections, 
we apply the general bound derived in this section 
to show that the $\epsilon$-expansion is controlled for
the $N=1$ and $N=2$ theories. 
%
%
%
In particular, 
we show that
 all half-skeleton diagrams that are not included in the present analysis are sub-leading,
 which automatically implies that non-half skeleton diagrams are also suppressed.

\subsection{Counterterms in the \texorpdfstring{$Z_2$}{Z2} Theory}
\label{app:Z2}

In this section,
we prove the control of the $\epsilon$-expansion in the $Z_2$ theory.
%
%
For each counter term $Z_{n,1}$, the relevant 
 half-skeleton diagrams are partitioned into sub-classes based on their kinematic properties.
Each class is analyzed one by one through the upper bound derived in the previous section.
Finally, we discuss potential logarithmic enhancements 
that are not captured in \ref{eq:GU}.
Before we begin, let us summarize the symbols introduced in the previous section and used in the following discussions.

\begin{itemize}

    \item 
    $\bar\delta$ :
    the maximal number of non-exclusive fermion propagators which are 
    independent of the exclusive fermion propagators 

    \item 
    $dK=\dim\ker\mathbb M_1$ :
    the number of mixed loops whose internal momenta are not fixed when all fermions are set to be on the Fermi surface and all bosons have vanishing $x$-momenta

    \item 
    $c^{\frac{E-2}{2}+V_u-L_b-dK} s^{\bar\delta+dK} y^{V_g/2} \kappa^{V_u}$ :
    the generic upper bound that sets the overall magnitude of a diagram

    \item 
    Generalized Fermion Vertex (GFV) : maximally connected subdiagrams containing only Yukawa vertices

    \item 
    Generalized Boson Vertex 
    (GBV) : maximally connected subdiagrams containing only four-boson vertices

    \item
    $\alpha$ :
    the number of GFVs that contain at least one external vertices

    \item 
    $\xi = E_{GBV}-E_{GFV}$ :
    the difference in the numbers of external legs attached to the generalized boson and fermion vertices 
    (only external legs contribute to $\xi$
    because internal lines contribute equally to
    $E_{GBV}$ and $E_{GFV}$)
    

    \item
    $V_u-L_b-dK \ge \lceil\alpha+\xi/4\rceil$ :
    the inequality that holds for all half-skeleton diagrams
    with $V_u>0$

\end{itemize}

\subsubsection{\tps{$Z_{6,1}$}{Z6}}


At the modified one-loop level,
there is only one contribution
(\fig{fig:Yukawa1Loop}) 
to $Z_{6,1}$,
which is of order $\mathcal O(ys)$.
Higher-order quantum corrections are 
partitioned into two sub-classes based on how the external boson legs are connected to the diagram: 
    $u$-type -- diagrams with the external boson attached to a boson vertex;
    and
    $g$-type -- diagrams with the external leg attached to a Yukawa vertex.
Beyond the modified one-loop level, diagrams of $g$-type must have $V_g\ge5$.

\begin{center}
\begin{forest}
    [$Z_{6,1}$,rectangle,draw
[ {\hyperref[controlZ2_baab]{(a)} $u$-type} ] 
[ {\hyperref[controlZ2_baac]{(b)} $g$-type, $V_g\ge5$} ]
    ]
\end{forest}
\end{center}

\paragraph{u-type.}\label{controlZ2_baab}

If the external boson leg is connected to the diagram through one four-boson vertex, then 
$\xi\equiv E_{GBV}-E_{GFV}=-1$. 
Moreover, since the external fermion legs must be continuously connected along a fermion line, this means that $\alpha=1$.
With $\alpha=1, \xi=-1$, Eq.~\eqref{eqDeltaRelation} implies that $V_u-L_b-dK\ge1$. 
With $\bar\delta\ge0$ and $dK\ge0$,
the counterterm scales as
\begin{align}
Z_{6,1}\sim\frac{1}{g}\Bigl(
    c^{\frac{E-2}{2}+V_u-L_b-dK}s^{\bar\delta+dK}y^{V_g/2}\kappa^{V_u}
    \Bigr)
    \le
    \frac{(yc)^{3/2}}{g}
    =yc.
\end{align}
This is suppressed by $c/s \ll 1$ compared to the modified one-loop correction.

\paragraph{g-type \& $V_g\ge5$.}\label{controlZ2_baac}

In this case,
the external boson leg is connected to the diagram through a Yukawa vertex.
Call the two fermion propagators adjacent to the external boson propagator $\alpha,\beta$. 
Any loop that goes through one of these propagators, must also go through the other. 
Moreover, the hot spot index of $\alpha$ and $\beta$ is different. 
We choose one of these propagators to be an exclusive propagator and the other to be a reserved propagator. 
%
Therefore, $\bar\delta\ge1$.
Let us also note, when the external boson is connected to a Yukawa vertex, $V_u-L_b-dK \geq 0$.\footnote{
For $V_u=0$, $L_b=dK=0$ 
from \eq{AlphaDk}
and the inequality trivially follows. 
For $V_u>0$, $\alpha\ge1$ and $\xi\equiv E_{GBV}-E_{GFV}=-3$, 
which implies 
$V_u-L_b-dK\ge\lceil \alpha+\xi/4\rceil \ge\lceil 1-3/4\rceil = 1$.}
Therefore, the counterterm scales as 
\begin{align}
    Z_{6,1}\sim\frac{1}{g}\Bigl(
    c^{\frac{E-2}{2}+V_u-L_b-dK}s^{\bar\delta+dK}y^{V_g/2}\kappa^{V_u}
    \Bigr)
    \le
    \frac{y^{5/2}c^{1/2}s}{g}
    = y^2 s.
\end{align}
This is suppressed by powers of $y$ compared to the modified one-loop level.

\subsubsection{\tps{$Z_{1,1}$}{Z1}}

The leading order contribution is generated from \fig{fig:FermionSE1Loop}, which is $Z_{1,1} \sim y$.
Diagrams beyond the modified one-loop order are divided into two sub-classes.

\begin{center}
\begin{forest}
    [$Z_1$,rectangle,draw
        [ {\hyperref[controlZ2_bb]{(a)} $V_u>0$} ] 
        [{\hyperref[controlZ2_bc]{(b)} $V_u=0$ \& $V_g\ge4$} ]
    ]
\end{forest}
\end{center}

\paragraph{
$V_u > 0$.}
\label{controlZ2_bb}

If $V_u > 0$, $\alpha=1$ and $\xi=-2$. 
Using \eq{eqDeltaRelation}, we find $V_u-L_b-dK\ge1$.
Then, together with $\bar\delta\ge0$ and $dK\ge0$, the generic upper bound takes the form:
\begin{align}
        Z_{1,1} \sim c^{\frac{E-2}{2}+V_u-L_b-dK} s^{\bar\delta+dK} y^{V_g/2} \kappa^{V_u} \le c.
\end{align}
This is suppressed by $c$ compared to the modified one-loop contribution.

\paragraph{
$V_u = 0$ \& $V_g\ge4$.
}\label{controlZ2_bc}

If $V_u=0$, $L_b=dK=0$. 
Since $\bar\delta\ge0$ and $V_g\ge4$ for all diagrams with more than one loop, the generic upper bound takes the form:
\begin{align}
       Z_{1,1} \sim c^{\frac{E-2}{2}+V_u-L_b-dK} s^{\bar\delta+dK} y^{V_g/2} \kappa^{V_u} \le y^2.
\end{align}
This is suppressed by powers of $y$ compared to the modified one-loop level.




\subsubsection{\tps{$Z_{2,1}, Z_{3,1}$}{Z2,Z3}}\label{appControlZ2_Z23}

At the modified one-loop order,
$Z_{2,1} = - Z_{3,1} \sim yc$. 
%
In the following, we analyze higher-order contributions to $Z_{2,1}$ and $Z_{3,1}$ first, and then analyze $Z_{2,1}-Z_{3,1}$ separately.
This is because the flow of $s$ within the line of quasi-fixed points is solely determined by 
$Z_{2,1}-Z_{3,1}$,
and the higher-order contributions to
$Z_{2,1}-Z_{3,1}$ are generally much smaller than individual $Z_{2,1}$ and $Z_{3,1}$.
When analyzing the higher-order contributions to $Z_{2,1}$ and $Z_{3,1}$ individually, we partition diagrams into three groups: 
{\hyperref[controlZ2_cb]{(a)}} $V_u>0$ and
{\hyperref[controlZ2_cc0]{(b)}} $V_u=0$ \& $V_g = 4$ \& $\bar\delta\ge1$,
{\hyperref[controlZ2_cc1]{(c)}} $V_u=0$ \& $V_g \ge 6$ \& $\bar\delta\ge1$.
The reason why $V_g < 4$ and $(V_g\ge4\;\&\;\bar\delta=0)$ diagrams are not considered is because they are either non-half-skeleton diagrams or coincide with diagrams already included at the modified one-loop level.
\footnote{ In other words, diagrams with vertex corrections have $\bar \delta > 0$.}
When analyzing $Z_{2,1}-Z_{3,1}$, we 
 further divide the case $(V_u=0\;\&\;V_g\ge4)$ into two sub-cases: {\hyperref[controlZ2_cf]{(e)}} $\bar\delta\ge2$ and 
{\hyperref[controlZ2_cg]{(f)}} $\bar\delta=1$.

\begin{center}
\begin{forest}
[${Z_{2,1},Z_{3,1}}$,rectangle,draw
    [{\hyperref[controlZ2_cb]{(a)} $V_u>0$}]
    [{$V_u=0$ \& $\bar\delta\ge1$}
        [{\hyperref[controlZ2_cc0]{(b)} $V_g=4$}]
        [{\hyperref[controlZ2_cc1]{(c)} $V_g\ge6$}]
    ]
]
\end{forest}
\begin{forest}
    [${Z_{2,1}-Z_{3,1}}$,rectangle,draw
       [{\hyperref[controlZ2_ce]{(d)} $V_u>0$}]
       [{$V_u=0$ \& $V_g\ge4$}
                [{\hyperref[controlZ2_cf]{(e)} $\bar\delta\ge2$}]
                [{\hyperref[controlZ2_cg]{(f)} $\bar\delta=1$}  ]
            ]]
\end{forest}
\end{center}

\paragraph{$Z_{2,1}, Z_{3,1}$: $V_u>0$.}\label{controlZ2_cb}
With $V_u > 0$, $\alpha=1$ and $\xi=-2$, 
\eq{eqDeltaRelation} implies $V_u-L_b-dK\ge1$.
For half-skeleton diagrams that do not have self-energy insertions, we have $V_g\ge4$.
%
%
Generically, $\mathcal F(v,c;\vec k)$ in \eq{eq:Fvc} does not give any additional suppression, and together with $\bar\delta\ge0$ and $dK\ge0$, the contribution to $Z_{2,1}$ and $Z_{3,1}$ individually scales as:
\begin{align}
    Z_{2,1},Z_{3,1} \sim \ub\sim c^{\frac{E-2}{2}+V_u-L_b-dK} s^{\bar\delta+dK} y^{V_g/2} \kappa^{V_u} \le cy^2.
\end{align}
This is suppressed by $y$ compared with the modified one-loop level.

\paragraph{$Z_{2,1}, Z_{3,1}$: $V_u=0$ \& $V_g=4$ \& $\bar\delta\ge1$.}\label{controlZ2_cc0}

There is only one diagram, \fig{fig:FSE_2Loop}, that satisfies these conditions. 
In App. \ref{app:LogZ2},
it is shown that 
$\mathcal F(k) \sim s k_y, s v k_x$ for this diagram.
%
Combining this with the generic upper bound, the counterterm scales as
\begin{align}
    Z_{2,1},Z_{3,1} \sim s \cdot \Bigl( c^{\frac{E-2}{2}+V_u-L_b-dK} s^{\bar\delta+dK} y^{V_g/2} \kappa^{V_u} \Bigr) 
    \le y^2s^2.
\end{align}
As is shown in Sec. \ref{app:LogZ2}, there is no additional enhancement by $\log (1/s)$ in this diagram.
Therefore, the two-loop correction is suppressed by a factor of $(\, \log(1/s)\, )^{-1/2}$ compared with the modified one-loop level near the fixed point.

\paragraph{$Z_{2,1}, Z_{3,1}$: $V_u=0$ \& $V_g\ge6$ \& $\bar\delta\ge1$.}\label{controlZ2_cc1}

For $V_u=0$, $L_b=dK=0$.
With $\bar\delta\ge1$ and $V_g\ge6$, 
the counterterm then generically scales as
\begin{align}
    Z_{2,1}, Z_{3,1} \sim
    c^{\frac{E-2}{2}+V_u-L_b-dK} s^{\bar\delta+dK} y^{V_g/2} \kappa^{V_u} \le y^3s.
    \label{z2z3_anomalous}
\end{align}
While these counterterms vanish at the proposed fixed point, these contributions are larger than the modified one-loop quantum correction.
In the main text, we show that including these quantum corrections do not alter the modified one-loop fixed point and its stability.
It is noted that the scaling in \eq{z2z3_anomalous} is saturated when the contribution from $\mathcal F(k)$ 
arises from $\varepsilon_1(k)$. 
This is because the external momentum can be directed to flow only through the boson propagators and the fermion propagators of hot spot $1+$.
Therefore, the leading contribution from this class of diagrams 
 gives $Z_{2,1} = Z_{3,1} = A y^3s$, where $A$ is a coefficient of order unity.

\paragraph{$Z_{2,1}-Z_{3,1}$: $V_u>0$.}\label{controlZ2_ce}

As is shown earlier, both $Z_{2,1}, Z_{3,1} \lesssim cy^2$, which is suppressed by $y$ compared with the modified one-loop level.

\paragraph{$Z_{2,1}-Z_{3,1}$: $V_u=0$ \& $V_g\ge4$ \& $\bar\delta\ge2$.}\label{controlZ2_cf}

The generic upper bound for this class of diagrams scale as $\ub \sim s^2$.
For $Z_{2,1}-Z_{3,1}$, $\mathcal F(k)$ contributes an additional factor of $s$
as is shown in \eq{eq:Z2m3}.
Combining this with the generic upper bound, the counterterm scales as:
\begin{align}
    Z_{2,1}-Z_{3,1} \sim
    s \cdot \Bigl(c^{\frac{E-2}{2}+V_u-L_b-dK} s^{\bar\delta+dK} y^{V_g/2} \kappa^{V_u}\Bigr) \le y^2 s^3.
\end{align}
This is suppressed by $y s$ compared with the modified one-loop level.

\paragraph{$Z_{2,1}-Z_{3,1}$: $V_u=0$ \& $V_g\ge4$ \& $\bar\delta=1$}\label{controlZ2_cg}

The generic upper bound for this class of diagrams scales as $\ub \sim y^2s$.
Combined with the extra suppression factor of $s$ from $\mathcal F(k)$, the counterterm scales as:
\begin{align}
    Z_{2,1}-Z_{3,1} \sim
    s \cdot \Bigl(c^{\frac{E-2}{2}+V_u-L_b-dK} s^{\bar\delta+dK} y^{V_g/2} \kappa^{V_u}\Bigr) \le y^2s^2.
\end{align}
As is shown in Sec. \ref{app:LogZ2}, there is no additional enhancement by $\log (1/s)$ in these diagrams.
Therefore, the higher-order corrections are suppressed by $y (\, \log(1/s)\, )^{-1/2}$ compared with the modified one-loop level near the proposed fixed point.

\subsubsection{\tps{$Z_{4,1}$}{Z4}}\label{appControlZ2_Z4}

\begin{figure}[htpb]
    \centering
    \begin{subfigure}{0.4\textwidth}
        \centering

\includegraphics[width=0.8\textwidth]{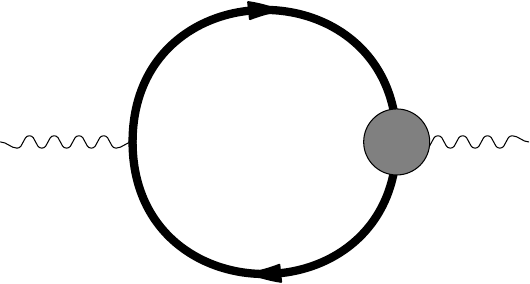}

        \caption{}
        \label{fig:BosonSkeletonPt1}
    \end{subfigure}
    \begin{subfigure}{0.4\textwidth}
        \centering

\includegraphics[width=0.8\textwidth]{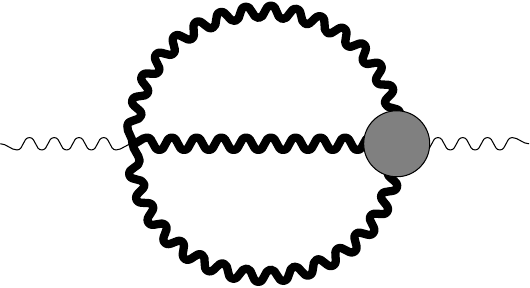}

        \caption{}
        \label{fig:BosonSkeletonPt2}
    \end{subfigure}
    \caption{
    Boson self-energies can be partitioned into two non-disjoint subsets.
    (a) and (b) correspond to the classes of diagrams where at least one external leg is connected to the diagram through the Yukawa vertex and the quartic vertex, respectively.
    }
    \label{fig:BosonSkeletonAllPt}
\end{figure}

The leading contribution to $Z_{4,1}$ comes from \fig{fig:BosonSE2LoopBoson},
which is of order $\mathcal O(\kappa^2) = \mathcal O(\epsilon)$.
Higher-order diagrams that contribute to $Z_{4,1}$ 
 is made of a non-disjoint union of $\mathcal S_1$ and $\mathcal S_2$. 
In this section, we use $\mathcal S_1$ to denote diagrams where there is at least one external Yukawa vertex (cf. \fig{fig:BosonSkeletonPt1}), namely, the diagrams where an external leg is attached to a Yukawa vertex.
We use $\mathcal S_2$ to denote diagrams with at least one external four-boson vertex (\fig{fig:BosonSkeletonPt2}).
In addition to the above partition, we shall partition diagrams in the class $\mathcal S_1$ into two more groups:
{\hyperref[controlZ2_db]{(a)} $V_u>0$} and
{\hyperref[controlZ2_dc]{(b)} $V_u=0\;\&\;V_g \ge 6$}.
%
Similarly, we partition the diagrams in class $\mathcal S_2$ into two sub-classes:
{\hyperref[controlZ2_dd]{(c)} $V_g>0$} and
{\hyperref[controlZ2_de]{(d)} $V_g=0$\;\&\;$V_u\ge3$}.
Note that the lower bound on $V_g$ in 
{\hyperref[controlZ2_dc]{(b)}}
and 
{\hyperref[controlZ2_de]{(d)}}
highlights the requirement on the number of Yukawa vertices for diagrams beyond the modified one-loop level.

\begin{center}
\begin{forest}
    [${Z_{4,1}}$,circle,draw
       [${\mathcal S_1}$
            [{\hyperref[controlZ2_db]{(a)} $V_u>0$}]
            [{\hyperref[controlZ2_dc]{(b)} $V_u=0 \;\&\;V_g \ge 6$}]]
       [${\mathcal S_2}$
            [{\hyperref[controlZ2_dd]{(c)} $V_g>0$}]
            [{\hyperref[controlZ2_de]{(d)} $V_g=0$\;\&\;$V_u\ge3$}]
        ]
    ]
\end{forest}
\end{center}

\paragraph{$\mathcal S_1$ \& $V_u>0$.}\label{controlZ2_db}
Diagrams in $\mathcal S_1$ have an external Yukawa vertex, 
 which ensures $\bar\delta\ge1$. 
Using \eq{eqDeltaRelation} with $\alpha\ge1$ and $\xi \ge -2$ gives 
$V_u-L_b-dK \ge 1$.
The counterterm then scales as:
\begin{align}
        Z_{4,1}\sim
        c^{\frac{E-2}{2}+V_u-L_b-dK} s^{\bar\delta+dK} y^{V_g/2} \kappa^{V_u}
        \le sc.
\end{align}
This is suppressed by $sc$ compared with the modified one-loop level.

\paragraph{$\mathcal S_1$ \& $V_u=0$ \& $V_g\ge6$.}\label{controlZ2_dc}
For $V_u=0$, $L_b=dK=0$ which implies that $(E-2)/2+V_u-L_b-dK=0$.
Since these diagrams have external Yukawa vertices, $\bar\delta\ge1$. 
Therefore, 
\begin{align}
        Z_{4,1}\sim
        c^{\frac{E-2}{2}+V_u-L_b-dK} s^{\bar\delta+dK} y^{V_g/2} \kappa^{V_u}
        \le y^3s.
\end{align}
This set of diagrams is suppressed by 
$s$ compared with the two-loop boson self-energy.

\paragraph{$\mathcal S_2$ \& $V_g>0$.}\label{controlZ2_dd}
Diagrams in $\mathcal S_2$ have an external boson vertex.
Because $V_g>0$, we 
 have either
$(\alpha,\xi) = (0,2)$\footnote{This will happen if there are no external Yukawa vertices.} or $(\alpha,\xi) = (1,0)$\footnote{This will happen if there is one external Yukawa vertex.}.
This implies 
$V_u-L_b-dK\ge1$.
Therefore,
\begin{align}
        Z_{4,1}\sim 
        c^{\frac{E-2}{2}+V_u-L_b-dK} s^{\bar\delta+dK} y^{V_g/2} \kappa^{V_u}
        \le c.
\end{align}
This is suppressed by $c$ compared with the modified one-loop level.

\paragraph{$\mathcal S_2$ \& $V_g=0$ \& $V_u\ge3$.}\label{controlZ2_de}
If $V_g=0$, then the entire diagram is of order $\kappa^{V_u}$.
%
With $V_u\ge3$, the counterterm scales as
\begin{align}
    Z_{4,1} \sim \kappa^{V_u} \le \kappa^3
\end{align}
This is suppressed at least by $\kappa$ compared with the modified one-loop level contribution to $Z_{5,1}$.\footnote{Recall, there are two diagrams, cf. \fig{fig:2LoopDiagrams}, that are necessary to include to tame the fixed point.}

\subsubsection{\tps{$Z_{5,1}$}{Z5}}

The leading contribution to $Z_{5,1}$ comes from \fig{fig:BosonSE2LoopFermion} which is of order $y$
at the fixed point.
We first classify higher-order diagrams based on the types of external vertices. 
There are three possibilities:
    $(g,g)$ -- two external Yukawa vertices;
    $(g,u)$ -- one external Yukawa vertex and one external four-boson vertex;
    $(u,u)$ -- two external four-boson vertices.
Type $(g,g)$ diagrams must have $\bar\delta\ge1$. 
Of type $(g,g)$,
there is only one half-skeleton diagram 
with $\bar\delta=1$ 
(\fig{fig:BosonSE1Loop}),
but it vanishes exactly. 
Therefore, we only need to consider diagrams with $\bar \delta \geq 2$.
For these diagrams, it follows that
$\bar\delta+dK \ge \bar\delta\ge2$. 
Moreover, $L\ge3$ beyond the modified one-loop level.
Diagrams of type $(u,u)$ 
 can be split into sub-classes based on $V_g=0$ and $V_g>0$. 
To go beyond modified one-loop, diagrams with $V_g=0$ must also have $V_u\ge3$.
Furthermore,
diagrams with $V_g>0$ must have an internal fermion loop,
which implies that $V_g\ge4$.\footnote{
Due to global momentum conservation, each fermion loop must contain an even number of internal fermion legs and thus an even number of Yukawa vertices.}
We may divide the $(u,u)$-class with $V_g\ge4$ diagrams further into two sub-cases:
    $c$-type -- diagrams where we can redirect the external momentum exclusively through boson propagators;
    $s$-type -- diagrams where the external momentum flows through some fermionic propagators.
Finally, the $s$-type diagrams can further be partitioned into cases with $dK=0$ and $dK>0$.

\begin{center}
\begin{forest}
    [${Z_{5,1}}$,circle,draw
    [{\hyperref[controlZ2_eb]{(a)} $(g,g)\;\&\;\bar\delta+dK\ge2\;\&\;L\ge3$}]
    [{\hyperref[controlZ2_ee]{(b)} $(g,u)$}]
       [{$(u,u)$}
            [{\hyperref[controlZ2_ef]{(c)} $V_g=0\;\&\;V_u\ge3$}] 
            [{$V_g\ge4$}
                [{\hyperref[controlZ2_eg]{(d)} $c$-type}]
                [{$s$-type}
                    [{\hyperref[controlZ2_eh]{(e)} $dK=0$}]
                    [{\hyperref[controlZ2_ei]{(f)} $dK>0$}]
                ]
            ]
        ]
    ]
\end{forest}
\end{center}

\paragraph{Type $(g,g)$ \&  $\bar\delta+dK\ge2$ \& $L\ge3$.}\label{controlZ2_eb}

%
Using the non-negativity of $V_u-L_b-dK$ and $\bar\delta+dK\ge2$
we find that the generic upper bound to be
\begin{align}
        \ub\sim
        c^{\frac{E-2}{2}+V_u-L_b-dK} s^{\bar\delta+dK} y^{V_g/2} \kappa^{V_u}
        \le
        s^2.
\end{align}
As is shown in \eq{eq:Fbosonself}, there is an additional factor of $s^2$ contributing to the counterterm from the momentum dependent, $\mathcal F(k)$.
Finally, with the additional factor of $c^{-2}$, due to the tree-level contribution, we find that this class of diagrams contribute to $Z_{5,1}$ as:
\begin{align}
    Z_{5,1}\lesssim s^4/c^2 \sim \frac{1}{\log(1/s)}.
\end{align}
As is shown in Sec. \ref{app:LogZ2}, there is no further logarithmic enhancment in $1/s$.
Consequently, higher-order diagrams are suppressed by $(\, \log(1/s)\, )^{-1}$ compared with the modified one-loop level.

\paragraph{Type $(g,u)$.}\label{controlZ2_ee}

Due to an external Yukawa vertex,
$\bar\delta\ge1$.
Due to the presence of both Yukawa and four-boson vertices, we can use 
\eq{eqDeltaRelation} with $\alpha=1,\xi=0$, which implies
$V_u-L_b-dK\ge1$.
Additionally, there is a factor of $s^2$ coming from $\mathcal F(k)$ (\eq{eq:Fbosonself}).
The counterterm then scales as
\begin{align}
    Z_{5,1} \sim \frac{1}{c^2}
    \cdot s^2 \cdot \Bigl( c^{\frac{E-2}{2}+V_u-L_b-dK} s^{\bar\delta+dK} y^{V_g/2} \kappa^{V_u} \Bigr)
    \le \frac{s^3}{c}.
\end{align}
%
This is suppressed by $s$ compared with the modified one-loop level.

\paragraph{Type $(u,u)$ \& $V_g=0$ \& $V_u\ge3$}\label{controlZ2_ef}

The contribution to $Z_{5,1}$ from these diagrams scales as:
\begin{align}
    Z_{5,1} \sim \kappa^{V_u} \le \kappa^3.
\end{align}
This is suppressed by $\kappa^2$ compared with the leading order contribution of the modified one-loop level.



\paragraph{Type $(u,u)$ \& $V_g\ge4$ \& $c$-type momentum redirection.}\label{controlZ2_eg}

This case corresponds to half-skeleton diagrams where both external legs are attached to quartic boson vertices.
Additionally, it is possible to redirect the external momentum only through the boson propagators.
This redirection gives rise to a factor of $c^2$ in $\mathcal F(k)$.
%
With $V_g\ge4$,
the counter-term scales as
\begin{align}
    Z_{5,1} \lesssim y^2. 
\end{align}
This is suppressed by $y$ compared to the modified one-loop order.

\paragraph{Type $(u,u)$ \& $V_g\ge4$ \& $s$-type momentum redirection \& $dK=0$.}\label{controlZ2_eh}

Applying \eq{eqDeltaSingleGBV}, and using $dK=0$, we find $V_u-L_b-dK=V_u-L_b\ge2$.\footnote{Note that in \eq{eqDeltaRelation} we find that for each individual GBV we have $V_u-L_b\ge1$. In $s$-type diagrams, there are at least two distinct GBVs from which the inequality follows.}
Since the dependence of the external momentum is only suppressed by $s^2$,
the counterterm scales as
        \begin{align}
        Z_{5,1} \sim 
        \frac{1}{c^2} \cdot s^2 \cdot
    \Bigl(c^{\frac{E-2}{2}+V_u-L_b-dK}s^{\bar\delta+dK}y^{V_g/2}\kappa^{V_u}\Bigr)
    \le
    s^2.
\end{align}
This is suppressed by $s^2$ compared with the modified one-loop level.

\paragraph{Type $(u,u)$ \& $V_g \geq 4$ \& $s$-type momentum redirection \& $dK>0$.}\label{controlZ2_ei}

%
Applying \eq{eqDeltaRelation}, we find $V_u-L_b-dK\ge1$ 
because $\alpha=0$ and $\xi=2$.
With $\bar\delta+dK \ge dK \ge 1$, the counterterm then scales as
\begin{align}
Z_{5,1} \sim
    \frac{1}{c^2} \cdot s^2 \cdot
    \Bigl(c^{\frac{E-2}{2}+V_u-L_b-dK}s^{\bar\delta+dK}y^{V_g/2}\kappa^{V_u}\Bigr)
    \le
    y^2s^3/c \sim \frac{y^2s}{\sqrt{\log(1/s)}}.
\end{align}
This is suppressed by $s$ compared with the modified one-loop level.

\subsubsection{\tps{$Z_{7,1}$}{Z7}}

The leading contribution to $Z_{7,1}$ comes from \fig{fig:Quartic1LoopBoson} which is of order $\kappa$.
Let us now partition the diagrams beyond the modified one-loop level
first  based on $V_u=0$ and $V_u>0$.
Diagrams with $V_u=0$ 
are further partitioned into
the ones with $\bar\delta\ge2$ and $\bar\delta=1$. 
However, the latter case with $\bar\delta=1$ are either contained in the modified one-loop level or are non-half-skeleton diagrams.\footnote{
Let us show that there is only one half-skeleton diagram with $V_u=0$ and $\bar\delta=1$ and it has already been included at the modified one-loop order.
First, all four external legs must be connected to the diagram through Yukawa vertices.
Let us call ``external fermion loops'' to be those loops which contain at least one external vertex.
If there is more than one external loop, then this means that $\bar\delta\ge2$. 
Thus, with $\bar\delta=1$, there must be only one such external fermion loop.
Beyond the one-loop level, diagrams in this class must have at least two Yukawa vertices in addition to the four external ones.
Since the external fermion loop contains four external Yukawa vertices, let us naturally call the segments of the loop between these vertices as ``branches.''
If it turns out that these branches are not connected amongst each other with internal boson propagators
then each non-trivial branch is a fermion self-energy insertion. 
Since we are disregarding diagrams with such insertions, we are left to consider diagrams in which there is a boson propagator connecting two distinct branches. 
By looking at this special propagator, together with the external fermion loop, we may construct two loops which partition the external Yukawa vertices and thus show that $\bar\delta\ge2$. 
Once again, since we are considering only diagrams with $\bar\delta=1$, this means that the only higher-loop diagrams that satisfy this condition are precisely the diagrams only having fermion self-energy insertions.
It is only at the one loop level, Fig.~(\ref{fig:Quartic1LoopFermion}) that we have a diagram without self-energy insertions which satisfies the conditions of this case.
}
For this reason we focus on the case $V_u=0$ and $\bar\delta\ge2$.
Diagrams with $V_u>0$ can 
 be further partitioned into ones with $V_g=0$ and $V_g>0$. 
Higher-order diagrams with $V_g=0$ necessarily have 
$V_u\ge3$,
and 
those with $V_g>0$ 
must have a fermion loop and it must also have at least four Yukawa vertices.
Continuing on, we can partition diagrams with $V_u>0$ and $V_g\ge4$ depending on $V_u-L_b-dK \ge 1$ or $V_u-L_b-dK=0$. 
In the latter case, according to the lower bound in \eq{eqDeltaRelation} we must also have $\alpha=1$ and $\xi=-4$, 
which implies that $\bar\delta\ge2$.\footnote{
Let us show here that four-boson vertex corrections with $V_u>0$ \& $\alpha=1,\xi=-4$ \& $\bar\delta+dK=1$ must contain self-energies. Since $\alpha\ge1$, it follows immediately that $\bar\delta\ge1$. From here we find that $dK=0$. With the above assumptions, the diagram must take the form of \fig{fig:Quartic1LoopFermion} with dressed fermion propagators. Since we are only considering half-skeleton diagrams this case does not contribute.
} 
Therefore, we shall partition this class of diagrams into two: $\bar\delta+dK =2$ and $\bar\delta+dK\ge3$.

\begin{center}
\begin{forest}
    [${Z_{7,1}}$,circle,draw
       [{\hyperref[controlZ2_fc]{(a)} $V_u=0$ \& $\bar\delta\ge2$}
       ]
       [{$V_u>0$} 
            [{\hyperref[controlZ2_fd]{(b)} $V_g=0$ \& $V_u\ge3$}] 
            [{$V_g\ge4$}
                [{\hyperref[controlZ2_ff]{(c)} $V_u-L_b-dK\ge1$}]
                [{$V_u-L_b-dK=0$}
                    [{\hyperref[controlZ2_fh]{(d)} $\bar\delta+dK=2$}]
                    [{\hyperref[controlZ2_fi]{(e)} $\bar\delta+dK\ge3$}]
                ]
            ]
        ]
    ]
\end{forest}
\end{center}

\paragraph{$V_u=0$ \& $\bar\delta\ge2$.}\label{controlZ2_fc}

For this class of diagrams, the counterterm scales as
\begin{align}
    Z_{7,1}\sim
    \frac{1}{u}\left( c^{\frac{E-2}{2}+V_u-L_b-dK} s^{\bar\delta+dK} y^{V_g/2} \kappa^{V_u}\right)
    \le \frac{1}{\kappa c^2}\left( c^1 s^2 y^{V_g/2} \right)
    \le \frac{s^2}{c}
    \le \frac{1}{\sqrt{\log(1/s)}}.
    \label{eq_fourboson_B24}
\end{align}
We use the fact that $y^{V_g/2}/\kappa \leq \epsilon^{1/2}$ for diagrams beyond the modified one-loop order with $V_g \geq 4$.
%
Diagrams with 
$L > \bar\delta+dK$
can not have further logarithmic
enhancements
as is shown in App. \ref{app:LogZ2},
and are suppressed by
$(\, \log(1/s)\, )^{-1/2}$ compared with the modified one-loop level.
There can be,
in principle,
logarithmic enhancements if $L = \bar\delta+dK$.
Since $V_u=0$, then $dK=0$\footnote{
This follows from 
Eq. \eqref{AlphaDk}
and
$N_{GFV}=\alpha=1$
for diagrams without quartic vertex. 
}.
The condition for logarithmic enhancements is then $L=\bar\delta$. 
Let us consider the cases with
$L=\bar\delta=2$ and $L=\bar\delta\ge3$ separately.
Diagrams in the former class are necessarily suppressed by an additional power of $s$ 
due to the presence of double pole for the integrations over the spatial momenta in the small $s$ limit 
as is shown in  App. \ref{app:QuarticTwoLoop1}.
Diagrams in the latter case are also suppressed by $s$ because $L \geq 3$:
$\ub \sim c s^L y^{V_g/2} \kappa^{V_u} \leq cs^3 $.

\paragraph{$V_u\ge3$ \& $V_g=0$.}\label{controlZ2_fd}

The counter terms for these diagrams scale as $Z_{7,1}\sim \kappa^{V_u-1}$. Since $V_u\ge3$, we have
\begin{align}
    Z_{7,1}\lesssim \kappa^2.
\end{align}
This is suppressed by $\kappa$ compared with the modified one-loop level.

\paragraph{$V_u>0$ \& $V_g\ge4$ \& $V_u-L_b-dK\ge1$.}\label{controlZ2_ff}

The counterterm then takes the form:
        \begin{align}
            Z_{7,1}\sim\frac{1}{u}\left( c^{\frac{E-2}{2}+V_u-L_b-dK} s^{\bar\delta+dK} y^{V_g/2} \kappa^{V_u} \right)
            \le y^2.
        \end{align}
This is suppressed by $\sqrt{\epsilon}$ compared with the modified one-loop level.


\paragraph{$V_u>0$ \& $V_g\ge4$ \& $V_u-L_b-dK=0$ \& $\bar\delta+dK=2$.}\label{controlZ2_fh}

The counterterm scales as:
\begin{align}
    Z_{7,1}\sim
    \frac{1}{u}\left( c^{\frac{E-2}{2}+V_u-L_b-dK} s^{\bar\delta+dK} y^{V_g/2} \kappa^{V_u} \right)
    \le \frac{y^2s^2}{c}
    \le \frac{y^2}{\sqrt{\log(1/s)}}.
\end{align}
Diagrams in this class can not have logarithmic enhancement as is shown in Sec. \ref{app:LogZ2}, and they are suppressed by $(\, \log(1/s)\, )^{-1/2}$ compared with the modified one-loop level.

\paragraph{$V_u>0$ \& $V_g\ge4$ \& $V_u-L_b-dK=0$ \& $\bar\delta+dK\ge3$.}\label{controlZ2_fi}

If $\bar\delta+dK\ge3$, then the counterterm will take the form of
\begin{align}
    Z_{7,1}\sim\frac{1}{u}\left( c^{\frac{E-2}{2}+V_u-L_b-dK} s^{\bar\delta+dK} y^{V_g/2} \kappa^{V_u} \right)
    \le \frac{y^2s^3}{c}
    \le y^2s.
\end{align}
This is suppressed by $s$ compared with the modified one-loop contribution.


\subsubsection{Logarithmic Enhancements in the \texorpdfstring{$Z_2$}{Z2} Theory}\label{app:LogZ2}

The general upper bound 
in \eq{eq:GU} captures power-law divergences in the small $c$ and $s$ limit.
In this section, we discuss extra enhancements by $\log 1/s$ that can arise in some diagrams. 
In particular, we show that counter terms that are logarithmically divergent in the cutoff do not have factors of $\log 1/s$.

%
We start with the simplest example that exhibits a logarithmic enhancement:
the two-loop boson self-energy in \fig{fig:BosonSE2LoopFermion}.
According to the generic upper bound, 
the two-loop boson self-energy gives
$Z_{5,1} \sim \mathcal O(y^2 s^4/c^2)$. 
A factor of $s^2$ originates from $\bar\delta=2$, 
an additional factor of $s^2$ from a momentum redirection, 
and a factor of $c^{-2}$ from matching to the $c^2$ suppression of the boson dispersion.
However, the explicit calculation of \fig{fig:BosonSE2LoopFermion} shows that $Z_{5,1} \sim y^2s^4\log(1/s)/c^2$.
To understand the 
 origin of the logarithmic enhancement,
let us rewrite \eq{eq:A40} 
for \fig{fig:BosonSE2LoopFermion} in terms of canonical variables,  
$E_1 = \varepsilon_1(p)$,
$E_2 = \varepsilon_3(p)$,
$E_3 = \varepsilon_1(q)$,
$E_4 = \varepsilon_3(q)$:
\begin{align}
\centering
\Upsilon_{(0,2),a}^{2L,\alpha,n=1}(\vec{k})
&= \frac{1}{v^2}
\int 
\dd{{\bf P}}\!\dd{{\bf Q}} 
\prod_i dE_i
\Tr \left[
\left(
  \frac{\mathbb{P_{+}}}{(E_1+\varepsilon_1(k)-i|\vb{P}|)(E_2-i|\vb{P}|)}+
  cc.
\right)
\left(
  \frac{\mathbb{Q_{+}}}{(E_3-i |\vb{Q}|)(E_4+\varepsilon_3(k)-i|\vb{Q}|)}+cc.
\right)
\right]
\nonumber\\
&\hspace{1cm}\times
\left[
(\vb Q-\vb P)^2 
+\frac14(sE_3+sE_4-sE_1-sE_2)^2
+\frac14(cE_3-cE_4-cE_1+cE_2)^2
\right]^{-1}.
\label{eq:B42}
\end{align}
With the mass renormalization subtracted and ${\bf K}$ set to be zero, the whole expression is proportional to $|\vec k|^2$ in $d=3$ with a logarithmically divergent coefficient.
$E_{i}$ with $i=1,\ldots,4$ represent the four-dimensional spatial momenta.
Since there is at least one propagator that decays as $1/E_i$ for each $E_i$, 
\eq{eq:B42} has no power-law divergence 
once $1/v^2$ is factored out.
It is tempting to set $c$ and $s$ zero in the rest of the integrand.
However, this is dangerous because 
the dispersion of the internal boson expressed in terms of the canonical variables is suppressed either by $s$ or $c$,
and setting $s=c=0$ inside the boson propagator leads to a factorization of integration into
\begin{align}
    \Upsilon_{(0,2),a}^{2L,\alpha,n=1}(\vec{k})
    -\Upsilon_{(0,2),a}^{2L,\alpha,n=1}(0)
    \sim
    \int \frac{\dd^{d-1}{(\vb Q-\vb P)}}{(\vb Q-\vb P)^2}
    \int \Bigl[\dd^{d-1}(\vb Q+\vb P) \prod_{i=1}^4\dd E_i\Bigr]
    \mathcal F(\vb P, \vb Q, E_1, \ldots, E_4).
    \label{eq:B43}
\end{align}
Because the boson propagator only depends on 
${\bf Q}-{\bf P}$, 
the rest of the integration exhibits a logarithmic divergence in $d=3$ 
 at fixed ${\bf Q}-{\bf P}$.
Together with the integration over 
${\bf Q}-{\bf P}$,
the diagram results in $\log^2 \Lambda$ divergence in $d=3$,
where $\Lambda$ is a UV cutoff.
While this is superficially similar to divergences of sub-diagrams, there are two crucial differences.
First, there is no lower order counter term that cancels this $\log^2 \Lambda$ divergence.
This is because the 
 one-loop boson self-energy supplemented with the counter-term for the one-loop vertex correction is independent of $|\vec k|^2$.
Second, this $\log^2 \Lambda$ divergence is an artifact of setting $c$ and $s$ to zero too early.
For a non-zero $s$,
the extra  logarithmic divergence in $\Lambda$ is converted to a $\log(1/s)$ enhancement as is shown through 
 the explicit calculation in \eq{eqA44}.
\footnote{This same behaviour was observed in the calculation of 
\fig{fig:2Lphi4}.}


This example reveals the mechanism for the logarithmic enhancement:
boson propagators that become independent of 
canonical variables in the small $s$ and $c$ limit are responsible for the  
extra $\log(1/s)$.
We refer to those boson propagators as {\it isolated} propagators.
The absence of such isolated boson propagators ensures that no propagator can be factored out of the integrations and  there is no superfluous $\log^2 \Lambda$ divergence  in the small $c$ and $s$ limit. 
In such cases, there is no extra enhancement of $\log 1/s$ for a non-zero $s$,
and
the generic upper bound in \eq{eq:GU} is valid as it is.

\begin{figure}[htpb]
    \centering
    \begin{subfigure}{0.4\linewidth}
        \includegraphics[width=0.7\textwidth]{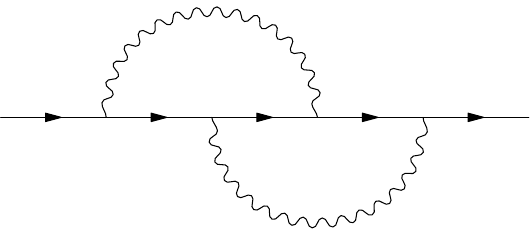}
        \caption{}
        \label{fig:FSE_2Loop}
    \end{subfigure}
    \begin{subfigure}{0.4\linewidth}
        \includegraphics[width=0.7\textwidth]{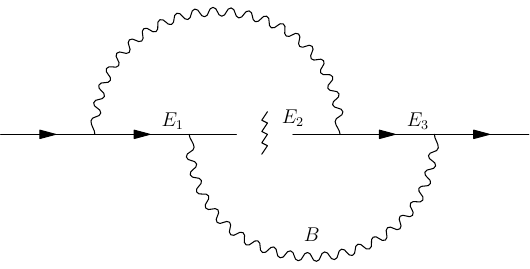}
        \caption{}
        \label{fig:FSE_2Loop_cut}
    \end{subfigure}
    \caption{
    The two-loop fermion self-energy in (a) has no logarithmic enhancement in $1/s$.
    This follows from the fact that the residual loop (b) obtained after cutting the reserved fermion line leaves no isolated boson propagator in the small $c$ and $s$ limit.}
\end{figure}
Before we generalize this, 
it is instructive to consider an example where there is no such logarithmic enhancement: the two-loop fermion self-energy in \fig{fig:FSE_2Loop}.
As is shown in page~\pageref{controlZ2_cg}, the contribution of this diagram to $Z_{2,1}-Z_{3,1}$, up to powers of $\epsilon$, 
is of order $\mathcal O(s^2/c) = \mathcal O(\,(\log(1/s))^{-1/2}\,)$.
If there was a logarithmic enhancement, this contribution would be non-negligible at the fixed point.
However, an explicit calculation shows that this graph does not exhibit 
a logarithmic enhancement.
Here, we show the absence of such logarithmic enhancement through a simpler argument.
The two-loop fermion self-energy takes the following form
\begin{align}
    \int dp dq\,
    \frac{1}{\vb P^2 + c^2 p_x^2 + c^2 p_y^2}
    \frac{1}{\vb Q^2 + c^2 q_x^2 + c^2 q_y^2}
    \frac{1}{(\varepsilon_3(p+k)-i|\vb P|)(\varepsilon_1(p+q+k)-i|\vb P+\vb Q|)(\varepsilon_3(q+k)-i|\vb Q|)},
    \label{eqB56}
\end{align}
where we set ${\bf K}=0$ as we consider the part of the self-energy that depends on the spatial momentum.
We now choose the following canonical variables:
$E_1 = \varepsilon_3(p)$, 
$E_2 = \varepsilon_1(p+q)$,
$E_3 = \varepsilon_3(q)$, 
$B_1 = cq_x$.
$E_1$ and $E_3$ are the 
 energies of the exclusive propagators that have been identified from the ELC procedure; they
 tame the $p_y$ and $q_y$, respectively.
$E_2$ corresponds to the energy of a reserved propagator, which 
tames the $p_x$ integration.
Finally, $B$ is the energy 
 associated with $q_x$ momentum that flows through one of the boson propagators.
There is one boson propagator that is not explicitly used in the construction of the canonical variables.
This propagator is a potential candidate for an isolated propagator.
However, it turns out that there is no isolated propagator in this diagram.
This can be understood once
 the two-loop fermion self-energy is written in terms of the canonical variables as
\begin{align}
    &\int dpdq\,
    \Bigl[\vb P^2 
    + (B-s(E_1+E_2+E_3)/2)^2 
    + (vB+c(E_1-E_2-E_3)/2)^2
    \Bigr]^{-1}
    \Bigl[\vb Q^2 + B^2 +  (vB - cE_3)^2\Bigr]^{-1}
    \nonumber\\
    &\hspace{1cm}\times
    \frac{1}{
    (E_1+\varepsilon_3(k)-i|\vb P|)(E_2+\varepsilon_1(k)-i|\vb P+\vb Q|)(E_3+\varepsilon_3(k)-i|\vb Q|)
    }.
    \label{eqB57}
\end{align}
There is no isolated propagator because all boson propagators end up being dependent on some canonical variables even in the small $s$ and $c$ limit. 
To understand why this is the case,
let us focus on the canonical variables
$E_2$ and $B$ associated with $x$-momenta.
First, we cut the exclusive propagator associated with $E_2$.
The remaining graph is a one-loop diagram that includes both of the boson propagators in the loop
as is shown in         \fig{fig:FSE_2Loop_cut}.
This guarantees that the variable $B$ that is associated with the $x$-momentum of the remaining loop must go through all boson propagators, leaving no isolated propagator.
As a side remark, we also note that the two-loop fermion self-energy exhibits a suppressed dependence on the external momentum.
If we shift
$(E_1,E_2,E_3)\to(E_1-\varepsilon_3(k), E_2-\varepsilon_1(k),E_3-\varepsilon_3(k))$, 
the external momentum appears only through
$s\varepsilon_1(k)$ and $ s\varepsilon_3(k)$.
Therefore, $\mathcal{F}(k) \sim s k_y, sv k_x$.

The above examples lead to 
 a general criterion for the presence of logarithmic enhancement.
Let a diagram have $\bar\delta+dK$ fermion propagators and 
$L-\bar\delta-dK$ 
boson propagators that are used to tame the $x$-momenta.
%
After cutting 
$\bar\delta+dK$ exclusive and reserved fermion propagators\footnote{
See below \eqref{eqnInitUpperBound} for its definition.}
in the diagram,
we are left with 
a connected diagram with $E+2(\bar\delta+dK)$ legs made of the original external legs and cut fermion propagators.
All boson propagators are still parts of the 
$(L-\bar\delta-dK)$-loop
connected diagram.
As long as the remaining connected diagram has non-zero loop,
every boson propagator must be traversed by at least one canonical variable associated with the $x$-loop momentum. 
Therefore, diagrams with $L-\bar\delta-dK > 0$ can not have any isolated propagator.
This is consistent with the examples we considered:
the two-loop boson self-energy, which has a logarithmic enhancement, has
$L-\bar\delta-dK = 0$
while the two-loop fermion self-energy, which has no logarithmic enhancement, 
has $L-\bar\delta-dK = 1$.
In summary, a logarithmic enhancement in $1/s$ can arise only if 
$L-\bar\delta-dK = 0$.
We conclude this section by showing that the cases in which a logarithmic enhancement would have spoiled the control of the $\epsilon$-expansion indeed do not have such an enhancement.

\paragraph*{Fermion self- energy.}\label{pglogFSE}

As is discussed on page~\pageref{controlZ2_cg}, 
the contribution of 
the fermion self energy with $V_u=0$ and $\bar\delta=1$ 
to $Z_{2,1}-Z_{3,1}$ 
is only suppressed by
$\mathcal O(s^2/c) \sim \mathcal O(\,(\log(1/s))^{-1/2}\,)$
relative to the modified one-loop order. 
%
If there was a logarithmic enhancement in the higher-order diagrams,
they can not be ignored at the fixed point.
Fortunately, there is no such logarithmic enhancement because the only half-skeleton diagram that satisfies
$V_u=0$, $\bar\delta=1$ and $\bar\delta+dK=L$ 
is the one-loop fermion self-energy which has already been included\footnote{
Since $V_u=0$, then $dK=0$ and so that $L = L_{fm}=\bar\delta=1$.}.
Therefore, higher-loop diagrams with $V_u=0$ and $\bar\delta=1$ are negligible at the fixed point.

\paragraph*{Boson self-energy.}

On page~\pageref{controlZ2_eb}, we find that the boson self-energies of type $(g,g)$ with $\bar\delta+dK=2$ is only suppressed by
$\mathcal O(s^4/c^2) = \mathcal O(1/\log(1/s))$.
If there was a logarithmic enhancement,
the higher-order diagrams would not be negligible.
Such a diagram
would satisfy $L = \bar\delta+dK = 2$
because
a logarithmic  enhancement can only arise for diagrams with $L-\bar\delta-dK=0$ and 
$\bar\delta+dK=2$ for the diagrams under consideration.
%
By using $L=2$, $V-I+L=1$ and $E+2I = 3V_g + 4V_u$,
we conclude $V_g + 2V_u = 4$.
Since $V_g\ge4$ in type $(g,g)$ diagrams with $L\ge2$, this immediately implies that the only diagram that possibly has a logarithmic enhancement is the two-loop boson self-energy, which is already included at the modified one-loop level.
Therefore, there is no additional diagram that is logarithmically enhanced in this class.

\paragraph*{Four-boson vertex correction.}

For $Z_{7,1}$, there are two cases in which
a logarithmic enhancement can potentially spoil the control of the expansion.
In case {\hyperref[controlZ2_fc]{(a)}}, 
this can happen for $L = \bar\delta=2$
because 
$dK = 0$ for this case
and
a logarithmic enhancement arises only if $L = \bar\delta+dK$.
There are only two diagrams that satisfy these conditions.
Although they indeed exhibit logarithmic enhancements,
an explicit calculation shows that they 
 are further suppressed by powers of $s$
 as is shown in App. \ref{app:QuarticTwoLoop1}.
This additional suppression arises because the integration over spatial momenta vanishes in the small $s$ limit due to a kinematic constraint that creates double pole in the complex plane of internal momenta.
In case {\hyperref[controlZ2_fh]{(d)}}, there is no logarithmic enhancement either.
Suppose there is a diagram in case {\hyperref[controlZ2_fh]{(d)}} that has a logarithmic enhancement. 
Then, $L=\bar\delta+dK=2$. 
Combining this with $V-I+L=1$ and $E+2I=3V_g+4V_u$, 
we have $6=V_g+2V_u$.
With $V_g\ge4$ and $V_u>0$, there is only one possibility: $V_g=4$ and $V_u=1$. 
For one-particle irreducible diagrams
without self-energy correction, diagrams with
$V_g=4$ and $V_u=1$ must have $dK=0$ 
because
the internal fermions participating in the Yukawa interactions must form a loop that is connected to some external legs.
Furthermore, there can not be a boson loop with $V_u=1$.
With $L_b=dK=0$, 
we have $V_u=V_u-L_b-dK=0$, which contradicts $V_u=1$.
Therefore, there are no higher order diagrams in case {\hyperref[controlZ2_fh]{(d)}} with logarithmic enhancements.

\subsection{Counterterms in the \texorpdfstring{$O(2)$}{O2} Theory}\label{app:O2}

In this section, we discuss the control of 
the $N=2$ theory.
We will see that the $O(2)$ algebra significantly restricts the number of non-vanishing Feynman diagrams, 
which significantly simplifies our analysis.
To see this extra kinematic constraint for the $O(2)$ theory,
we start by rewriting the action for the $O(2)$ theory by replacing
$\Phi(q) = \phi^x(q)\tau^x + \phi^y(q)\tau^y$ with  
$\Phi_\pm(q) 
= \frac12 (\phi^x(q)\pm i\phi^y(q))(\tau^x\mp i \tau^y) 
\equiv \frac12 \phi^\mp(q) \tau^\mp$.
With
$    \frac12 \sum_{\alpha\in\{x,y\}}
    \phi^\alpha(q)\phi^\alpha(-q)
    = \frac14\Tr
    \left(
        \Phi_+(q)\Phi_-(-q) +
        \Phi_-(q)\Phi_+(-q)
    \right)$,
    $\sum_{\alpha\in\{x,y\}}
    \phi^\alpha(q)\tau^\alpha = \Phi_+(q)+\Phi_-(q)$
    and
    $
    \sum_{\alpha,\beta\in\{x,y\}}
    \phi^\alpha(p_1)\phi^\alpha(p_2)\phi^\beta(p_3)
    \phi^\beta(p_4)=
    \frac14 
    \Tr\left(
        \Phi_+(p_1)\Phi_-(p_2) +
        \Phi_-(p_1)\Phi_+(p_2)
    \right)
    \Tr\left(
        \Phi_+(p_3)\Phi_-(p_4) +
        \Phi_-(p_3)\Phi_+(p_4)
    \right)$,
%
the action that depends on the critical boson is rewritten as
\begin{subequations}
\begin{align}
S_b &= \frac14\sum_\alpha\int \frac{d^{d+1}q}{(2\pi)^{d+1}}
\left(\textbf{Q}^2 + c^2 |\vec q|^2\right) 
\Tr\left(
        \Phi_+(q)\Phi_-(-q) +
        \Phi_-(q)\Phi_+(-q)
    \right),\\
S_{g} &= \frac{g\mu^{(3-d)/2}}{\sqrt{N_f}}
\sum_{j=1}^{N_f}
\sum_{n=1}^4 \sum_{\sigma,\sigma'=\uparrow,\downarrow} \int \frac{d^{d+1}k}{(2\pi)^{d+1}} \frac{d^{d+1}q}{(2\pi)^{d+1}}
\overline\Psi_{\bar n,\sigma,j}(k+q)
\Bigl(\Phi_+(q)+\Phi_-(q)\Bigr)_{\sigma,\sigma'}
(i\gamma_{d-1})
\Psi_{n,\sigma',j}(k),
\\
S_{u} &= 
\frac{u \mu^{3-d}}{4}
\int \prod_{j=1}^4 dp_j 
\delta\left(\sum_{i=1}^4p_i\right)
\Tr\left(
        \Phi_+(p_1)\Phi_-(p_2) +
        \Phi_-(p_1)\Phi_+(p_2)
\right)
\Tr\left(
        \Phi_+(p_3)\Phi_-(p_4) +
        \Phi_-(p_3)\Phi_+(p_4)
\right).
\end{align}
\end{subequations}
It is noted that the Yukawa interaction can be rewritten in terms of the raising and lowering fields, $\Phi_\pm$ as
\begin{align}
S_{g} &= \frac{g\mu^{(3-d)/2}}{\sqrt{N_f}}
\sum_{j=1}^{N_f}
\sum_{n=1}^4 \int \frac{d^{d+1}k}{(2\pi)^{d+1}}     \frac{d^{d+1}q}{(2\pi)^{d+1}}
\Bigl(
\overline\Psi_{\bar n,\uparrow,j}(k+q)
\phi^+(q)
(i\gamma_{d-1})
\Psi_{n,\downarrow,j}(k)
+\overline\Psi_{\bar n,\downarrow,j}(k+q)
\phi^-(q)
(i\gamma_{d-1})
\Psi_{n,\uparrow,j}(k)
\Bigr).
\end{align}
This action is symmetric under the $SO(2)=U(1)$ symmetry,
\begin{align}
    \phi^\pm \to e^{\pm i \theta} \phi^\pm,\;\;\;
    \Psi_{n,\sigma,j} \to e^{i(\delta_{\sigma,\uparrow}-\delta_{\sigma,\downarrow})\theta/2}\Psi_{n,\sigma,j},
\end{align}
where $\phi^\pm$ carries spin $\pm1$ while the fermion carries  
spin $\pm1/2$. 
%
%

\begin{figure}[htpb!]
    \centering
    \includegraphics[width=0.25\textwidth]{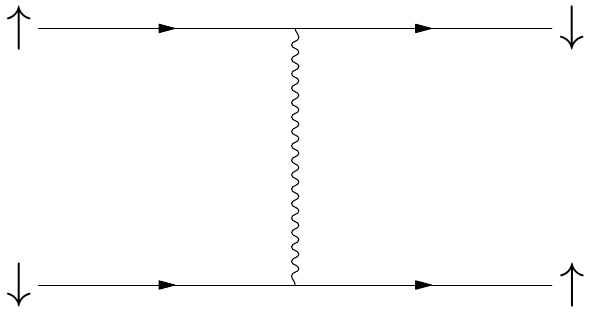}
    \caption{
    In $O(2)$ theory, 
    the critical boson mediates only a specific two-body interaction where
    there are always a pair of incoming fermions with spin $\uparrow$ and $\downarrow$ and an outgoing pair of fermions with flipped spins. }
    \label{fig:conservationofspin}
\end{figure}

\begin{figure}[htpb!]
    \centering
    \includegraphics[width=0.5\textwidth]{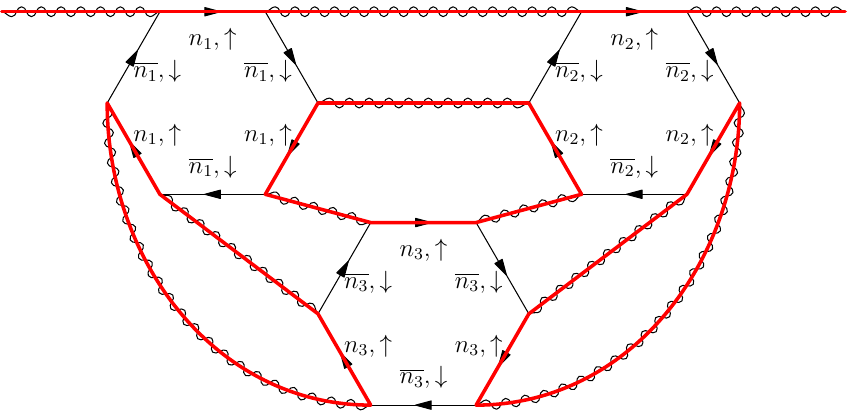}
    \caption{
A boson self-energy with three fermion loops which are labelled by $i=1,2,3$.
The fermion propagators in fermion loop $i$ are labelled with hot spot indices ($n_i$ and $\bar{n}_i$) and spins.
All thick (red) lines carry the external momentum $k$, and fermion loop $i$ carries momentum $q_i$.
The kinematic constraint of the $O(2)$ theory guarantees that 
in each fermion loop
$k$ appears in all fermions of one type of hot spots but not in fermions of the other type. 
This allows us to shift the fermion loop momenta as
$\varepsilon_{n_i}(q_i+k)\to \varepsilon_{n_i}(k)$ and
$\varepsilon_{\bar{n}_i}(q_i)\to \varepsilon_{\bar n_i}(q_i)$ 
(or 
$\varepsilon_{n_i}(q_i)\to \varepsilon_{n_i}(k)$ and
$\varepsilon_{\bar{n}_i}(q_i+k)\to \varepsilon_{\bar n_i}(q_i)$ 
).
This makes sure that $k$ appears only in the boson propagators through $c \vec k$.
}
    \label{fig:demonstrationO2bosonSE}
\end{figure}

The SO(2) spin conservation at each Yukawa vertex combined with the momentum conservation leads to 
 the following kinematic constraints that is crucial for our proof of the control: 
\begin{enumerate}
\item
An internal boson, if connected to two Yukawa interactions, must always have the form shown in \fig{fig:conservationofspin}, 
    where two fermions of different spin enter and get flipped. 
This follows from the fact that bosonic propagator is between $\phi^+$ and $\phi^-$, which translate to a $\tau^+ \otimes \tau^-$ or $\tau^- \otimes \tau^+$ effective interaction for the fermions.
Therefore, it is not possible to have two spin-$\uparrow$ fermions entering a two-body effective interaction as shown in \fig{fig:conservationofspin}.e
For this reason,
diagrams such as \fig{fig:BosonSE2LoopFermion} identically vanish. 
\item 
    In the remaining diagrams that survive,
    all Yukawa interactions scatter 
        spin-$\uparrow$ fermions to 
        spin-$\downarrow$ fermions or vice-versa. Therefore,
    the external momentum can always be directed to flow along fermion propagators of spin-$\uparrow$ in the same direction of propagation as the fermion. 
\item 
All other spin-$\uparrow$ fermions,
which does not carry the external momentum, 
are parts of some loops that can be formed out of only spin-$\uparrow$ fermions and bosons. 
This is due to the conservation of spin. 
Spin $+1/2$ carried by those spin-$\uparrow$ fermions can only be transferred to internal bosons which then transfer the spin to 
 spin-$\downarrow$ fermions to flip its spin to $\uparrow$.
 The chains of those spin-$\uparrow$ fermions and bosons form loops.
Those loops do not include those fermion propagators of spin-$\uparrow$ that already carry the external momentum.
This is illustrated in \fig{fig:demonstrationO2bosonSE}.
Now, we shift those loop momenta by the external momentum so that all spin-$\uparrow$ fermions carry external momentum.
\item 
The above constraints have a significant consequence for the boson self-energy.
For the diagrams that contribute to the boson self-energy, all fermion propagators are parts of fermion loops.
Within each fermion loop, all fermion propagators of spin $\uparrow$ must be of one type of hot spot (say hot spot $n$) because the spin alternate together with the hot spot index within fermion loops.
Therefore, in each fermion loop, 
fermions with spin $\uparrow$ and $\downarrow$ depend on the external momentum $k$ and the fermion loop momentum $q_i$ through
$\epsilon_n(\vec q_i+\vec k)$ and
$\epsilon_{\bar n}(\vec q_i)$, respectively.
Now, we can shift $\vec q_i$ within each fermion loop to remove $\vec k$ in all fermion propagators in the loop  because $\epsilon_n$ and $\epsilon_{\bar{n}}$ are linearly independent. 
Since this can be done in each fermion loop,
$\vec k$ drops in all fermion propagators and only appears in the boson propagators through $c\vec k$. 
Therefore,
    $\mathcal F(k) \sim c^2 |\vec k|^2$ for the boson self-energy
    and
    $Z_{5,1} \sim \ub$.
\end{enumerate}
In the rest of the section, we go over each $Z_{n,1}$ and show that the
$\epsilon$-expansion is controlled. The following discussion has the same
structure as the previous section on the $Z_2$ theory.



\subsubsection{\tps{$Z_{4,1},Z_{5,1}$}{Z4,Z5}}


The leading order contribution to $Z_{4,1}+Z_{5,1}$ comes from \fig{fig:BosonSE2LoopBoson} and is of order $\kappa^2 \sim \epsilon$.
The leading order contribution to $Z_{4,1}-Z_{5,1}$ comes from \fig{fig:BosonSE1Loop} and is of order $y\sim (c\epsilon)^{1/3}$.
Now, let us recall the beta functions for ${c,s,y,\kappa}$:
\begin{align}
    \begin{pmatrix}
        \frac{1}{zc}\beta_c\\
        \frac{1}{zs}\beta_s\\
        \frac{1}{zy}\beta_y\\
        \frac{1}{z\kappa}\beta_\kappa
    \end{pmatrix}
    =
    \begin{pmatrix}
1&0&-1&-1/2&1/2&0&0\\
1&-1&0&-1/2&1/2&0&0\\
-1&0&-1&-1/2&-1/2&2&0\\
0&0&0&-1&-1&0&1
    \end{pmatrix}
    \begin{pmatrix}
        Z_{1,1}'\\
        Z_{2,1}'\\
        Z_{3,1}'\\
        Z_{4,1}'\\
        Z_{5,1}'\\
        Z_{6,1}'\\
        Z_{7,1}'
    \end{pmatrix}
    +
    \begin{pmatrix}
        0\\0\\-\epsilon\\-\epsilon
    \end{pmatrix}.
\end{align}
We see here that the flow of $c,s,y,\kappa$ depends on the difference and addition of $Z_{4,1}$ and $Z_{5,1}$ rather than them separately.
For this reason we shall analyze the contribution to $Z_{4,1} \pm Z_{5,1}$ rather than $Z_{4,1}$ and $Z_{5,1}$ individually. 
Let us now partition higher-order diagrams depending on the number of Yukawa vertices. 
Note that there are no higher-loop half-skeleton diagrams with $V_g=2$ because the only half-skeleton boson self-energy with $V_g=2$ is \fig{fig:BosonSE1Loop}. 
For this reason, we consider two classes $V_g=0$ and $V_g\ge4$.
To go beyond the modified one-loop level, diagrams with $V_g=0$ must also have $V_u\ge3$.

\begin{center}
\begin{forest}
    [{$Z_{4,1} \pm Z_{5,1}$},rectangle,draw
        [ {\hyperref[controlO2_bb]{(a)} $V_g=0$ \& $V_u\ge3$}] 
        [ {\hyperref[controlO2_bd]{(b)} 
 $V_g\ge4$}] 
    ]
\end{forest}
\end{center}


\paragraph{$V_g=0$ \& $V_u\ge3$.}\label{controlO2_bb}

Diagrams with $V_g=0$ only contribute to $Z_{4,1} + Z_{5,1}$. The counterterm then scales as:
\begin{align}
    Z_{4,1}+Z_{5,1}\sim \kappa^{V_u} \le \kappa^3
\end{align}
This is suppressed by $\kappa$ compared with the modified one-loop level.

\paragraph{$V_g\ge4$.}\label{controlO2_bd}

For diagrams with $V_g\ge4$, the generic upper bound is given by
\begin{align}
    Z_{4,1} \pm Z_{5,1}\sim \ub \sim c^{(E-2)/2+V_u-L_b-dK}
    s^{\bar\delta+dK}
    y^{V_g/2}
    \kappa^{V_u}
    \le y^2 \sim c^{2/3}.
\end{align}
For $Z_{4,1}+Z_{5,1}$, this is suppressed by a factor of $c^{2/3}$ compared with the modified one-loop level.
For $Z_{4,1}-Z_{5,1}$, this is suppressed by a factor of $c^{1/3}$ compared with the modified one-loop level.

\subsubsection{\tps{$Z_{1,1}$}{Z1}}


At the modified one-loop level, the leading order contribution to $Z_{1,1}$ comes from \fig{fig:FermionSE1Loop} and is of order $y \sim \epsilon^{1/3}c^{1/3}$.
Diagrams beyond the modified one-loop must have $V_g\ge4$. For such diagrams, the counterterm scales as:
\begin{align}
    Z_{1,1}\sim c^{\frac{E-2}{2}+V_u-L_b-dK} s^{\bar\delta+dK} y^{V_g/2} 
    \kappa^{V_u}
    \le  y^2.
\end{align}
This is suppressed by a factor of $y$ compared with the modified one-loop level.






\subsubsection{\tps{$Z_{2,1},Z_{3,1}$}{Z2,Z3}}

The contribution to $Z_{2,1}$ and $Z_{3,1}$ at the modified one-loop order comes from \fig{fig:FermionSE1Loop} and is of order $yc \sim \epsilon^{1/3}c^{4/3}$.
Let us partition higher-order diagrams into those which have fermion loops (`FL') and those that do not (`No FL').
We further partition the diagrams in `No FL' depending on $V_u$ and $V_g$. 
For $V_u=0$ and $V_g=2,4$, there are no diagrams beyond the modified one-loop level that are non-skeleton.
Therefore we consider $(V_u=0,V_g=6)$, $(V_u=0,V_g\ge8)$, and $V_u>0$.

\begin{center}
\begin{forest}
    [{$Z_{2,1}, Z_{3,1}$},rectangle,draw
        [{\hyperref[controlO2_cb]{(a)} FL}]
        [No FL
            [{$V_u=0$}
                [{\hyperref[controlO2_ce]{(b)} $V_g=6$}]
                [{\hyperref[controlO2_cf]{(c)} $V_g\ge8$}]
            ]
            [{\hyperref[controlO2_cg]{(d)} $V_u>0$}]
        ]
    ]
\end{forest}
\end{center}

\paragraph{FL.}\label{controlO2_cb}

If there is at least one fermion loop in a diagram that contributes to $Z_{2,1}$ or $Z_{3,1}$, $V_g\ge8$. 
%
Using \eq{eq:Ffermionself}, we find the counterterm scales as:
\begin{align}
    Z_{2,1},Z_{3,1}\sim
    c^{\frac{E-2}{2}+V_u-L_b-dK} s^{\bar\delta+dK} y^{V_g/2} \kappa^{V_u}
    \le
    y^4
    \sim c^{4/3}\epsilon^{4/3}.
\end{align}
This is suppressed by $\epsilon$ compared with the modified one-loop level.

\paragraph{No FL \& $V_u=0$ \& $V_g=6$.}\label{controlO2_ce}

\begin{figure}[htpb]
    \centering
\includegraphics[scale=0.5]{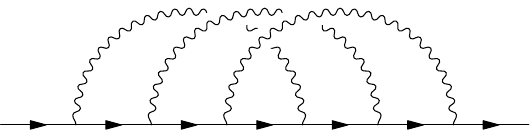}
    \caption{
A three-loop fermion self-energy with 
no fermion loops, $V_u=0$, and $V_g=6$.
    }
    \label{fig:output-4}
\end{figure}

There is only one non-vanishing half-skeleton diagram in this class  as is shown in \fig{fig:output-4}.
The generic upper bound for this diagram scales as
\begin{align}
    \ub \sim 
    c^{\frac{E-2}{2}+V_u-L_b-dK} s^{\bar\delta+dK} y^{V_g/2} \kappa^{V_u}
    = y^3 
    \sim c\epsilon.
\end{align}
Let us now consider the dependence on external momentum.
Without loss of generality, let us assume that the external fermion lies on hot spot $1+$, and suppose that the momentum of the boson propagators is $p_1$, $p_2$, and $p_3$, 
so that the momentum of the first four fermions from left to right are $k,k+p_1,k+p_1+p_2, k+p_1+p_2+p_3$.
%
Direct the external momentum to go through the main fermion line.
If we shift the $y$-components of the internal momenta, 
$p_{1,y}\to p_{1,y}+\varepsilon_3(k)$,
$p_{2,y}\to p_{2,y}-\varepsilon_3(k)$,
$p_{3,y}\to p_{3,y}+\varepsilon_3(k)$,
then this diagram will manifestly depend on the external momentum only through
$\Delta_1 =  \varepsilon_1(k)$,
$\Delta_2 = c\varepsilon_3(k)$.
%
Combining the generic upper bound and the dependence on the external momentum, we can conclude the following.
\begin{itemize}
    \item 
    The contribution 
    of \fig{fig:output-4} to
    $Z_{2,1}$ and $Z_{3,1}$ 
    is of order $\mathcal O(y^3) \sim \mathcal O(c\epsilon)$,
    which is nominally larger than the modified one-loop contribution.
    Recall at the modified one-loop level, the contribution to $Z_{2,1}$ and $Z_{3,1}$ is of order $\mathcal O(c^{4/3}\epsilon^{1/3})$. 
    However, they all vanish at the fixed point.
    \item 
    The contribution 
    of \fig{fig:output-4} to
    $Z_{2,1}-Z_{3,1}$ is additionally suppressed by a factor of $c$ because the part of the quantum correction proportional to $\Delta_1=\varepsilon_1(k)$ does not contribute to $Z_{2,1}-Z_{3,1}$ and only the part proportional to $\Delta_2 = c\varepsilon_3(k)$ contributes.
\end{itemize}
In summary, this higher-order fermion self-energy gives rise to a counterterm,
\begin{align}
    Z_{2,1} = Z_{3,1} = A y^3.
\end{align}
In the main text, it is shown that these contributions do not affect the fixed point and its stability.

\paragraph{No FL \& $V_u=0$ \& $V_g\ge8$.}\label{controlO2_cf}

The counterterm scales as:
\begin{align}
    Z_{2,1},Z_{3,1} \sim \ub \sim
    c^{\frac{E-2}{2}+V_u-L_b-dK} s^{\bar\delta+dK} y^{V_g/2} \kappa^{V_u}
    \le
    y^4
    \sim c^{4/3}\epsilon^{4/3}.
\end{align}
This is suppressed by $\epsilon$ compared with the modified one-loop level.


\paragraph{No FL \& $V_u>0$.}\label{controlO2_cg}

\begin{figure}[htpb]
    \centering
\includegraphics[scale=0.3]{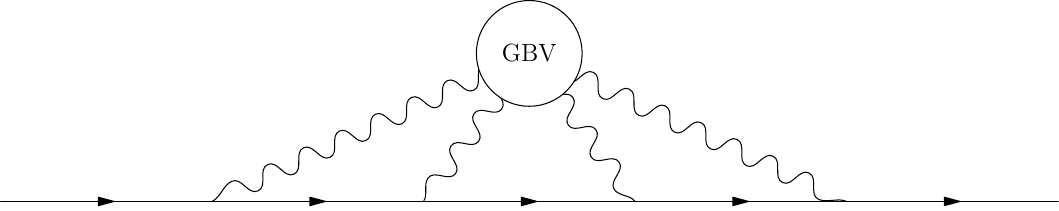}
        \caption{ A class of diagrams that contribute to the fermion self-energy which do not have fermion loops and $V_u>0$. }
    \label{fig:output-3}
\end{figure}

\fig{fig:output-3}
is an example of diagrams in this class.

\noindent
We can use \eq{eqDeltaSingleGBV} to show that $V_u-L_b\ge1$ in this case. 
Additionally, we have $dK=0$, $V_g\ge4$.
Therefore, 
the counterterm scales as
\begin{align}
    Z_{2,1},Z_{3,1}\sim\ub\sim
    c^{\frac{E-2}{2}+V_u-L_b-dK} s^{\bar\delta+dK} y^{V_g/2} \kappa^{V_u}
    \le
    y^2c
    \le c^{5/3}.
\end{align}
This is suppressed by $c^{1/3}$ compared with the modified one-loop level.

\subsubsection{\tps{$Z_{7,1}$}{Z7}}

The leading modified one-loop contributions to $Z_{7,1}$ arise from \fig{fig:Quartic1LoopBoson} and \fig{fig:2Lphi4}. 
Their magnitudes are $\mathcal O(\kappa)$ and $\mathcal O(s^2y^3/\kappa c)$ both of which are of order $\sqrt\epsilon$.
Going beyond the modified one-loop level, let us partition the diagrams according to $V_u$.
If $V_u=0$, we must have $V_g\ge8$ beyond the modified one-loop level.\footnote{
The modified one-loop order with $V_u=0$ takes into account all diagrams with $V_g\le6$. 
The 1PI diagrams with $V_g \le 6$ can have 
only one fermion loop, like in \fig{fig:2Lphi4}. 
There is one diagram with $V_g=4$, \fig{fig:Quartic1LoopFermion}, and one diagram with $V_g=6$, \fig{fig:2Lphi4}.}
Diagrams with $V_u>0$ are further partitioned depending on the type of fermion loops in the diagram.
If there are no fermion loops (`No FL'), then we must have $V_u\ge3$.
If there are fermion loops, then either $(\alpha,\xi) = (1,-4)$ or $(\alpha,\xi) \neq (1,-4)$.

\begin{center}
\begin{forest}
    [{$Z_{7,1}$},rectangle,draw
        [{\hyperref[controlO2_db]{(a)} $V_u=0$ \& $V_g\ge8$}]
        [{$V_u>0$}
[{\hyperref[controlO2_dc]{(b)} No FL \& $V_u\ge3$}]
[{\hyperref[controlO2_dd]{(c)} FL \& $(\alpha,\xi)\ne(1,-4)$}]
[{\hyperref[controlO2_de]{(d)} FL  \& $(\alpha,\xi)=(1,-4)$}]
        ]
    ]
\end{forest}
\end{center}

\paragraph{$V_u=0$ \& $V_g\ge8$.}\label{controlO2_db}

The counterterm scales as:\footnote{$L_b=dK=0$ due to $V_u=0$.}
\begin{align}
    Z_{7,1}\sim\frac{1}{\kappa c^2}\left(
        c^{(E-2)/2+V_u-L_b-dK} s^{\bar\delta+dK} c^{\delta-dK} y^{V_g/2} \kappa^{V_u}
        \right)
    \le \frac{cy^4}{\kappa c^2}
    \sim c^{1/3}\epsilon^{7/6}.
\end{align}
This is suppressed by $c^{1/3}$ compared with the modified one-loop level.


\paragraph{$V_u\ge3$ \& No FL}\label{controlO2_dc}

If there are no fermion loops, the diagram contains only boson propagators and four-boson vertices.
This means that the counterterm scales as:
\begin{align}
    Z_{7,1} \sim \kappa^{V_u-1}
\end{align}
As discussed in the $Z_2$ theory, they are order of $\kappa^{V_u-1}$ because $V_u-L_b=1$ and $E=4$.
Higher-order graphs with $V_u\ge3$ are suppressed by powers of $\kappa\sim\epsilon^{1/2}$.

\paragraph{$V_u>0$ \&
FL \&
$(\alpha,\xi)\ne(1,-4)$.}\label{controlO2_dd}

This class breaks into three sub-classes:
$(\alpha,\xi) = (0,4)$;
$\alpha\ge1$, $\xi\ge-2$;
$\alpha>1$, $\xi=-4$.
In all three cases,
$V_u-L_b-dK\ge1$.
Moreover, since there should be at least one GFV, this means that $V_g\ge4$. 
Diagrams with $V_g = 2$ with fermion loops necessarily have self-energy corrections and are not half-skeleton diagrams.
The counterterm scales as
\begin{align}
    Z_{7,1}\sim\frac{1}{u}\left( c^{\frac{E-2}{2}+V_u-L_b-dK} s^{\bar\delta+dK} y^{V_g/2} \kappa^{V_u} \right)
    \sim
    \frac{1}{\kappa c^2}\left( c^2 y^2 \kappa^{V_u} \right)
    \le y^2\kappa^{V_u-1}
    \le c^{2/3}.
\end{align}
This is suppressed by $c^{2/3}$ compared with the modified one-loop level.

\paragraph{$V_u>0$ \& 
FL \& $\alpha=1,\xi=-4$.}\label{controlO2_de}

The constraints ensure that $V_g\ge8$.
This means that the counterterm scales as
\begin{align}
    Z_{7,1}\sim\frac{1}{u}\left( c^{\frac{E-2}{2}+V_u-L_b-dK} s^{\bar\delta+dK} y^{V_g/2} \kappa^{V_u} \right)
    \lesssim
    \frac{1}{\kappa c^2}\left( c y^4 \kappa^{V_u} \right)
    \sim
    c^{1/3}.
\end{align}
This is suppressed by $c^{1/3}$ compared with modified one-loop level.

\subsubsection{\tps{$Z_{6,1}$}{Z6}}

At the modified one-loop level, $Z_{6,1}$ vanishes exactly.
Non-vanishing higher-order corrections 
scale as
\begin{align}
    Z_{6,1}\sim\frac{1}{g}\left( c^{\frac{E-2}{2}+V_u-L_b-dK} s^{\bar\delta+dK} y^{V_g/2} \kappa^{V_u} \right)
    =
    \frac{1}{(yc)^{1/2}}\left( c^{1/2} y^{V_g/2} \kappa^{V_u} \right)c^{V_u-L_b-dK}
    \le
    y^{(V_g-1)/2}.
\end{align}
%
Beyond the modified one-loop order, we have $V_g\ge5$, which implies that 
\begin{align}Z_{6,1} \lesssim y^2 \sim (c\epsilon)^{2/3}.\end{align}
As is shown in the main text, 
this higher-order correction do not affect the fixed point 
 and it stability.

\section{
Enhancement of superconducting and charge density wave fluctuations
}

\label{app:Instabilities}

In this appendix, we compute the anomalous dimension of superconducting and charge density wave order parameters to understand fluctuations that are enhanced at the quantum critical point.
For this, we turn on an infinitesimally small source that couples to fermion bi-linears,
\begin{equation}
\mathcal{S}_{\rho}[V_{\rho},\Theta]=V_{\rho} \mu \int\dd{k}\tilde{\Psi}^{(\rho)}_{n,\sigma,j}(k) \Theta_{n,\sigma,j;m,\tilde{\sigma},l}\Psi_{m,\tilde{\sigma},l}(k) + h.c. ,
\end{equation}
where we have used Einstein summation convention. 
$ V_\rho$  with $\rho=\SC$ or $\CDW$ is the dimensionless source for the superconducting and charge density wave order parameters, respectively.
$\tilde\Psi^{(\rho)}$ are defined as  $ \tilde{\Psi}^{(\CDW)}_{n,\sigma,j}(k) =\bar{\Psi}_{n,\sigma,j}(k) $ and $ \tilde{\Psi}^{(\SC)}_{n,\sigma,j}(k) ={\Psi}^{\top}_{n,\sigma',j}(-k) 
(i\tau_{\sigma'\sigma}^{y})$. 
$ \hat{\Omega}$
is a $2 \times 2$ matrix that acts on the spinor indices. Each $
\Theta_{n,\sigma,j;m,\tilde{\sigma},l} $ is a linear combination of Gamma matrices. 
$n=1,2,3,4$,
$\sigma=1,2$ 
and
$j=1,2,\dots,N_f$ 
are the hotspot, 
spin and 
flavour indices, respectively. 

UV divergences
coming from this insertion is cancelled by counter terms of the form $
\mathcal{S}_{\rho,\mathcal{CT}}=V_{\rho}\mu\int\dd{k}
\tilde{\Psi}^{\rho}{(k)}\mathcal{A}_{\rho}[\Theta]{\Psi{(k)}} $ with $ \mathcal{A}_{\rho}[\Theta]=\sum_m
{Z}_{\rho,m}[\Theta] \epsilon^{-m}$,
where ${Z}_{\rho,m}[\Theta]$ is a linear functions of $\Theta$. 
The bare action is given by the sum of the
original action and the counter term action. 
If
$\mathcal{A}_\rho[\Theta] = \mathcal{A}_\rho \times \Theta$,
the bare interaction strength is written as
$
V_{\rho,B}=\frac{\mu
\mathcal{Z}_{\rho}}{\mathcal{Z}_{\tau}^{d-1}\mathcal{Z}_{\psi}}V_{\rho}$, where $
\mathcal{Z}_\rho=1+\mathcal{A}_\rho $. 
Requiring that the bare source is
independent of $ \mu $, 
we obtain the flow equation 
for the renormalized source,
$
\dv{V_{\rho}}{l}=V_{\rho}(1+\gamma_{\rho}). $ Here $ \gamma_{\rho}=z\left(
\frac{g}{2}\partial_g + u\partial_{u} \right) \left(Z_{3,1}-Z_{\rho,1}\right)$
is the anomalous dimension of the source. 

\begin{figure}[t]
	\centering
	\begin{subfigure}[t]{0.2\textwidth}
		\centering
		\includegraphics[width=\textwidth]{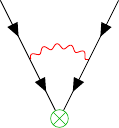}
		\caption{}
		\label{fig:SC_pairing}
	\end{subfigure}%
 \hspace{1cm}
	\begin{subfigure}[t]{0.2\textwidth}
		\centering
		\includegraphics[width=\textwidth]{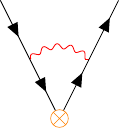}
		\caption{}
		\label{fig:CDW_pairing}
	\end{subfigure}%
	\caption{\label{fig:PairingChannels}\small{One loop diagrams that contribute to the anomalous dimension of the fermion bi-linear in the (a) particle-particle and  (b) particle-hole channels.}}
\end{figure}

To the leading order in $\epsilon$, the counter-terms can be obtained by evaluating the diagrams in Fig.~\ref{fig:PairingChannels}. 
Here, we consider the channels where the momentum of fermion bilinear is either $0$ or $2k_F$.
In this case, we can focus on
$\Theta$s that are diagonal in the hot-spot indices,
\begin{equation}
    \Theta_{n,\sigma,j; n',\sigma',j'} = b_n \d_{n,n'} [\vec{a}\cdot \vec{\tau}]_{\s,\s'} \delta_{j,j'}\hat{\Omega},
\end{equation}
where $\vec{b}= (b_1,b_2,b_3,b_4)$ is a vector of real numbers, $\vec{a}=(a_0,a_x,a_y,a_z)$ and $\vec{\tau} = (1,\tau^x,\tau^y,\tau^z)$. The insertion becomes
\begin{equation}
\mathcal{S}_{\rho}(\vec{a},\vec{b},\hat{\Omega})=V_{\rho} \mu \sum_{n,\sigma,{\sigma}',j}\int\!\dd{k} b_n\tilde\Psi^{(\rho)}_{n,\sigma,j}(k) [\vec{a}\cdot \vec{\tau}]_{\s,\s'}\hat{\Omega}\Psi_{n,\sigma',j}(k)+h.c..
\end{equation}
When $\rho=\SC$, the matrix $\tau^y(\vec{a}\cdot\vec{\tau}) \otimes \hat{\Omega}$ must be anti-symmetric to not vanish due to the anticommuting property of the fermions.

\subsection{Particle-particle channel
}

In the superconducting channel,
the one-loop contribution to the quantum effective action reads 
\begin{align}
\delta\mathcal{S}_{\mathrm{SC}}(A,\hat{\Omega},b)&= \sum_{n,\sigma,\tilde\sigma}\frac{V_\SC g^2}{N_f\mu^{d-4}}\!\!\int\!\dd{k} b_{\bar n}\Psi^{\top}_{n,\sigma,j}(-k) (i\tau^y)\left(\sum_{\alpha\in S_G} -\tau^\alpha(\vec{a}\cdot \vec{\tau})\tau^\alpha\right)_{\sigma,\tilde\sigma}\Upsilon_{\mathrm{SC}}^{(n)}(k,\hat{\Omega})\Psi_{n,\tilde{\sigma},j}(k) +h.c.,
\label{eq:deltaSsc}
\end{align}
where 
\begin{align}
\Upsilon_{\mathrm{SC}}^{(n)}(k,\hat{\Omega})&=\!\int\!\dd{q} D(q-k) \left(G_{\bar{n}}(-q)i\gamma_{d-1} \right)^{\top} \hat{\Omega} \left(G_{\bar{n}}(q)i\gamma_{d-1} \right), \notag\\
&=-\!\int\!\dd{q} \frac{\left(\varepsilon_{\bar{n}}(\vec{q}\,)\gamma_{d-1}-\vb{\Gamma}\cdot\vb{Q} \right)^{\top}\!\!\gamma_{d-1}^{\top}\hat{\Omega}\gamma_{d-1} \left( \varepsilon_{\bar{n}}(\vec{q}\,)\gamma_{d-1}-\vb{\Gamma}\cdot\vb{Q} \right)}{\left(c^2(\vec{q}-\vec{k}\,)^2+|\vb{Q-K}|^2\right)\left(\varepsilon_{\bar{n}}(\vec{q}\,)^2+|\vb{Q}|^2\right)^2},  \notag\\
&=-\int\frac{\dd{q'}}{v_F^2} \frac{\left(q_\parallel \gamma_{d-1}-q_\mu \gamma^{\mu}+k_0 \gamma_0\right)\gamma_{d-1}\hat{\Omega}\gamma_{d-1}\left(q_{\parallel} \gamma_{d-1}-\vb{\Gamma \cdot Q}\right)}{\left(\hat{c}^2(q-k)_\parallel^2+\hat{c}^2(q-k)_\perp^2+|\vb{Q-K}|^2 \right)\left(q_\parallel^{2}+|\vb{Q}|^2\right)^2}.
\label{eq:C5}
\end{align}
We have used $ \gamma_0^{\top}=-\gamma_0 $ and $ \gamma_i^{\top}=+\gamma_i $ for $ i=1,\dots,d-1 $. 
The change of variables performed in the last equality is the same as in the calculation of the 
one-loop fermion self-energy in Appendix \ref{app:FermionSE}. 
In the last expression, we have used the Einstein summation convention with $ \mu=1,\dots,d-2.$  
To extract $ 1/\epsilon $  pole, we can set $ \vec{k}=0 $ and keep $ \vb{K}\neq0 $ as an IR regulator. 
Because of parity, we only need to consider integrals where the numerators are either $q_\parallel^2$ or $\vb{Q}_a \vb{Q}_b$
(the integration of each factor is done in  App.~\ref{app:evaluationDiagrams}). 
The result can be written as
\begin{equation}\label{eq:ValOmegaprimeSC}
\Upsilon_{\mathrm{SC}}^{(n)}(k,\hat{\Omega})=\frac{-1}{16\pi v_F c \epsilon}\!\left[h_{6}^{f}(v,c)\hat\Omega'+h_{6}^{s}(v,c)\hat{\Omega}+\mathcal{O}\left(\epsilon \right)\right].
\end{equation}
Here $h_{6}^f(v,c)$ and $h_{6}^s(v,c)$ are defined in Eq.~\ref{eq:hSCfreq} and Eq.~\ref{eq:hSCspace}, respectively. 
They satisfy $h_{6}^f(0,0)=h_{6}^s(0,0)=1$.
For each  $\hat \Omega$,
$\hat \Omega'$ generated from the quantum correction is given by
$\hat{\Omega}'= \frac{1}{2}\left(
\left[\sum_{\mu=1}^{d-2}\gamma_\mu\gamma_{\scriptscriptstyle{d-1}}\hat{\Omega}\gamma_{\scriptscriptstyle{d-1}}\gamma_\mu\right]
-\gamma_0\gamma_{\scriptscriptstyle{d-1}}\hat{\Omega}\gamma_{\scriptscriptstyle{d-1}}\gamma_0\right)$.
In the $v,c \to 0 $ limit,
we obtain 
\begin{equation}
\Upsilon_{\mathrm{SC}}^{(n)}(k,\hat{\Omega})=\frac{-1}{16\pi v_F c \epsilon}\!\left[\Lambda_{\SC}[\hat\Omega]\hat{\Omega}+\mathcal{O}\left(\epsilon,v,c \right)\right],
\end{equation}
where 
\begin{equation}
    \begin{array}{|c|c|c|c|c|}\hline
        \hat{\Omega} &  1 & \gamma_0 &\gamma_{d-1} & \gamma_{\mu}\\\hline
       \Lambda_{\SC}[\Omega]& 1 & 2 & 1 &0\\
       \hline
    \end{array}.
\end{equation}
It is noted that $\hat \Omega$ determines
the total momentum of Cooper pairs and the charge under the $\pi$-rotation.
For $\hat\Omega=1$, the total momentum is $2k_F$, while for $\hat\Omega=\gamma_{d-1},\gamma_0$ the total momentum is zero. 
Under the $\pi$-rotation, $\Psi_{n}(k) \to \pm \gamma_{d-1}\Psi_{n}(R_{\pi}k) $,  
$\hat \Omega$ is transformed into
$\gamma_{d-1}^{\top}\hat\Omega \gamma_{d-1} = (-1)^l \hat\Omega $. 
The spin-dependence in \eq{eq:deltaSsc} is determined from
$\tilde{\tau}^A=-\sum_{\alpha\in S_G}(\tau^\alpha \tau^A \tau^\alpha)
 = \Lambda_{s,G}[\tau^A] \tau^A$,
where the eigenvalue
$\Lambda_{s,G}[\tau^A]$ is summarized as
\begin{equation}
    \begin{array}{|c|c|c|c|c|} 
    \hline
   G & 1 & \tau^x & \tau^y & \tau^z\\ \hline
   Z_2 & -1 & 1 & 1 & -1 \\
   O(2) & -2 & 0 & 0 & 2\\
   O(3) & -3 & 1 & 1 & 1 \\\hline
\end{array}.
\end{equation}
The factor of the eigenvalue associated with the hot-spot wavefunction is $\Lambda_l[\vec{b}] = b_n/b_{\bar{n}} = \pm 1$.
The net anomalous dimension can be written as
\begin{equation}
    \lambda = -\frac{g^2}{16\pi c  N_f} \Lambda_l[\vec{b}]\times \Lambda_{\SC}[\hat\Omega] \times \Lambda_{s,G}[\tau^A].
\end{equation}
The pairing vertex that is most enhanced is determined to be the one with the largest $\lambda$
under the constraint that $(\tau^y \tau^ A)\otimes\hat\Omega$ is anti-symmetric: 
\begin{itemize}
 \item For $G=Z_2$, the most enhanced channels are $(\tau^A,b_n/b_{\bar{n}},\hat{\Omega})=(+\tau^z,+1,\gamma_0),(+\tau^{x/y},-1,\gamma_0)$.
\item For $G=O(2)$, the most enhanced channels are 
$(\tau^A,b_n/b_{\bar{n}},\hat{\Omega})=(+\tau^z,-1,\gamma_0)$.
    \item For $G=O(3)$, the most enhanced channels are  $(\tau^A,b_n/b_{\bar{n}},\hat{\Omega})=(+1,+1,1),(+1,+1,\gamma_{d-1})$.
\end{itemize}
An additional degeneracy can arise from the relative sign between $b_1$ and $b_2$.
The relative sign between $b_1$ and $b_2$ determines the transformation properties under $\pi/2$ rotations. 
The quantum number under the spin rotation $\Psi \to \exp(i(\theta/2) \vec{n}\cdot\vec{\tau}) \Psi$ is determined by $\tau^A$. 
Note that some of the $d$-wave channels are still enhanced in the $G=Z_2$ and $O(2)$ theories. 
For instance the channel $(\tau^A,b_n/b_{\bar{n}},\hat{\Omega})=(1,1,1)$ have a positive $\lambda$ for all theories. 

\subsection{Particle-hole channel }
\def\PH{\mathrm{PH}}

In the particle-hole channel, 
the one-loop contribution to the quantum action becomes 
\begin{align*}
\delta\mathcal{S}_{\mathrm{PH}}(A,\hat{\Omega},b)&= \frac{V_\rho g^2}{N_f\mu^{d-4}}\!\!\sum_{n,\sigma,\tilde\sigma}\int\!\dd{k} b_{\bar n}\bar{\Psi}_{n,\sigma,j}(k) \left[\sum_{\a\in S_G}\tau^\alpha(\vec{a}\cdot\vec{\tau})\tau^\alpha\right]_{\sigma,\tilde\sigma}\Upsilon_{\mathrm{PH}}^{(n)}(k,\hat{\Omega})\Psi_{n,\tilde{\sigma},j}(k),
\end{align*}
where
\begin{align*}
\Upsilon_{\mathrm{PH}}^{(n)}(k,\hat{\Omega})&=\!\int\!\dd{q} D(q-k) i\gamma_{d-1} G_{\bar{n}}(q)\hat{\Omega} G_{\bar{n}}(q)i\gamma_{d-1} , \notag\\
&=\!\int\!\dd{q} \frac{\gamma_{d-1}\left(\varepsilon_{\bar{n}}(\vec{q}\,)\gamma_{d-1}+\vb{\Gamma}\cdot\vb{Q} \right)\hat{\Omega} \left( \varepsilon_{\bar{n}}(\vec{q}\,)\gamma_{d-1}+\vb{\Gamma}\cdot\vb{Q} \right)\gamma_{d-1}}{\left(c^2(\vec{q}-\vec{k}\,)^2+|\vb{Q-K}|^2\right)\left(\varepsilon_{\bar{n}}(\vec{q}\,)^2+|\vb{Q}|^2\right)^2},  \notag\\
&=\int\frac{\dd{q'}}{v_F^2} \frac{\gamma_{d-1}\left(q_\parallel \gamma_{d-1}+\vb{\Gamma \cdot Q}\right)\hat{\Omega}\left(q_{\parallel} \gamma_{d-1}+\vb{\Gamma \cdot Q}\right)\gamma_{d-1}}{\left(\hat{c}^2(q-k)_\parallel^2+\hat{c}^2(q-k)_\perp^2+|\vb{Q-K}|^2 \right)\left(q_\parallel^{2}+|\vb{Q}|^2\right)^2}.
\end{align*}
The change of variables performed in the last equality is the same as for the SC channels. 
We also set $\vec{k}=0$.
The computation of 
$\Upsilon_{\mathrm{PH}}^{(n)}(k,\hat{\Omega})$
is parallel to that of 
$\Upsilon_{\mathrm{SC}}^{(n)}(k,\hat{\Omega})$,
\begin{equation}
\Upsilon_{\mathrm{PH}}^{(n)}(k,\hat{\Omega})=\frac{1}{16\pi v_F c \epsilon}\!\left[
h_{6}^{f}(v,c)
\hat{\Omega}''+
h_{6}^{s}(v,c)
\hat{\Omega}+\mathcal{O}\left(\epsilon \right)\right],
\end{equation}
where
$\hat{\Omega}''=\frac{1}{2}\left[\sum_{\mu=0}^{d-2}\gamma_\mu\gamma_{d-1}{\Omega}\gamma_{d-1}\gamma_\mu\right]$.
The anomalous dimension is determined by the eigenvalue
$\Lambda_{\PH}[\hat\Omega]$
defined through $\hat\Omega''=\Lambda_{\PH}[\hat\Omega]\hat\Omega$ with
\begin{equation}
    \begin{array}{|c|c|c|c|c|}\hline
        \hat{\Omega} &  1 & \gamma_0 &\gamma_{d-1} & \gamma_{\mu}\\\hline
       \Lambda_{\PH}[\hat\Omega]& 2&1 & 0&1\\
       \hline
    \end{array}
\end{equation}
The net anomalous dimension becomes
\begin{equation}
    \lambda = 
    -\frac{g^2}{16\pi c  N_f} \Lambda_l[\vec{b}]\times \Lambda_{\PH}[\hat\Omega] \times \Lambda_{s,G}[\tau^A].
\end{equation}
Unlike the particle-particle channel, there is no constraint on $\hat \Omega$. 
For each spin symmetry group,
the dominant charge-density wave vertex is 
\begin{itemize}
    \item For $G=Z_2$, the most enhanced channels are $(\tau^A,b_n/b_{\bar{n}},\hat{\Omega})=(1/\tau^z, +1, 1), (\tau^x/\tau^y, -1, 1)$.
    \item For $G=O(2)$, the most enhanced channels are $(\tau^A,b_n/b_{\bar{n}},\hat{\Omega})=(1,1,1),(+\tau^z,-1,1)$.
    \item For $G=O(3)$, the most enhanced channels are $(\tau^A,b_n/b_{\bar{n}},\hat{\Omega})=(+1,+1,1)$.
\end{itemize}

\subsection{Evaluation of diagrams }\label{app:evaluationDiagrams}

Here, we show the details for 
 the computation of  \eq{eq:C5}.
For convenience, 
we divide \eq{eq:C5} into the part that is proportional to 
$\vb{Q}_{a}\vb{Q}_{b}$ 
and the other part proportional to
$q_\parallel^2$.

\subsubsection{
\texorpdfstring{$\vb{Q}_{a}\vb{Q}_{b}$}{QaQb} part
}

The term that is proportional to 
$\vb{Q}_{a}\vb{Q}_{b}$ 
in \eq{eq:C5} is written as
\begin{align}
I_{a,b}^{f}\left(\vb{K}\right)&=\int\frac{\dd{q}}{v_F^2} \frac{
\vb{Q}_{a}\vb{Q}_{b}
}{\left(\hat{c}^2q_\parallel^2+\hat{c}^2q_\perp^2+|\vb{Q-K}|^2 \right)\left(q_\parallel^{2}+|\vb{Q}|^2\right)^2} \notag \\
&=\frac{1}{2v_Fc}\int\frac{\dd{q_{\parallel}}\dd{\vb{Q}}}{(2\pi)^d} \frac{\frac{\delta_{a,b}}{(d-1)}|\vb{Q}|^2}{\left(q_\parallel^{2}+|\vb{Q}|^2\right)^2\left(\hat{c}^2q_\parallel^2+|\vb{Q-K}|^2 \right)^{1/2}} \notag\\
&=\frac{\delta_{a,b}}{2v_{F} c(d-1)}\int_{\mathcal{D}_2}\!\!\dd{x_1}\dd{x_2}  \int\frac{\dd{q_{\parallel}}\dd{\vb{Q}}}{(2\pi)^d}\frac{x_1\,x_2^{-\frac{1}{2}}}{B\left(\frac{1}{2},2\right)}  \frac{|\vb{Q}|^2}{\left(|\vb{Q}-x_2\vb{K}|^2+(x_1+x_2 \hat{c}^2)q_\parallel^2+x_1 x_2 |\vb{K}|^2\right)^{\frac{5}{2}}},
\end{align}
where $\int_{\mathcal{D}_2}\!\!\dd{x_1}\dd{x_2} = \int_{0}^1\dd{x_1}\int_0^1\dd{x_2}\delta(x_1+x_2-1)$. 
With a shift 
$ \vb{Q}\rightarrow\vb{Q}+x_2\vb{K}$, 
a rescaling $q_\parallel\rightarrow \frac{q_\parallel}{\sqrt{x_1+x_2 \hat{c}^2}} $, 
and discarding terms that are non-singular, we obtain
\begin{align}
I_{a,b}^{f}\left(\vb{K}\right)&=\frac{\delta_{a,b}}{2v_{F} c d }\int_{\mathcal{D}_2}\!\!\frac{\dd{x_1}\dd{x_2}}{B\left(\frac{1}{2},2\right)}\frac{x_1\,x_2^{-\frac{1}{2}}}{\sqrt{x_1+\hat{c}^2x_2}}\!\!\int\!\!\frac{\dd{\vb{X}}}{(2\pi)^d}
\frac{|\vb{X}|^2}{\left(|\vb{X}|^2+x_1 x_2 |\vb{K}|^2\right)^{\frac{5}{2}}}, \notag\\
&=\frac{\delta_{a,b}}{2 v_F c}\frac{\Gamma\left(\frac{3-d}{2}\right)|\vb{K}|^{d-3}}{(4\pi)^\frac{d+1}{2}}\int_{\mathcal{D}_2}\!\!{\dd{x_1}\dd{x_2}}\frac{x_1\,x_2^{-\frac{1}{2}}}{\sqrt{x_1+\hat{c}^2x_2}}\frac{1}{(x_1 x_2)^{3-d}}, \notag\\
&=\frac{\delta_{a,b}}{32\pi v_F c\,\epsilon}\left(h_{6}^{f}(v,c) +\mathcal{O}(\epsilon) \right),
\end{align}
where 
the integration variables are combined into a d-dimensional vector $ \vb{X}=(\vb{Q},q_\parallel)$,
the spherical symmetry is used to replace $ |\vb{Q}|^2$ with $\frac{d-1}{d} |\vb{X}|^2 $,
and
\begin{equation}\label{eq:hSCfreq}
h_{6}^f(v,c)1=\frac{2}{\pi}\int_{\mathcal{D}_2}\!\!{\dd{x_1}\dd{x_2}}\frac{x_1\,x_2^{-\frac{1}{2}}}{\sqrt{x_1+\hat{c}^2x_2}}=\frac{8}{3\pi} \,_2F_1\left(\frac{1}{2},\frac{1}{2};\frac{5}{2};1-\hat{c}^2\right).
\end{equation}
In the small $ \hat{c} $ limit, $ h_{6}^{f}(v,c)= 1-\frac{\hat{c}^2}{2}+\mathcal{O}\left(\hat{c}^3\right)$. 

\subsubsection{
\texorpdfstring{$q_\parallel^2$}{Square q parallel} part }
The term that is proportional to 
$q_\parallel^2$ 
in \eq{eq:C5} reads
\begin{align}\allowdisplaybreaks
I^{s}\left(\vb{K}\right)&=\int\frac{\dd{q'}}{v_F^2} \frac{q_\parallel^2}{\left(\hat{c}^2q_\parallel^2+\hat{c}^2q_\perp^2+|\vb{Q-K}|^2 \right)\left(q_\parallel^{2}+|\vb{Q}|^2\right)^2}, \notag\\
&=\frac{1}{2v_{F} c}\int_{\mathcal{D}_2}\!\!\dd{x_1}\dd{x_2}  \int\frac{\dd{q_{\parallel}}\dd{\vb{Q}}}{(2\pi)^d}  \frac{x_1\,x_2^{-\frac{1}{2}}}{B\left(\frac{1}{2},2\right)}  \frac{q_\parallel^2}{\left(|\vb{Q}-x_2\vb{K}|^2+(x_1+x_2 \hat{c}^2)q_\parallel^2+x_1 x_2 |\vb{K}|^2\right)^{\frac{5}{2}}}.
\end{align} \allowdisplaybreaks[0]
We follow the same steps as before with one difference in replacing $ q_\parallel^2$ with $\frac{1}{d}|\vb{X}|^2 $ to obtain
\begin{align}
I^{s}\left(\vb{K}\right)&=\frac{1}{2v_{F} c d }\int_{\mathcal{D}_2}\!\!\frac{\dd{x_1}\dd{x_2}}{B\left(\frac{1}{2},2\right)}\frac{x_1\,x_2^{-\frac{1}{2}}}{\left(x_1+\hat{c}^2x_2\right)^{\frac{3}{2}}}\!\!\int\!\!\frac{\dd{\vb{X}}}{(2\pi)^d}
\frac{|\vb{X}|^2}{\left(|\vb{X}|^2+x_1 x_2 |\vb{K}|^2\right)^{\frac{5}{2}}} \notag\\
&=\frac{1}{2 v_F c}\frac{\Gamma\left(\frac{3-d}{2}\right)|\vb{K}|^{d-3}}{(4\pi)^\frac{d+1}{2}}\int_{\mathcal{D}_2}\!\!{\dd{x_1}\dd{x_2}}\frac{x_1\,x_2^{-\frac{1}{2}}}{\left(x_1+\hat{c}^2x_2\right)^{\frac{3}{2}}}\frac{1}{(x_1 x_2)^{3-d}} \notag\\
&=\frac{1}{16\pi v_F c\,\epsilon}\left(h_{6}^{s}(v,c) +\mathcal{O}(\epsilon) \right),
\end{align}
where 
\begin{align}\label{eq:hSCspace}
h_{6}^{s}(v,c)&=\frac{1}{\pi}\int_{\mathcal{D}_2}\!\!{\dd{x_1}\dd{x_2}}\frac{x_1\,x_2^{-\frac{1}{2}}}{{\left(x_1+\hat{c}^2x_2\right)^{\frac{3}{2}} }}=\frac{4}{3\pi} \,_2F_1\left(\frac{1}{2},\frac{3}{2};\frac{5}{2};1-\hat{c}^2\right)\nn
&=\frac{2}{\pi}\frac{d_0(\hat{c})-\hat{c}}{1-\hat{c}^2}=1-\frac{4}{\pi}\hat{c}+\frac{3}{2}\hat{c}^2+\mathcal{O}\left(\hat{c}^3\right).
\end{align}

\section*{Acknowledgement}

We thank Joseph Maciejko for discussions.
This research was supported by the Natural Sciences 
and Engineering Research Council of
Canada. Research at the Perimeter Institute is supported in part by the
Government of Canada through Industry Canada, and by the Province of
Ontario through the Ministry of Research and Information. 
VC acknowledges the support of the Perimeter Scholars International (PSI) program during the 2017-2018 academic year at the Perimeter Institute. 
Portions of this work were conducted for VC's PSI essay.

\bibliographystyle{apsrev4-1-title}
\bibliography{ref_try}

\begin{thebibliography}{75}%
\makeatletter
\providecommand \@ifxundefined [1]{%
 \@ifx{#1\undefined}
}%
\providecommand \@ifnum [1]{%
 \ifnum #1\expandafter \@firstoftwo
 \else \expandafter \@secondoftwo
 \fi
}%
\providecommand \@ifx [1]{%
 \ifx #1\expandafter \@firstoftwo
 \else \expandafter \@secondoftwo
 \fi
}%
\providecommand \natexlab [1]{#1}%
\providecommand \enquote  [1]{``#1''}%
\providecommand \bibnamefont  [1]{#1}%
\providecommand \bibfnamefont [1]{#1}%
\providecommand \citenamefont [1]{#1}%
\providecommand \href@noop [0]{\@secondoftwo}%
\providecommand \href [0]{\begingroup \@sanitize@url \@href}%
\providecommand \@href[1]{\@@startlink{#1}\@@href}%
\providecommand \@@href[1]{\endgroup#1\@@endlink}%
\providecommand \@sanitize@url [0]{\catcode `\\12\catcode `\$12\catcode
  `\&12\catcode `\#12\catcode `\^12\catcode `\_12\catcode `\%12\relax}%
\providecommand \@@startlink[1]{}%
\providecommand \@@endlink[0]{}%
\providecommand \url  [0]{\begingroup\@sanitize@url \@url }%
\providecommand \@url [1]{\endgroup\@href {#1}{\urlprefix }}%
\providecommand \urlprefix  [0]{URL }%
\providecommand \Eprint [0]{\href }%
\providecommand \doibase [0]{http://dx.doi.org/}%
\providecommand \selectlanguage [0]{\@gobble}%
\providecommand \bibinfo  [0]{\@secondoftwo}%
\providecommand \bibfield  [0]{\@secondoftwo}%
\providecommand \translation [1]{[#1]}%
\providecommand \BibitemOpen [0]{}%
\providecommand \bibitemStop [0]{}%
\providecommand \bibitemNoStop [0]{.\EOS\space}%
\providecommand \EOS [0]{\spacefactor3000\relax}%
\providecommand \BibitemShut  [1]{\csname bibitem#1\endcsname}%
\let\auto@bib@innerbib\@empty
\bibitem [{\citenamefont {Hooft}(1980)}]{Hooft1980}%
  \BibitemOpen
  \bibfield  {author} {\bibinfo {author} {\bibfnamefont {G.}~\bibnamefont
  {Hooft}},\ }\enquote {\bibinfo {title} {Naturalness, chiral symmetry, and
  spontaneous chiral symmetry breaking},}\ in\ \href {\doibase
  10.1007/978-1-4684-7571-5_9} {\emph {\bibinfo {booktitle} {Recent
  Developments in Gauge Theories}}},\ \bibinfo {editor} {edited by\ \bibinfo
  {editor} {\bibfnamefont {G.}~\bibnamefont {Hooft}}, \bibinfo {editor}
  {\bibfnamefont {C.}~\bibnamefont {Itzykson}}, \bibinfo {editor}
  {\bibfnamefont {A.}~\bibnamefont {Jaffe}}, \bibinfo {editor} {\bibfnamefont
  {H.}~\bibnamefont {Lehmann}}, \bibinfo {editor} {\bibfnamefont {P.~K.}\
  \bibnamefont {Mitter}}, \bibinfo {editor} {\bibfnamefont {I.~M.}\
  \bibnamefont {Singer}}, \ and\ \bibinfo {editor} {\bibfnamefont
  {R.}~\bibnamefont {Stora}}}\ (\bibinfo  {publisher} {Springer US},\ \bibinfo
  {address} {Boston, MA},\ \bibinfo {year} {1980})\ pp.\ \bibinfo {pages}
  {135--157}\BibitemShut {NoStop}%
\bibitem [{\citenamefont {Seiberg}(1993)}]{SEIBERG1993469}%
  \BibitemOpen
  \bibfield  {author} {\bibinfo {author} {\bibfnamefont {N.}~\bibnamefont
  {Seiberg}},\ }\bibfield  {title} {\enquote {\bibinfo {title} {Naturalness
  versus supersymmetric non-renormalization theorems},}\ }\href {\doibase
  10.1016/0370-2693(93)91541-T} {\bibfield  {journal} {\bibinfo  {journal}
  {Physics Letters B}\ }\textbf {\bibinfo {volume} {318}},\ \bibinfo {pages}
  {469 } (\bibinfo {year} {1993})}\BibitemShut {NoStop}%
\bibitem [{\citenamefont {Lee}(2018{\natexlab{a}})}]{SUNGSIKREVIEW}%
  \BibitemOpen
  \bibfield  {author} {\bibinfo {author} {\bibfnamefont {S.-S.}\ \bibnamefont
  {Lee}},\ }\bibfield  {title} {\enquote {\bibinfo {title} {Recent developments
  in non-fermi liquid theory},}\ }\href {\doibase
  10.1146/annurev-conmatphys-031016-025531} {\bibfield  {journal} {\bibinfo
  {journal} {Annu. Rev. of Condens. Matter Phys.}\ }\textbf {\bibinfo {volume}
  {9}},\ \bibinfo {pages} {227} (\bibinfo {year}
  {2018}{\natexlab{a}})}\BibitemShut {NoStop}%
\bibitem [{\citenamefont {Hooft}(1974)}]{THOOFT}%
  \BibitemOpen
  \bibfield  {author} {\bibinfo {author} {\bibfnamefont {G.}~\bibnamefont
  {Hooft}},\ }\bibfield  {title} {\enquote {\bibinfo {title} {A planar diagram
  theory for strong interactions},}\ }\href {\doibase
  https://doi.org/10.1016/0550-3213(74)90154-0} {\bibfield  {journal} {\bibinfo
   {journal} {Nuclear Physics B}\ }\textbf {\bibinfo {volume} {72}},\ \bibinfo
  {pages} {461 } (\bibinfo {year} {1974})}\BibitemShut {NoStop}%
\bibitem [{\citenamefont {Lee}(2009)}]{SSLEE}%
  \BibitemOpen
  \bibfield  {author} {\bibinfo {author} {\bibfnamefont {S.-S.}\ \bibnamefont
  {Lee}},\ }\bibfield  {title} {\enquote {\bibinfo {title} {Low-energy
  effective theory of fermi surface coupled with u(1) gauge field in $2+1$
  dimensions},}\ }\href {\doibase 10.1103/PhysRevB.80.165102} {\bibfield
  {journal} {\bibinfo  {journal} {Phys. Rev. B}\ }\textbf {\bibinfo {volume}
  {80}},\ \bibinfo {pages} {165102} (\bibinfo {year} {2009})}\BibitemShut
  {NoStop}%
\bibitem [{\citenamefont {Abanov}\ and\ \citenamefont
  {Chubukov}(2000)}]{ABANOV1}%
  \BibitemOpen
  \bibfield  {author} {\bibinfo {author} {\bibfnamefont {A.}~\bibnamefont
  {Abanov}}\ and\ \bibinfo {author} {\bibfnamefont {A.~V.}\ \bibnamefont
  {Chubukov}},\ }\bibfield  {title} {\enquote {\bibinfo {title} {{Spin-Fermion
  Model near the Quantum Critical Point: One-Loop Renormalization Group
  Results}},}\ }\href {\doibase 10.1103/PhysRevLett.84.5608} {\bibfield
  {journal} {\bibinfo  {journal} {Phys. Rev. Lett.}\ }\textbf {\bibinfo
  {volume} {84}},\ \bibinfo {pages} {5608} (\bibinfo {year}
  {2000})}\BibitemShut {NoStop}%
\bibitem [{\citenamefont {Abanov}\ \emph {et~al.}(2003)\citenamefont {Abanov},
  \citenamefont {Chubukov},\ and\ \citenamefont {Schmalian}}]{ABANOV3}%
  \BibitemOpen
  \bibfield  {author} {\bibinfo {author} {\bibfnamefont {A.}~\bibnamefont
  {Abanov}}, \bibinfo {author} {\bibfnamefont {A.~V.}\ \bibnamefont
  {Chubukov}}, \ and\ \bibinfo {author} {\bibfnamefont {J.}~\bibnamefont
  {Schmalian}},\ }\bibfield  {title} {\enquote {\bibinfo {title}
  {{Quantum-critical theory of the spin-fermion model and its application to
  cuprates: Normal state analysis}},}\ }\href {\doibase
  10.1080/0001873021000057123} {\bibfield  {journal} {\bibinfo  {journal} {Adv.
  Phys.}\ }\textbf {\bibinfo {volume} {52}},\ \bibinfo {pages} {119} (\bibinfo
  {year} {2003})}\BibitemShut {NoStop}%
\bibitem [{\citenamefont {Abanov}\ and\ \citenamefont
  {Chubukov}(2004)}]{ABANOV2}%
  \BibitemOpen
  \bibfield  {author} {\bibinfo {author} {\bibfnamefont {A.}~\bibnamefont
  {Abanov}}\ and\ \bibinfo {author} {\bibfnamefont {A.}~\bibnamefont
  {Chubukov}},\ }\bibfield  {title} {\enquote {\bibinfo {title} {{Anomalous
  Scaling at the Quantum Critical Point in Itinerant Antiferromagnets}},}\
  }\href {\doibase 10.1103/PhysRevLett.93.255702} {\bibfield  {journal}
  {\bibinfo  {journal} {Phys. Rev. Lett.}\ }\textbf {\bibinfo {volume} {93}},\
  \bibinfo {pages} {255702} (\bibinfo {year} {2004})}\BibitemShut {NoStop}%
\bibitem [{\citenamefont {Hartnoll}\ \emph {et~al.}(2011)\citenamefont
  {Hartnoll}, \citenamefont {Hofman}, \citenamefont {Metlitski},\ and\
  \citenamefont {Sachdev}}]{HARTNOLL}%
  \BibitemOpen
  \bibfield  {author} {\bibinfo {author} {\bibfnamefont {S.~A.}\ \bibnamefont
  {Hartnoll}}, \bibinfo {author} {\bibfnamefont {D.~M.}\ \bibnamefont
  {Hofman}}, \bibinfo {author} {\bibfnamefont {M.~A.}\ \bibnamefont
  {Metlitski}}, \ and\ \bibinfo {author} {\bibfnamefont {S.}~\bibnamefont
  {Sachdev}},\ }\bibfield  {title} {\enquote {\bibinfo {title} {Quantum
  critical response at the onset of spin-density-wave order in two-dimensional
  metals},}\ }\href {\doibase 10.1103/PhysRevB.84.125115} {\bibfield  {journal}
  {\bibinfo  {journal} {Phys. Rev. B}\ }\textbf {\bibinfo {volume} {84}},\
  \bibinfo {pages} {125115} (\bibinfo {year} {2011})}\BibitemShut {NoStop}%
\bibitem [{\citenamefont {Abrahams}\ and\ \citenamefont
  {W\"olfe}(2012)}]{ABRAHAMS}%
  \BibitemOpen
  \bibfield  {author} {\bibinfo {author} {\bibfnamefont {E.}~\bibnamefont
  {Abrahams}}\ and\ \bibinfo {author} {\bibfnamefont {P.}~\bibnamefont
  {W\"olfe}},\ }\bibfield  {title} {\enquote {\bibinfo {title} {{Critical
  quasiparticle theory applied to heavy fermion metals near an
  antiferromagnetic quantum phase transition}},}\ }\href {\doibase
  10.1073/pnas.1200346109} {\bibfield  {journal} {\bibinfo  {journal} {Proc.
  Natl. Acad. Sci.}\ }\textbf {\bibinfo {volume} {109}},\ \bibinfo {pages}
  {3238} (\bibinfo {year} {2012})}\BibitemShut {NoStop}%
\bibitem [{\citenamefont {Lee}\ \emph {et~al.}(2013)\citenamefont {Lee},
  \citenamefont {Strack},\ and\ \citenamefont {Sachdev}}]{PhysRevB.87.045104}%
  \BibitemOpen
  \bibfield  {author} {\bibinfo {author} {\bibfnamefont {J.}~\bibnamefont
  {Lee}}, \bibinfo {author} {\bibfnamefont {P.}~\bibnamefont {Strack}}, \ and\
  \bibinfo {author} {\bibfnamefont {S.}~\bibnamefont {Sachdev}},\ }\bibfield
  {title} {\enquote {\bibinfo {title} {Quantum criticality of reconstructing
  fermi surfaces in antiferromagnetic metals},}\ }\href {\doibase
  10.1103/PhysRevB.87.045104} {\bibfield  {journal} {\bibinfo  {journal} {Phys.
  Rev. B}\ }\textbf {\bibinfo {volume} {87}},\ \bibinfo {pages} {045104}
  (\bibinfo {year} {2013})}\BibitemShut {NoStop}%
\bibitem [{\citenamefont {de~Carvalho}\ and\ \citenamefont
  {Freire}(2014)}]{DECARVALHO}%
  \BibitemOpen
  \bibfield  {author} {\bibinfo {author} {\bibfnamefont {V.~S.}\ \bibnamefont
  {de~Carvalho}}\ and\ \bibinfo {author} {\bibfnamefont {H.}~\bibnamefont
  {Freire}},\ }\bibfield  {title} {\enquote {\bibinfo {title} {Evidence of a
  short-range incommensurate d-wave charge order from a fermionic two-loop
  renormalization group calculation of a 2d model with hot spots},}\ }\href
  {\doibase 10.1016/j.aop.2014.05.009} {\bibfield  {journal} {\bibinfo
  {journal} {Annals of Physics}\ }\textbf {\bibinfo {volume} {348}},\ \bibinfo
  {pages} {32 } (\bibinfo {year} {2014})}\BibitemShut {NoStop}%
\bibitem [{\citenamefont {Patel}\ \emph {et~al.}(2015)\citenamefont {Patel},
  \citenamefont {Strack},\ and\ \citenamefont {Sachdev}}]{PATEL}%
  \BibitemOpen
  \bibfield  {author} {\bibinfo {author} {\bibfnamefont {A.~A.}\ \bibnamefont
  {Patel}}, \bibinfo {author} {\bibfnamefont {P.}~\bibnamefont {Strack}}, \
  and\ \bibinfo {author} {\bibfnamefont {S.}~\bibnamefont {Sachdev}},\
  }\bibfield  {title} {\enquote {\bibinfo {title} {Hyperscaling at the spin
  density wave quantum critical point in two-dimensional metals},}\ }\href
  {\doibase 10.1103/PhysRevB.92.165105} {\bibfield  {journal} {\bibinfo
  {journal} {Phys. Rev. B}\ }\textbf {\bibinfo {volume} {92}},\ \bibinfo
  {pages} {165105} (\bibinfo {year} {2015})}\BibitemShut {NoStop}%
\bibitem [{\citenamefont {Patel}\ and\ \citenamefont {Sachdev}(2014)}]{PATEL2}%
  \BibitemOpen
  \bibfield  {author} {\bibinfo {author} {\bibfnamefont {A.~A.}\ \bibnamefont
  {Patel}}\ and\ \bibinfo {author} {\bibfnamefont {S.}~\bibnamefont
  {Sachdev}},\ }\bibfield  {title} {\enquote {\bibinfo {title} {dc resistivity
  at the onset of spin density wave order in two-dimensional metals},}\ }\href
  {\doibase 10.1103/PhysRevB.90.165146} {\bibfield  {journal} {\bibinfo
  {journal} {Phys. Rev. B}\ }\textbf {\bibinfo {volume} {90}},\ \bibinfo
  {pages} {165146} (\bibinfo {year} {2014})}\BibitemShut {NoStop}%
\bibitem [{\citenamefont {Varma}(2015)}]{VARMA2}%
  \BibitemOpen
  \bibfield  {author} {\bibinfo {author} {\bibfnamefont {C.~M.}\ \bibnamefont
  {Varma}},\ }\bibfield  {title} {\enquote {\bibinfo {title} {Quantum
  criticality in quasi-two-dimensional itinerant antiferromagnets},}\ }\href
  {\doibase 10.1103/PhysRevLett.115.186405} {\bibfield  {journal} {\bibinfo
  {journal} {Phys. Rev. Lett.}\ }\textbf {\bibinfo {volume} {115}},\ \bibinfo
  {pages} {186405} (\bibinfo {year} {2015})}\BibitemShut {NoStop}%
\bibitem [{\citenamefont {Maier}\ and\ \citenamefont {Strack}(2016)}]{MAIER}%
  \BibitemOpen
  \bibfield  {author} {\bibinfo {author} {\bibfnamefont {S.~A.}\ \bibnamefont
  {Maier}}\ and\ \bibinfo {author} {\bibfnamefont {P.}~\bibnamefont {Strack}},\
  }\bibfield  {title} {\enquote {\bibinfo {title} {Universality in
  antiferromagnetic strange metals},}\ }\href {\doibase
  10.1103/PhysRevB.93.165114} {\bibfield  {journal} {\bibinfo  {journal} {Phys.
  Rev. B}\ }\textbf {\bibinfo {volume} {93}},\ \bibinfo {pages} {165114}
  (\bibinfo {year} {2016})}\BibitemShut {NoStop}%
\bibitem [{\citenamefont {Varma}\ \emph {et~al.}(2018)\citenamefont {Varma},
  \citenamefont {Gannon}, \citenamefont {Aronson}, \citenamefont
  {Rodriguez-Rivera},\ and\ \citenamefont {Qiu}}]{VARMA3}%
  \BibitemOpen
  \bibfield  {author} {\bibinfo {author} {\bibfnamefont {C.~M.}\ \bibnamefont
  {Varma}}, \bibinfo {author} {\bibfnamefont {W.~J.}\ \bibnamefont {Gannon}},
  \bibinfo {author} {\bibfnamefont {M.~C.}\ \bibnamefont {Aronson}}, \bibinfo
  {author} {\bibfnamefont {J.~A.}\ \bibnamefont {Rodriguez-Rivera}}, \ and\
  \bibinfo {author} {\bibfnamefont {Y.}~\bibnamefont {Qiu}},\ }\bibfield
  {title} {\enquote {\bibinfo {title} {Quantum critical singularities in
  two-dimensional metallic xy ferromagnets},}\ }\href {\doibase
  10.1103/PhysRevB.97.085134} {\bibfield  {journal} {\bibinfo  {journal} {Phys.
  Rev. B}\ }\textbf {\bibinfo {volume} {97}},\ \bibinfo {pages} {085134}
  (\bibinfo {year} {2018})}\BibitemShut {NoStop}%
\bibitem [{\citenamefont {Metlitski}\ and\ \citenamefont
  {Sachdev}(2010{\natexlab{a}})}]{MAX2}%
  \BibitemOpen
  \bibfield  {author} {\bibinfo {author} {\bibfnamefont {M.~A.}\ \bibnamefont
  {Metlitski}}\ and\ \bibinfo {author} {\bibfnamefont {S.}~\bibnamefont
  {Sachdev}},\ }\bibfield  {title} {\enquote {\bibinfo {title} {Quantum phase
  transitions of metals in two spatial dimensions. ii. spin density wave
  order},}\ }\href {\doibase 10.1103/PhysRevB.82.075128} {\bibfield  {journal}
  {\bibinfo  {journal} {Phys. Rev. B}\ }\textbf {\bibinfo {volume} {82}},\
  \bibinfo {pages} {075128} (\bibinfo {year} {2010}{\natexlab{a}})}\BibitemShut
  {NoStop}%
\bibitem [{\citenamefont {Berg}\ \emph
  {et~al.}(2012{\natexlab{a}})\citenamefont {Berg}, \citenamefont {Metlitski},\
  and\ \citenamefont {Sachdev}}]{MAX1}%
  \BibitemOpen
  \bibfield  {author} {\bibinfo {author} {\bibfnamefont {E.}~\bibnamefont
  {Berg}}, \bibinfo {author} {\bibfnamefont {M.}~\bibnamefont {Metlitski}}, \
  and\ \bibinfo {author} {\bibfnamefont {S.}~\bibnamefont {Sachdev}},\
  }\bibfield  {title} {\enquote {\bibinfo {title} {Sign-problem--free quantum
  monte carlo of the onset of antiferromagnetism in metals},}\ }\href {\doibase
  10.1126/science.1227769} {\bibfield  {journal} {\bibinfo  {journal}
  {Science}\ }\textbf {\bibinfo {volume} {338}},\ \bibinfo {pages} {1606}
  (\bibinfo {year} {2012}{\natexlab{a}})}\BibitemShut {NoStop}%
\bibitem [{\citenamefont {Li}\ \emph {et~al.}(2016)\citenamefont {Li},
  \citenamefont {Wang}, \citenamefont {Yao},\ and\ \citenamefont
  {Lee}}]{LIHAI2}%
  \BibitemOpen
  \bibfield  {author} {\bibinfo {author} {\bibfnamefont {Z.-X.}\ \bibnamefont
  {Li}}, \bibinfo {author} {\bibfnamefont {F.}~\bibnamefont {Wang}}, \bibinfo
  {author} {\bibfnamefont {H.}~\bibnamefont {Yao}}, \ and\ \bibinfo {author}
  {\bibfnamefont {D.-H.}\ \bibnamefont {Lee}},\ }\bibfield  {title} {\enquote
  {\bibinfo {title} {What makes the tc of monolayer fese on srtio3 so high: a
  sign-problem-free quantum monte carlo study},}\ }\href {\doibase
  10.1007/s11434-016-1087-x} {\bibfield  {journal} {\bibinfo  {journal}
  {Science Bulletin}\ }\textbf {\bibinfo {volume} {61}},\ \bibinfo {pages} {925
  } (\bibinfo {year} {2016})}\BibitemShut {NoStop}%
\bibitem [{\citenamefont {Schattner}\ \emph {et~al.}(2016)\citenamefont
  {Schattner}, \citenamefont {Gerlach}, \citenamefont {Trebst},\ and\
  \citenamefont {Berg}}]{SCHATTNER2}%
  \BibitemOpen
  \bibfield  {author} {\bibinfo {author} {\bibfnamefont {Y.}~\bibnamefont
  {Schattner}}, \bibinfo {author} {\bibfnamefont {M.~H.}\ \bibnamefont
  {Gerlach}}, \bibinfo {author} {\bibfnamefont {S.}~\bibnamefont {Trebst}}, \
  and\ \bibinfo {author} {\bibfnamefont {E.}~\bibnamefont {Berg}},\ }\bibfield
  {title} {\enquote {\bibinfo {title} {Competing orders in a nearly
  antiferromagnetic metal},}\ }\href {\doibase 10.1103/PhysRevLett.117.097002}
  {\bibfield  {journal} {\bibinfo  {journal} {Phys. Rev. Lett.}\ }\textbf
  {\bibinfo {volume} {117}},\ \bibinfo {pages} {097002} (\bibinfo {year}
  {2016})}\BibitemShut {NoStop}%
\bibitem [{\citenamefont {Gerlach}\ \emph {et~al.}(2017)\citenamefont
  {Gerlach}, \citenamefont {Schattner}, \citenamefont {Berg},\ and\
  \citenamefont {Trebst}}]{GERLACH}%
  \BibitemOpen
  \bibfield  {author} {\bibinfo {author} {\bibfnamefont {M.~H.}\ \bibnamefont
  {Gerlach}}, \bibinfo {author} {\bibfnamefont {Y.}~\bibnamefont {Schattner}},
  \bibinfo {author} {\bibfnamefont {E.}~\bibnamefont {Berg}}, \ and\ \bibinfo
  {author} {\bibfnamefont {S.}~\bibnamefont {Trebst}},\ }\bibfield  {title}
  {\enquote {\bibinfo {title} {Quantum critical properties of a metallic
  spin-density-wave transition},}\ }\href {\doibase 10.1103/PhysRevB.95.035124}
  {\bibfield  {journal} {\bibinfo  {journal} {Phys. Rev. B}\ }\textbf {\bibinfo
  {volume} {95}},\ \bibinfo {pages} {035124} (\bibinfo {year}
  {2017})}\BibitemShut {NoStop}%
\bibitem [{\citenamefont {Li}\ \emph {et~al.}(2017)\citenamefont {Li},
  \citenamefont {Wang}, \citenamefont {Yao},\ and\ \citenamefont
  {Lee}}]{LIHAI}%
  \BibitemOpen
  \bibfield  {author} {\bibinfo {author} {\bibfnamefont {Z.-X.}\ \bibnamefont
  {Li}}, \bibinfo {author} {\bibfnamefont {F.}~\bibnamefont {Wang}}, \bibinfo
  {author} {\bibfnamefont {H.}~\bibnamefont {Yao}}, \ and\ \bibinfo {author}
  {\bibfnamefont {D.-H.}\ \bibnamefont {Lee}},\ }\bibfield  {title} {\enquote
  {\bibinfo {title} {Nature of the effective interaction in electron-doped
  cuprate superconductors: A sign-problem-free quantum monte carlo study},}\
  }\href {\doibase 10.1103/PhysRevB.95.214505} {\bibfield  {journal} {\bibinfo
  {journal} {Phys. Rev. B}\ }\textbf {\bibinfo {volume} {95}},\ \bibinfo
  {pages} {214505} (\bibinfo {year} {2017})}\BibitemShut {NoStop}%
\bibitem [{\citenamefont {Wang}\ \emph {et~al.}(2018)\citenamefont {Wang},
  \citenamefont {Wang}, \citenamefont {Schattner}, \citenamefont {Berg},\ and\
  \citenamefont {Fernandes}}]{WANG2}%
  \BibitemOpen
  \bibfield  {author} {\bibinfo {author} {\bibfnamefont {X.}~\bibnamefont
  {Wang}}, \bibinfo {author} {\bibfnamefont {Y.}~\bibnamefont {Wang}}, \bibinfo
  {author} {\bibfnamefont {Y.}~\bibnamefont {Schattner}}, \bibinfo {author}
  {\bibfnamefont {E.}~\bibnamefont {Berg}}, \ and\ \bibinfo {author}
  {\bibfnamefont {R.~M.}\ \bibnamefont {Fernandes}},\ }\bibfield  {title}
  {\enquote {\bibinfo {title} {Fragility of charge order near an
  antiferromagnetic quantum critical point},}\ }\href {\doibase
  10.1103/PhysRevLett.120.247002} {\bibfield  {journal} {\bibinfo  {journal}
  {Phys. Rev. Lett.}\ }\textbf {\bibinfo {volume} {120}},\ \bibinfo {pages}
  {247002} (\bibinfo {year} {2018})}\BibitemShut {NoStop}%
\bibitem [{\citenamefont {Sur}\ and\ \citenamefont
  {Lee}(2015{\natexlab{a}})}]{SHOUVIK}%
  \BibitemOpen
  \bibfield  {author} {\bibinfo {author} {\bibfnamefont {S.}~\bibnamefont
  {Sur}}\ and\ \bibinfo {author} {\bibfnamefont {S.-S.}\ \bibnamefont {Lee}},\
  }\bibfield  {title} {\enquote {\bibinfo {title} {Quasilocal strange metal},}\
  }\href {\doibase 10.1103/PhysRevB.91.125136} {\bibfield  {journal} {\bibinfo
  {journal} {Phys. Rev. B}\ }\textbf {\bibinfo {volume} {91}},\ \bibinfo
  {pages} {125136} (\bibinfo {year} {2015}{\natexlab{a}})}\BibitemShut
  {NoStop}%
\bibitem [{\citenamefont {Lunts}\ \emph
  {et~al.}(2017{\natexlab{a}})\citenamefont {Lunts}, \citenamefont {Schlief},\
  and\ \citenamefont {Lee}}]{LUNTS}%
  \BibitemOpen
  \bibfield  {author} {\bibinfo {author} {\bibfnamefont {P.}~\bibnamefont
  {Lunts}}, \bibinfo {author} {\bibfnamefont {A.}~\bibnamefont {Schlief}}, \
  and\ \bibinfo {author} {\bibfnamefont {S.-S.}\ \bibnamefont {Lee}},\
  }\bibfield  {title} {\enquote {\bibinfo {title} {Emergence of a control
  parameter for the antiferromagnetic quantum critical metal},}\ }\href
  {\doibase 10.1103/PhysRevB.95.245109} {\bibfield  {journal} {\bibinfo
  {journal} {Phys. Rev. B}\ }\textbf {\bibinfo {volume} {95}},\ \bibinfo
  {pages} {245109} (\bibinfo {year} {2017}{\natexlab{a}})}\BibitemShut
  {NoStop}%
\bibitem [{\citenamefont {Schlief}\ \emph {et~al.}(2017)\citenamefont
  {Schlief}, \citenamefont {Lunts},\ and\ \citenamefont {Lee}}]{SCHLIEF}%
  \BibitemOpen
  \bibfield  {author} {\bibinfo {author} {\bibfnamefont {A.}~\bibnamefont
  {Schlief}}, \bibinfo {author} {\bibfnamefont {P.}~\bibnamefont {Lunts}}, \
  and\ \bibinfo {author} {\bibfnamefont {S.-S.}\ \bibnamefont {Lee}},\
  }\bibfield  {title} {\enquote {\bibinfo {title} {Exact critical exponents for
  the antiferromagnetic quantum critical metal in two dimensions},}\ }\href
  {\doibase 10.1103/PhysRevX.7.021010} {\bibfield  {journal} {\bibinfo
  {journal} {Phys. Rev. X}\ }\textbf {\bibinfo {volume} {7}},\ \bibinfo {pages}
  {021010} (\bibinfo {year} {2017})}\BibitemShut {NoStop}%
\bibitem [{\citenamefont {Sur}\ and\ \citenamefont
  {Lee}(2016{\natexlab{a}})}]{SHOUVIK3}%
  \BibitemOpen
  \bibfield  {author} {\bibinfo {author} {\bibfnamefont {S.}~\bibnamefont
  {Sur}}\ and\ \bibinfo {author} {\bibfnamefont {S.-S.}\ \bibnamefont {Lee}},\
  }\bibfield  {title} {\enquote {\bibinfo {title} {Anisotropic non-fermi
  liquids},}\ }\href {\doibase 10.1103/PhysRevB.94.195135} {\bibfield
  {journal} {\bibinfo  {journal} {Phys. Rev. B}\ }\textbf {\bibinfo {volume}
  {94}},\ \bibinfo {pages} {195135} (\bibinfo {year}
  {2016}{\natexlab{a}})}\BibitemShut {NoStop}%
\bibitem [{\citenamefont {Hertz}(1976)}]{HERTZ}%
  \BibitemOpen
  \bibfield  {author} {\bibinfo {author} {\bibfnamefont {J.~A.}\ \bibnamefont
  {Hertz}},\ }\bibfield  {title} {\enquote {\bibinfo {title} {Quantum critical
  phenomena},}\ }\href {\doibase 10.1103/PhysRevB.14.1165} {\bibfield
  {journal} {\bibinfo  {journal} {Phys. Rev. B}\ }\textbf {\bibinfo {volume}
  {14}},\ \bibinfo {pages} {1165} (\bibinfo {year} {1976})}\BibitemShut
  {NoStop}%
\bibitem [{\citenamefont {Millis}(1993)}]{MILLIS}%
  \BibitemOpen
  \bibfield  {author} {\bibinfo {author} {\bibfnamefont {A.~J.}\ \bibnamefont
  {Millis}},\ }\bibfield  {title} {\enquote {\bibinfo {title} {Effect of a
  nonzero temperature on quantum critical points in itinerant fermion
  systems},}\ }\href {\doibase 10.1103/PhysRevB.48.7183} {\bibfield  {journal}
  {\bibinfo  {journal} {Phys. Rev. B}\ }\textbf {\bibinfo {volume} {48}},\
  \bibinfo {pages} {7183} (\bibinfo {year} {1993})}\BibitemShut {NoStop}%
\bibitem [{\citenamefont {Varma}\ \emph
  {et~al.}(1989{\natexlab{a}})\citenamefont {Varma}, \citenamefont
  {Littlewood}, \citenamefont {Schmitt-Rink}, \citenamefont {Abrahams},\ and\
  \citenamefont {Ruckenstein}}]{VARMALI}%
  \BibitemOpen
  \bibfield  {author} {\bibinfo {author} {\bibfnamefont {C.~M.}\ \bibnamefont
  {Varma}}, \bibinfo {author} {\bibfnamefont {P.~B.}\ \bibnamefont
  {Littlewood}}, \bibinfo {author} {\bibfnamefont {S.}~\bibnamefont
  {Schmitt-Rink}}, \bibinfo {author} {\bibfnamefont {E.}~\bibnamefont
  {Abrahams}}, \ and\ \bibinfo {author} {\bibfnamefont {A.~E.}\ \bibnamefont
  {Ruckenstein}},\ }\bibfield  {title} {\enquote {\bibinfo {title}
  {Phenomenology of the normal state of cu-o high-temperature
  superconductors},}\ }\href {\doibase 10.1103/PhysRevLett.63.1996} {\bibfield
  {journal} {\bibinfo  {journal} {Phys. Rev. Lett.}\ }\textbf {\bibinfo
  {volume} {63}},\ \bibinfo {pages} {1996} (\bibinfo {year}
  {1989}{\natexlab{a}})}\BibitemShut {NoStop}%
\bibitem [{\citenamefont {Polchinski}(1994)}]{POLCHINSKI2}%
  \BibitemOpen
  \bibfield  {author} {\bibinfo {author} {\bibfnamefont {J.}~\bibnamefont
  {Polchinski}},\ }\bibfield  {title} {\enquote {\bibinfo {title} {Low-energy
  dynamics of the spinon-gauge system},}\ }\href {\doibase
  10.1016/0550-3213(94)90449-9} {\bibfield  {journal} {\bibinfo  {journal}
  {Nuclear Physics B}\ }\textbf {\bibinfo {volume} {422}},\ \bibinfo {pages}
  {617} (\bibinfo {year} {1994})}\BibitemShut {NoStop}%
\bibitem [{\citenamefont {Lee}(1989)}]{PLEE1}%
  \BibitemOpen
  \bibfield  {author} {\bibinfo {author} {\bibfnamefont {P.~A.}\ \bibnamefont
  {Lee}},\ }\bibfield  {title} {\enquote {\bibinfo {title} {Gauge field,
  aharonov-bohm flux, and high-${T}_{c}$ superconductivity},}\ }\href {\doibase
  10.1103/PhysRevLett.63.680} {\bibfield  {journal} {\bibinfo  {journal} {Phys.
  Rev. Lett.}\ }\textbf {\bibinfo {volume} {63}},\ \bibinfo {pages} {680}
  (\bibinfo {year} {1989})}\BibitemShut {NoStop}%
\bibitem [{\citenamefont {Altshuler}\ \emph {et~al.}(1994)\citenamefont
  {Altshuler}, \citenamefont {Ioffe},\ and\ \citenamefont
  {Millis}}]{ALTSHULER}%
  \BibitemOpen
  \bibfield  {author} {\bibinfo {author} {\bibfnamefont {B.~L.}\ \bibnamefont
  {Altshuler}}, \bibinfo {author} {\bibfnamefont {L.~B.}\ \bibnamefont
  {Ioffe}}, \ and\ \bibinfo {author} {\bibfnamefont {A.~J.}\ \bibnamefont
  {Millis}},\ }\bibfield  {title} {\enquote {\bibinfo {title} {{Low-energy
  properties of fermions with singular interactions}},}\ }\href {\doibase
  10.1103/PhysRevB.50.14048} {\bibfield  {journal} {\bibinfo  {journal} {Phys.
  Rev. B}\ }\textbf {\bibinfo {volume} {50}},\ \bibinfo {pages} {14048}
  (\bibinfo {year} {1994})}\BibitemShut {NoStop}%
\bibitem [{\citenamefont {Kim}\ \emph {et~al.}(1994)\citenamefont {Kim},
  \citenamefont {Furusaki}, \citenamefont {Wen},\ and\ \citenamefont
  {Lee}}]{YBKIM}%
  \BibitemOpen
  \bibfield  {author} {\bibinfo {author} {\bibfnamefont {Y.~B.}\ \bibnamefont
  {Kim}}, \bibinfo {author} {\bibfnamefont {A.}~\bibnamefont {Furusaki}},
  \bibinfo {author} {\bibfnamefont {X.-G.}\ \bibnamefont {Wen}}, \ and\
  \bibinfo {author} {\bibfnamefont {P.~A.}\ \bibnamefont {Lee}},\ }\bibfield
  {title} {\enquote {\bibinfo {title} {Gauge-invariant response functions of
  fermions coupled to a gauge field},}\ }\href {\doibase
  10.1103/PhysRevB.50.17917} {\bibfield  {journal} {\bibinfo  {journal} {Phys.
  Rev. B}\ }\textbf {\bibinfo {volume} {50}},\ \bibinfo {pages} {17917}
  (\bibinfo {year} {1994})}\BibitemShut {NoStop}%
\bibitem [{\citenamefont {Metlitski}\ and\ \citenamefont
  {Sachdev}(2010{\natexlab{b}})}]{MAX0}%
  \BibitemOpen
  \bibfield  {author} {\bibinfo {author} {\bibfnamefont {M.~A.}\ \bibnamefont
  {Metlitski}}\ and\ \bibinfo {author} {\bibfnamefont {S.}~\bibnamefont
  {Sachdev}},\ }\bibfield  {title} {\enquote {\bibinfo {title} {Quantum phase
  transitions of metals in two spatial dimensions. i. ising-nematic order},}\
  }\href {\doibase 10.1103/PhysRevB.82.075127} {\bibfield  {journal} {\bibinfo
  {journal} {Phys. Rev. B}\ }\textbf {\bibinfo {volume} {82}},\ \bibinfo
  {pages} {075127} (\bibinfo {year} {2010}{\natexlab{b}})}\BibitemShut
  {NoStop}%
\bibitem [{\citenamefont {Nayak}\ and\ \citenamefont {Wilczek}(1994)}]{NAYAK2}%
  \BibitemOpen
  \bibfield  {author} {\bibinfo {author} {\bibfnamefont {C.}~\bibnamefont
  {Nayak}}\ and\ \bibinfo {author} {\bibfnamefont {F.}~\bibnamefont
  {Wilczek}},\ }\bibfield  {title} {\enquote {\bibinfo {title} {Renormalization
  group approach to low temperature properties of a non-fermi liquid metal},}\
  }\href {\doibase https://doi.org/10.1016/0550-3213(94)90158-9} {\bibfield
  {journal} {\bibinfo  {journal} {Nuclear Physics B}\ }\textbf {\bibinfo
  {volume} {430}},\ \bibinfo {pages} {534 } (\bibinfo {year}
  {1994})}\BibitemShut {NoStop}%
\bibitem [{\citenamefont {Stewart}(2001)}]{STEWART}%
  \BibitemOpen
  \bibfield  {author} {\bibinfo {author} {\bibfnamefont {G.~R.}\ \bibnamefont
  {Stewart}},\ }\bibfield  {title} {\enquote {\bibinfo {title}
  {Non-fermi-liquid behavior in $d$- and $f$-electron metals},}\ }\href
  {\doibase 10.1103/RevModPhys.73.797} {\bibfield  {journal} {\bibinfo
  {journal} {Rev. Mod. Phys.}\ }\textbf {\bibinfo {volume} {73}},\ \bibinfo
  {pages} {797} (\bibinfo {year} {2001})}\BibitemShut {NoStop}%
\bibitem [{\citenamefont {Schofield}(1999)}]{SCHOFIELD}%
  \BibitemOpen
  \bibfield  {author} {\bibinfo {author} {\bibfnamefont {A.~J.}\ \bibnamefont
  {Schofield}},\ }\bibfield  {title} {\enquote {\bibinfo {title} {Non-fermi
  liquids},}\ }\href {\doibase 10.1080/001075199181602} {\bibfield  {journal}
  {\bibinfo  {journal} {Contemporary Physics}\ }\textbf {\bibinfo {volume}
  {40}},\ \bibinfo {pages} {95} (\bibinfo {year} {1999})}\BibitemShut {NoStop}%
\bibitem [{\citenamefont {Dalidovich}\ and\ \citenamefont
  {Lee}(2013{\natexlab{a}})}]{DENNIS}%
  \BibitemOpen
  \bibfield  {author} {\bibinfo {author} {\bibfnamefont {D.}~\bibnamefont
  {Dalidovich}}\ and\ \bibinfo {author} {\bibfnamefont {S.-S.}\ \bibnamefont
  {Lee}},\ }\bibfield  {title} {\enquote {\bibinfo {title} {Perturbative
  non-fermi liquids from dimensional regularization},}\ }\href {\doibase
  10.1103/PhysRevB.88.245106} {\bibfield  {journal} {\bibinfo  {journal} {Phys.
  Rev. B}\ }\textbf {\bibinfo {volume} {88}},\ \bibinfo {pages} {245106}
  (\bibinfo {year} {2013}{\natexlab{a}})}\BibitemShut {NoStop}%
\bibitem [{\citenamefont {Senthil}(2008)}]{SENTHIL}%
  \BibitemOpen
  \bibfield  {author} {\bibinfo {author} {\bibfnamefont {T.}~\bibnamefont
  {Senthil}},\ }\bibfield  {title} {\enquote {\bibinfo {title} {Critical fermi
  surfaces and non-fermi liquid metals},}\ }\href {\doibase
  10.1103/PhysRevB.78.035103} {\bibfield  {journal} {\bibinfo  {journal} {Phys.
  Rev. B}\ }\textbf {\bibinfo {volume} {78}},\ \bibinfo {pages} {035103}
  (\bibinfo {year} {2008})}\BibitemShut {NoStop}%
\bibitem [{\citenamefont {Mross}\ \emph {et~al.}(2010)\citenamefont {Mross},
  \citenamefont {McGreevy}, \citenamefont {Liu},\ and\ \citenamefont
  {Senthil}}]{MROSS}%
  \BibitemOpen
  \bibfield  {author} {\bibinfo {author} {\bibfnamefont {D.~F.}\ \bibnamefont
  {Mross}}, \bibinfo {author} {\bibfnamefont {J.}~\bibnamefont {McGreevy}},
  \bibinfo {author} {\bibfnamefont {H.}~\bibnamefont {Liu}}, \ and\ \bibinfo
  {author} {\bibfnamefont {T.}~\bibnamefont {Senthil}},\ }\bibfield  {title}
  {\enquote {\bibinfo {title} {Controlled expansion for certain
  non-fermi-liquid metals},}\ }\href {\doibase 10.1103/PhysRevB.82.045121}
  {\bibfield  {journal} {\bibinfo  {journal} {Phys. Rev. B}\ }\textbf {\bibinfo
  {volume} {82}},\ \bibinfo {pages} {045121} (\bibinfo {year}
  {2010})}\BibitemShut {NoStop}%
\bibitem [{\citenamefont {Fitzpatrick}\ \emph {et~al.}(2013)\citenamefont
  {Fitzpatrick}, \citenamefont {Kachru}, \citenamefont {Kaplan},\ and\
  \citenamefont {Raghu}}]{FITZPATRICK}%
  \BibitemOpen
  \bibfield  {author} {\bibinfo {author} {\bibfnamefont {A.~L.}\ \bibnamefont
  {Fitzpatrick}}, \bibinfo {author} {\bibfnamefont {S.}~\bibnamefont {Kachru}},
  \bibinfo {author} {\bibfnamefont {J.}~\bibnamefont {Kaplan}}, \ and\ \bibinfo
  {author} {\bibfnamefont {S.}~\bibnamefont {Raghu}},\ }\bibfield  {title}
  {\enquote {\bibinfo {title} {Non-fermi-liquid fixed point in a wilsonian
  theory of quantum critical metals},}\ }\href {\doibase
  10.1103/PhysRevB.88.125116} {\bibfield  {journal} {\bibinfo  {journal} {Phys.
  Rev. B}\ }\textbf {\bibinfo {volume} {88}},\ \bibinfo {pages} {125116}
  (\bibinfo {year} {2013})}\BibitemShut {NoStop}%
\bibitem [{\citenamefont {Sur}\ and\ \citenamefont {Lee}(2014)}]{SHOUVIK2}%
  \BibitemOpen
  \bibfield  {author} {\bibinfo {author} {\bibfnamefont {S.}~\bibnamefont
  {Sur}}\ and\ \bibinfo {author} {\bibfnamefont {S.-S.}\ \bibnamefont {Lee}},\
  }\bibfield  {title} {\enquote {\bibinfo {title} {Chiral non-fermi liquids},}\
  }\href {\doibase 10.1103/PhysRevB.90.045121} {\bibfield  {journal} {\bibinfo
  {journal} {Phys. Rev. B}\ }\textbf {\bibinfo {volume} {90}},\ \bibinfo
  {pages} {045121} (\bibinfo {year} {2014})}\BibitemShut {NoStop}%
\bibitem [{\citenamefont {Else}\ \emph {et~al.}(2021)\citenamefont {Else},
  \citenamefont {Thorngren},\ and\ \citenamefont
  {Senthil}}]{PhysRevX.11.021005}%
  \BibitemOpen
  \bibfield  {author} {\bibinfo {author} {\bibfnamefont {D.~V.}\ \bibnamefont
  {Else}}, \bibinfo {author} {\bibfnamefont {R.}~\bibnamefont {Thorngren}}, \
  and\ \bibinfo {author} {\bibfnamefont {T.}~\bibnamefont {Senthil}},\
  }\bibfield  {title} {\enquote {\bibinfo {title} {Non-fermi liquids as ersatz
  fermi liquids: General constraints on compressible metals},}\ }\href
  {\doibase 10.1103/PhysRevX.11.021005} {\bibfield  {journal} {\bibinfo
  {journal} {Phys. Rev. X}\ }\textbf {\bibinfo {volume} {11}},\ \bibinfo
  {pages} {021005} (\bibinfo {year} {2021})}\BibitemShut {NoStop}%
\bibitem [{\citenamefont {Debbeler}\ and\ \citenamefont
  {Metzner}(2023)}]{PhysRevB.107.165152}%
  \BibitemOpen
  \bibfield  {author} {\bibinfo {author} {\bibfnamefont {L.}~\bibnamefont
  {Debbeler}}\ and\ \bibinfo {author} {\bibfnamefont {W.}~\bibnamefont
  {Metzner}},\ }\bibfield  {title} {\enquote {\bibinfo {title} {{Non-Fermi
  liquid behavior at flat hot spots from quantum critical fluctuations at the
  onset of charge- or spin-density wave order}},}\ }\href {\doibase
  10.1103/PhysRevB.107.165152} {\bibfield  {journal} {\bibinfo  {journal}
  {Phys. Rev. B}\ }\textbf {\bibinfo {volume} {107}},\ \bibinfo {pages}
  {165152} (\bibinfo {year} {2023})}\BibitemShut {NoStop}%
\bibitem [{\citenamefont {Berg}\ \emph {et~al.}(2019)\citenamefont {Berg},
  \citenamefont {Lederer}, \citenamefont {Schattner},\ and\ \citenamefont
  {Trebst}}]{berglederer}%
  \BibitemOpen
  \bibfield  {author} {\bibinfo {author} {\bibfnamefont {E.}~\bibnamefont
  {Berg}}, \bibinfo {author} {\bibfnamefont {S.}~\bibnamefont {Lederer}},
  \bibinfo {author} {\bibfnamefont {Y.}~\bibnamefont {Schattner}}, \ and\
  \bibinfo {author} {\bibfnamefont {S.}~\bibnamefont {Trebst}},\ }\bibfield
  {title} {\enquote {\bibinfo {title} {Monte carlo studies of quantum critical
  metals},}\ }\href {\doibase
  https://doi.org/10.1146/annurev-conmatphys-031218-013339} {\bibfield
  {journal} {\bibinfo  {journal} {Annual Review of Condensed Matter Physics}\
  }\textbf {\bibinfo {volume} {10}},\ \bibinfo {pages} {63} (\bibinfo {year}
  {2019})}\BibitemShut {NoStop}%
\bibitem [{\citenamefont {Borges}\ \emph {et~al.}(2023)\citenamefont {Borges},
  \citenamefont {Borissov}, \citenamefont {Singh}, \citenamefont {Schlief},\
  and\ \citenamefont {Lee}}]{BORGES2023169221}%
  \BibitemOpen
  \bibfield  {author} {\bibinfo {author} {\bibfnamefont {F.}~\bibnamefont
  {Borges}}, \bibinfo {author} {\bibfnamefont {A.}~\bibnamefont {Borissov}},
  \bibinfo {author} {\bibfnamefont {A.}~\bibnamefont {Singh}}, \bibinfo
  {author} {\bibfnamefont {A.}~\bibnamefont {Schlief}}, \ and\ \bibinfo
  {author} {\bibfnamefont {S.-S.}\ \bibnamefont {Lee}},\ }\bibfield  {title}
  {\enquote {\bibinfo {title} {{Field-theoretic functional renormalization
  group formalism for non-Fermi liquids and its application to the
  antiferromagnetic quantum critical metal in two dimensions}},}\ }\href
  {\doibase https://doi.org/10.1016/j.aop.2023.169221} {\bibfield  {journal}
  {\bibinfo  {journal} {Annals of Physics}\ }\textbf {\bibinfo {volume}
  {450}},\ \bibinfo {pages} {169221} (\bibinfo {year} {2023})}\BibitemShut
  {NoStop}%
\bibitem [{\citenamefont {Kukreja}\ \emph {et~al.}(2024)\citenamefont
  {Kukreja}, \citenamefont {Besharat},\ and\ \citenamefont
  {Lee}}]{2024arXiv240509450K}%
  \BibitemOpen
  \bibfield  {author} {\bibinfo {author} {\bibfnamefont {S.}~\bibnamefont
  {Kukreja}}, \bibinfo {author} {\bibfnamefont {A.}~\bibnamefont {Besharat}}, \
  and\ \bibinfo {author} {\bibfnamefont {S.-S.}\ \bibnamefont {Lee}},\
  }\bibfield  {title} {\enquote {\bibinfo {title} {Projective fixed points for
  non-fermi liquids: A case study of the ising-nematic quantum critical
  metal},}\ }\href {\doibase 10.1103/PhysRevB.110.155142} {\bibfield  {journal}
  {\bibinfo  {journal} {Phys. Rev. B}\ }\textbf {\bibinfo {volume} {110}},\
  \bibinfo {pages} {155142} (\bibinfo {year} {2024})}\BibitemShut {NoStop}%
\bibitem [{\citenamefont {Landau}(1957)}]{LANDAU}%
  \BibitemOpen
  \bibfield  {author} {\bibinfo {author} {\bibfnamefont {L.}~\bibnamefont
  {Landau}},\ }\bibfield  {title} {\enquote {\bibinfo {title} {The theory of a
  fermi liquid},}\ }\href {http://www.jetp.ac.ru/cgi-bin/dn/e_003_06_0920.pdf}
  {\bibfield  {journal} {\bibinfo  {journal} {Sov. Phys. JETP}\ }\textbf
  {\bibinfo {volume} {3}},\ \bibinfo {pages} {920} (\bibinfo {year}
  {1957})}\BibitemShut {NoStop}%
\bibitem [{\citenamefont {Benfatto}\ and\ \citenamefont
  {Gallavotti}(1990)}]{PhysRevB.42.9967}%
  \BibitemOpen
  \bibfield  {author} {\bibinfo {author} {\bibfnamefont {G.}~\bibnamefont
  {Benfatto}}\ and\ \bibinfo {author} {\bibfnamefont {G.}~\bibnamefont
  {Gallavotti}},\ }\bibfield  {title} {\enquote {\bibinfo {title}
  {{Renormalization-group approach to the theory of the Fermi surface}},}\
  }\href {\doibase 10.1103/PhysRevB.42.9967} {\bibfield  {journal} {\bibinfo
  {journal} {Phys. Rev. B}\ }\textbf {\bibinfo {volume} {42}},\ \bibinfo
  {pages} {9967} (\bibinfo {year} {1990})}\BibitemShut {NoStop}%
\bibitem [{\citenamefont {Shankar}(1994)}]{SHANKAR}%
  \BibitemOpen
  \bibfield  {author} {\bibinfo {author} {\bibfnamefont {R.}~\bibnamefont
  {Shankar}},\ }\bibfield  {title} {\enquote {\bibinfo {title}
  {Renormalization-group approach to interacting fermions},}\ }\href {\doibase
  10.1103/RevModPhys.66.129} {\bibfield  {journal} {\bibinfo  {journal} {Rev.
  Mod. Phys.}\ }\textbf {\bibinfo {volume} {66}},\ \bibinfo {pages} {129}
  (\bibinfo {year} {1994})}\BibitemShut {NoStop}%
\bibitem [{\citenamefont {Polchinski}(1992)}]{POLCHINSKI1}%
  \BibitemOpen
  \bibfield  {author} {\bibinfo {author} {\bibfnamefont {J.}~\bibnamefont
  {Polchinski}},\ }\bibfield  {title} {\enquote {\bibinfo {title} {{Effective
  Field Theory and the Fermi Surface}},}\ }\href@noop {} {\bibfield  {journal}
  {\bibinfo  {journal} {ArXiv High Energy Physics - Theory e-prints}\ }
  (\bibinfo {year} {1992})},\ \Eprint {http://arxiv.org/abs/hep-th/9210046}
  {hep-th/9210046} \BibitemShut {NoStop}%
\bibitem [{\citenamefont {Helm}\ \emph {et~al.}(2010)\citenamefont {Helm},
  \citenamefont {Kartsovnik}, \citenamefont {Sheikin}, \citenamefont
  {Bartkowiak}, \citenamefont {Wolff-Fabris}, \citenamefont {Bittner},
  \citenamefont {Biberacher}, \citenamefont {Lambacher}, \citenamefont {Erb},
  \citenamefont {Wosnitza},\ and\ \citenamefont {Gross}}]{Cuprates2010Example}%
  \BibitemOpen
  \bibfield  {author} {\bibinfo {author} {\bibfnamefont {T.}~\bibnamefont
  {Helm}}, \bibinfo {author} {\bibfnamefont {M.~V.}\ \bibnamefont
  {Kartsovnik}}, \bibinfo {author} {\bibfnamefont {I.}~\bibnamefont {Sheikin}},
  \bibinfo {author} {\bibfnamefont {M.}~\bibnamefont {Bartkowiak}}, \bibinfo
  {author} {\bibfnamefont {F.}~\bibnamefont {Wolff-Fabris}}, \bibinfo {author}
  {\bibfnamefont {N.}~\bibnamefont {Bittner}}, \bibinfo {author} {\bibfnamefont
  {W.}~\bibnamefont {Biberacher}}, \bibinfo {author} {\bibfnamefont
  {M.}~\bibnamefont {Lambacher}}, \bibinfo {author} {\bibfnamefont
  {A.}~\bibnamefont {Erb}}, \bibinfo {author} {\bibfnamefont {J.}~\bibnamefont
  {Wosnitza}}, \ and\ \bibinfo {author} {\bibfnamefont {R.}~\bibnamefont
  {Gross}},\ }\bibfield  {title} {\enquote {\bibinfo {title} {Magnetic
  breakdown in the electron-doped cuprate superconductor
  ${\mathrm{nd}}_{2\ensuremath{-}x}{\mathrm{ce}}_{x}{\mathrm{cuo}}_{4}$: The
  reconstructed fermi surface survives in the strongly overdoped regime},}\
  }\href {\doibase 10.1103/PhysRevLett.105.247002} {\bibfield  {journal}
  {\bibinfo  {journal} {Phys. Rev. Lett.}\ }\textbf {\bibinfo {volume} {105}},\
  \bibinfo {pages} {247002} (\bibinfo {year} {2010})}\BibitemShut {NoStop}%
\bibitem [{\citenamefont {Hashimoto}\ \emph {et~al.}(2012)\citenamefont
  {Hashimoto}, \citenamefont {Cho}, \citenamefont {Shibauchi}, \citenamefont
  {Kasahara}, \citenamefont {Mizukami}, \citenamefont {Katsumata},
  \citenamefont {Tsuruhara}, \citenamefont {Terashima}, \citenamefont {Ikeda},
  \citenamefont {Tanatar}, \citenamefont {Kitano}, \citenamefont {Salovich},
  \citenamefont {Giannetta}, \citenamefont {Walmsley}, \citenamefont
  {Carrington}, \citenamefont {Prozorov},\ and\ \citenamefont
  {Matsuda}}]{IronPnic2012Example}%
  \BibitemOpen
  \bibfield  {author} {\bibinfo {author} {\bibfnamefont {K.}~\bibnamefont
  {Hashimoto}}, \bibinfo {author} {\bibfnamefont {K.}~\bibnamefont {Cho}},
  \bibinfo {author} {\bibfnamefont {T.}~\bibnamefont {Shibauchi}}, \bibinfo
  {author} {\bibfnamefont {S.}~\bibnamefont {Kasahara}}, \bibinfo {author}
  {\bibfnamefont {Y.}~\bibnamefont {Mizukami}}, \bibinfo {author}
  {\bibfnamefont {R.}~\bibnamefont {Katsumata}}, \bibinfo {author}
  {\bibfnamefont {Y.}~\bibnamefont {Tsuruhara}}, \bibinfo {author}
  {\bibfnamefont {T.}~\bibnamefont {Terashima}}, \bibinfo {author}
  {\bibfnamefont {H.}~\bibnamefont {Ikeda}}, \bibinfo {author} {\bibfnamefont
  {M.~A.}\ \bibnamefont {Tanatar}}, \bibinfo {author} {\bibfnamefont
  {H.}~\bibnamefont {Kitano}}, \bibinfo {author} {\bibfnamefont
  {N.}~\bibnamefont {Salovich}}, \bibinfo {author} {\bibfnamefont {R.~W.}\
  \bibnamefont {Giannetta}}, \bibinfo {author} {\bibfnamefont {P.}~\bibnamefont
  {Walmsley}}, \bibinfo {author} {\bibfnamefont {A.}~\bibnamefont
  {Carrington}}, \bibinfo {author} {\bibfnamefont {R.}~\bibnamefont
  {Prozorov}}, \ and\ \bibinfo {author} {\bibfnamefont {Y.}~\bibnamefont
  {Matsuda}},\ }\bibfield  {title} {\enquote {\bibinfo {title} {A sharp peak of
  the zero-temperature penetration depth at optimal composition in
  bafe2(as1{\textendash}xpx)2},}\ }\href {\doibase 10.1126/science.1219821}
  {\bibfield  {journal} {\bibinfo  {journal} {Science}\ }\textbf {\bibinfo
  {volume} {336}},\ \bibinfo {pages} {1554} (\bibinfo {year}
  {2012})}\BibitemShut {NoStop}%
\bibitem [{\citenamefont {Park}\ \emph {et~al.}(2006)\citenamefont {Park},
  \citenamefont {Ronning}, \citenamefont {Yuan}, \citenamefont {Salamon},
  \citenamefont {Movshovich}, \citenamefont {Sarrao},\ and\ \citenamefont
  {Thompson}}]{HeavyFermion2006Example}%
  \BibitemOpen
  \bibfield  {author} {\bibinfo {author} {\bibfnamefont {T.}~\bibnamefont
  {Park}}, \bibinfo {author} {\bibfnamefont {F.}~\bibnamefont {Ronning}},
  \bibinfo {author} {\bibfnamefont {H.~Q.}\ \bibnamefont {Yuan}}, \bibinfo
  {author} {\bibfnamefont {M.~B.}\ \bibnamefont {Salamon}}, \bibinfo {author}
  {\bibfnamefont {R.}~\bibnamefont {Movshovich}}, \bibinfo {author}
  {\bibfnamefont {J.~L.}\ \bibnamefont {Sarrao}}, \ and\ \bibinfo {author}
  {\bibfnamefont {J.~D.}\ \bibnamefont {Thompson}},\ }\bibfield  {title}
  {\enquote {\bibinfo {title} {Hidden magnetism and quantum criticality in the
  heavy fermion superconductor {CeRhIn$_5$}},}\ }\href
  {https://doi.org/10.1038/nature04571} {\bibfield  {journal} {\bibinfo
  {journal} {Nature}\ }\textbf {\bibinfo {volume} {440}},\ \bibinfo {pages}
  {65} (\bibinfo {year} {2006})}\BibitemShut {NoStop}%
\bibitem [{\citenamefont {Lunts}\ \emph {et~al.}(2023)\citenamefont {Lunts},
  \citenamefont {Albergo},\ and\ \citenamefont {Lindsey}}]{Lunts:2023un}%
  \BibitemOpen
  \bibfield  {author} {\bibinfo {author} {\bibfnamefont {P.}~\bibnamefont
  {Lunts}}, \bibinfo {author} {\bibfnamefont {M.~S.}\ \bibnamefont {Albergo}},
  \ and\ \bibinfo {author} {\bibfnamefont {M.}~\bibnamefont {Lindsey}},\
  }\bibfield  {title} {\enquote {\bibinfo {title} {{Non-Hertz-Millis scaling of
  the antiferromagnetic quantum critical metal via scalable Hybrid Monte
  Carlo}},}\ }\href {\doibase 10.1038/s41467-023-37686-4} {\bibfield  {journal}
  {\bibinfo  {journal} {Nature Communications}\ }\textbf {\bibinfo {volume}
  {14}},\ \bibinfo {pages} {2547} (\bibinfo {year} {2023})}\BibitemShut
  {NoStop}%
\bibitem [{\citenamefont {Berg}\ \emph
  {et~al.}(2012{\natexlab{b}})\citenamefont {Berg}, \citenamefont {Metlitski},\
  and\ \citenamefont {Sachdev}}]{berg2012signfreemontecarlo}%
  \BibitemOpen
  \bibfield  {author} {\bibinfo {author} {\bibfnamefont {E.}~\bibnamefont
  {Berg}}, \bibinfo {author} {\bibfnamefont {M.~A.}\ \bibnamefont {Metlitski}},
  \ and\ \bibinfo {author} {\bibfnamefont {S.}~\bibnamefont {Sachdev}},\
  }\bibfield  {title} {\enquote {\bibinfo {title} {Sign-problem–free quantum
  monte carlo of the onset of antiferromagnetism in metals},}\ }\href {\doibase
  10.1126/science.1227769} {\bibfield  {journal} {\bibinfo  {journal}
  {Science}\ }\textbf {\bibinfo {volume} {338}},\ \bibinfo {pages} {1606}
  (\bibinfo {year} {2012}{\natexlab{b}})}\BibitemShut {NoStop}%
\bibitem [{\citenamefont {Bauer}\ \emph {et~al.}(2020)\citenamefont {Bauer},
  \citenamefont {Schattner}, \citenamefont {Trebst},\ and\ \citenamefont
  {Berg}}]{PhysRevResearch.2.023008}%
  \BibitemOpen
  \bibfield  {author} {\bibinfo {author} {\bibfnamefont {C.}~\bibnamefont
  {Bauer}}, \bibinfo {author} {\bibfnamefont {Y.}~\bibnamefont {Schattner}},
  \bibinfo {author} {\bibfnamefont {S.}~\bibnamefont {Trebst}}, \ and\ \bibinfo
  {author} {\bibfnamefont {E.}~\bibnamefont {Berg}},\ }\bibfield  {title}
  {\enquote {\bibinfo {title} {{Hierarchy of energy scales in an O(3) symmetric
  antiferromagnetic quantum critical metal: A Monte Carlo study}},}\ }\href
  {\doibase 10.1103/PhysRevResearch.2.023008} {\bibfield  {journal} {\bibinfo
  {journal} {Phys. Rev. Res.}\ }\textbf {\bibinfo {volume} {2}},\ \bibinfo
  {pages} {023008} (\bibinfo {year} {2020})}\BibitemShut {NoStop}%
\bibitem [{\citenamefont {Xu}\ \emph {et~al.}(2019)\citenamefont {Xu},
  \citenamefont {Liu}, \citenamefont {Pan}, \citenamefont {Qi}, \citenamefont
  {Sun},\ and\ \citenamefont {Meng}}]{Xu_2019}%
  \BibitemOpen
  \bibfield  {author} {\bibinfo {author} {\bibfnamefont {X.~Y.}\ \bibnamefont
  {Xu}}, \bibinfo {author} {\bibfnamefont {Z.~H.}\ \bibnamefont {Liu}},
  \bibinfo {author} {\bibfnamefont {G.}~\bibnamefont {Pan}}, \bibinfo {author}
  {\bibfnamefont {Y.}~\bibnamefont {Qi}}, \bibinfo {author} {\bibfnamefont
  {K.}~\bibnamefont {Sun}}, \ and\ \bibinfo {author} {\bibfnamefont {Z.~Y.}\
  \bibnamefont {Meng}},\ }\bibfield  {title} {\enquote {\bibinfo {title}
  {{Revealing fermionic quantum criticality from new Monte Carlo
  techniques}},}\ }\href {\doibase 10.1088/1361-648X/ab3295} {\bibfield
  {journal} {\bibinfo  {journal} {Journal of Physics: Condensed Matter}\
  }\textbf {\bibinfo {volume} {31}},\ \bibinfo {pages} {463001} (\bibinfo
  {year} {2019})}\BibitemShut {NoStop}%
\bibitem [{\citenamefont {Teixeira}\ \emph {et~al.}(2023)\citenamefont
  {Teixeira}, \citenamefont {P\'epin},\ and\ \citenamefont
  {Freire}}]{2023arXiv230506421T}%
  \BibitemOpen
  \bibfield  {author} {\bibinfo {author} {\bibfnamefont {R.~M.~P.}\
  \bibnamefont {Teixeira}}, \bibinfo {author} {\bibfnamefont {C.}~\bibnamefont
  {P\'epin}}, \ and\ \bibinfo {author} {\bibfnamefont {H.}~\bibnamefont
  {Freire}},\ }\bibfield  {title} {\enquote {\bibinfo {title} {Strange
  metallicity in an antiferromagnetic quantum critical model: A
  sign-problem-free quantum monte carlo study},}\ }\href {\doibase
  10.1103/PhysRevB.108.085131} {\bibfield  {journal} {\bibinfo  {journal}
  {Phys. Rev. B}\ }\textbf {\bibinfo {volume} {108}},\ \bibinfo {pages}
  {085131} (\bibinfo {year} {2023})}\BibitemShut {NoStop}%
\bibitem [{\citenamefont {Sur}\ and\ \citenamefont
  {Lee}(2015{\natexlab{b}})}]{Sur2015Quasilocal}%
  \BibitemOpen
  \bibfield  {author} {\bibinfo {author} {\bibfnamefont {S.}~\bibnamefont
  {Sur}}\ and\ \bibinfo {author} {\bibfnamefont {S.-S.}\ \bibnamefont {Lee}},\
  }\bibfield  {title} {\enquote {\bibinfo {title} {Quasilocal strange metal},}\
  }\href {\doibase 10.1103/PhysRevB.91.125136} {\bibfield  {journal} {\bibinfo
  {journal} {Phys. Rev. B}\ }\textbf {\bibinfo {volume} {91}},\ \bibinfo
  {pages} {125136} (\bibinfo {year} {2015}{\natexlab{b}})}\BibitemShut
  {NoStop}%
\bibitem [{\citenamefont {Halbinger}\ \emph {et~al.}(2019)\citenamefont
  {Halbinger}, \citenamefont {Pimenov},\ and\ \citenamefont
  {Punk}}]{PhysRevB.99.195102}%
  \BibitemOpen
  \bibfield  {author} {\bibinfo {author} {\bibfnamefont {J.}~\bibnamefont
  {Halbinger}}, \bibinfo {author} {\bibfnamefont {D.}~\bibnamefont {Pimenov}},
  \ and\ \bibinfo {author} {\bibfnamefont {M.}~\bibnamefont {Punk}},\
  }\bibfield  {title} {\enquote {\bibinfo {title} {{Incommensurate $2{k}_{F}$
  density wave quantum criticality in two-dimensional metals}},}\ }\href
  {\doibase 10.1103/PhysRevB.99.195102} {\bibfield  {journal} {\bibinfo
  {journal} {Phys. Rev. B}\ }\textbf {\bibinfo {volume} {99}},\ \bibinfo
  {pages} {195102} (\bibinfo {year} {2019})}\BibitemShut {NoStop}%
\bibitem [{\citenamefont {S\'ykora}\ and\ \citenamefont
  {Metzner}(2021)}]{PhysRevB.104.125123}%
  \BibitemOpen
  \bibfield  {author} {\bibinfo {author} {\bibfnamefont {J.}~\bibnamefont
  {S\'ykora}}\ and\ \bibinfo {author} {\bibfnamefont {W.}~\bibnamefont
  {Metzner}},\ }\bibfield  {title} {\enquote {\bibinfo {title} {{Fluctuation
  effects at the onset of $2{k}_{F}$ density wave order with two pairs of hot
  spots in two-dimensional metals}},}\ }\href {\doibase
  10.1103/PhysRevB.104.125123} {\bibfield  {journal} {\bibinfo  {journal}
  {Phys. Rev. B}\ }\textbf {\bibinfo {volume} {104}},\ \bibinfo {pages}
  {125123} (\bibinfo {year} {2021})}\BibitemShut {NoStop}%
\bibitem [{\citenamefont {Dalidovich}\ and\ \citenamefont
  {Lee}(2013{\natexlab{b}})}]{Lee2013DimReg}%
  \BibitemOpen
  \bibfield  {author} {\bibinfo {author} {\bibfnamefont {D.}~\bibnamefont
  {Dalidovich}}\ and\ \bibinfo {author} {\bibfnamefont {S.-S.}\ \bibnamefont
  {Lee}},\ }\bibfield  {title} {\enquote {\bibinfo {title} {Perturbative
  non-fermi liquids from dimensional regularization},}\ }\href {\doibase
  10.1103/PhysRevB.88.245106} {\bibfield  {journal} {\bibinfo  {journal} {Phys.
  Rev. B}\ }\textbf {\bibinfo {volume} {88}},\ \bibinfo {pages} {245106}
  (\bibinfo {year} {2013}{\natexlab{b}})}\BibitemShut {NoStop}%
\bibitem [{\citenamefont {Wilson}\ and\ \citenamefont
  {Fisher}(1972)}]{WilsonFisher1972WFBoson}%
  \BibitemOpen
  \bibfield  {author} {\bibinfo {author} {\bibfnamefont {K.~G.}\ \bibnamefont
  {Wilson}}\ and\ \bibinfo {author} {\bibfnamefont {M.~E.}\ \bibnamefont
  {Fisher}},\ }\bibfield  {title} {\enquote {\bibinfo {title} {Critical
  exponents in 3.99 dimensions},}\ }\href {\doibase 10.1103/PhysRevLett.28.240}
  {\bibfield  {journal} {\bibinfo  {journal} {Phys. Rev. Lett.}\ }\textbf
  {\bibinfo {volume} {28}},\ \bibinfo {pages} {240} (\bibinfo {year}
  {1972})}\BibitemShut {NoStop}%
\bibitem [{\citenamefont {Poland}\ \emph {et~al.}(2019)\citenamefont {Poland},
  \citenamefont {Rychkov},\ and\ \citenamefont {Vichi}}]{RevModPhys.91.015002}%
  \BibitemOpen
  \bibfield  {author} {\bibinfo {author} {\bibfnamefont {D.}~\bibnamefont
  {Poland}}, \bibinfo {author} {\bibfnamefont {S.}~\bibnamefont {Rychkov}}, \
  and\ \bibinfo {author} {\bibfnamefont {A.}~\bibnamefont {Vichi}},\ }\bibfield
   {title} {\enquote {\bibinfo {title} {{The conformal bootstrap: Theory,
  numerical techniques, and applications}},}\ }\href {\doibase
  10.1103/RevModPhys.91.015002} {\bibfield  {journal} {\bibinfo  {journal}
  {Rev. Mod. Phys.}\ }\textbf {\bibinfo {volume} {91}},\ \bibinfo {pages}
  {015002} (\bibinfo {year} {2019})}\BibitemShut {NoStop}%
\bibitem [{\citenamefont {Varma}\ \emph
  {et~al.}(1989{\natexlab{b}})\citenamefont {Varma}, \citenamefont
  {Littlewood}, \citenamefont {Schmitt-Rink}, \citenamefont {Abrahams},\ and\
  \citenamefont {Ruckenstein}}]{Varma1996MarginalFermiLiquid}%
  \BibitemOpen
  \bibfield  {author} {\bibinfo {author} {\bibfnamefont {C.~M.}\ \bibnamefont
  {Varma}}, \bibinfo {author} {\bibfnamefont {P.~B.}\ \bibnamefont
  {Littlewood}}, \bibinfo {author} {\bibfnamefont {S.}~\bibnamefont
  {Schmitt-Rink}}, \bibinfo {author} {\bibfnamefont {E.}~\bibnamefont
  {Abrahams}}, \ and\ \bibinfo {author} {\bibfnamefont {A.~E.}\ \bibnamefont
  {Ruckenstein}},\ }\bibfield  {title} {\enquote {\bibinfo {title}
  {Phenomenology of the normal state of cu-o high-temperature
  superconductors},}\ }\href {\doibase 10.1103/PhysRevLett.63.1996} {\bibfield
  {journal} {\bibinfo  {journal} {Phys. Rev. Lett.}\ }\textbf {\bibinfo
  {volume} {63}},\ \bibinfo {pages} {1996} (\bibinfo {year}
  {1989}{\natexlab{b}})}\BibitemShut {NoStop}%
\bibitem [{\citenamefont {Sur}\ and\ \citenamefont
  {Lee}(2016{\natexlab{b}})}]{Sur2016Anisotropic}%
  \BibitemOpen
  \bibfield  {author} {\bibinfo {author} {\bibfnamefont {S.}~\bibnamefont
  {Sur}}\ and\ \bibinfo {author} {\bibfnamefont {S.-S.}\ \bibnamefont {Lee}},\
  }\bibfield  {title} {\enquote {\bibinfo {title} {Anisotropic non-fermi
  liquids},}\ }\href {\doibase 10.1103/PhysRevB.94.195135} {\bibfield
  {journal} {\bibinfo  {journal} {Phys. Rev. B}\ }\textbf {\bibinfo {volume}
  {94}},\ \bibinfo {pages} {195135} (\bibinfo {year}
  {2016}{\natexlab{b}})}\BibitemShut {NoStop}%
\bibitem [{\citenamefont {Lee}(2018{\natexlab{b}})}]{Lee2018Review}%
  \BibitemOpen
  \bibfield  {author} {\bibinfo {author} {\bibfnamefont {S.-S.}\ \bibnamefont
  {Lee}},\ }\bibfield  {title} {\enquote {\bibinfo {title} {Recent developments
  in non-fermi liquid theory},}\ }\href {\doibase
  10.1146/annurev-conmatphys-031016-025531} {\bibfield  {journal} {\bibinfo
  {journal} {Annual Review of Condensed Matter Physics}\ }\textbf {\bibinfo
  {volume} {9}},\ \bibinfo {pages} {227} (\bibinfo {year}
  {2018}{\natexlab{b}})}\BibitemShut {NoStop}%
\bibitem [{\citenamefont {Lunts}\ \emph
  {et~al.}(2017{\natexlab{b}})\citenamefont {Lunts}, \citenamefont {Schlief},\
  and\ \citenamefont {Lee}}]{Lunts2017SU2ControlParameter}%
  \BibitemOpen
  \bibfield  {author} {\bibinfo {author} {\bibfnamefont {P.}~\bibnamefont
  {Lunts}}, \bibinfo {author} {\bibfnamefont {A.}~\bibnamefont {Schlief}}, \
  and\ \bibinfo {author} {\bibfnamefont {S.-S.}\ \bibnamefont {Lee}},\
  }\bibfield  {title} {\enquote {\bibinfo {title} {Emergence of a control
  parameter for the antiferromagnetic quantum critical metal},}\ }\href
  {\doibase 10.1103/PhysRevB.95.245109} {\bibfield  {journal} {\bibinfo
  {journal} {Phys. Rev. B}\ }\textbf {\bibinfo {volume} {95}},\ \bibinfo
  {pages} {245109} (\bibinfo {year} {2017}{\natexlab{b}})}\BibitemShut
  {NoStop}%
\bibitem [{\citenamefont {Schlief}\ \emph {et~al.}(2018)\citenamefont
  {Schlief}, \citenamefont {Lunts},\ and\ \citenamefont {Lee}}]{SCHLIEF2}%
  \BibitemOpen
  \bibfield  {author} {\bibinfo {author} {\bibfnamefont {A.}~\bibnamefont
  {Schlief}}, \bibinfo {author} {\bibfnamefont {P.}~\bibnamefont {Lunts}}, \
  and\ \bibinfo {author} {\bibfnamefont {S.-S.}\ \bibnamefont {Lee}},\
  }\bibfield  {title} {\enquote {\bibinfo {title} {Noncommutativity between the
  low-energy limit and integer dimension limits in the $\ensuremath{\epsilon}$
  expansion: A case study of the antiferromagnetic quantum critical metal},}\
  }\href {\doibase 10.1103/PhysRevB.98.075140} {\bibfield  {journal} {\bibinfo
  {journal} {Phys. Rev. B}\ }\textbf {\bibinfo {volume} {98}},\ \bibinfo
  {pages} {075140} (\bibinfo {year} {2018})}\BibitemShut {NoStop}%
\bibitem [{\citenamefont {Halbinger}\ and\ \citenamefont
  {Punk}(2021)}]{PhysRevB.103.235157}%
  \BibitemOpen
  \bibfield  {author} {\bibinfo {author} {\bibfnamefont {J.}~\bibnamefont
  {Halbinger}}\ and\ \bibinfo {author} {\bibfnamefont {M.}~\bibnamefont
  {Punk}},\ }\bibfield  {title} {\enquote {\bibinfo {title} {{Quenched disorder
  at antiferromagnetic quantum critical points in two-dimensional metals}},}\
  }\href {\doibase 10.1103/PhysRevB.103.235157} {\bibfield  {journal} {\bibinfo
   {journal} {Phys. Rev. B}\ }\textbf {\bibinfo {volume} {103}},\ \bibinfo
  {pages} {235157} (\bibinfo {year} {2021})}\BibitemShut {NoStop}%
\bibitem [{\citenamefont {Jang}\ and\ \citenamefont
  {Kim}(2023)}]{JANG2023169164}%
  \BibitemOpen
  \bibfield  {author} {\bibinfo {author} {\bibfnamefont {I.}~\bibnamefont
  {Jang}}\ and\ \bibinfo {author} {\bibfnamefont {K.-S.}\ \bibnamefont {Kim}},\
  }\bibfield  {title} {\enquote {\bibinfo {title} {{Effects of general
  non-magnetic quenched disorder on a spin-density-wave quantum critical
  metallic system in two spatial dimension}},}\ }\href {\doibase
  https://doi.org/10.1016/j.aop.2022.169164} {\bibfield  {journal} {\bibinfo
  {journal} {Annals of Physics}\ }\textbf {\bibinfo {volume} {448}},\ \bibinfo
  {pages} {169164} (\bibinfo {year} {2023})}\BibitemShut {NoStop}%
\bibitem [{\citenamefont {Lee}(2014)}]{lee:2014uf}%
  \BibitemOpen
  \bibfield  {author} {\bibinfo {author} {\bibfnamefont {S.-S.}\ \bibnamefont
  {Lee}},\ }\bibfield  {title} {\enquote {\bibinfo {title} {Quantum
  renormalization group and holography},}\ }\href {\doibase
  10.1007/JHEP01(2014)076} {\bibfield  {journal} {\bibinfo  {journal} {Journal
  of High Energy Physics}\ }\textbf {\bibinfo {volume} {2014}},\ \bibinfo
  {pages} {76} (\bibinfo {year} {2014})}\BibitemShut {NoStop}%
\end{thebibliography}%

\end{document}